ESA/SCI(2020)1
November 2020

# Ariel

## Atmospheric Remote-sensing Infrared Exoplanet Large-survey

## Enabling planetary science across light-years

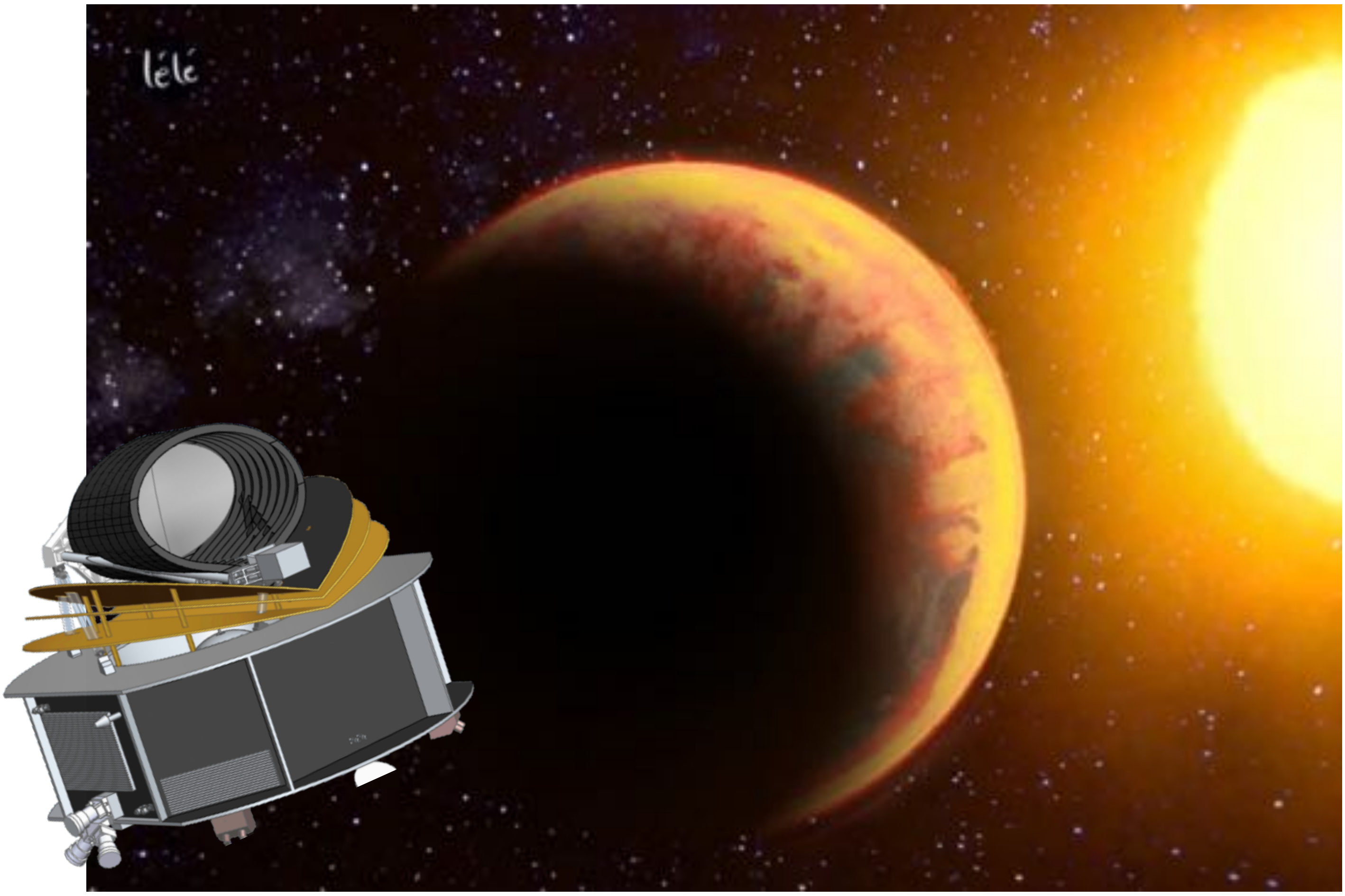

Cover image credit: Léa Changeat

Definition Study Report

**European Space Agency**

---------- PAGE INTENTIONALLY LEFT BLANK ----------

| **Ariel Mission Summary** | |
|---|---|
| **Key Science Questions to be Addressed** | • *What are the physical processes shaping planetary atmospheres?*<br>• *What are exoplanets made of?*<br>• *How do planets and planetary systems form and evolve?* |
| **Science Objectives** (Chapter 2) | • Detect and determine the composition and structure of a large number of planetary atmospheres<br>• Constrain planetary interiors by removing degeneracies in the interpretation of mass-radius diagrams<br>• Constrain planetary formation and evolution models through measurements of the elemental composition (evidence for migration)<br>• Determine the energy budget of planetary atmospheres (e.g. albedo, vertical and horizontal temperature structure, weather/temporal variations)<br>• Identify and constrain chemical processes at work (e.g thermochemistry, photochemistry, transport, quenching)<br>• Constrain the properties of clouds (e.g. cloud type, particle size, distribution, patchiness)<br>• Investigate the impact of stellar and planetary environment on exoplanet properties<br>• Identification of different populations of planets and atmospheres (e.g. through colour- colour diagrams)<br>• Capacity to do a population study and go into a detailed study of select planets |
| **Ariel Core Survey** (Chapter 3) | • Survey of ~1000 transiting exoplanets from gas giants to rocky planets, in the hot to temperate zones of A to M-type host stars<br>• Target selection before launch based on ESA science team and community inputs<br>• Delivery of a homogeneous catalogue of planetary spectra, yielding refined molecular abundances, chemical gradients and atmospheric structure; diurnal and seasonal variations; presence of clouds and measurement of albedo. |
| **Observational Strategy** (Chapter 3, 7) | • Transit, eclipse, phase-curve spectroscopy with broad (0.5-7.8 µm), instantaneous, uninterrupted spectra<br>• High photometric stability on transit timescales<br>• Large instantaneous sky coverage<br>• Focus on planets around relatively nearby thus relatively bright stellar hosts.<br>• Four-tiered approach: 4 different samples are observed at optimised spectral resolutions and SNRs.<br>• Required SNR obtained by performing multiple observations. |
| **Payload** (Chapter 4) | • Off-axis Cassegrain telescope, 1.1 m × 0.73 m elliptical M1; diffraction limited at 3µm. Mirrors, optical bench and telescope all manufactured from Aluminium alloy for isothermal design with minimal thermo-elastic deformation.<br>• Ariel InfraRed Spectrometer (AIRS) provides low/medium resolution (R=30–200) spectroscopy in 1.95-7.8 µm range.<br>• FGS includes 3 photometric channels (two used for guiding as well as science) between 0.5-1.1 µm + low resolution NIR spectrometer for 1.1-1.95 µm range.<br>• Thermal: Warm SVM, cryogenic PLM cooled passively to ~55 K with the thermal shield assembly. Active cooler (Neon JT) included to ensure AIRS detector operating temperature of ~42 K<br>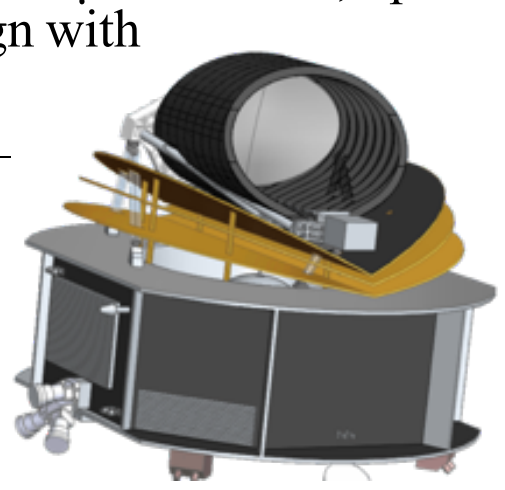 |
| **Spacecraft** (Chapter 5) | • Spacecraft budgets: ~1.5 t launch mass, ~1 kW power<br>• Communications: X-band, 236 Gbit / week of science and housekeeping data<br>• Fine pointing requirements across instrument LoS (3 sigma): APE ≤ 1"; RPE ≤ 230 mas up to 90 s; PDE ≤ 70 mas up to 10 hrs for integrations of 90 s<br>• AOCS: Fine Guidance Sensor and Reaction Wheels on dampers to minimise micro-vibrations |
| **Launch and Operations** (Chapter 5, 6) | • Ariane 6.2 launch from Kourou planned in 2029 towards the large amplitude orbit around L2, in dual launch configuration together with the Comet-Interceptor (F1) mission<br>• 4 years nominal lifetime (goal 6 years extended mission)<br>• MOC at ESOC, SOC at ESAC, and Instrument Operations and Science Data Centre (IOSDC) distributed across consortium member states<br>• 14 hrs split in 3 ground contacts per week with 35 m ESTRACK ground stations |
| **Data Policy** (Chapter 9) | • It is recognized that Ariel data will be of great interest to the general astronomical and exoplanet community, and Ariel wants to embrace the general community.<br>• Ariel will offer community involvement in target selection, and a very open data policy.<br>• Regular timely public releases of high quality data products at various processing levels will be provided throughout the mission, with Science Demonstration Phase data, as well as Tier 1 data throughout the mission provided immediately after quality control has been completed. |

# Foreword

The concept of a mission devoted to atmospheric characterization of exoplanets through transit spectroscopy was first considered in Europe in 2007, shortly after the DARWIN proposal submitted to ESA for the first Cosmic Vision call for L-class missions was rejected because of the need for further scientific and technical developments. Following the decision, both ESA (EP-RAT panel report, October 2010) and the Exoplanetary Community (Blue Dot Team – Barcelona conference, September 2009) started a discussion to define a roadmap for exoplanetary research.

Both groups concluded that an intermediate step was needed, both scientifically and technically, before the characterisation of Earth-like planets could be tackled, and recommended a transit spectroscopy mission as a first step to atmospheric characterisation. A short study was undertaken at ESTEC in the context of the ExoPlanet Roadmap Advisory Team mandate under the name ESM (Exoplanet Spectroscopy Mission). Following this study the *Exoplanet Characterisation Observatory* (EChO) was proposed and accepted for assessment phase study in the context of the ESA Cosmic Vision 2015-2025 programme M3 medium class mission opportunity. Although eventually not selected, the EChO study[1] allowed further development of the technical building blocks and the science case for an eventual transit spectroscopy mission.

In response to the call for the next medium class opportunity, Cosmic Vision M4, a proposal was submitted in January 2015: the *Atmospheric Remote-sensing InfraRed Large-survey* (ARIEL). The mission was one of the three selected in June 2015 for study in a Phase 0/A, a competitive assessment phase[2]. ARIEL was eventually selected as M4 in March 2018, and went into Phase B1, the definition study phase. The name of the mission has been changed to Ariel after selection.

During Phase B1, the science case was studied in depth and consolidated under auspices of the Science Advisory Team, the bulk of the work being performed in a large number of science working groups in the Ariel Mission Consortium (AMC). The ESA Study Team and AMC reviewed the mission requirements, the technical design and analysis of the complete payload module (including telescope, instruments, guidance system and supporting infrastructure). The AMC developed an end-to-end performance simulator of the complete system. Two industrial contractors (Airbus Defence and Space, France and ThalesAlenia Space, France) reviewed the mission requirements, the technical design and analysis of the s/c and performed a programmatic analysis of the mission. Dedicated iterations were done in conjunction with both industrial and payload studies to harmonise the interfaces between the s/c and the payload, and to consolidate the payload accommodation. Recently the ESA Mission Adoption Review has successfully been concluded.

This definition study report presents a summary of the very large body of work that has been undertaken on the Ariel mission over the 30-month period of the Ariel definition phase. As such, it represents the contributions of a large number of parties (ESA, industry, institutes and universities from 17 ESA member states, NASA CASE team), encompassing a very large number of people.

The successful public Ariel: Science, Mission & Community 2020 workshop was held in ESTEC, Noordwijk, on 14-16 January 2020 (https://www.cosmos.esa.int/web/ariel/conference-2020). Over 200 participants from 19 countries attended the conference which had the objectives to present the mission and its science as proposed for mission adoption, and involve the planetary and astrophysical community at large in the mission. Presentations and discussions addressed how Ariel can work in conjunction with other ground-based and space-based observatories to best further our knowledge of exoplanetary science.

In the six years since Ariel was first conceived in 2014, the number of confirmed exoplanets has increased from ~1000 to over 4300, providing an ever more tantalising prospect of looking beyond our solar system and enabling planetary science across light years.

ESA Ariel Science Advisory Team – September 2020

[1] EChO – *Experimental Astronomy*, Special Issue, Volume 40, issue 2-3, December 2015; ESA EChO Assessment Study Report

[2] Ariel – *Experimental Astronomy*, Special Issue, Volume 46, issue 1, November 2018; ESA Ariel Assessment Study Report

# Authorship, acknowledgements

This report has been prepared by:

| ESA Science Advisory Team (SAT) | | |
|---|---|---|
| *Name* | *Affiliation* | *City, Country* |
| Giovanna Tinetti | University College London | London, UK |
| Carole Haswell | The Open University | Milton Keynes, UK |
| Jérémy Leconte | Université de Bordeaux | Bordeaux, France |
| Pierre-Olivier Lagage | CEA | Saclay, France |
| Giusi Micela | INAF / Observatory of Palermo | Palermo, Italy |
| Michel Min | SRON | Utrecht, The Netherlands |
| Leonardo Testi | ESO | Garching, Germany |
| Diego Turrini | INAF / IAPS | Rome, Italy |
| Bart Vandenbussche | KU Leuven | Leuven, Belgium |
| Maria Rosa Zapatero Osorio | INTA | Madrid, Spain |
| **Supported by** | | |
| Paul Eccleston (Ariel PM) | RAL Space | Harwell, UK |
| Mark Swain (CASE PI) | Jet Propulsion Laboratory | Pasadena, CA, US |

The ESA Team supporting the activities comprises:

<table>
<tr><th colspan="3">ESA Study Team</th></tr>
<tr><td>Ludovic Puig (Study Manager)</td><td rowspan="11">ESTEC</td><td rowspan="11">Noordwijk, The Netherlands</td></tr>
<tr><td>Göran Pilbratt (Study Scientist)</td></tr>
<tr><td>Theresa Lüftinger (Project Scientist)</td></tr>
<tr><td>Francesco Ratti (Payload Manager)</td></tr>
<tr><td>Carsten Scharmberg (Payload Manager)</td></tr>
<tr><td>Jean-Christophe Salvignol (Project Manager)</td></tr>
<tr><td>Nathalie Boudin (Payload Engineer)</td></tr>
<tr><td>Jean-Philippe Halain (PRODEX Officer)</td></tr>
<tr><td>Martin Haag (PRODEX Officer)</td></tr>
<tr><td>Pierre-Elie Crouzet (Detector Engineer)</td></tr>
<tr><td></td></tr>
<tr><td>Ralf Kohley (SOC)</td><td>ESAC</td><td>Madrid, Spain</td></tr>
<tr><td>Kate Symonds (MOC)</td><td rowspan="2">ESOC</td><td rowspan="2">Darmstadt, Germany</td></tr>
<tr><td>Florian Renk (Mission Analysis)</td></tr>
<tr><th colspan="3">ESA Coordinators</th></tr>
<tr><td>Luigi Colangeli</td><td rowspan="2">ESTEC</td><td rowspan="2">Noordwijk, The Netherlands</td></tr>
<tr><td>Paul McNamara</td></tr>
</table>

The Ariel Mission Consortium is supported by their respective national funding agencies. The team would like to thank the agencies for their support during the Definition Study. Consortium team members contributing to the Definition Study are listed in the next page.

We would like to thank all the ESA colleagues in D/SCI, D/TEC, D/OPS, and D/IPL who supported the study.

We also acknowledge an atmosphere of fruitful collaboration with our industrial contractors:

- Airbus Defence and Space, France
- ThalesAlenia Space, France

Cover image credit: Léa Changeat

**Consortium Co-PIs**

Giovanna Tinetti, UCL, UK; Anna Aret, U. of Tartu, Estonia; Jean-Philippe Beaulieu, IAP, France; Lars Buchhave, DTU, Denmark; Martin Ferus, J. Heyrovský Institute of Physical Chemistry CAS, Czech Republic; Matt Griffin, U. of Cardiff, UK; Manuel Guedel, U. of Vienna, Austria; Paul Hartogh, Max Planck Sonnnensystem, Germany; Pierre-Olivier Lagage, CEA, France; Pedro Machado, Institute of Astrophysics and Space Sciences, Portugal; Giuseppe Malaguti, INAF-OAS Bologna, Italy; Giusi Micela, INAF – Osservatorio Astronomico di Palermo, Italy; Michiel Min, SRON Netherlands Institute for Space Research, the Netherlands; Enric Pallé, IAC, Spain; Mirek Rataj, Space Research Centre, Polish Academy of Science, Poland; Tom Ray, Dublin Institute for Advanced Studies, Ireland; Ignasi Ribas, IEEC–CSIC, Spain; Robert Szabó, Konkoly Observatory, Hungary; Mark Swain, JPL, US; Jonathan Tan, Chalmers U., Sweden; Bart Vandenbussche, KU Leuven, Belgium; Stephanie Werner, U. of Oslo, Norway;

**Consortium Management Team and Consortium System Team**

Paul Eccleston, RAL Space, UK; Andrew Caldwell, RAL Space, UK; Manuel Abreu, IA, Portugal; Gustavo Alonso, UPM, Spain; Jerome Amiaux, CEA, France; Michel Berthé, CEA, France; Georgia Bishop, RAL Space, UK; Neil Bowles, University of Oxford, UK; Manuel Carmona, IEEC-UB, Spain; Deirdre Coffey, DIAS, Ireland; Josep Colomé, IEEC-CSIC, Spain; Martin Crook, RAL TD, UK; Lucile Désjonqueres, RAL Space, UK; José J. Díaz, IAC, Spain; Rachel Drummond, RAL Space, UK; Mauro Focardi, INAF-OAA, Italy; Jose M. Gómez, IEEC, Spain; Warren Holmes, JPL, US; Matthijs Krijger, SRON, Netherlands; Zsolt Kovacs, Admatis, Hungary; Tom Hunt, MSSL, UK; Richardo Machado, ActiveSpace, Portugal; Gianluca Morgante, INAF – OAS, Bologna, Italy; Marc Ollivier, IAS Paris, France; Roland Ottensamer, U. of Vienna, Austria; Emanuele Pace, U. di Firenze, Italy; Teresa Pagano, RUAG, Switzerland; Enzo Pascale, La Sapienza U. of Rome; Chris Pearson, RAL Space, UK; Søren Møller Pedersen, DTU, Denmark; Moshe Pniel, JPL, US; Stéphane Roose, CSL, Belgium; Giorgio Savini, UCL, UK; Richard Stamper, RAL Space, UK; Peter Szirovicza, Admatis, Hungary; Janos Szoke, Admatis, Hungary; Ian Tosh, RAL Space, UK; Francesc Vilardell, IEEC, Spain;

**Consortium Science Team Coordinators**

Joanna Barstow, OU, UK; Luca Borsato, U. of Padova, Italy; Sarah Casewell, U. of Leicester, UK; Quentin Changeat, UCL, UK; Benjamin Charnay, LESIA, France; Svatopluk Civiš, J. Heyrovský Institute of Physical Chemistry CAS, Czech Republic; Vincent Coudé du Foresto, LESIA, France; Athena Coustenis, LESIA, France; Nicolas Cowan, McGill U., Canada; Camilla Danielski, CSED, UK; Olivier Demangeon, IA, Portugal; Pierre Drossart, IAP, France; Billy N. Edwards, UCL, UK; Gabriella Gilli, IA, Portugal; Therese Encrenaz, LESIA, France; Csaba Kiss, Konkoly Observatory, Hungary, Anastasia Kokori, Royal Museums Greenwich, UK; Masahiro Ikoma, U. of Tokyo, Japan; Jérémy Leconte, U. de Bordeaux, France; Juan Carlos Morales, IEEC, Spain; João Mendonça, DTU, Denmark; Andrea Moneti, IAP, France; Lorenzo Mugnai, La Sapienza, Italy; Antonio García Muñoz, Technische Universität Berlin, Germany; Ravit Helled, U. of Zurich, Switzerland; Mihkel Kama, U. of Tartu/UCL, Estonia/UK; Yamila Miguel, SRON, the Netherlands; Nikos Nikolaou, UCL, UK; Isabella Pagano, INAF, Italy; Olja Panic, Leeds, UK; Miriam Rengel, MPS, Germany; Hans Rickman, PAS Space Research Centre, Poland; Marco Rocchetto, Konica Minolta; Subhajit Sarkar, U. of Cardiff, UK; Franck Selsis, U. de Bordeaux, France; Jonathan Tennyson, UCL, UK; Angelos Tsiaras, UCL, UK; Diego Turrini, IAPS, Italy; Olivia Venot, LISA, France; Krisztián Vida, Konkoly Observatory, Hungary; Ingo P. Waldmann, UCL, UK; Sergey Yurchenko, UCL, UK; Gyula Szabó, U. of Eötvös Loránd, Hungary; Rob Zellem, JPL, US.

**Scientists and Engineers contributing to Ariel Definition Study Report**

Ahmed Al-Refaie, UCL, UK; Javier Perez Alvarez, UPM Madrid, Spain; Lara Anisman, UCL, UK; Axel Arhancet, CEA, FR; Jaume Ateca, IEEC – UB, SP; Robin Baeyens, KU Leuven, Belgium; John R. Barnes, OU, UK; Taylor Bell, McGill U., CA; Serena Benatti, OAPa, IT; Katia Biazzo, OA Roma, IT; Maria Błęcka, CBK, PL; Aldo Stefano Bonomo, OATo, IT; José Bosch, IEEC – UB, SP; Diego Bossini, U. do Porto, PT; Jeremy Bourgalais, LATMOS, FR; Daniele Brienza, INAF-IAPS, IT; Anna Brucalassi, OAA, IT; Giovanni Bruno, OACa, IT; Hamish Caines, UCL, UK; Simon Calcutt, University of Oxford, UK; Tiago Campante, U. do Porto, PT; Rodolfo Canestrari, INAF-IASFPa, IT; Nick Cann, RAL Space, UK; Giada Casali, OAA, IT; Albert Casas, IEEC – UB, SP; Giuseppe Cassone, ICPF-CNR, IT; Christophe Cara, CEA, France; Manuel Carmona, Uni Barcelona, Spain; Ludmila Carone, MPIA, DE; Nathalie Carrasco, LATMOS, FR; Quentin Changeat, UCL, UK; Paolo Chioetto, IFN-CNR, Italy; Fausto Cortecchia, INAF-OAS Bologna, IT; Markus

Czupalla, Uni Aachen / CBK, Poland; Katy L. Chubb, SRON, NL; Angela Ciaravella, INAF-OAPa, IT; Antonio Claret, IAA, SP; Riccardo Claudi, INAF OAPd, IT; Claudio Codella, OAA, IT; Maya Garcia Comas, IAA-CSIC, SP; Gianluca Cracchiolo, U. di Palermo, IT; Patricio Cubillos, SRI, Austria; Vania Da Peppo, CNR-IFN, Italy; Leen Decin, KU Leuven, Belgium; Clemence Dejabrun, IAS, FR; Elisa Delgado-Mena, U. do Porto, PT; Anna Di Giorgio, INAF, Italy; Emiliano Diolaiti, INAF-OAS Bologna, IT; Caroline Dorn, U. of Zurich, CH; Vanessa Doublier, IAP, FR; Eric Doumayrou, CEA, FR; Georgina Dransfield, U. of Birmingham, UK; Luc Dumaye, CEA, FR; Emma Dunford, UCL, UK; Antonio Jimenez Escobar, INAF – OAPa, IT; Vincent Van Eylen, MSSL, UK; Maria Farina, INAF-IAPS, IT; Davide Fedele, OAA, IT; Alejandro Fernández, UPM, SP; Benjamin Fleury, JPL, US; Sergio Fonte, IAPS, IT; Jean Fontignie, CEA, FR; Luca Fossati, IWF, AU; Bernd Funke, IAA-CSIC, SP; Camille Galy, CSL, Belgium; Zoltán Garai, MTA-ELTE, HU; Andrés García, UPM, SP; Alberto García-Rigo, IEEC, SP; Antonio Garufi, OAA, IT; Giuseppe Germano Sacco, OAA, IT; Paolo Giacobbe, OATo, IT; Alejandro Gómez, UPM, SP; Arturo Gonzalez, UPM Madrid, Spain; Francisco Gonzalez-Galindo, IAA-CSIC, SP; Davide Grassi, IAPS-INAF, IT; Caitlin Griffith, UoA, US; Mario Giuseppe Guarcello, OAPa, IT; Audrey Goujon, CEA, FR; Amélie Gressier, IAP, FR; Aleksandra Grzegorczyk, CBK, Poland; Tristan Guillot, OCA, FR; Gloria Guilluy, OATo, IT; Peter Hargrave, U. of Cardiff, UK; Marie-Laure Hellin, CSL, Belgium; Enrique Herrero, IEEC, SP; Matt Hills, RAL Technology Department, UK; Benoit Horeau, CEA, FR; Yuichi Ito, UCL, UK; Niels Christian Jessen, DTU, Denmark; Petr Kabath, AA, Czech Republic; Szilárd Kálmán, U. of Szeged, HU; Yui Kawashima, SRON, NL; Tadahiro Kimura, U. of Tokyo, JP; Antonín Knížek, JHI CAS, CZE; Laura Kreidberg, MPIA, DE; Ronald Kruid, JPL, US; Diederik J. M. Kruijssen, U. of Heidelberg, DE; Petr Kubelík, JHI CAS, CZ; Luisa Lara, IAA-CSIC, SP; Sebastien Lebonnois, LMD, FR; David Lee; UK ATC, UK; Maxence Lefevre, LESIA, FR; Tim Lichtenberg, U. of Oxford, UK; Daniele Locci, INAF-OAPa, IT; Matteo Lombini, INAF-OAS Bologna, IT; Alejandro Sanchez Lopez, IAA-CSIC, SP; Andrea Lorenzani, INAF-OAA, IT; Ryan MacDonald, Cornell U., US; Laura Magrini, OAA, IT; Jesus Maldonado, OAPa, IT; Emmanuel Marcq, LATMOS, FR; Alessandra Migliorini, IAPS-INAF, IT; Darius Modirrousta-Galian, U. of Palermo, IT; Karan Molaverdikhani, MPIA, DE; Sergio Molinari, IAPS, IT; Paul Mollière, MPIA, DE; Vincent Moreau, CEA, FR; Giuseppe Morello, CEA, FR; Gilles Morinaud, IAS, FR; Mario Morvan, UCL, UK; Julianne I. Moses, SSI, US; Salima Mouzali, CEA, FR; Nariman Nakhjiri, IEEC – CSIC, SP; Luca Naponiello, U. di Firenze, IT; Norio Narita, U. of Tokyo, JP; Valerio Nascimbeni, OAP, IT; Athanasia Nikolaou, La Sapienza, IT; Vladimiro Noce, U. di Firenze, IT; Fabrizio Oliva, IAPS, IT; Pietro Palladino, U. Palermo, IT; Andreas Papageorgiou, U. of Cardiff, UK; Vivien Parmentier, U. of Oxford, UK; Giovanni Peres, U. of Palermo, IT; Javier Pérez, UPM, SP; Santiago Perez-Hoyos UPV/EHU, SP; Manuel Perger, IEEC – CSIC, SP; Cesare Cecchi Pestellini, INAF-OAPa, IT; Antonino Petralia, INAF-OAPa, IT; Anne Philippon, IAS, FR; Arianna Piccialli, BIRA-IASB, Belgium; Marco Pignatari, U. of Hull, UK; Giampaolo Piotto, OAP, IT; Linda Podio, OAA, IT; Gianluca Polenta, ASI/SSDC, IT; Giampaolo Preti, U. di Firenze, IT; Theodor Pribulla, MTA-ELTE, HU; Manuel Lopez Puertas, IAA-CSIC, SP; Monica Rainer, OAA, IT; Jean-Michel Reess, LESIA, FR; Paul Rimmer U. of Cambridge, UK; Séverine Robert, BIRA-IASB, Belgium; Albert Rosich, IEEC – CSIC, SP; Loic Rossi, LATMOS, FR; Duncan Rust, MSSL, UK; Ayman Saleh, UCL, UK; Nicoletta Sanna, OAA, IT; Eugenio Schisano, IAPS, IT; Laura Schreiber, INAF-OAS Bologna, IT; Victor Schwartz, CEA, FR; Antonio Scippa, Università di Firenze, IT; Bálint Seli, Konkoly Obs., HU; Sho Shibata, U. of Tokyo, JP; Caroline Simpson, RAL Space, UK; Oliver Shorttle, U. of Cambridge, UK; N. Skaf, LESIA, FR; Konrad Skup, CBK, Poland; Mateusz Sobiecki, CBK, Poland; Sergio Sousa, U. do Porto, PT; Alessandro Sozzetti, OATo, IT; Judit Šponer, BFÚ CAS, CZ; Lukas Steiger, Toptec – IPP, Czech Rep; Paolo Tanga, OCA, FR; Paul Tackley, ETH-Zurich, CH; Jake Taylor, U. of Oxford, UK; Matthias Tecza, University of Oxford, UK; Luca Terenzi, INAF-OAS, IT; Pascal Tremblin, CEA, FR; Andrea Tozzi, INAF-OAA, IT; Amaury Triaud, U. of Birmingham, UK; Loïc Trompet, BIRA-IASB, Belgium; Shang-Min Tsai, U. of Oxford, UK; Maria Tsantaki, OAA, IT; Diana Valencia, U. of Toronto, CA; Ann Carine Vandaele, BIRA-IASB, Belgium; Mathieu Van der Swaelmen, OAA, IT; Adibekyan Vardan, U. do Porto, PT; Gautam Vasisht, JPL, US; Allona Vazan, Hebrew U. of Jerusalem, IL; Ciro Del Vecchio, INAF-OAA, IT; Dave Waltham, RH, UK; Piotr Wawer, CBK, Poland; Thomas Widemann, LESIA, FR; Paulina Wolkenberg, IAPS, IT; Gordon Hou Yip, UCL, UK; Yuk Yung, Caltech, US; Mantas Zilinskas, U. of Leiden, NL; Tiziano Zingales, U. de Bordeaux, FR; Paola Zuppella, IFN – CNR, Italy.

# Table of contents

# 1 Executive Summary

The Ariel mission will address the fundamental questions on what exoplanets are made of and how planetary systems form and evolve by investigating the atmospheres of many hundreds of diverse planets orbiting different types of stars. This large and unbiased survey will contribute to answering the first of the four ambitious topics listed in the ESA's Cosmic Vision: "*What are the conditions for planet formation and the emergence of life?*". Thousands of exoplanets have now been discovered with a huge range of masses, sizes and orbits: from rocky Earth-like planets to large gas giants grazing the surface of their host star. There is no known, discernible pattern linking the presence, size, or orbital parameters of a planet to the nature of its parent star. We have little idea whether the chemistry of a planet's surface and atmosphere is linked to its formation environment, or whether the type of host star drives the physics and chemistry of the planet's birth and evolution.

Ariel will observe around a thousand transiting planets, including gas giants, Neptunes, super-Earths and Earth-size planets around a range of host star types. This comprehensive approach will underpin statistical understanding generating robust conclusions which are simply not possible with smaller samples or patchy coverage of the relevant parameter space. Ariel will use transit spectroscopy in the 1.1-7.8 μm spectral range and photometry in multiple narrow bands covering the optical and near-infrared (NIR). We will focus on warm and hot planets to take advantage of their well-mixed atmospheres which should show minimal condensation and sequestration of high-Z materials and thus reveal their bulk elemental composition (especially C, O, N, S, Si). Observations of these warm/hot exoplanets will drive understanding of the early stages of planetary and atmospheric formation during the nebular phase and the following few million years. Ariel will thus provide a complete picture of the chemical nature of the exoplanets and relate this directly to the planetary parameters (see e.g. Figure 1-1) and the type and chemical environment of the host star.

For this ambitious scientific programme, Ariel is designed as a dedicated survey mission for transit and eclipse spectroscopy, capable of observing a large and well-defined planet sample within its 4-year mission lifetime. Transit, eclipse and phase-curve spectroscopy methods, whereby the signal from the star and planet are differentiated using knowledge of the planetary ephemerides, allow us to measure atmospheric signals from the planet at levels of 10-50 part per million (ppm) relative to the star. Given the brightness of the target host stars more sophisticated techniques, such as eclipse mapping, will also be used to give deeper insights. *These observations require a specifically designed, stable payload and satellite platform with broad, instantaneous wavelength coverage to detect many molecular species, probe the thermal structure, identify clouds and monitor the stellar activity*. The wavelength range covered by Ariel includes all the expected major atmospheric gases from e.g. $H_2O$, $CO_2$, $CH_4$ $NH_3$, HCN, $H_2S$ through to the more exotic metallic compounds, such as TiO, VO, and condensed species.

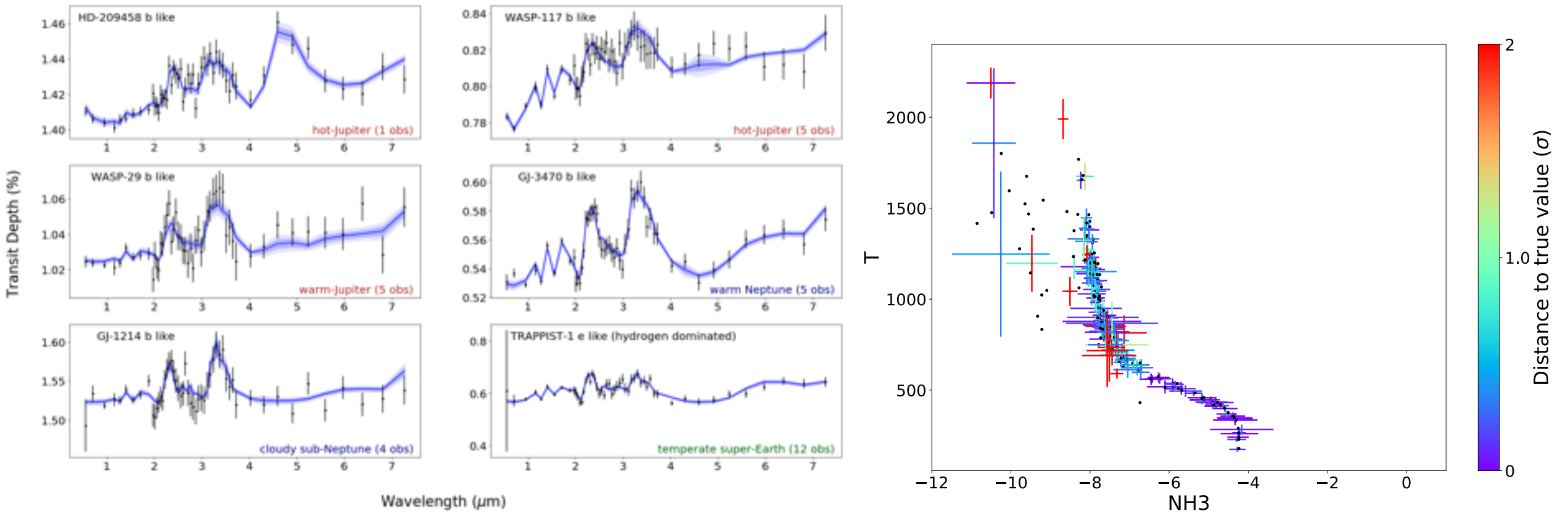

*Figure 1-1: Retrieval outcomes of the simulated Ariel yields (Changeat et al. 2020a). Left: Examples of observed spectra (black) and the best fit retrieved spectra (blue) for six planets out of 150+ used in the simulations. The planets, with bulk parameters from Edwards et al. (2019), are assumed to host an hydrogen dominated envelope, with randomised clouds and with trace gases following thermo-chemical equilibrium (Agundez et al. 2012). The errors on signal were computed with ArielRad (Mugnai et al. 2020) and observations combined to reach Tier 2 requirements (see Table 3-1). Right: Temperature correlation map obtained for $NH_3$ from the same run. X axis: retrieved logarithmic abundance of $NH_3$, y axis: planetary equilibrium temperature in K. The bars indicate the 1-σ error on the retrieved parameter, while the colours indicate their distance (in unit of σ) to the true value (black dots). The imposed equilibrium chemistry trend is accurately recovered from the retrieval analysis of the full Ariel sample. Our results demonstrate the ability of Ariel Tier-2 survey to reveal trends between the chemistry and associated planetary parameters.*

**Performance Evaluation** – Detailed performance simulations of Ariel observations have been performed using time domain simulators which use conservative estimates of payload, spacecraft and ground processing performance with a full model of all significant noise sources in the measurements. Using a list of potential targets incorporating the latest available statistics on exoplanets, these simulations show that Ariel will be able to observe between 500 and 1000 exoplanets during the nominal mission lifetime with the exact number to be dependent on the exact details of the survey strategy which will evolve (under the guidance of the science team) up to, and indeed during, flight.

**Ariel Data policy** – Ariel data and data products will be of great interest and utility for the entire exoplanet community, beyond those directly involved in the mission. In response, the data policy has been designed to embrace the astronomy community in general and the exoplanet community in particular. The intention is to provide high quality data in a timely manner and to have a continuous dialogue with the wider community, maximising the science and impact that can be achieved by the mission.

The target list will be drawn from a larger list of all potential Ariel targets. This larger list will be made public and maintained publicly. Inputs will be solicited from the general community through formal time-bounded processes such as whitepapers and meetings. Moreover the list will be presented and maintained in an interactive environment enabling direct input from members of the community. The target list will be drawn from the larger list through scientific priorities and other guidelines through scheduling exercises performed under the responsibility of the Ariel Science team. The community will be kept informed about the status of the target list, as will the ESA Advisory Bodies whose feedback will be solicited.

A Science Demonstration Phase (SDP) will be conducted as the final step before routine science phase operations commence, validating the foreseen observing plan and demonstrating the data quality achievable by Ariel. The SDP is foreseen to provide approximately one month's worth of data observed in the manner planned for the core survey. These data will be made public on a timescale of about a month, in conjunction with organisation of a major public workshop.

Regular timely public releases of high quality data products at various processing Levels will be provided throughout the mission, it is foreseen that after the SDP data release, there will be data releases every six months. The data 'raw telemetry files' (Level 0) will be unpacked and uncompressed into 'uncalibrated photometric/spectral science frames' (Level 1), and pipeline processed to 'calibrated photometric/spectral images' (Level 1.5), and to 'target light-curves' (Level 2) where 'target' comprises star + planet, and 'exoplanet spectra' (Level 3), respectively. The lower data product levels and lower tiers will be released quicker than higher levels and tiers, but the objective is timely release of all levels to maximise the science return of Ariel.

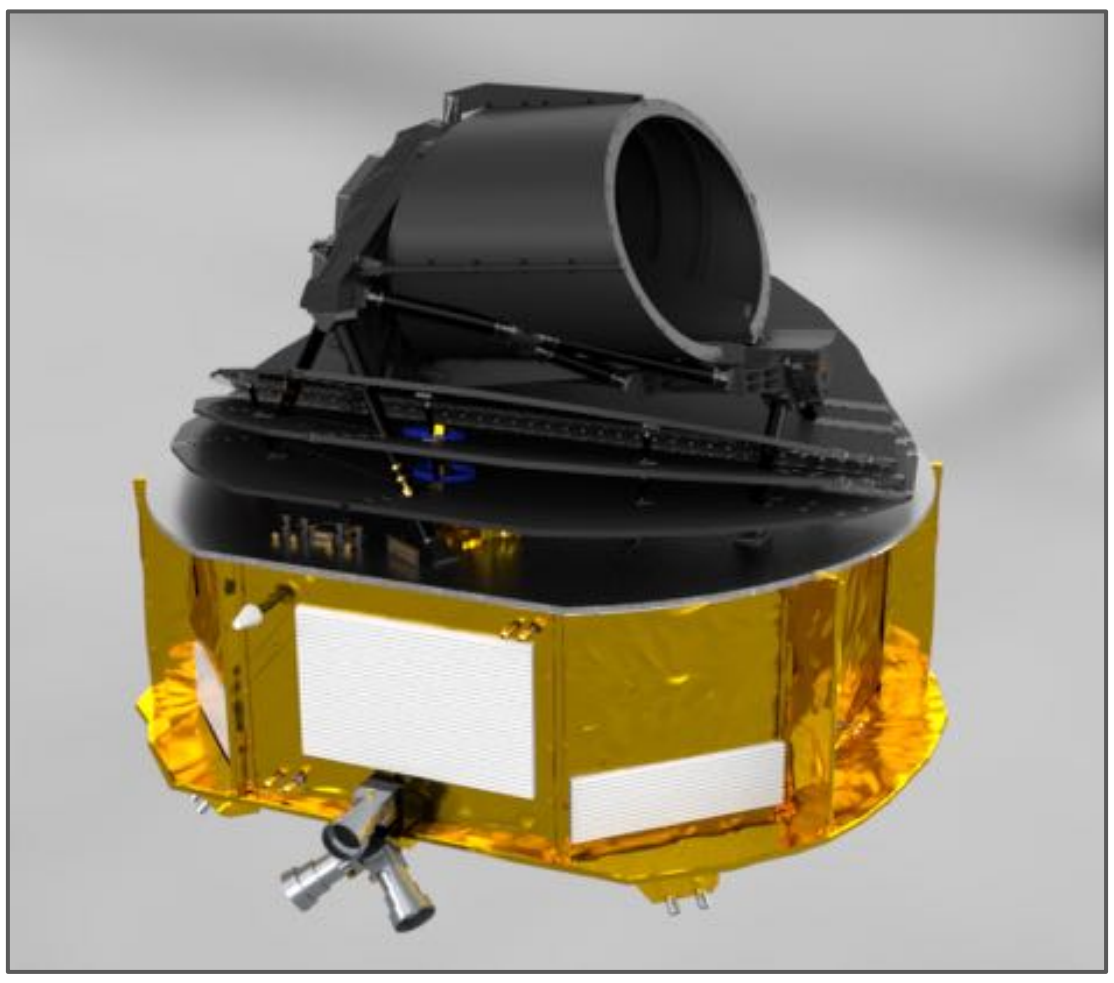

**Mission and spacecraft design** –Ariel is planned to be launched from Kourou (FR) on board an Ariane 6.2 in 2029, in a dual launch configuration with Comet-Interceptor. Ariel will sit underneath the Dual Launch Structure, while Comet-I will ride on the top. The nominal operations orbit is a large amplitude orbit around the Sun-Earth L2 point. This orbit provides a stable environment, along with a large instantaneous field of regard, both of which are key to allowing Ariel to meet its science objectives. The spacecraft is designed in a modular way, with a service module (SVM) and a payload module (PLM) that can be procured and tested in parallel. The SVM contains all the units required to operate the spacecraft and maintain the payload in its nominal operating conditions. The spacecraft has a wet mass of ~1.5 t and a power generation capability of ~1 kW. 236 Gbit of science data are generated every week, and are down-linked in three ground contacts totalling 14 hours / week using an X-band system and the 35 m ESTRACK ground stations. The pointing requirements achieved by the AOCS system in the fine pointing mode are (3 sigma): APE ≤ 1"; RPE (within 0.1s) ≤ 180 mas; RPE (up to 90 s) of the MPE (on 0.1 s) ≤ 130 mas; PDE ≤ 70 mas up to 10 hours for integrations of 90 seconds. This is achieved with a Fine Guidance Sensor (FGS, part of the payload instrument suite) and reaction wheels only as the sole actuators (accommodated on dampers to minimise any micro-vibrations).

The PLM design is optimised to fulfil the science requirements while keeping the technical risks and costs within the M4 programmatic constraints of the payload consortium. The baseline integrated payload consists of an all-Aluminium off-axis Cassegrain telescope (primary mirror 1100 mm × 730 mm ellipse) with a re-focussing mechanism accommodated behind the M2 mirror that allows correction for any misalignment generated during the telescope assembly or launch and cool down. The telescope feeds a collimated beam, split in wavelengths by dichroics, into two separate instruments with coincident fields of view. The FGS contains three photometric channels (VISPhot [0.50-0.60 μm], FGS#1 [0.6-0.8 μm], FGS#2 [0.80-1.10 μm]) and a coarse resolution spectrometer (NIRSpec [1.10-1.95 μm] with R ≥ 15). FGS#1 and FGS#2 will also be used as a redundant system for providing guidance and closed-loop feedback to the high stability pointing of the spacecraft. The FGS provides simultaneous information on the photometric stability of the target stars, and NIRSpec is optimized for cloud characterisation. AIRS contains two channels (AIRS-Ch0 [1.95-3.9 μm] with R ≥ 100, AIRS-Ch1 [3.9-7.8 μm] with R ≥ 30). The payload module is passively cooled to ~55 K by isolation from the spacecraft bus via a series of V-Groove radiators; the detectors for the AIRS are the only items that require active cooling to < 42 K via an active Ne JT cooler.

The instrument design uses only technologies with a high degree of technical maturity. The observation method of transit spectroscopy means requires no significant angular resolution and detailed performance studies show that a telescope collecting area of 0.64 $m^2$ is sufficient to achieve the necessary observations on all the ARIEL targets within the mission lifetime. The satellite is best placed into an L2 orbit to provide high thermal stability and maximum field of regard which leads to high observational efficiency on the time bounded observations of transits. ARIEL is complementary to other international facilities (such as TESS, launched in 2018) and will build on the success of ESA exoplanet missions such as CHEOPSs and PLATO.

**Ground segment** – The Ariel ground segment will be implemented and operated by ESA and the Ariel Mission Consortium (AMC) in collaboration. The mission operations will be conducted by the Mission Operations Centre (MOC) at ESOC under ESA responsibility, and ESA will also provide the ground stations. The science operations will be conducted by the Science Ground Segment consisting of the Science Operations Centre (SOC) at ESAC provided by ESA, and the AMC-provided nationally funded Instrument Operations and Science Data Centre (IOSDC). The provision of the target list will be an Ariel Science Team responsibility, involving the community, and review by the ESA Advisory Bodies. The actual observation scheduling will be a joint MOC-SOC-IOSDC activity with scientific guidance provided by the Ariel Science Team.

The SOC will host the Ariel archive comprising the mission data base and science data archive. A range of data products will be produced at the SOC and used to populate the science archive, employing data processing pipelines delivered to the SOC by the IOSDC. Final exoplanet spectra will be produced using state-of-the-art tools developed at the IOSDC, and delivered to the SOC for ingestion into the Ariel science archive. The pipeline, as well as scripts with critical parameters used to generate the final exoplanet spectra, will be clearly and thoroughly documented and made available to users to enable reprocessing of data taken with Ariel.

**Communication and Outreach** – Beyond the science community, Ariel's mission to characterise distant worlds offers an immense opportunity to capture the public imagination and inspire the next generation of scientists and engineers. Through the provision of enquiry-based educational programmes and citizen science platforms, school students and members of the public will have the opportunity to participate directly in the analysis of Ariel.

Humans have been speculating about the cosmos and the possible existence of other worlds throughout recorded history, and probably for longer. Since the discovery of exoplanets in the 1990s this field of astronomy and planetary science has exploded, being one of the most exciting and dynamic. Currently we know over 4300 exoplanets orbiting in excess of 3000 host stars. We know the sizes of about ¾ of these, and for about one fifth we have mass information. Only for a few percent of all known exoplanets do we have both, enabling rudimentary conclusions about their bulk compositions, and comparison with models of planetary structure. For even fewer do we have atmospheric information.

We now stand at the threshold of a revolution in our understanding of our place in the Universe: Ariel is the next step. Only by performing a chemical census of a large, diverse sample of about a thousand exoplanets, studying each as a world in its own right, and as a member of a (sub-)population, can we hope to understand how our own Solar System, and our own planet Earth, fit in the big picture of innumerable other worlds. We are the first generation to know that the ancient hypothesis about planets around other stars is true. We are also the first to be capable of studying these worlds. Ariel will seize this unique opportunity and enable transformative science in this exciting field.

# 2 Scientific Objectives

## 2.1 Ariel as Part of Cosmic Vision

The first major theme of ESA's Cosmic Vision program (ESA BR-247, 2005) poses the questions of *how do planets form and what are the conditions that might make them (or their moons) habitable*. Even within the limits of our current observational capabilities, studies of extrasolar planets have provided a unique contribution to improving our understanding of these subjects and provided us with a clearer view of the place that the Solar System and the Earth occupy in the galactic context.

This new perspective is showcased, for instance, by the discovery that planets with radii between those of the Earth and Neptune (1.0-3.9 $R_\oplus$) are by far the most common in the sample identified by the NASA Kepler mission (e.g. Howard et al. 2012; Fressin et al. 2013), while our Solar System has no example of these intermediate planets. These discoveries have opened important questions about the nature, composition and formation history of planets classified as "transitional" in Figure 2-1. Transitional planets may either have a rocky core enveloped in a $H_2$-He gaseous envelope or contain a significant amount of $H_2O$-dominated ices/fluids (Valencia et al. 2013; Zeng et al. 2019), but the knowledge of the planetary radius alone is not enough to unveil the actual nature of these planets. This question remains unanswered, in most cases, even when the planetary mass, determined through radial velocity measurements, is added to the picture (Figure 2-2 right). The atmospheric composition, if known, would be an important discriminant of the actual nature of these planets.

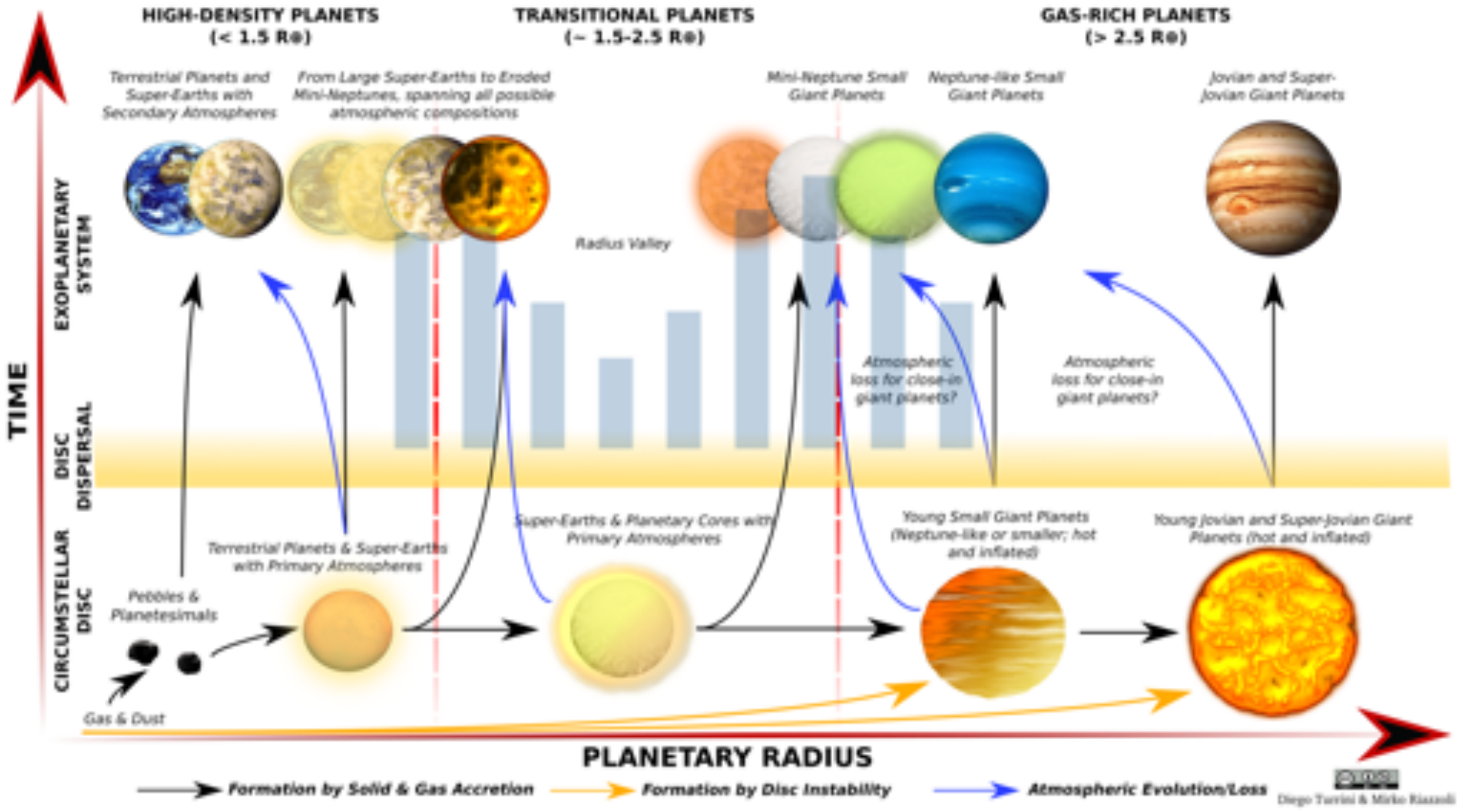

*Figure 2-1: Schematic representation of our current understanding of the formation and evolutionary paths that, starting from the gas and dust in circumstellar discs (the bottom left corner of the diagram), create the different kinds of planets currently observed. Black and orange arrows indicate the paths linked to the formation process (e.g. disc instability, solid accretion, gas capture) while blue arrows indicate the paths shaped by atmospheric evolution (e.g. atmospheric escape, atmospheric erosion, outgassing). Planets are divided into three broad categories: high-density planets (mainly composed by Si, Mg, Fe, C, O), gas-rich planets (for which H and He represent a significant fraction of their mass) and transitional planets (encompassing the transition between the largest high-density planets and the smallest gas-rich planets). The Solar System offers us examples of high-density planets and gas-rich planets but not of transitional planets, for which we need to look to exoplanets. The light blue histogram shows the frequency rate of planets across the transition region as a function of planetary radii from Fulton & Petigura (2018). Figure updated from Turrini et al. (2018).*

Tracing the atmospheric composition is important not only for planets in the transitional region or smaller, but also for the larger, gas-rich ones. In particular, the formation of the gas giants is a long-standing problem in planetary science. New observational evidence coming from Solar System (e.g. Morbidelli et al. 2016; Desch et al. 2018; Pirani et al. 2019; Öberg & Wordworth 2019; Bosman et al. 2019) and ALMA observations of complex structures and evolution even in very young protoplanetary discs (e.g. ALMA Partnership 2015; Isella et al. 2016; Fedele et al. 2017, 2018; Long et al. 2018; Andrews et al. 2018), seems to indicate that giant planets can form very rapidly, possibly as fast as ~1 Myr, over a wider range of distances from the host star than previously thought before the gas in the circumstellar disc dissipates. By contrast, the accretion of the cores of giant planets via the collision and mergers of planetesimals and pebbles is much slower according to models, particularly in the outer regions of circumstellar discs. Different formation and migration scenarios explored in the literature to address this issue might have left their imprints in the relative abundance of heavier elements present in the atmospheres. Today we can only speculate about how the "mature state" of a planet might be linked to its formation environment, or about how the type of host star drives the physics and chemistry of the planet's birth and evolution (Figure 2-1).

Progress with the (exo)planet-related science questions highlighted in ESA's Cosmic Vision and elsewhere demands a large, comprehensive survey of the chemical composition and structure of exoplanetary atmospheres. The Ariel mission has been conceived and designed to conduct such a survey and, through it, to identify the key factors affecting the formation and evolution of planetary systems. It is for these reasons that Ariel has been selected as ESA M4 mission in March 2018.

### 2.1.1 Exoplanet demographics in 2020

The discovery of the first planet around a main sequence star (Mayor & Queloz 1995) was recognised in 2019 with the Nobel prize for Physics. Since their discovery in the early 1990s planets have been found around every type of star, including pulsars (Wolszczan & Frail 1992), binaries (e.g. Doyle et al. 2011; Welsh et al. 2012), ultra-cool dwarfs (Gillon et al. 2017) and pre-main sequence stars (Plavchan et al. 2020). Planets appear to be rather ubiquitous. Current statistical estimates generated with different discovery techniques indicate that, on average, every star in our galaxy hosts at least one planetary companion (Cassan et al. 2012; Batalha et al. 2014; Howard et al. 2010) and therefore billions of planets should exist just in our Milky Way. The number of planets discovered is still far from representing an unbiased sample of the billions mentioned above, and as a result, a great deal of effort has been and will be spent to increase the number of known extrasolar planets (4300+ at the time of writing) and overcome the limits imposed by the incomplete sample (Figure 2-2 left).

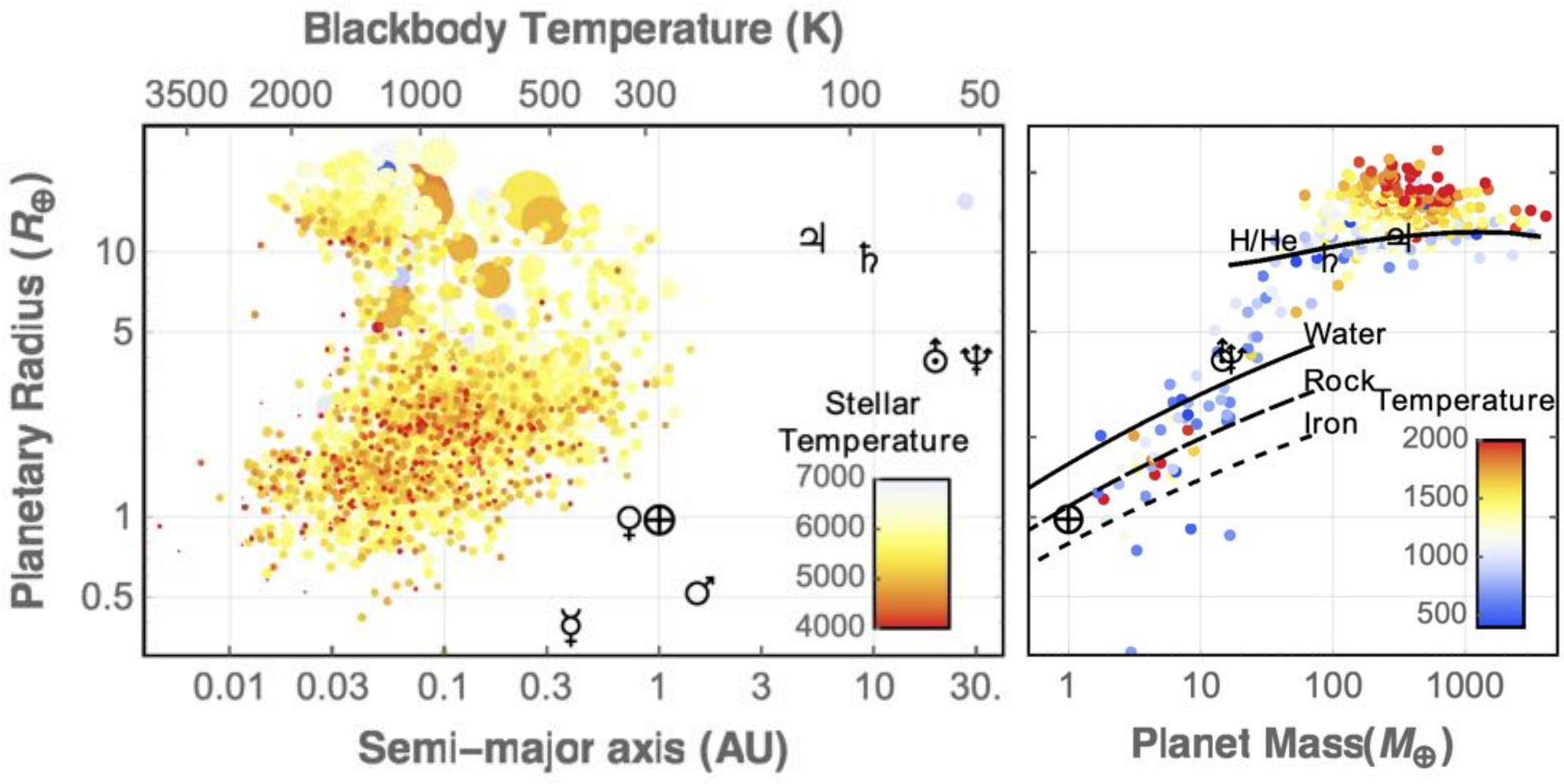

*Figure 2-2: Left: currently known exoplanets, plotted as a function of distance to the star (up to 30 au) and planetary radii. Known planets span several orders of magnitude in both parameters. The planets in the Solar System are shown for reference, each identified by its astronomical symbol. On the same plot we show the conversion of the orbital period into planetary equilibrium temperature, i.e. the temperature that a black body would have at the given distance of a Sun-like star. Notice that in general a planet is not a black body: the albedo and atmospheric greenhouse effect, which are currently unknown for most exoplanets, have a great impact on its real temperature. Data from exoplanets.org (accessed March 2020). Right: Masses and radii of known transiting exoplanets. Black lines show mass-radius relations for a variety of internal compositions: the models cannot fully capture the variety of cases and break the degeneracies in the interpretation of the bulk composition. Data from exoplanets.org (accessed March 2020)*

The ESA Gaia mission's full data release, foreseen in 2024, will enable the discovery of tens of thousands new planets with the astrometry method (Perryman et al. 2014). In addition, the ongoing release of results from the NASA Kepler and K2 missions (e.g. Batalha 2014; Borucki 2016; Mayo et al. 2018; Petigura et al. 2018; Ethan et al. 2019) and ground-based surveys (e.g. Pepe et al. 2018; Gillon et al. 2018; Batista et al. 2018; Nielsen et al. 2019; Boccaletti et al. 2020) will add to the current ground and space based efforts (see Table 3-4 and Table 3-5). The NASA Transiting Exoplanet Survey Satellite (TESS, Ricker et al. 2016), launched in April 2018, is anticipated to find several thousand transiting planets around stars that are 30-100 times brighter than those surveyed with Kepler/K2 (Barclay et al. 2018). Of these, several hundred are expected to be Earth-sized and super-Earth-sized planets and an extensive ground-based program aims to measure masses for many of these. Already, 2000+ TESS candidates have been found and tens of planets have been confirmed (e.g. Huang et al. 2018; Dragomir et al. 2019; Vanderspek et al. 2019). The mission was recently funded for a two-year extension that will see the spacecraft survey the ecliptic as well as re-observe portions of the sky covered in the primary mission (https://tess.mit.edu/observations/).

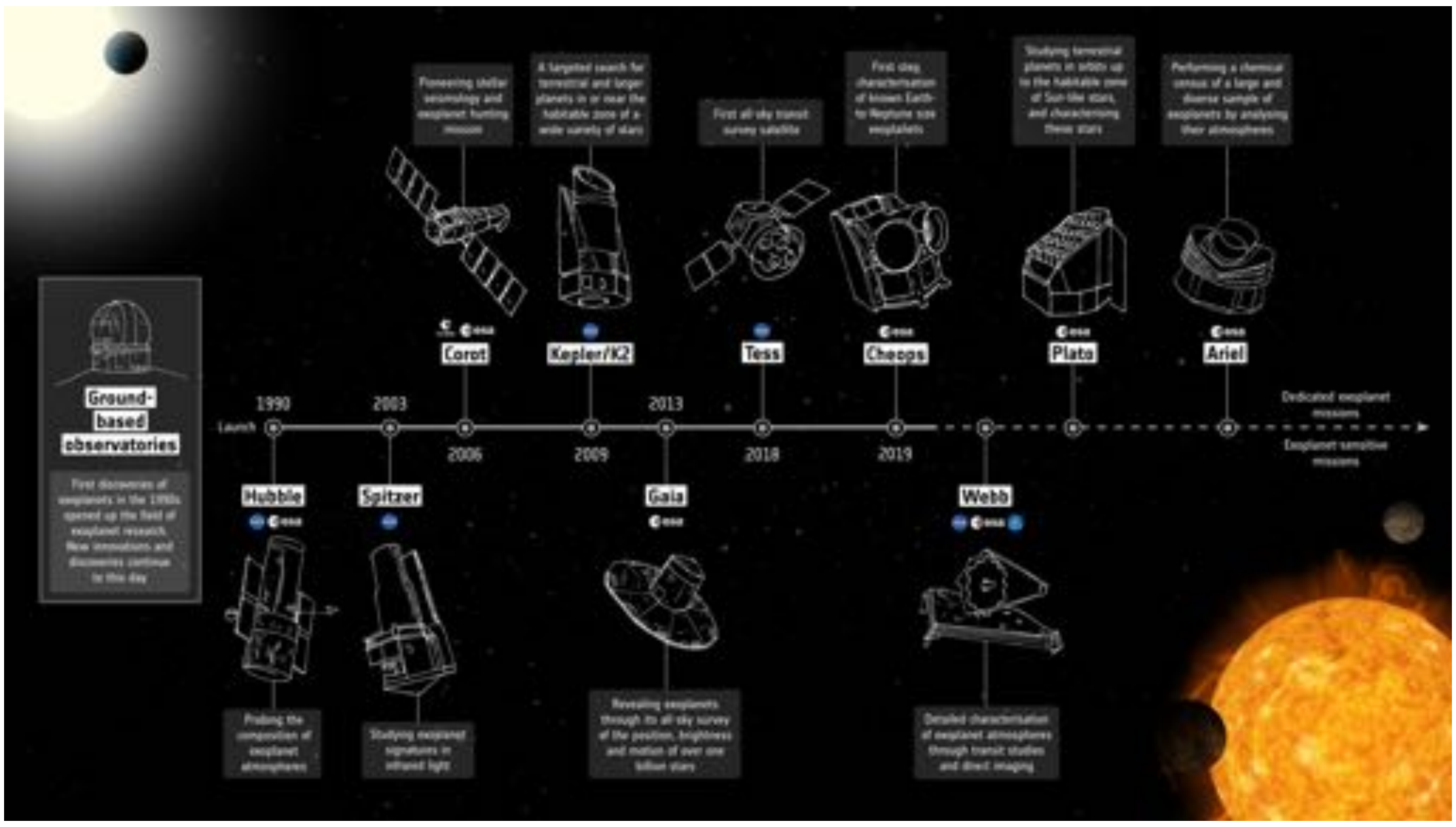

*Figure 2-3: Infographic highlighting the main space-based contributors to the exoplanetary field, including not only exoplanet-dedicated missions, but also exoplanet-sensitive missions, past, present and future. Credit ESA.*

The launch of ESA CHEOPS in 2019 and the selection of PLATO as ESA's M3 mission, to be launched in the 2026 timeframe, are important goals in the ESA's Cosmic Vision program to address the current observational bias in the available sample of planets (Figure 2-2 left). CHEOPS has the scientific goal to search for transits of exoplanets by means of high-precision photometry on bright stars already known to host planets (Benz et al. 2020). In the future, we can look forward to many more discoveries from the PLATO mission, especially small transiting planets in the habitable zones of solar-like stars (Rauer & Heras, 2018).

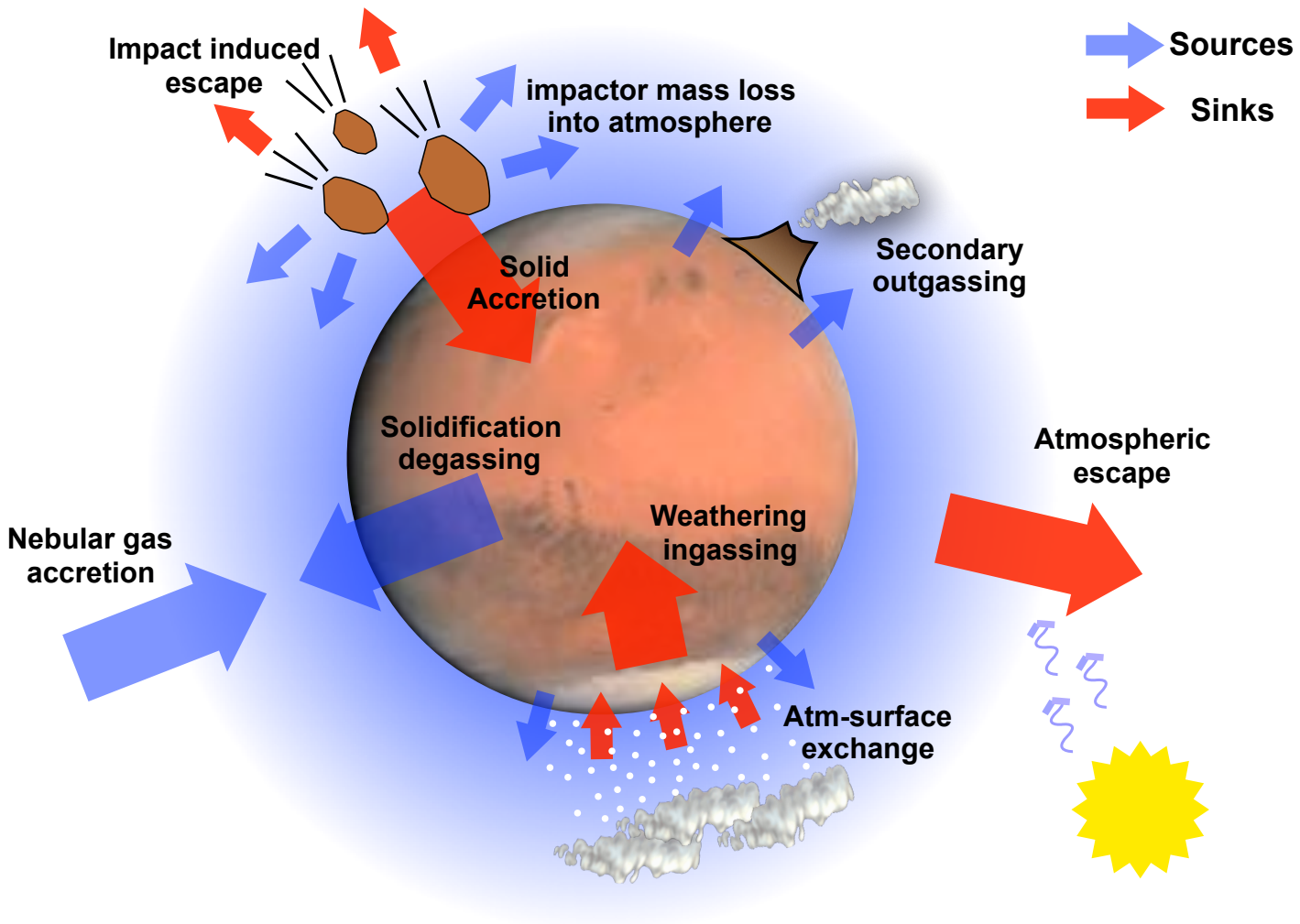

*Figure 2-4: Key physical processes influencing the composition and structure of a planetary atmosphere. Red arrows represent mechanisms removing gas from the atmosphere while blue arrows are source mechanisms. While the analysis of a single planet cannot establish the relative impact of all these processes on the atmosphere, by expanding observations to a large number of very diverse exoplanets, we can use the information obtained to disentangle the various effects.*

The search for new worlds in our galaxy in the past couple of decades has been highly successful and the prospects for the next decade are even brighter (see Figure 2-3). The information provided by the currently planned efforts, however, mainly deals with the orbits and the basic planetary parameters – e.g. mass, size – of the discovered planets. In the next decade, emphasis in the field of exo-planetary science will have to expand towards the understanding of the nature of the exo-planetary bodies and their formation and evolutionary history. This goal is achievable through the remote sensing observation of a large sample of exoplanetary atmospheres. Knowing what exoplanets are made of is essential not only to clarify their individual histories (see Figure 2-4) but also to seek the underlying physical and chemical properties of planetary systems. Most critically, this information can then be used to understand key mechanisms that govern planetary evolution at different time scales.

By performing a large survey of chemical composition, Ariel will open a new vista in exoplanet demographics.

## 2.1.2 Current observations of exo-atmospheres

### *2.1.2.1 Spectroscopic measurements of exoplanetary atmospheres*

In recent years a small yet steadily increasing number of exoplanetary atmospheres have been characterised with space- and ground-based observatories. Ultraviolet, optical, and infrared spectra, recorded through transit, eclipse, high-dispersion, and direct imaging techniques, have offered glimpses of the atmospheric structure and composition of exotic worlds orbiting other stars.

Transit and eclipse spectroscopy have formed the kernel of exoplanet atmospheric characterisation. Pioneering results have been obtained using transit spectroscopy with Hubble, Spitzer and ground-based facilities, enabling the detection of a few of the most abundant ionic (e.g. Linsky et al. 2010), atomic (e.g. Redfield et al. 2008) and molecular species (e.g. Swain et al. 2009), hazes and condensates (Pont et al. 2008) and constraints to be placed on the planet's thermal structure (Knutson et al. 2007). Information on the stability of the atmospheres of transiting planets has been collected through UV observations with Hubble (e.g. Vidal-Madjar et al. 2003; Fossati et al. 2010; Bourrier et al. 2020): hydrodynamic escape processes are likely to occur for most of the planets orbiting too close to their parent star. The infrared (IR) range, on the contrary, offers the possibility of probing the neutral atmospheres of exoplanets and exploring their thermal structure (e.g. Majeau et al. 2012; Line et al. 2016). In the IR the molecular bands are more intense and broader than in the visible (Tinetti et al. 2013) and less perturbed by small particle clouds, and are hence easier to detect. On a large scale, the IR transit and eclipse spectra of hot-Jupiters seem to be dominated by the signature of water vapour (e.g. Swain et al. 2009; Tinetti et al. 2007b; Deming et al. 2013; Danielski et al. 2014; Wakeford et al. 2017; Mikal-Evans et al. 2020; Pluriel et al. 2020). Other carbon-bearing molecules, metal hydrides and oxides (e.g. TiO, VO, FeH, AlO), $H^-$, have been suggested being present in a few of these atmospheres (e.g. Swain et al. 2009; Haynes et al. 2015; Evans et al. 2016; Gandhi & Madhusudhan, 2019; von Essen et al. 2019; Edwards et al. 2020b; Skaf et al. 2020; Changeat et al 2020e; Chubb et al. 2020a, Yip et al. 2020b).

High-contrast imaging photometry and spectroscopy in the infrared have been obtained for a number of young gaseous planets using ground-based dedicated instruments (Macintosh et al. 2015; Barman et al. 2015; Gravity Collaboration 2019, see Figure 2-5), such as VLT (SPHERE), Gemini (GPI), Subaru (SCExAO), VLTI (GRAVITY). The comparison of the chemical composition of these young gaseous objects with the composition of their migrated siblings probed through transit will help to clarify the role played by migration and by extreme irradiation on gaseous planets .

High-dispersion spectroscopic measurements from the ground have been successfully applied for both exoplanet transmission (Snellen et al. 2010) and emission spectroscopy (Brogi et al. 2012). Metallicity measurements via the detection of atomic and ionized species – e.g. Fe, Fe+ – and atomic and ionized Ca have already been detected in the atmospheres of ultra hot-Jupiter planets (Hoeijmakers et al. 2018; Yan et al. 2019). Atmospheric lines crucial to understand evaporation were also measured, like the Hα line and the Balmer series in the visible range (Casasayas et al. 2019) and the He I triplet in the mid-infrared (Nortmann et al. 2018, see Figure 2-5). In most recent years, the study of exoplanetary atmospheres has shifted from the investigation of individual planets to the characterisation of populations (e.g. Iyer et al. 2016; Sing et al. 2016; Barstow et al. 2017, Tsiaras et al. 2018, Fu et al. 2018; Fisher et al. 2018; Pinhas et al. 2019; Garhart et al. 2020). So far, population studies of gaseous planetary atmospheres have focused mainly on the abundance of water vapour and on the presence/absence of condensates/hazes. While available data are not yet good enough to support conclusive results on how the atmospheric chemistry correlates to other planetary and atmospheric parameters (e.g. Tsiaras et al. 2018), these studies are a remarkable step forward in our understanding of exoplanetary atmospheres.

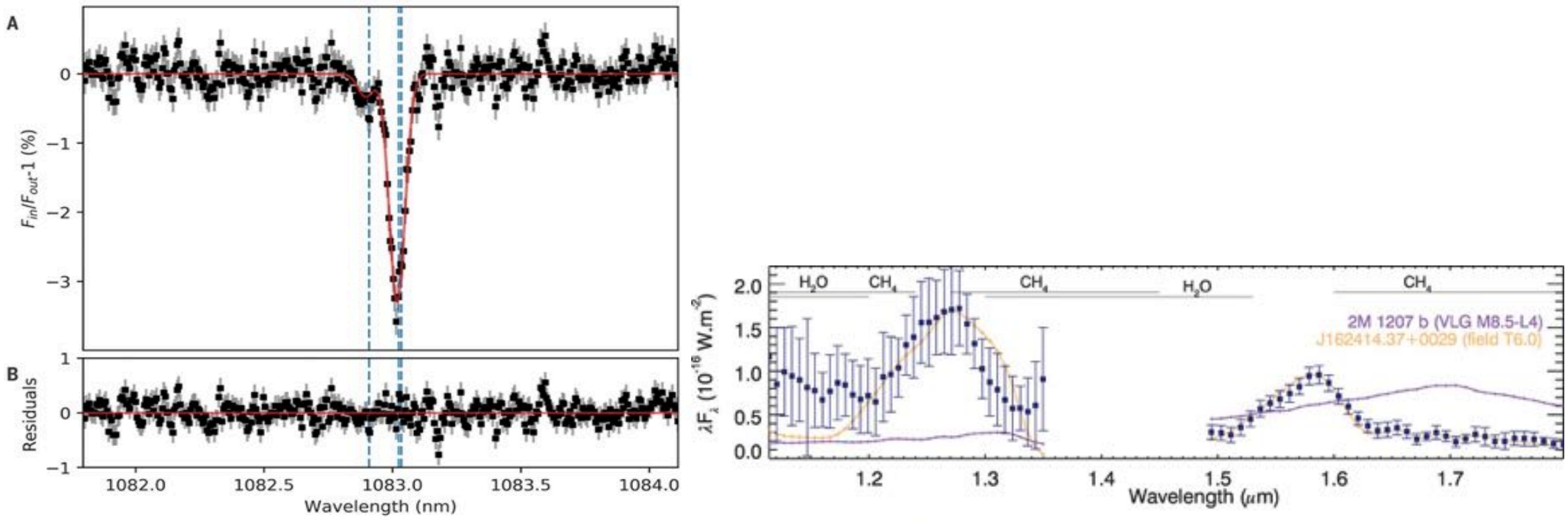

*Figure 2-5: Left: Ground-based detection of an extended helium atmosphere in the Saturn-mass exoplanet WASP-69b (Nortmann et al. 2018). The planetary absorption in the He I triplet at 1083 nm was recoded with the CARMENES spectrograph. Right: J- and H-band spectra for 51 Eri b from GPI data (Macintosh et al. 2015).*

While the rapid development of the study of gaseous exoplanets has revealed more and more about their atmospheric properties, the study of planets smaller than Neptune has, so far, remained very limited. Current observations of the atmospheres of these small worlds have not yet allowed us to infer precise constraints on their nature. The transitional planet GJ-1214 b was observed multiple times with Hubble (Kreidberg et al. 2014) and other instruments. Current observations are suggestive of a cloudy or hazy world, which has motivated an extensive literature (e.g. Valencia et al. 2013; Kataria et al. 2014; Charnay et al. 2015; Gao & Benneke 2018; Lavvas et al. 2019; Miguel et al. 2019; Venot et al. 2020b). Based on its bulk density, we speculate its atmosphere as being hydrogen dominated; however so far this has not been confirmed by the direct measurement of its atmosphere. An interesting case is 55 Cnc e, a very hot transitional planet orbiting around its star in less than one Earth day. Most recent observations with Spitzer/Hubble suggest a very strong day-night thermal gradient with a volatile atmosphere around it (e.g. Tsiaras et al. 2016b; Demory et al. 2016; Angelo & Hu, 2017, see Figure 2-7). However, further observations are needed to constrain current models of the atmospheric composition and stability (Ito et al. 2015; Hammond & Pierrehumbert 2017; Modirrousta-Galian et al. 2020). Hubble observations of the TRAPPIST-1 planetary system (Guillon et al. 2017) did not reveal the atmospheric composition of these rocky worlds (de Wit et al. 2016, 2018; Ducrot et al. 2018). Recent observations with HST of the transitional planet K2-18b, orbiting within the habitable-zone of its host star, have revealed the presence of water vapour in its atmosphere (Tsiaras et al. 2019; Benneke et al. 2019). However additional observations at longer wavelengths are needed to understand the atmospheric thickness and the water to hydrogen ratio in the atmosphere (Changeat et al. 2020d; Madhusudhan et al. 2020)

Despite some early successes, current data remain sparse – in particular, there is insufficient wavelength coverage and most observations were not made simultaneously. Because an absolute calibration at the level of 10–100 ppm is not guaranteed by available instruments, great caution is needed when one combines multiple datasets at different wavelengths which were not recorded simultaneously (Yip et al. 2020a). Because the data are sparse, their interpretation is rarely unique. To explore the degeneracy, reliability, and correlations among the atmospheric parameters extracted from the data, the past decade has seen a surge in spectral retrieval models developed by many teams (e.g. Terrile et al. 2008; Irwin et al. 2008; Madhusudhan & Seager 2009; Line et al. 2013; Waldmann et al. 2015b; Cubillos et al. 2016; Lavie 2017; MacDonald & Madhusudhan 2017a,b; Gandhi & Madhusudhan 2018; Goyal et al. 2018, Al-Refaie et al. 2020; Min et al. 2020).

New and better data of uniform calibration and quality are essential for the characterisation of exoplanet atmospheres; most importantly, we need the data for a large population of objects: both objectives can be achieved with a dedicated space mission like Ariel. Figure 2-6, bottom panels, illustrate the capabilities of Ariel for recording high quality, broad wavelength – simultaneously covering 0.5-7.8 µm – spectra.

Ariel will deliver large, homogeneous datasets of visible to mid-infrared spectroscopy and phase-curves. This builds on the legacy of the current state of the art, while mitigating key issues with more fragmented datasets.

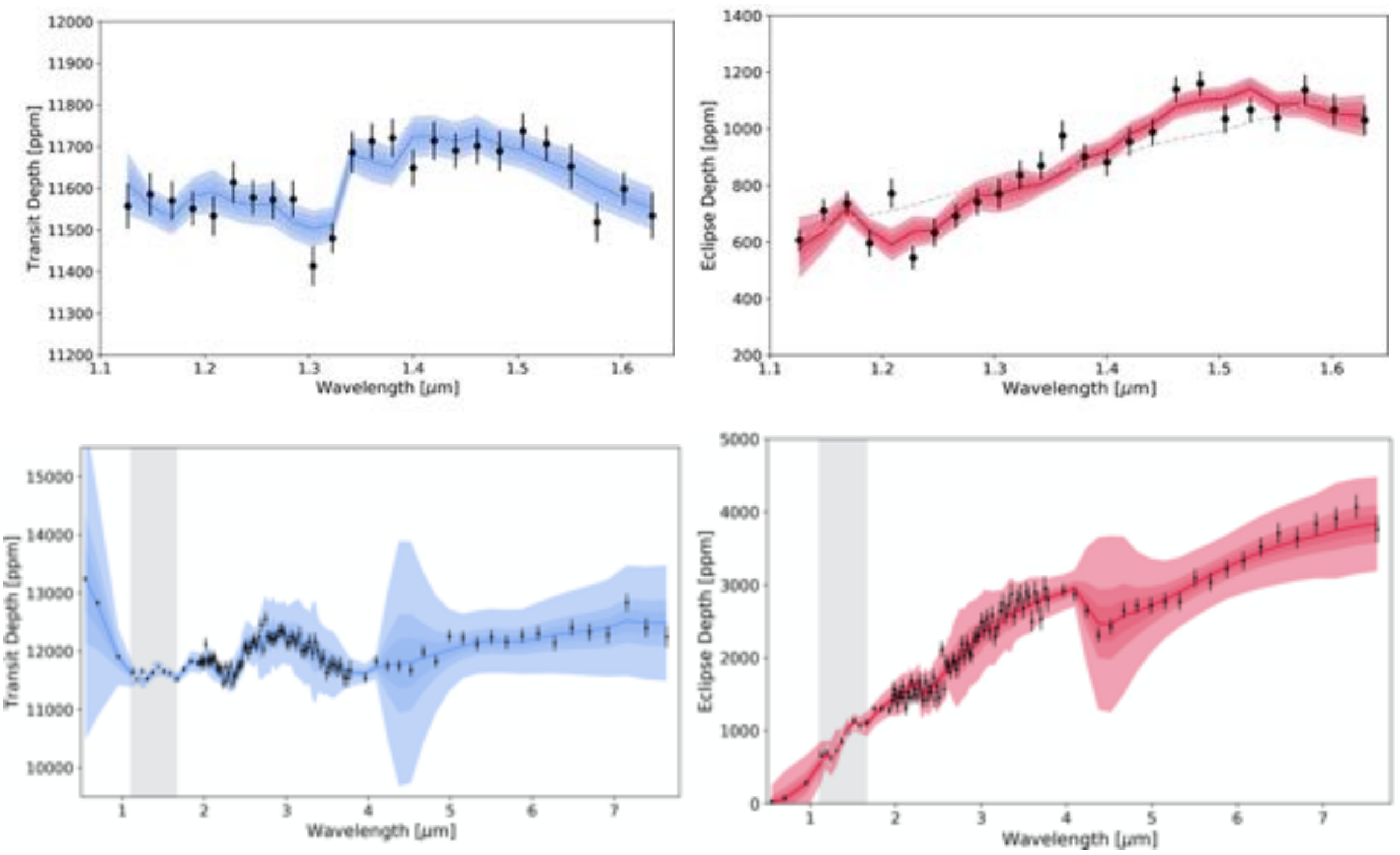

*Figure 2-6: Figures from Edwards et al. (2020b). Top: HST-WFC3 observations of WASP-76 b. Left: transit spectrum, right: eclipse spectrum. The interpretation of these spectra through spectral retrieval models suggests the presence of $H_2O$ and possibly TiO. Bottom: Simulated Ariel observations of the same planet. The Ariel spectra (black points) are for a single observation at the native resolution of the instrumentation (i.e., Tier 3, see Chapter 3). The grey box indicates the wavelength range covered by the Hubble WFC3 G141 grism and the data points from WFC3 are shown in white. Blue and red areas: 1-3σ uncertainties from the WFC3 data.*

### 2.1.2.2 Phase-curves

Phase-curves have so far been published for more than two dozen planets (e.g. Knutson et al. 2007; Laughlin et al. 2009; Esteves et al. 2015; Kreidberg et al. 2019b) and additional observations are available for as many planets. The majority of these phase-curves were obtained with the Spitzer Space Telescope (Werner et al. 2004), primarily as part of the post-cryogenic "warm" mission. Due to its far-Earth orbit, Spitzer was capable (as Ariel will be) of continuous monitoring for an entire planetary orbit. Continuous Spitzer photometric phase-curves of transiting (and eclipsing) planets have been the most useful. While multi-epoch phase-curves and/or phase-curves of non-transiting planets are often affected by instrument systematics which are difficult to correct for (Harrington et al. 2006; Cowan et al. 2007; Krick et al. 2016), out-of-eclipse baseline can well constrain the phase variations of short-period planets, as demonstrated by Wong et al. (2014).

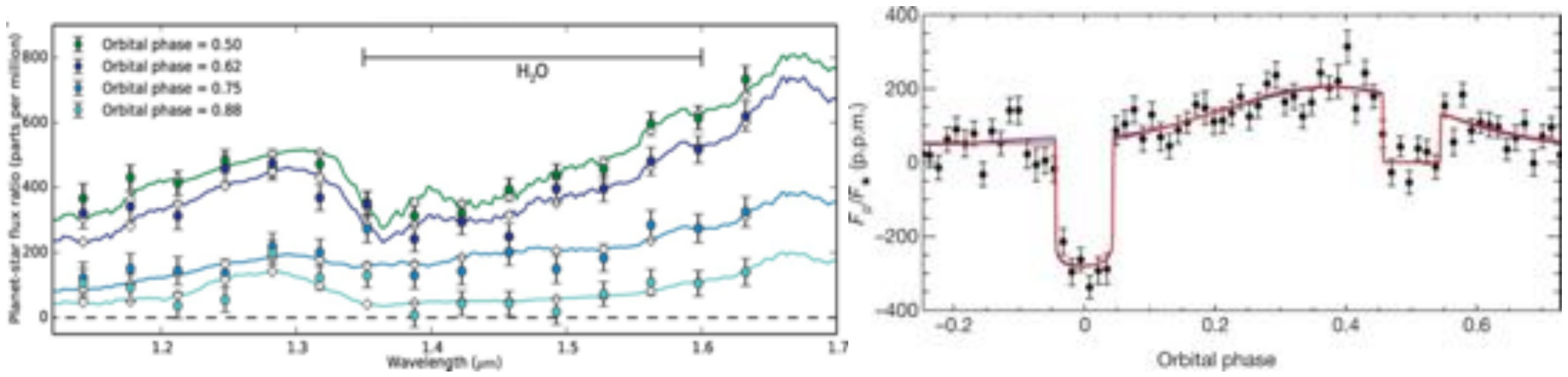

*Figure 2-7: Left: Phase-resolved emission spectra of WASP-43b relative to the stellar flux from Stevenson et al. (2014). Planet-to-star flux ratios at four planet phases are displayed: full (0.5, eclipse), waning gibbous (0.62), half (0.75), and waning crescent (0.88). Right: longitudinal thermal brightness map of the nearby transiting super-Earth 55 Cancri e (Demory et al. 2016) revealing highly asymmetric dayside thermal emission and a strong day–night temperature contrast. Eight photometric data sets obtained with Spitzer at 4.5 µm were combined and folded. The black filled circles represent the relative flux ($F_p/F_\star$) variation in phase and are data binned per 15 min.*

More recently, a few near infrared (NIR) spectroscopic phase-curves have been measured with Hubble. To date, HST phase-curves have been published for the hot-Jupiters WASP-43b (Stevenson et al. 2014, see Figure 2-7, WASP-103b (Kreidberg et al. 2018), and WASP-18b (Arcangeli et al. 2019). These spectroscopic phase-curves allowed for the first detection of water vapour and probed the atmospheric temperature-pressure profile as a function of longitude. The interpretation of the data has improved through time to include the effect of planetary reflected light, which can be significant at these NIR wavelengths (e.g. Stevenson et al. 2014; Keating & Cowan, 2017; Mendonça et al. 2018a).

HST has offered us a glimpse of what exoplanetary characterisation will be like in the era of Ariel, but the reduction and interpretation of these phase-curves has not been without problems. Hubble's low Earth orbit makes it impossible to obtain continuous measurements throughout a planet's orbit. Full-orbit phase-curves are therefore stitched together, which requires a model of telescope and detector systematics, and this has improved with time. Ariel will operate in L2 and therefore will have continuous visibility of the target throughout the orbit.

Optical phase-curves have been observed for a handful of planets with Kepler (e.g. Armstrong et al. 2016) and TESS. For the hottest planets, these phase-curves still primarily probe thermal emission from the planet, while for cooler planets they probe reflected starlight (Parmentier et al. 2016). If optical phase-variations can be attributed to reflected light, then they can be inverted to make a longitudinal albedo map of the planet's day-side (Cowan et al. 2009, 2013; Demory et al. 2013). We refer to Cowan & Fujii (2018) for a review of all aspects of planet mapping. Reflected planetary light is only detectable in short-period systems where thermal emission is not negligible, so eclipses or phase-curves at longer wavelengths are required to interpret correctly optical phase curves. This will not be an issue for Ariel because of its ability to probe simultaneously optical, NIR and MIR wavelengths.

## 2.2 Key Science Questions Addressed by Ariel

Ariel will address three fundamental questions:

- *What are the physical processes shaping planetary atmospheres?*

  The thermal structure of the atmosphere and its chemical and elemental composition relies on our understanding of how a variety of atmospheric processes interact in a wide diversity of environments.

- *What are exoplanets made of?*

  Although mass-radius measurements are essential to constrain the bulk enrichment and elemental composition of many planets, these constraints are highly degenerate: only direct measurements of the atmospheric composition can lift these degeneracies.

- *How do planets and planetary systems form and evolve?*

  In addition to the wealth of information on planetary systems' orbital architectures provided by ongoing surveys, having accurate constraints on the *bulk* elemental composition of many, diverse exoplanets will be key in advancing planet formation theory.

These three questions are intrinsically linked: understanding planetary atmospheres will inform us of the planet compositions which, in turn, will tell us about their formation.

Progress in the three key exoplanetary science areas (atmospheres; composition; formation and evolution) will depend on a spectroscopic survey of hundreds of transiting planets, spanning different planetary sizes, a range of equilibrium temperatures and orbiting a variety of stellar types. This is the goal of Ariel.

In the following sections, we discuss more detailed science cases identified by the Ariel Mission Consortium Science Working Groups, including recent work performed during Phase B. Addressing those cases will provide responses to the three fundamental questions described in the previous paragraph. Quantitative estimates of the feasibility of specific observations or of the accuracy/precision with which atmospheric parameters can be constrained by Ariel data will be presented in Chapters 3 and 7.

Throughout this report, we will follow the schematic approach adopted in Figure 2-1 and divide planets into three broad categories:

- Gas-rich planets, i.e. objects with radii larger than Neptune's. Current formation and structure models agree that a significant fraction of the planet must be composed of primordial Hydrogen/Helium gas.
- High-density or terrestrial planets, i.e. objects which lost their primary Hydrogen/Helium envelope. This category would encompass objects roughly resembling the four inner planets of the Solar System as well as equivalents to the icy satellites of the giant planets (sometimes called ocean planets), or very irradiated magma ocean planets. They may or may not have retained a secondary atmosphere. Based on theoretical arguments, we expect those planets to be smaller than 1.5 Earth radii, although any clear classification will have to wait a proper atmospheric characterization.
- Transitional planets, i.e. objects whose bulk composition is dominated by heavy elements while their atmospheres managed to retain a significant fraction of hydrogen and helium. In between the two aforementioned categories, they play a crucial role in our understanding of planet formation as they are the most numerous in the galaxy while having no counterpart in the Solar System.

### 2.2.1 How Ariel will place the Solar System into a broader context

For centuries the Solar System has been our only example of planetary system and the planets within it our only template for planetary bodies in our Galaxy. Thanks to the level of detail with which we have been able to study them, Solar System planets have enabled great progress in understanding planet formation. However, our own planetary system is only one among many, with peculiarities – some of these listed below – that cannot be extended to other planets in the Milky Way.

#### *Solar System difficulty 1 – Condensation*

Compared to many exoplanets discovered to date, our four gas-rich planets are cold: their atmospheric composition, our window into their bulk composition, is thus extremely affected by condensation and removal processes. For instance, the relative abundance of the most common heavy element, oxygen, cannot be measured directly through remote sensing observations because its main molecular carrier, water, condenses in the atmosphere and is removed from the observable region (see e.g. Taylor et al. 2004). This information is so crucial in the study of the Solar System's origins that it justifies no less than two missions in situ (NASA's Juno and ESA's JUICE) with the objective to constrain the bulk composition of Jupiter alone.

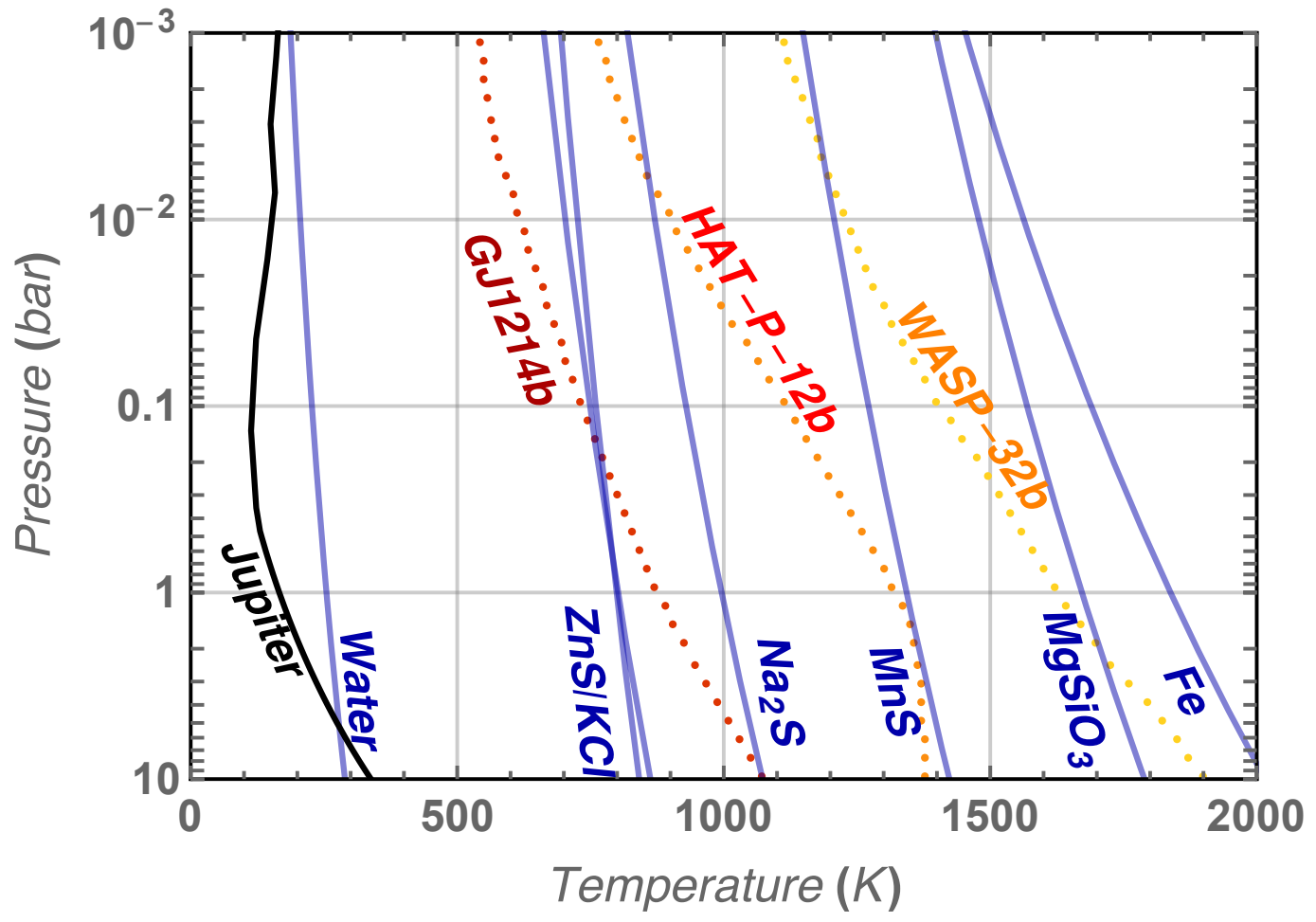

*Figure 2-8: Atmospheric temperature profiles (dotted lines) compared to the condensation curve of selected condensing species in an atmosphere with Solar elemental abundances (solid blue curves). The thermal profiles are taken from Mollière et al. (2017) for prototypical planets with an equilibrium temperature of 500K (GJ 1214 b; red), 1000K (HAT-P-12 b; Orange), and 1500K (WASP-32 b; yellow). The thermal profiler corded by Galileo for Jupiter is in black. The condensation of a given species occurs when the temperature profile crosses the corresponding condensation curve; the atmosphere is depleted of that species above that point. While Jupiter is too cold to keep a large amount of water vapour aloft, even the coldest planets in our sample ($T_{eq}$~500K) should not experience water condensation. As other species (e.g. CO, $CH_4$, $NH_3$, $CO_2$) condense at even lower temperatures, the case is even more compelling for these species. However, we do expect Ariel will probe a long list of exotic clouds in the atmospheres of planets at different equilibrium temperatures.*

As described in Figure 2-8, even for the coldest objects in the Ariel sample, the atmospheric temperatures will be well above the condensation temperature for water. Most of the other main reservoirs of oxygen, carbon, and nitrogen (e.g. CO, $CH_4$, $CO_2$, $NH_3$, $N_2$) condense at even lower temperatures, so the case is even stronger for these molecules. For some of the hottest objects, even more refractory elements (silicates, metals) can be detected, allowing us to constrain almost all of the major constituents of a potential planetary core. Observations of hot gaseous exoplanets can therefore provide unique access to their elemental composition (especially C, O, N, S) and enable the understanding of the early stage of planetary and atmospheric formation during the nebular phase and the following few million years (see §2.2.2).

***Solar System difficulty 2 – Chemical (dis)equilibrium***

In the Solar System the condensation of some key molecules deep down implies that any inference about the deep composition relies on modelling. In general, disequilibrium chemical models are used to link the deep elemental abundances to the abundance of some specific trace molecules in the observable atmosphere (Wang et al. 2016). This challenge makes the results model dependent (Cavalié et al. 2017). This will not be the case for Ariel targets, as it is discussed extensively in the following sections.

***Solar System difficulty 3 – There are only 8 planets and 1 star in the Solar System...***

The Solar System's planets do not sample all possible outcomes of planet formation processes! We now know that the most abundant class of planets, i.e. the transitional planets in Figure 2-1, is not even represented in the Solar system. Even in terms of climate and atmospheric evolution, Solar System's planets all share a common host star, and have evolved for the same amount of time.

While the knowledge of the planets in our Solar System is growing and becoming more accurate due to the ambitious exploration programs by ESA, NASA and other agencies, the statistics will always be too small to draw conclusions about the general properties of planets and planetary systems. A larger population of planets covering a broader parameter space in terms of size, mass, orbital characteristics, and stellar host is needed to progress in our understanding of the general principles underlying planet formation and evolution. Interestingly, such a general understanding will also inevitably shed a new light on the origin of our own planetary system.

### 2.2.2 What are the physical processes shaping planetary atmospheres?

Ariel will observe many planets in very different atmospheric regimes, aiming at identifying the most important physical processes at play and how they interact. We discuss in this section a few specific questions that Ariel will address in this area.

#### *2.2.2.1 What is the thermal structure of planetary atmospheres?*

Vertical structure and global energy budget: planetary temperatures are determined by the balance between the absorption of the stellar radiation and the atmospheric thermal emission. This equilibrium is regulated by the molecules absorbing and emitting in the atmosphere. Measuring the bulk temperature of the atmosphere and, when possible, its vertical structure (e.g. from secondary eclipses) is important to constrain the atmospheric composition. For example:

- A *higher-than-expected* temperature or a temperature inversion suggests the presence of an atmospheric absorber, even if said absorber cannot be identified with Ariel (e.g. UV absorber).
- A *lower-than-expected* temperature points to the presence of either a strong thermal emitter (which should be detectable) or very reflective aerosols.

Current observations suggest that many exoplanets are totally or partially covered by reflective aerosols (e.g. Sing et al. 2016; Kreidberg et al. 2014). However, we do not know the nature of these aerosols, whether they are condensate **clouds** or photochemical **hazes**. We expect, though, the thickness and composition of clouds to be strongly correlated to the bulk planetary temperature, as it is for the L-T transition in brown dwarfs (Saumon, Marley 2008, Morley et al. 2012, Parmentier et al. 2016, Charnay et al. 2018, see also Figure 2-11). Ariel Tier 2 and Tier 3 surveys (see Section 3.1) will measure the bulk temperature of the atmosphere and its vertical structure for many planets. For a subsample, Ariel will measure or constrain the geometric and Bond **Albedo** (Ariel Phase Curve WG report, 2020). The albedo of an exoplanet is determined by the optical properties and cloud/haze cover: testing whether there are albedo transitions and at which temperatures they happen, will be important to identify the condensed species. For a rocky planet with a thin atmosphere the Bond Albedo can constrain the composition of the surface (Kreidberg et al. 2019b). More in general, the albedo and the energy budget are essential ingredients to understand the climate and thermal evolution of exoplanets.

Three-dimensional structure and heat redistribution: a peculiarity of transiting exoplanets compared to their Solar System siblings is that most of them are expected to be tidally synchronized, i.e. with one hemisphere permanently facing the host star. The thermal gradients between day-side, terminator, and night-side of transiting planets are thus expected to be much larger than those measured in our closest neighbours. The thermal gradients depend on the atmospheric redistribution of energy from the day side (Komacek et al. 2016). While there are several 3D Global Climate models developed to study the circulation in the atmospheres of tidally locked planets (Showman et al. 2002; 2009; Leconte et al. 2013; Mendonça et al. 2018b; Sainsbury-Martinez et al. 2019), the dynamical regimes present in those atmospheres have never been observed and observational constraints are sorely needed to understand them. The temperatures in the day-side can be measured through eclipse observations, while the temperatures at the terminator through transit observations. Additional constraints on the heat redistribution will be provided by the optical albedo and the thermal emission of the day side, simultaneously measured by Ariel during the eclipse. Direct constraints on the 3D thermal structure of the atmosphere will come from eclipse-mapping and phase-curve observations (see Chapter 3). The difficulties with all past phase-curve observations (see Section 2.1.2.2) are that: *i*) optical data alone do not allow to disentangle the reflected starlight part –which encodes information on the cloud structure as well as the composition and particle size of the condensates – from the emitted part; *ii*) IR data alone are also incomplete: it is impossible to know whether a low brightness is due to a cool atmosphere or to high altitude clouds which hide the deep atmosphere. Ariel's high-quality, simultaneous data across the optical and infrared will thus be key to disentangle all those contributions and decipher the 3D thermal structure of the atmosphere (Ariel Phase Curve WG report, 2020; Figure 2-9).

Ariel will perform phase-curve observations on a sizable sample of planets with different rotation rates and temperatures, allowing to probe different circulation regimes and to study the underlying mechanisms at the origin of atmospheric dynamics and structures.

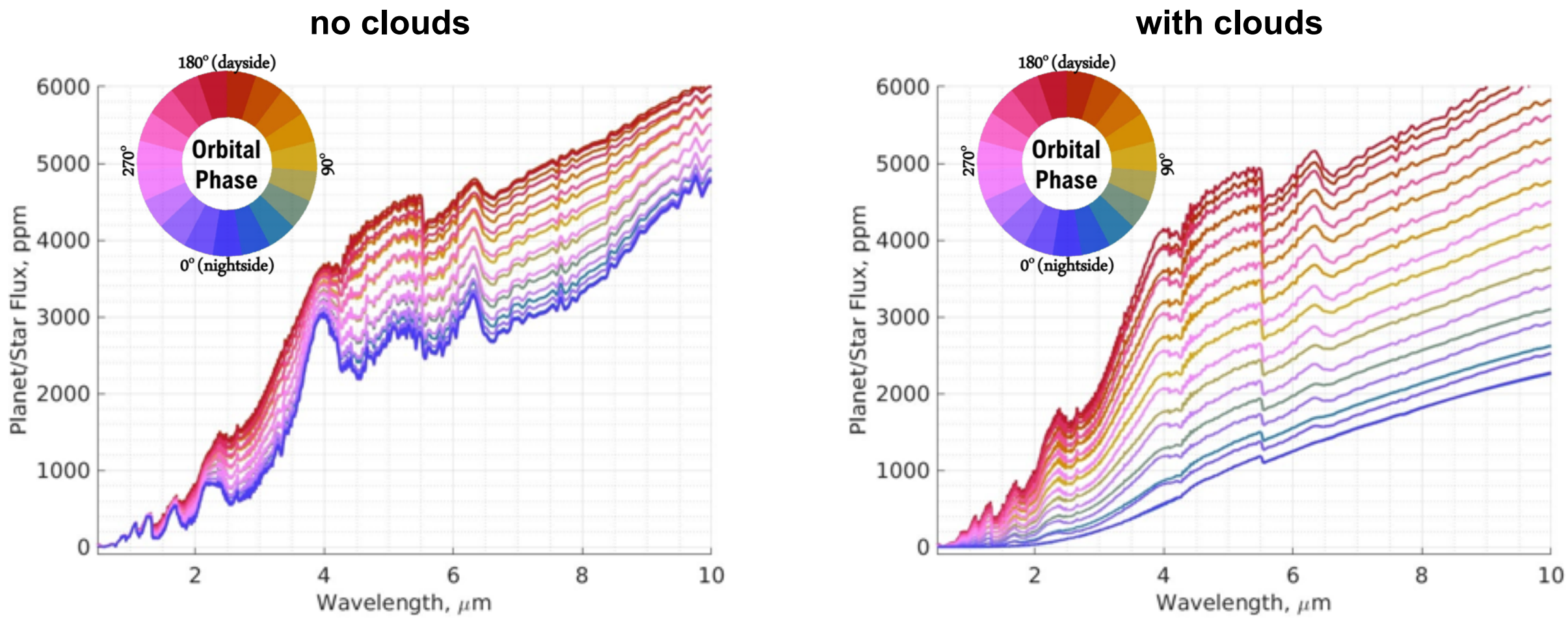

*Figure 2-9: Simulated spectra as emitted from hot-Jupiter WASP-43 b at different orbital phases. The simulations were obtained with a 3D Global Climate Model (THOR: Mendonça et al. 2016). The circular map indicates the colours for the different orbital phases: the transit occurs at orbital phase 0° (blue), the eclipse at 180° (red). Left: clear sky atmosphere; right: cloudy atmosphere in the night-side. The removal of spectral features, characteristic of clouds, only at some orbital phases is key to locating these clouds and constraining their composition. Here the condensing species have a sublimation temperature below the day-side temperature.*

Finally, all observed Solar System planets and many brown dwarfs show some sort of temporal variability in their atmospheres (Artiguau et al. 2009). Currently, there are almost no constraints on temporal variability in the atmospheres of extrasolar planets, partly because it is difficult to re-observe targets with a non-dedicated observatory. Theoretical models predict that temporal variability could be induced by shocks (Fromang et al. 2016), atmospheric waves (Komacek & Showman, 2020), and magneto-hydrodynamical effects (Rogers et al. 2017). By observing at length and through time select targets in diverse environments, Ariel will allow to test whether and how specific physical processes may influence planetary atmospheres.

#### 2.2.2.2 Where does non-equilibrium chemistry play an important role?

The atmospheric temperatures of short-period exoplanets are very hot and therefore one might expect the chemical composition of these atmospheres being close to the thermochemical equilibrium, as high temperatures may lead to very fast chemical kinetics. These were the assumptions made by the first atmospheric models used to study these planets (e.g. Burrows et al. 1999; Seager et al. 2000). However, it was soon realised that many warm and hot atmospheres could be far from equilibrium for several reasons. Firstly, as it happens on Earth, photochemistry can break some otherwise stable molecules in the upper atmosphere. Secondly, in gas giants and brown dwarfs, vertical turbulent mixing transports gas upward faster than the timescale needed by chemistry to reach the equilibrium (vertical quenching). As kinetics slows down with decreasing temperature and pressure, this scenario almost always happens at some atmospheric pressure, with the exception of the atmospheres of ultra-hot planets (T > 2000 K). In close-in planets, horizontal quenching can become predominant. Here the strong thermal gradient between the day and night side coupled to the large-scale atmospheric motions transporting the gas around the planet can also cause a departure from the equilibrium. Where quenching dominates, chemical abundances are expected to be more uniform, therefore facilitating the job of spectral retrieval models. By contrast, where chemistry is fast, we expect large chemical inhomogeneities in the atmosphere. The information that temperature and composition must be linked is helpful to interpret those cases. The study of more complex cases at the transitions between these regimes will be very important to improve our understanding of chemistry in hot atmospheres.

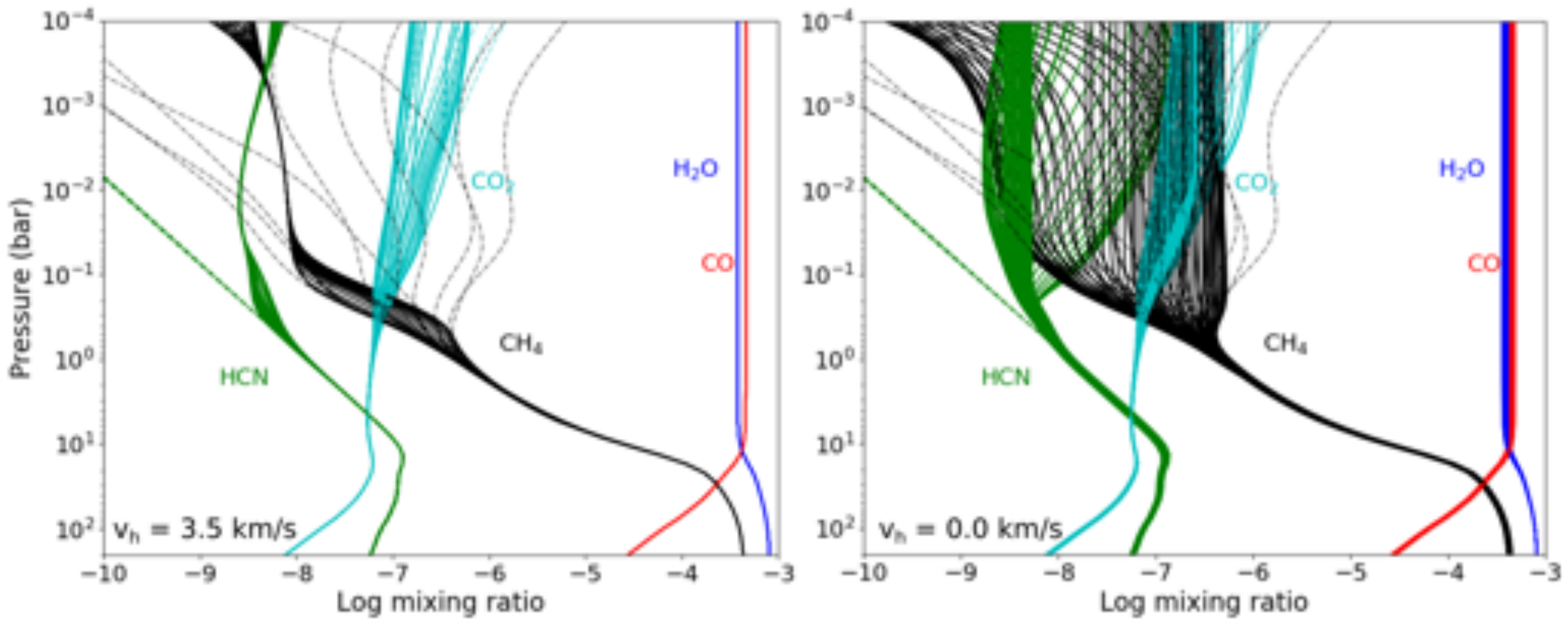

*Figure 2-10: Simulated chemical profiles in the atmosphere of the hot-Jupiter HD 209458 b. Individual species are colour-coded, with equatorial profiles plotted for various longitudes. The atmosphere exhibits small longitudinal differences when including horizontal quenching (left), whereas longitudinal differences are pronounced when horizontal quenching is ignored (right). The corresponding dash-dotted profiles represent the chemical composition at the equilibrium (Ariel Chemistry WG report 2020).*

Significant efforts in the exoplanet community and in the Ariel Chemistry WG (see the Ariel Chemistry WG report (2020) for references) have been made to understand where non-equilibrium chemistry plays an important role. This work uses a comprehensive hierarchy of models, from 0D “chemistry in a box” models to fully fledged 3D simulations. The emerging picture is the following:

- For planets colder than ~750K, vertical and horizontal quenching dominate kinetics and the gas is completely out of equilibrium. As a result, the atmosphere is mostly homogeneous, except for the upper parts that can be altered by photochemistry (Venot et al. 2015).
- For planets with equilibrium temperatures between ~750 and 1500K, the temperatures on the day side become sufficiently hot to gradually restore the chemical equilibrium there, although vertical quenching is still significant (Shulyak et al. 2020). The very efficient day-night circulation, however, generates an important horizontal quenching, so that the composition of the limb and the night-side are completely determined by the conditions on the day-side (Agundez et al. 2014; Mendonça et al. 2018b; Drummond et al. 2020; Figure 2-10).
- For planets hotter than ~1500-2000K, the atmospheric temperatures are so hot that even the night-side is close to the chemical equilibrium (Parmentier et al. 2018b).

The transition temperatures mentioned above are indicative and need to be constrained by observations. This basic picture can be influenced by additional parameters, such as the dynamical regime, which is controlled by the rotation rate and the planet size, and the elemental composition. It has been recently shown, for example, that the transition from equilibrium to disequilibrium may occur at different temperatures in a carbon enriched atmosphere (Ariel Chemistry WG report 2020). Given these premises, one could ask: is the departure from the chemical equilibrium a hindrance to the determination of the bulk elemental abundance of a planet? The answer is no: while the condensation of some key molecules deep down in the atmospheres of the Solar System planets makes any inference on their deep composition model dependent, this will not be the case for the Ariel targets. As long as the most abundant molecules carrying a given element – for example, $H_2O$, CO, $CO_2$ for oxygen – can be constrained, the deep elemental abundances can be determined.

The broad spectral coverage of Ariel will be an important asset to determine the deep elemental abundances of gas-rich exoplanets, as it will allow to detect directly the key molecular carriers of those elements, e.g. $H_2O$, CO, $CO_2$, $NH_3$, $CH_4$, HCN, $H_2S$, $C_2H_2$, $PH_3$.

Some elements are carried by undetectable molecular species, e.g. nitrogen in the form of $N_2$. To constrain these elements, the use of chemical models will be needed. Fortunately, thermochemistry at the warm to high temperatures, pressures and compositions relevant to Ariel is well-served by huge databases used in industry to model combustion in engines (Bounaceur et al. 2007; Wakelam et al. 2015; Venot et al. 2020a; see Section 7.1).

#### *2.2.2.3 How do Clouds and Hazes form?*

As mentioned before, a growing body of evidence suggests aerosols are present in the atmosphere of exoplanets (e.g. Sing et al. 2016, Parmentier et al. 2016). The nature of these aerosols is however unknown. Possible candidates are mineral condensate clouds (Morley et al. 2013; Powell et al. 2018) and hazes based on Sulphur or hydrocarbon precursors (Gao et al. 2017; Lavvas & Koskinen 2017; Hörst et al. 2018; Kawashima & Ikoma 2019).

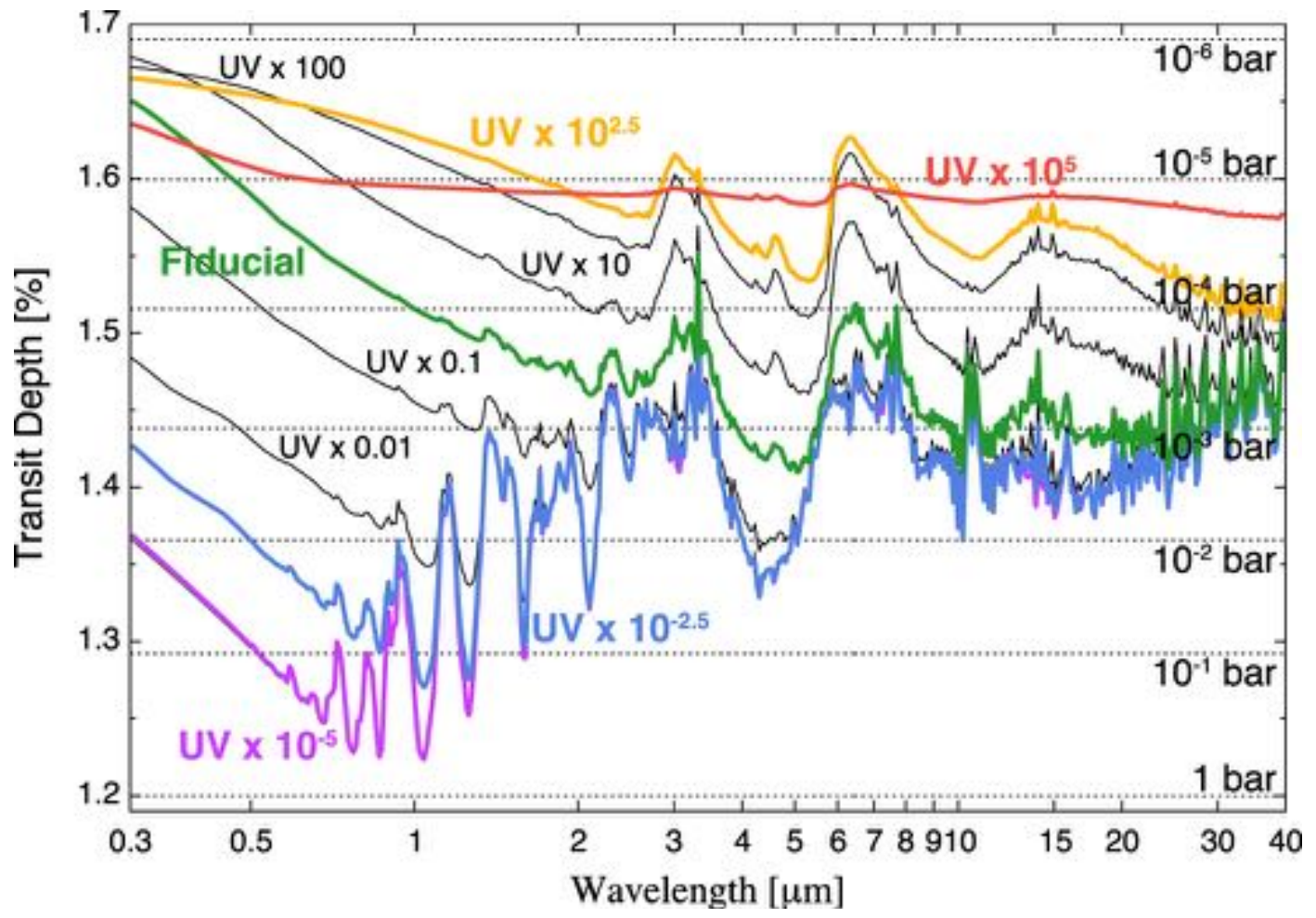

*Figure 2-11: Simulated transmission spectra of GJ 1214b-like planets with solar metallicity and hydrocarbon hazes triggered by different UV irradiation intensity. The spectra are dominated by aerosol properties in the optical and by molecular features in the IR. The concentration of the aerosols is proportional to the UV flux received by the planet. Figure from Kawashima & Ikoma (2019).*

By simultaneously covering 0.5-7.8 μm, Ariel can record the spectral energy distribution from the optical, where aerosols produce strong extinction, to the infrared, where spectral features are less obscured (see Figure 2-11). This behaviour will allow us to constrain the microphysical properties of aerosols, such as their particle sizes, distribution and optical thicknesses for individual planets. In some cases, the spectral features of some condensate species will allow us to directly infer the composition of the particles. As this information will be available for many planets, we will be able to understand what are the main mechanisms for aerosol formation as a function of the planetary parameters.

Condensate clouds and hazes are expected to follow very different trends: while the formation of mineral clouds is expected to be primarily determined by the temperature of the atmosphere and its composition (see Figure 2-11 or Parmentier et al. (2016) for details), haze formation is expected to be strongly linked to the level of UV radiation received from the star (Kawashima & Ikoma 2019).

#### *2.2.2.4 Non-Local Thermodynamic Equilibrium emissions*

Non local thermodynamic equilibrium processes (non-LTE) play a crucial role in the energy budget of low density upper atmospheric layers (Lopez-Puertas & Taylor 2001). They have to be considered when interpreting strongly irradiating exoplanets. $CO_2$ and CO non-LTE fluorescences are common in telluric atmospheres, and CO emission has been also detected in Neptune (Fletcher et al. 2010). Non-LTE $CH_4$ emission could explain the unexpected observed features around 3.3 μm in the atmosphere of HD 189733b (Swain et al. 2010; Waldman et al. 2012).

More recently, non-LTE effects have opened a new window into the physics of the upper atmosphere of strongly irradiated gaseous planets through the absorption of a metastable state of Helium. The detection of extended atmospheres thanks to non-LTE He lines in several planets (Spake et al. 2018; Allart et al. 2018; Lampon et al. 2020) has enable the study of interesting regimes of atmospheric escape. While He absorption is not directly accessible to Ariel, non-LTE emissions from $CO_2$, CO, $H_2O$, and $CH_4$, which are expected to peak at 2-5 μm, are detectable with Ariel.

#### *2.2.2.5 How does the high-energy environment created by the star affect the atmosphere?*

Planets orbiting stars of different stellar types and in various evolutionary stages do face very different conditions and are often embedded in an environment that is far from being as quiet as the present solar neighbourhood. Their host stars often expose them to extreme environments, which are predominantly shaped by the stellar magnetic activity level. The stellar activity level and the associated wind output, in turn, are directly related to stellar interior structure and rotational evolution. Surface magnetic activity signatures like flares, coronal mass ejections (CMEs), and ionized particle winds are driven by a convection-rotation dynamo. This dynamo also leads to magnetic field structures of changing complexity, once again linked to the stellar rotational evolution. Closely related to the underlying solar/stellar magnetic field is the output level of short-wavelength radiation in X-ray and extreme ultraviolet radiation (XUV). Models suggest that the interaction between the stellar and planetary magnetospheres may enhance the stellar activity (e.g. McIvor et al. 2006; Lanza 2008, 2013; Cohen et al. 2011a,b), even producing non-stationary phenomena.

Activity and wind outputs are crucial for the evolution of planetary atmospheres since they drive thermal and non-thermal atmospheric mass-loss processes (e.g. Bourrier et al. 2013; Koskinen et al. 2013a,b; Modirrousta-Galian et al. 2020), and upper-atmosphere heating, ionization, and chemistry (Roble et al. 1987; Chassefière 1996).

Phenomena related to stellar activity all evolve significantly over the star's lifetime. The detailed early history of a given star depends on its disc lifetime. If the star loses its disc quickly, it begins to rapidly spin up due to pre-main sequence contraction. All pre-main sequence stars are very energetic XUV sources emitting at some maximum (saturated) level. If the disc disappears early and planets are therefore exposed early to this radiation, their primordial atmospheres are more affected by evaporation. In extreme cases, close-in planets may evaporate completely or lose a substantial fraction of their mass, as supported by the recent observation of a possible remnant core of a giant planet (Armstrong et al. 2020) .

Consequently, very young stars span a wide range of rotational velocities. Once they reach the main sequence, the stellar rotation slows down gradually due to magnetic braking induced by the wind channelled by the magnetic fields, so the star becomes less active (Kraft 1967). The braking is more effective for rapidly rotating stars, therefore their rotation periods eventually converge and, after a few hundred Myr, stars exhibit age-dependent increases in rotation periods, which depend on their spectral type rather than on their initial rotation velocities. The early rotational evolution paths therefore depend sensitively on the initial conditions (Gallet & Bouvier 2013; Matt et al. 2015, Johnstone et al. 2015c) and the structure of the magnetic fields (Réville et al. 2015), while the late evolution depends only on spectral type.

It is hardly possible and challenging to observe stellar winds and measure the related mass-loss rates. For a limited number of Sun-like stars, the interaction of the stellar wind with the astrosphere has been observed in Lyα profiles providing possible constraints on their mass loss rates (Wood et al. 2004). Radio observations (Fichtinger et al. 2017) also provide valuable upper limits for a few Sun-like stars. However, the most unambiguous and well-constrained observational signature of stellar winds is probably the stellar spin-down on the main sequence. Many stellar wind models have been developed and applied, including theoretical models that consider energy balance in the transition region (Cranmer & Saar 2011), hydrodynamic models of the solar wind scaled to the winds of other stars (Johnstone et al. 2015a,b), and 3D MHD models based on observed magnetic field geometries (Vidotto et al. 2015). Recently, more physically based MHD models have been applied that involve heating and accelerating the wind using Alfvén waves (van der Holst et al. 2014; Airapetian & Usmanov 2016; Cohen et al. 2017; Boro Saikia et al. 2020), though there are still uncertainties and free parameters in these models.

Energetic radiation and particles emitted by the star produce photochemical and charge exchange effects in the upper layers of the planetary atmospheres. The affected atmospheric layers depend on the detailed energy of the radiation and particles, and on the atmospheric composition (Locci et al. 2019). *Ariel data will allow us to test the predictions of these interactions obtained through models and laboratory experiments* (Ariel Chemistry WG report 2020; Bourgalais et al. 2020). *The Ariel large-scale survey, will further observe how these interactions change along the stellar evolutionary path or around different stellar types.* The effects will be variable on several time scales, from the very long stellar evolutionary times to the very short time scales of flares (Venot et al. 2016). Interestingly, for the short time scales, *Ariel's ability to measure simultaneously the stellar activity in the visible and molecular features in the infrared will permit to correlate the level of activity to its atmospheric response*. Modelling and laboratory experiments suggest that Ariel has strong potential in detecting the tracers of this interaction in planets around the most active stars (Bourgalais et al. 2020).

Planetary atmospheres as a whole can be strongly affected by XUV radiation due to evaporation (as already mentioned for primordial H/He dominated atmospheres), and the loss rate depends on atmospheric composition and density. High levels of XUV irradiation may lead to rapid thermal (Tian et al. 2008) and, combined with a strong stellar wind, non-thermal (Lichtenegger et al. 2010) escape of nitrogen-dominated atmospheres. These effects may be counteracted by the presence of high levels of cooling $CO_2$ like in present-day Venus (Johnstone et al. 2019). Similarly, water-rich atmospheres can be photo-lysed by XUV and can disappear in a relatively short time (Johnstone 2020). By combining the knowledge of the present atmospheric composition and the star's activity evolutionary path we can constrain the initial atmospheric composition and its atmospheric evolution (Johnstone 2020).

### 2.2.3 How can Ariel help to overcome degeneracies in the study of the exoplanet interiors?

While the knowledge of planetary masses and radii helps to constrain the planetary interior composition, important degeneracies are left which may only be resolved by adding independent constraints. In the case of Solar System planets, gravitational moments offer an important insight into the planetary interior structure, but these cannot be measured for exoplanets. Even if the measurements of the mass and radius were obtained at any arbitrary precision – a statement that could almost be considered true in the Solar System – large degeneracies would still remain in determining the actual bulk composition of the interior (Figure 2-12).

The composition and the thermal structure of the atmosphere, as determined through Ariel observations, will mitigate or remove some of the key model assumptions that are necessary today to interpret the bulk composition of a planet. This additional information will result in a more accurate determination of the exoplanets' internal structure and composition for individual exoplanets. Thanks to the improved constraints on the interior of many individual planets, Ariel will address several new questions about the interiors of exoplanets, as a population.

See Table 2-1 below and also Table 1 in Helled et al. (2020).

*Table 2-1: Limitations on deducing the bulk composition of exoplanets and how Ariel will mitigate them.*

| Degeneracies in the study of exoplanet interiors with only density constraints | Ariel's contribution to addressing the degeneracies |
|---|---|
| Many different combinations of materials can yield the same mean planetary density (Baraffe et al. 2008; Valencia et al. 2013). | The atmospheric composition measured with Ariel constrains the elemental abundance in the deep gaseous envelope. Although this composition could be different from that of the core, if the core exists, this information reduces the number of plausible options to consider in modelling the interior. |
| The partitioning of heavy elements between the gaseous envelope and a potential core is unknown. The heavy elements are often assumed to be confined into a well-defined core with a given composition (e.g. pure water, silicate, or iron), which biases the inferred bulk enrichment. | The atmospheric enrichment observable with Ariel provides a lower limit for the enrichment of the deep gaseous envelope. This important information can be used in models of the interior. |
| For any planet with a significant atmosphere (>~0.01 per cent in weight) the temperature of the atmosphere and of the gaseous envelope strongly affect the planetary density (Leconte & Chabrier 2012; Turbet et al. 2019). For the planets in the Solar System, the problem of the unknown internal temperature is mitigated by the knowledge of the atmospheric temperature, which serves as an upper boundary condition that can be integrated downward to yield the whole thermal profile. For exoplanets, these temperatures are currently estimated through models of the thermal evolution: said models have to assume a plausible atmospheric composition to quantify the opacities. Also, all the processes heating the interior are assumed to be known in these models, an assumption that we already know being false: a glaring example is the issue of the large unexplained radii measured for many transiting hot-Jupiters (the so-called radius anomaly, see e.g. Guillot 2008). | Ariel spectroscopic measurements will constrain the atmospheric temperature profile for exoplanets down to the bar level. By providing the albedo, the thermal structure and the atmospheric composition, Ariel observations will significantly improve current estimates of the rate at which the interior can cool down and contract (see Figure 2-12). |
| At present, any composition inference based on the sole knowledge of the mass and radius relies on strong additional assumptions needed to break these degeneracies. | As it will constrain both the composition and the thermal structure of the atmosphere, Ariel observations will mitigate or remove some of the key assumptions that are necessary today to determine the bulk composition of a planet. |

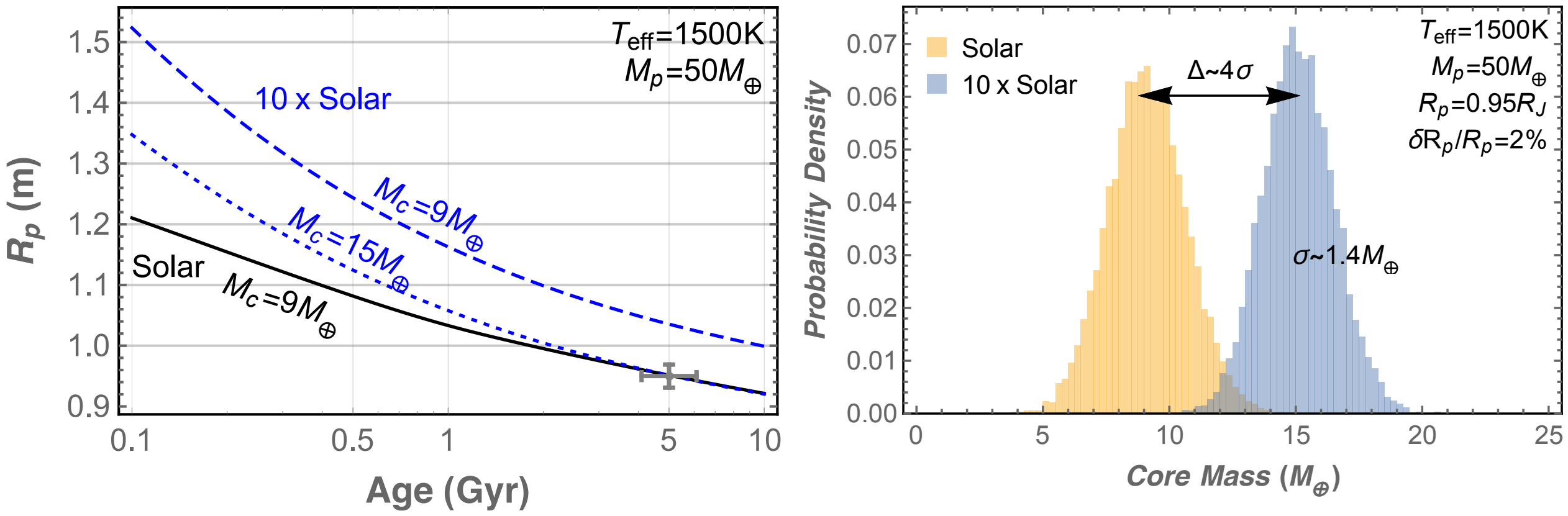

*Figure 2-12: Benefits of knowing the atmospheric composition to estimate the core mass of a gaseous planet. Left: Radius evolution tracks for a half Saturn mass planet with two different atmospheric compositions (black: Solar; Blue: 10 times solar) illustrating the core inference process (models from Leconte & Chabrier, 2013). For the solar composition, a nine Earth-mass core is sufficient to explain the observed radius (grey cross). A super-solar atmosphere, being more opaque, slows down the cooling, hence the contraction, of the planet. A larger — here 15 Earth-mass — core is thus needed to explain the measured radius. Not knowing the composition can lead to a 70% bias on the core mass inferred. Right: the calculations are repeated many times, randomizing the measured radius (Thorngren et al. 2016). The radius uncertainty was assumed to be as small as 2%, in line with the precisions expected from future missions (Rauer & Heras 2018). For each atmospheric composition, a histogram shows the probability for the core to have a given mass: a 2% radial uncertainty entails 1 sigma = 1.4 Earth mass uncertainty for the core, but this is not accurate. The difference (or bias) on the average core mass inferred in the two atmospheric scenarios is four times larger. By constraining the atmospheric composition, Ariel will guarantee more accurate core mass predictions.*

#### 2.2.3.1 How well are heavy elements distributed inside giant exoplanets?

Based on Solar System studies, we currently believe that the smaller the planets, the more enriched in heavy elements they are. This hypothesis has recently been partly verified by Thorngren et al. 2016, although their study is limited to planets colder than ~1000K to avoid the bias of the inflated radii measured for the heavily irradiated planets (e.g. Guillot 2008). An outstanding question is whether these heavy elements are sequestered into a dense core or mixed in the gaseous envelope (Leconte & Chabrier 2012; Vazan et al. 2015; Debras & Chabrier 2019). As outlined in Section 2.2.1, it is difficult to answer this question in the Solar System because the atmospheric composition is not representative of that of the envelope and gravitational moments are not sufficient to lift the degeneracy (Leconte & Chabrier 2012). In addition, all current loose constraints on atmospheric metallicity are limited to hot planets (>1000 K), which are easier to observe (e.g. Sing et al. 2016). As a result, it is currently impossible to compare the atmospheric and bulk metallicity for any given planet.

Ariel will target many planets in the 500-1000K regime, hot enough to avoid water condensation and cold enough to avoid the bias of the inflated radii. This sub-sample will enable a direct comparison of the atmospheric and bulk metallicity. Ariel measurements of the atmospheric temperature and metallicity will also help to extrapolate more accurately the bulk enrichment.

Determining the interior and atmospheric enrichment for a large number of planets is key to understanding interior mixing processes (Thorngren & Fortney 2019). As described in Section 2.2.4, understanding how bulk metallicity correlates with various parameters, in particular the planet's mass, will provide essential constraints to formation theories.

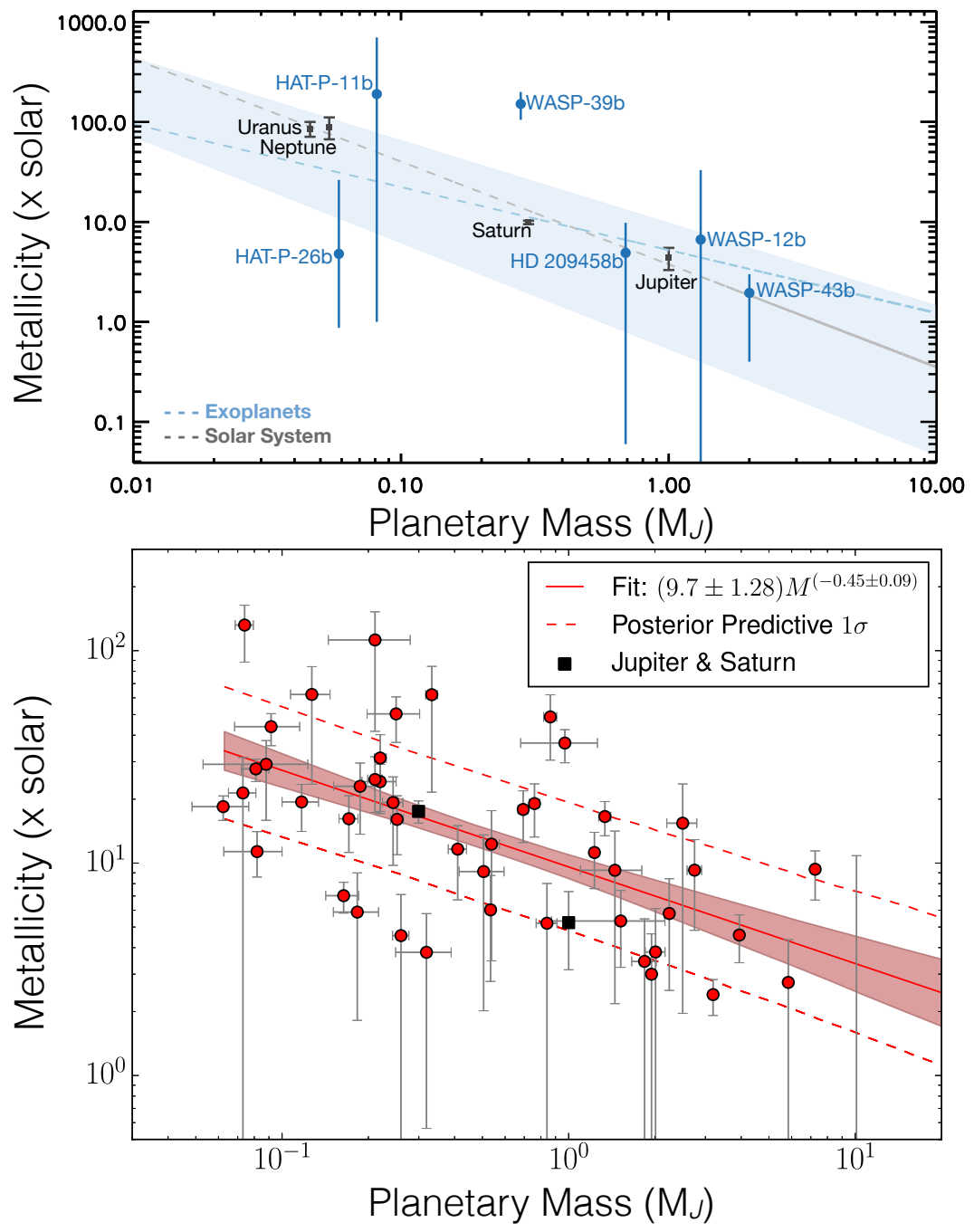

*Figure 2-13: Current constraints on the planetary mass vs. planetary metallicity. Solar System data are indicated in black. Top: atmospheric metallicity of individual exoplanets inferred from transit spectroscopy (blue dots). Large uncertainties are still present for exoplanets. Figure adapted from Sing et al. (2018). Bottom: bulk metallicity inferred from mass radius determination and interior modelling of individual planets (red dots). Figure adapted from Thorngren et al. (2016).*

*Ariel will provide more accurate and precise measurements of the atmospheric metallicity for a large number of planets in the 500-1000 K regime, for which also bulk metallicities can be inferred as discussed in 2.2.3.1. Having both atmospheric and bulk metallicities may provide unprecedented constraints on atmosphere/interior mixing processes.*

### 2.2.3.2 *What is the true nature of transitional planets?*

The important constraints that Ariel will provide to the study of the internal structure of gas-rich exoplanets (see Table 2-1) will also be applicable to the transitional planets (see Figure 2-1). It is worth emphasizing that for these planets, Ariel will enable a transformational shift in our understanding of their nature.

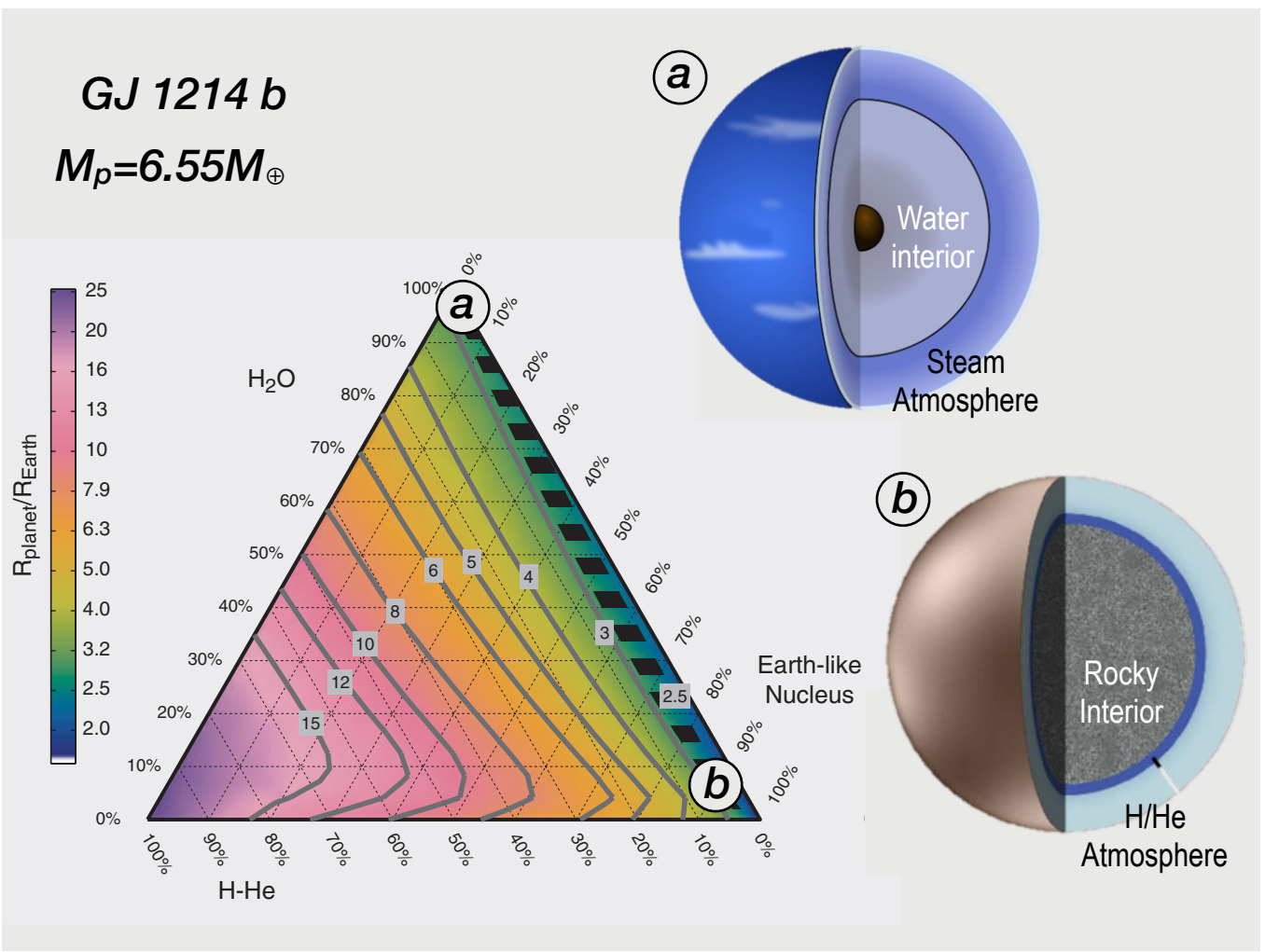

*Figure 2-14: Degeneracy in the internal composition of a planet when only the mean density is known. The ternary diagram relates the composition in terms of Earth-like nucleus fraction, water+ices fraction, and H/He fraction to the total mass and to the radius (colour coded). The models have assumed the mass of GJ 1214b. Each vertex corresponds to 100%, and the opposite side to 0% of a particular component. Grey lines: curves where the radius is constant. In our example, the available mass-radius data constrain GJ 1214b to the black dashed band on the right side of the ternary diagram. An almost pure water composition (case a) and a 90% rocky core with a 10% envelope of mixed water and H/He (case b) are both consistent with the present data. Only a further characterization of the gaseous envelope can remove this degeneracy. Adapted from Valencia et al. 2013.*

As noted very early (Adams et al. 2008; Valencia et al. 2013), for transitional planets, very different bulk compositions yield the same bulk density and thus cannot be discriminated at the moment, as illustrated by Figure 2-14. The presence of a massive hydrogen envelope today can only be inferred for the largest planets, and only in a statistical sense (Rogers 2015). These studies conclude that a robust determination of the composition of the upper atmosphere of these planets will be key to infer their bulk composition. The recent detection of water vapour and hydrogen in the atmosphere of the transitional planet K2-18b with the HST/WFC3 has demonstrated the potential of such observations, even though the poor spectral coverage of WFC3 could provide only limited constraints (Tsiaras et al. 2019; Benneke et al. 2019). We explain in the next section how the atmospheric composition of a large number of transitional planets is essential to understand how these planets formed.

## 2.2.4 How do planets and planetary systems form and evolve?

### *2.2.4.1 How can we track down the dynamical history and formation environment of giant exoplanets ?*

Giant planets are the class of planets for which the link between composition and formation process is expected to be the most direct, which is why a number of efforts are being devoted by the international exoplanetary community to understand this link and how to take advantage of it. Results obtained both within and outside the Ariel Mission Consortium confirm that this link can be optimally exploited thanks to the long list of molecules detectable by Ariel.

The diverse and representative sample of planets observed by Ariel will allow to perform population studies: these studies will be key to explore at statistical level the connection between the formation regions and the migration histories of giant planets and their atmospheric composition. This knowledge will prove essential to refine our theoretical understanding of the formation and migration histories of individual planets.

The first and most obvious challenge concerns the information carried by the different elements and their main molecular carriers about how and where the planets formed. On the one hand, this challenge is being addressed by observational campaigns of protoplanetary discs, which are providing increasingly refined constraints on the composition of the birth environment of giant planets both at molecular (e.g. Favre et al. 2018; van 't Hoff et al. 2018; Lee et al. 2019; Bianchi et al. 2019; Drozdovskaya et al. 2019) and elemental (e.g. Dutrey et al. 2011; Favre et al. 2013; McClure et al. 2016; Kama et al. 2016a, 2016b, 2019; Trapman et al. 2017; Semenov et al. 2018; Cleeves et al. 2018; Zhang et al. 2020) levels, in particular high-Z elements like carbon (C), oxygen (O), nitrogen (N) and sulphur (S) (see the Ariel Planetary Formation WG report 2020 for a more detailed discussion). On the other hand, a growing number of exoplanetary studies have been exploring the link between the relative concentration of the two most abundant high-Z elements, C and O, and the planetary formation process (see e.g. Madhusudhan et al. 2016, and references therein for a recent overview).

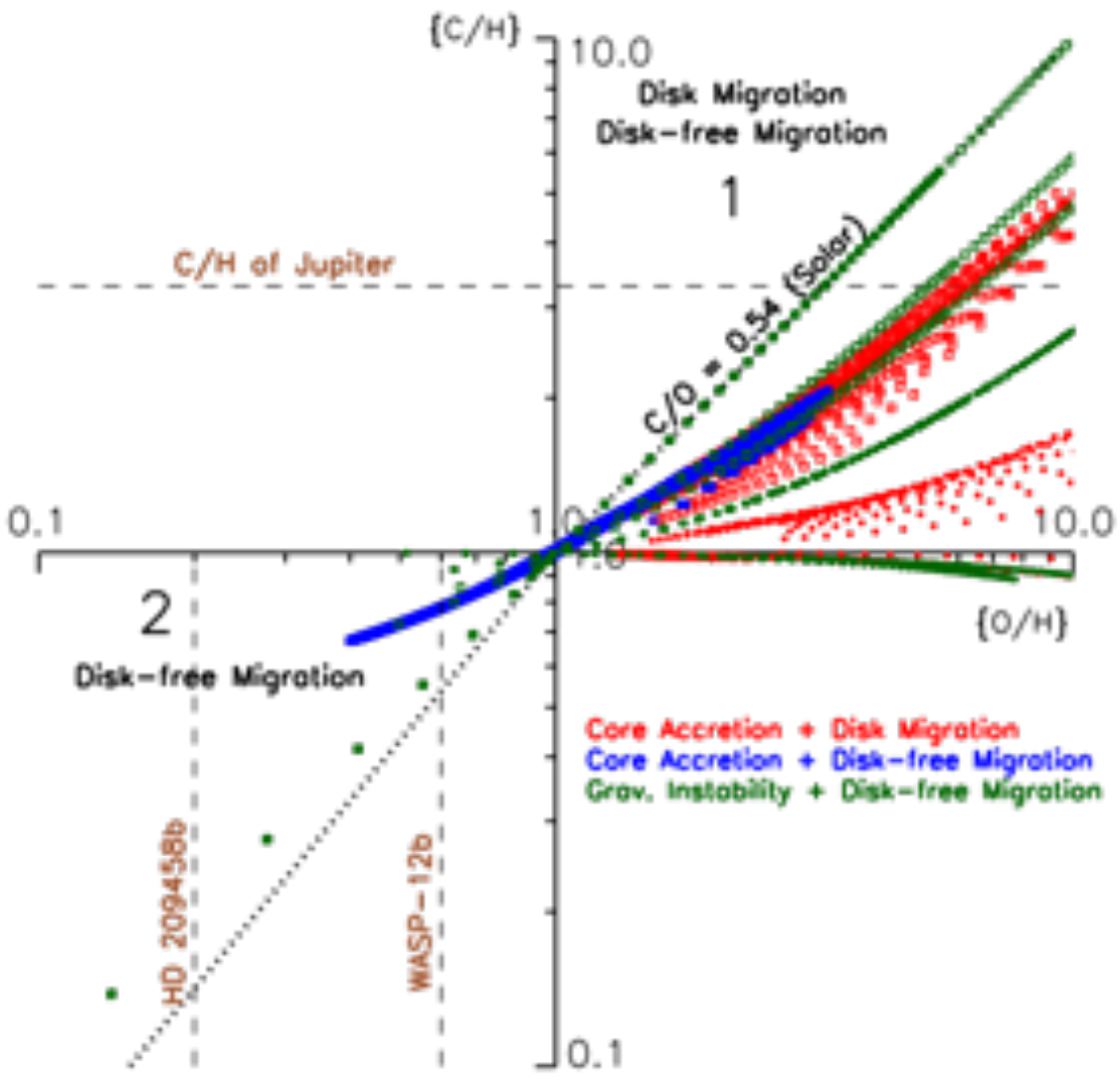

*Figure 2-15: Example of the predicted C and O abundances (in units of solar abundances) for giant planets forming at different locations in protoplanetary discs, through different formation mechanisms (core accretion or gravitational instability), and experiencing different types of migrations (disc-driven or disc-free, e.g. resulting from planet-planet scattering). The different formation environments and mechanisms produce distinct signatures in the C/O ratio of the giant planets, with high metallicity giant planets being associated with sub-stellar C/O ratios (top right quadrant) and low metallicity giant planets being associated with super-stellar C/O ratios (bottom left quadrant). Figure from Madhusudhan et al. (2016).*

The general expectation since the early results from Öberg et al. (2011) is for *low metallicity giant planets*, where the bulk of C and O are accreted from the gas, to be characterized by *super-stellar C/O ratios*. By contrast, *high-metallicity giant planets*, where C and O are dominated by the capture of solids, are expected to feature *sub-stellar C/O ratios*. Studies across both the exoplanetary and the astrochemical international communities confirm the possibility of using the C/O ratio as a powerful proxy to constrain the formation region of planets and their early dynamical evolution (see e.g. Figure 2-15 and Madhusudhan et al. 2016, and references therein for a recent overview; Mordasini et al. 2016 and Cridland et al. 2019 for an example of recent results from planetary synthesis models and Figure 2-16 for results on the link between planetary formation, migration and C/O ratio obtained within the Ariel Mission Consortium).

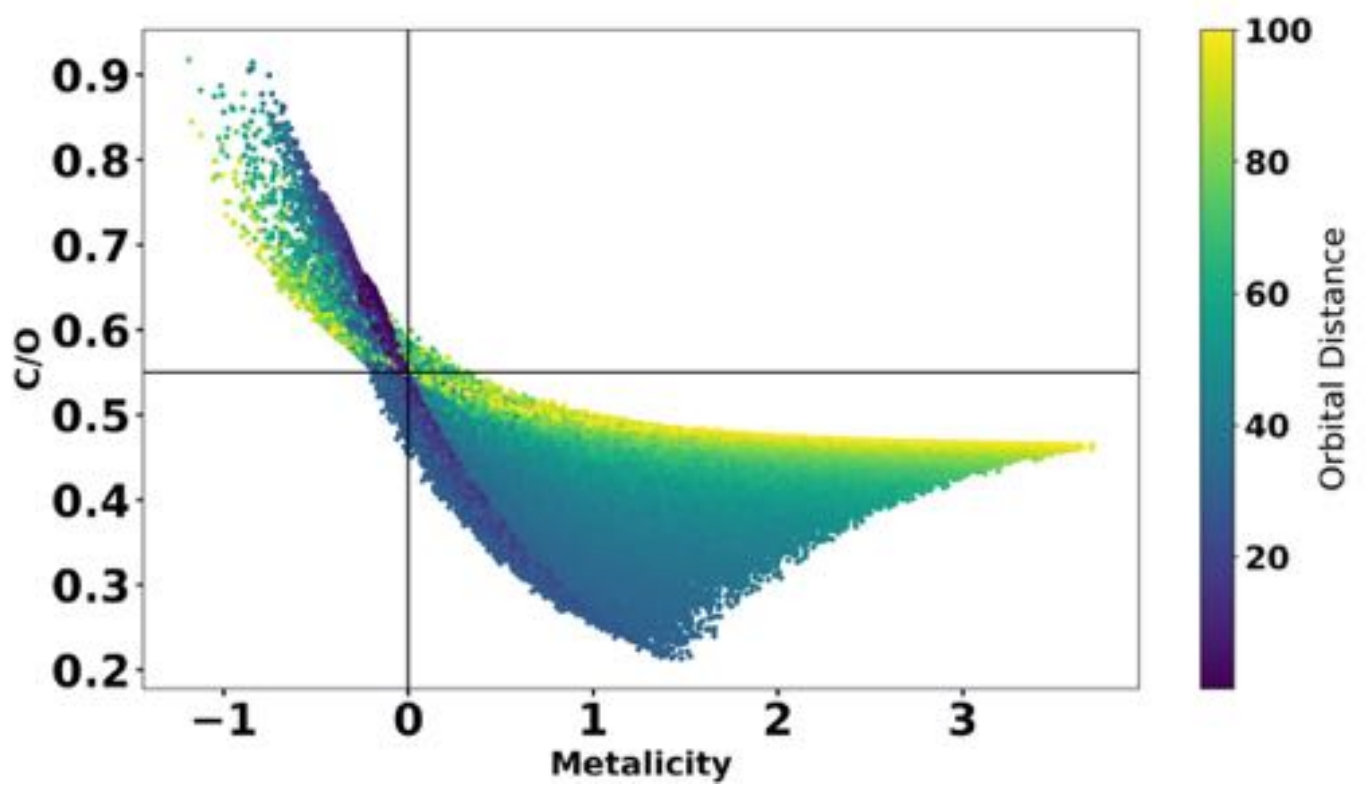

*Figure 2-16: Figure from Khorshid et al. (2020). Example of the relationship between planetary C/O ratio and metallicity of giant planets (specifically, Jovian analogues) in response to different protoplanetary disc environments and migration tracks. Each point represents a different choice for the disc parameters (e.g. planetesimal fraction, dust grain size distribution, and core mass at which the planet starts migrating). The colour scale indicates the location where the planet starts its journey, which ends at the same orbital distance for all planets.*

The main obstacles to a straightforward use of the C/O ratio are the still poorly understood details of the evolving chemical environment of protoplanetary discs in which the giant planets form. Examples include the extension of the planet-forming region (see e.g. ALMA Partnership et al. 2015; Isella et al. 2016; Fedele et al. 2017, 2018; Long et al. 2018, and Andrews et al. 2018 for an example of the wide range of orbital distances at which giant planets are observed in discs) and the partitioning of C and O across the different solid phases (rocks, organics, and ices) and molecules (e.g. $H_2O$, $CO_2$, CO, $CH_4$, $CH_3OH$, PAH) in protoplanetary discs (see e.g. Palme et al. 2014; Thiabaud et al. 2014; Marboeuf et al. 2014a,b; Mordasini et al. 2016; Eistrup et al. 2016, 2018; Cridland et al. 2019; Bianchi et al. 2019; Drozdovskaya et al. 2019; Doyle et al. 2019 for a perspective of this rapidly evolving field). These uncertainties can result in degeneracy in the interpretation of the C/O ratio if this parameter is used alone.

The broad coverage of Ariel in terms of molecules offers a straightforward way out of the possible limitations in the diagnostic use of the C/O, i.e. the use of *multiple elemental ratios* (e.g. N/O, C/N, Turrini et al. 2018).

An illustrative example is supplied by Figure 2-12, where we show the results obtained in the simulations performed within the Ariel Mission Consortium (Ariel Planetary Formation WG report 2020; Turrini et al. 2020a) when considering also nitrogen as a tracing element. The simulations focus on the formation and migration of a Jupiter-like planet in a protoplanetary disc surrounding a Sun-like star. Different starting positions are assumed for the seed of the giant planets with respect to the main snow lines. Also, wide orbital regions are sampled, in particular where giant planets have been suggested by ALMA observations.

The joint use of C, N and O allows us to compute two additional elemental ratios: C/N and N/O. As shown in Figure 2-17, the N/O ratio grows with migration for low metallicity giant planets and decreases for high metallicity ones: this happens because the bulk of O is trapped into solid (rocks and water ice) already after the first few au from the host star while the bulk of N remains in gas form until the outermost regions of protoplanetary discs. The C/N behaves in the opposite way for the same reasons. Sensitivity tests performed with different retrieval tools produced within and outside the Ariel's Consortium confirm Ariel's capability to observe the main carriers of C, O and N in exoplanetary atmospheres in the abundance ranges indicated by the planetary formation models (Ariel Spectral Retrieval WG report 2020).

The slopes of the curves shown in Figure 2-17 might change depending on the specific characteristics of the protoplanetary discs in which different giant planets formed: e.g. the different partitioning of N between the two main molecular carriers $NH_3$ and $N_2$. Yet the behaviour of the two elemental ratios is fundamentally determined by the solidly established volatility of N. As a result, the farther the giant planets start their migration from the host star, the more their C/N and N/O ratios will diverge from the stellar values. Furthermore, due to the different behaviours of C, O and N, solar values for the three ratios (C/O, N/O, C/N) can only be obtained in giant planets if their metallicity originates from a solar composition source. This only happens if the three elements are all in volatile phase (i.e. a purely solar gas) or if they are all in condensed phase (i.e. trapped in solids beyond the $N_2$ snow line). Similar considerations on the volatility of N and the meaning of a solar C/N ratio applied to the study of Jupiter's formation in the Solar System (Öberg & Wordsworth 2019; Bosman et al. 2019) have recently led to the suggestion that of Jupiter could have formed near or beyond the $N_2$ snow line, a scenario further supported from a dynamical point of view also by recent studies of the origins of Jupiter's Trojan asteroids (Pirani et al. 2019). The second main challenge to understand the link between planetary formation and composition is represented by the interpretation of the compositional data in the context of the evolution of the planet and of its host system. Also from this point of view, studies performed both within and outside the Ariel Mission Consortium are already supplying the tools and frameworks to address successfully this future challenge. One example is provided by the relationship between the displacement of giant planets due to disc-driven orbital planetary migration and the planetary metallicity.

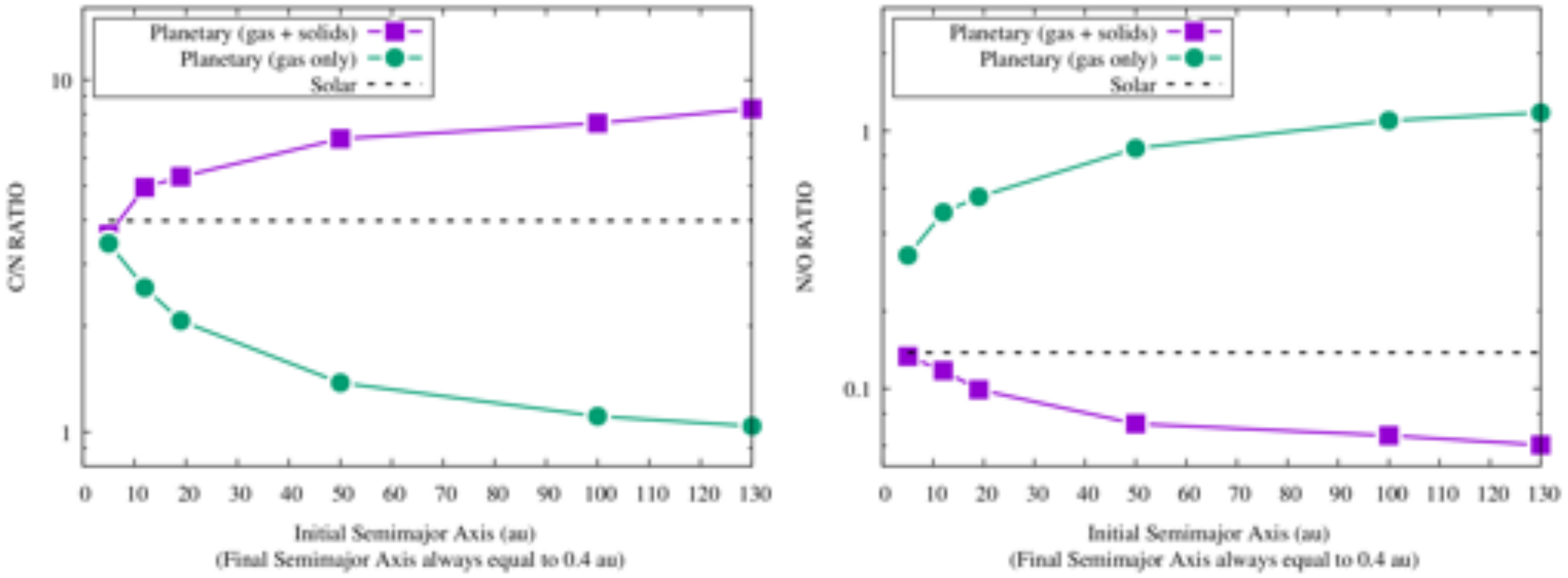

*Figure 2-17: C/N and N/O ratios for giant planets starting their formation process at different distances from a Sun-like star and migrating to 0.4 AU in 3 Myr in the simulations performed by the Ariel Mission Consortium (Ariel Planetary Formation WG report 2020; Turrini et al. 2020c). The green line and symbols indicate the behaviour of the two elemental ratios for giant planets accreting only gas from the protoplanetary disc (low metallicity case) while the purple ones indicate the behaviour of the two ratios for giant planets whose metallicity is dominated by the accretion of solids (high metallicity case). The filled squares and circles mark the initial positions of the giant planets in both cases (i.e. 5, 12, 19, 50, 100 and 130 au). The horizontal dashed lines indicate the stellar elemental ratios, assuming a solar composition for the host star and its protoplanetary disc. Figure from Turrini et al. (2020c).*

Shibata et al. (2020) showed that migrating giant planets capture planetesimals with total masses of several tens of Earth masses only if the planets start their formation at tens of AU in compact and relatively massive discs. A similar result has been obtained in a parallel study within the Ariel Mission Consortium (Ariel Planetary Formation WG report 2020; Turrini et al. 2020c) focusing on extended but not exceedingly massive discs if the giant planets have formation regions comparable to those recently observed by ALMA (from a few tens to about a hundred au, e.g. Andrews et al. 2018).

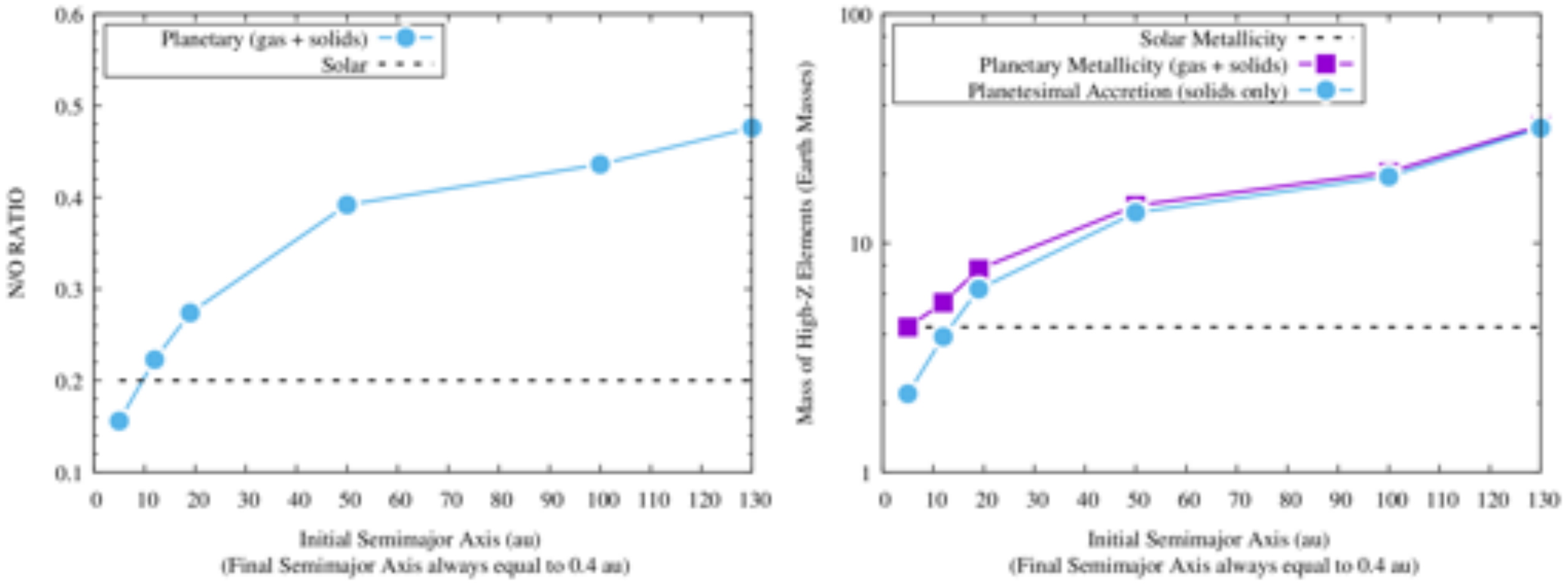

*Figure 2-18: S/N ratio (left) and heavy-elements enrichment (right) for giant planets starting their formation process at different distances from their host, Sun-like star and migrating to 0.4 au in 3 Myr in the simulations performed by the Ariel Mission Consortium (Ariel Planetary Formation WG report 2020; Turrini et al. 2020c). The purple and blue curves in the right-hand plot show the mass of high-Z material contributing to the metallicity of the giant planet as supplied by both gas and solids and by solids only, respectively. The filled squares and circles mark the initial positions of the giant planets in both cases (i.e. 5, 12, 19, 50, 100 and 130 au). The horizontal dashed lines indicated the solar S/N ratio and the enrichment associated with a solar metallicity. Figure from Turrini et al. (2020c).*

Both the above studies showed that the total captured planetesimal mass increases with increasing migration distances. In particular they both showed that, as mean motion resonances trapping and aerodynamic gas drag put a cap to the planetesimal capture efficiency of migrating planets, *large-scale disc-driven migration* is required to explain the enrichment of giant planets with several tens Earth masses of heavy elements as reported by exoplanetary population studies (Thorngren et al. 2016; Wakefield et al. 2017) and the giant planets in the Solar System (e.g. Atreya et al. 2018 and references therein). As a result, estimates of the metallicity/heavy-elements enrichment of giant planets would allow to constrain the dynamical framework of their formation process. While metallicity can be indirectly estimated through the mass-radius relationship of giant planets, the broad spectral and molecular coverage of Ariel provides an additional way to probe this quantity by constraining the sulphur over nitrogen (S/N) ratio. Given that the bulk of S is efficiently trapped into refractory solids (e.g. Lodders 2010; Kama et al. 2019) while the bulk of N remains in gas phase as $N_2$ for most of the extension of discs (e.g. Pollack et al. 1994; Eistrup et al. 2016; Öberg & Wordsworth 2019; Bosman et al. 2019), this ratio is correlated – even if not linearly proportional – to planetary metallicity as shown in Figure 2-18 (Ariel Planetary Formation WG report 2020).

As mentioned previously, disc-driven migration is not the only dynamical process capable of delivering giant planets from their formation regions to the orbital distances where Ariel will observe them. As supported by recent population studies of multi-planet exoplanetary systems, *chaos and planet-planet scattering* can cause episodes of late migration (see Figure 2-19) and also play an important role in shaping the architecture of known exoplanets and migrating giant planets to the Ariel's observational region (Limbach & Turner 2015; Zinzi & Turrini 2017; Laskar & Petit 2017; Turrini et al. 2020a). Ariel's compositional data will help to address this degeneracy in the dynamical history of individual planets (see also Turrini et al. 2018). However, as part of the efforts of the Ariel Mission Consortium, it was shown that the use of metrics linked to the angular momentum deficit of planetary systems (Turrini et al. 2020b) allows to extract information from the architectures of the systems themselves and to constrain the intensity and timing of the possible chaotic phases. The combination of Ariel's compositional data on individual giant planets with the information provided by the architecture of their host systems will provide unprecedented insight into the history of both.

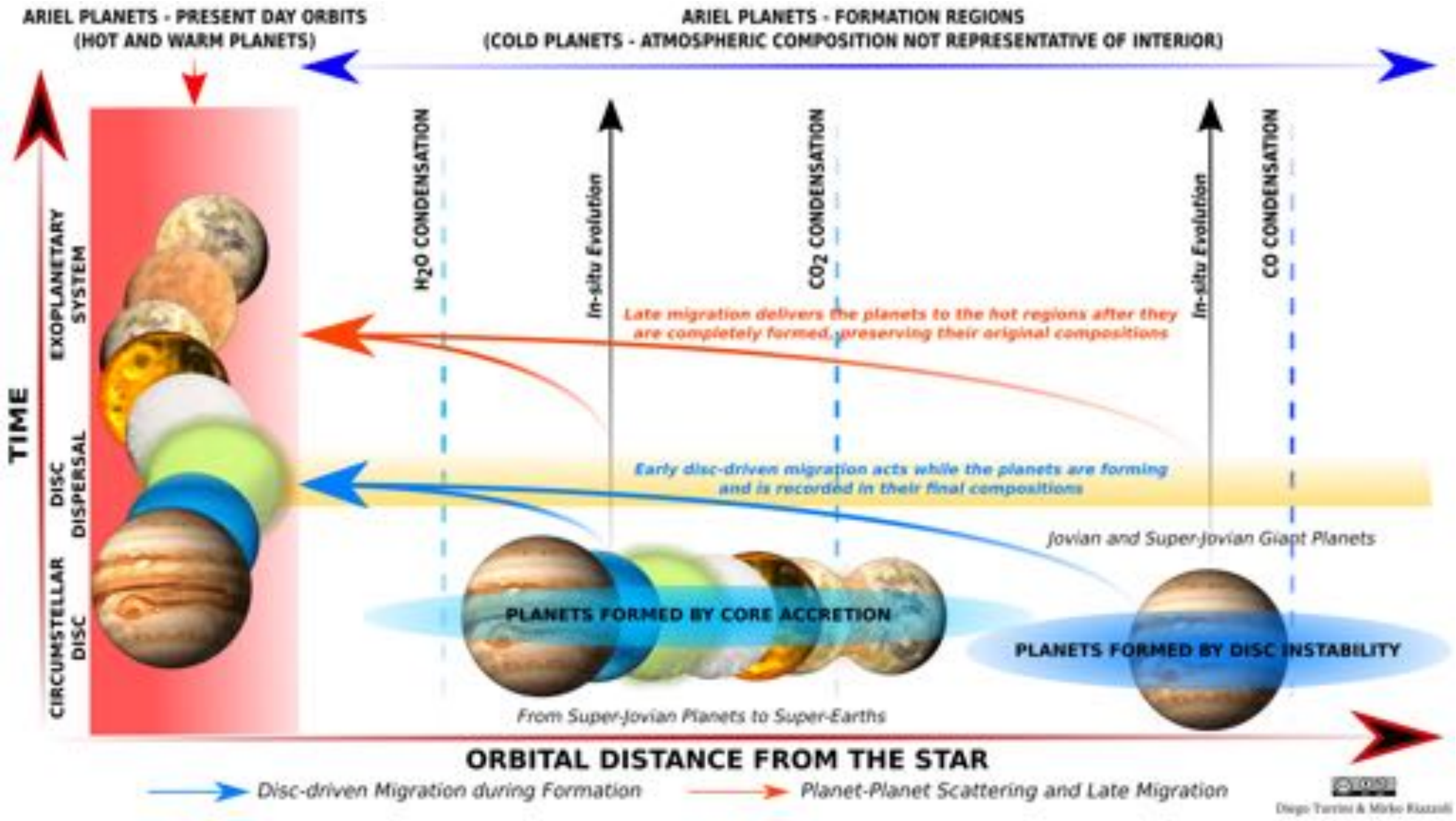

*Figure 2-19: Schematic representation of the dynamical pathways that can deliver planets from their formation regions to orbital regions where Ariel will observe them. While embedded in the native protoplanetary discs, planets can experience early disc-driven migration due to their interactions with the surrounding gas. Dynamical chaos and planet-planet scattering events can also affect newly formed planetary systems and cause their planets to experience late migration. Early migration acts while planets are still forming, while late migration occurs after: as a result, the two mechanisms are associated with different compositional signatures of the planets. Adapted from Turrini et al. (2018).*

Finally, the link between the planetary formation process and the chemical environments of protoplanetary discs, which Ariel will explore by observing the atmospheric composition of giant exoplanets, indicates that Ariel's observations can also have major implications beyond the field of exoplanets. The chemical environment of protoplanetary discs, in fact, is influenced by a number of factors linked to the stellar (e.g. stellar mass and metallicity) and galactic (e.g. stellar birth environment, galactic chemical evolution) environments (see Ariel Planetary Formation WG report 2020 and references therein for an overview of these subjects). These factors might provide a source of uncertainty in the interpretation of Ariel's early data, and will be therefore taken into account during the selection of Ariel's observational sample. However, the same argument also means that Ariel's observations have the potential for impacting in a major way our understanding of the role played by these environmental factors in setting the stage for the planetary formation process. This is particularly true if we consider an extended mission and the even larger and more diverse observational sample it will bring. Consequently, Ariel is uniquely suited not only to explore in unprecedented details and from different angles the outcomes of the planetary formation process, but also their connection with the star formation process.

#### 2.2.4.2 *Is the divide in the population of transitional planets due to atmospheric evolution or a legacy from formation?*

The range of planetary sizes spanning the super-Earths and the sub-Neptunian population is the one we understand the least, as no example of such planets exists in the Solar System. We know that the transition between these two families of planets should occur somewhere between 1 and 3 Earth radii, but to date we still lack a clear understanding of how this transition occurs, e.g. whether it is continuous or a sharp and abrupt change. The Kepler mission revealed a gap in the population of sub-Neptune planets with orbital periods below 100 days (Owen & Wu 2013): there is a factor of two deficit in the occurrence rate of planets between 1.5 and 2 times the size of the Earth (Fulton et al. 2017). This seems to divide exoplanets in the transitional region in two populations (Figure 2-20): slightly larger, low density planets that are believed to have accreted and kept significant H/He in their atmospheres, and smaller, high density ones that did not.

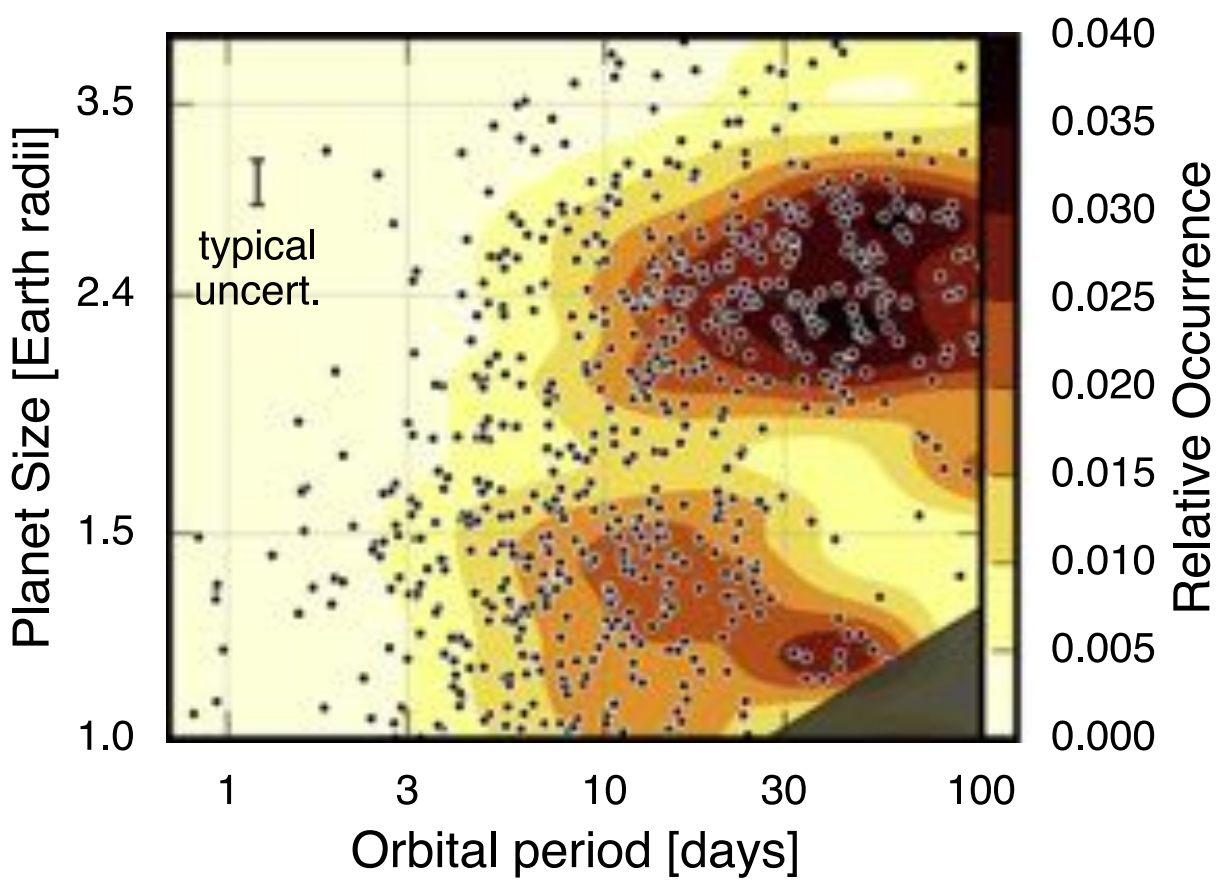

*Figure 2-20: Distribution of planets as a function of planet size and orbital period. Dots are observed planets. The colour shading is the relative occurrence rate of planets per star at a given location in this parameter space corrected for various observational biases. Two populations (islands with darker shading) are visible. For planets in the gap (lighter shading), membership of one or the other of these two populations cannot be determined without further atmospheric characterization. Figure from Fulton et al. (2018).*

However, the reasons for this divide remain uncertain. Photoevaporation of the atmosphere occurring after the dispersal of the protoplanetary nebula has been proposed to be the key mechanism (Owen & Wu 2017; Jin & Mordasini 2018), as this process is very efficient to strip an atmosphere from a planet below some threshold, thus creating a "photoevaporation valley" (Owen & Wu 2013; Lopez & Fortney 2013). However, other scenarios involving differences in the formation pathway have been proposed (Ginzburg et al. 2018; Gupta & Schlichting 2019; Kite et al. 2020). To make things more complex, recent observations by K2 have shown that the location and shape of the gap changes with the type of the host star, suggesting that all these processes could contribute to varying degrees in sculpting these populations (Fulton & Petigura 2018; Cloutier & Menou 2020).

In the "radius vs equilibrium temperature" parameter space, the limits of these two populations are blurred by the fact that both gaseous envelopes and cores may have very different enrichments and compositions from one planet to another. As also discussed previously, this makes planetary radius and density very unreliable probes of the nature of planets in this size range, prompting the need for additional constraints.

Figure 2-21 shows the predicted mass, radius, interior ice fraction and atmospheric fraction of transitional planets, such as super-Earths and sub-Neptunes around 0.3-$M_\odot$ stars, which were generated by the planetesimal-based population synthesis models (Tian & Ida 2015) accounting for the effects of water

production via atmosphere-magma interaction (Kimura & Ikoma 2020). The synthesized planets were sampled according to their transit probability, meaning that many of those planets are close to their host star. The colour-coding indicates the ice/water content of the planetary interior; specifically, red symbols are rocky planets formed in regions interior to the snowline, blue ones are ice-dominated planets that grew beyond the snowline, followed by orbital migration, and green ones are water-rich rocky planets coming from near the snowline. The symbol size indicates the mass fraction of the atmosphere, which also contains water. It is found that symbols with various colours and sizes are mixed in the radius range between 1.5 and 4 $R_\oplus$. This indicates that the synthesized planets are diverse in bulk composition and origin. Although not shown here, the fractions of rocky and icy planets depend on poorly-understood initial conditions and processes such as the size and distribution of building blocks, orbital migration, etc. Thus, the bulk composition of transitional planets places a crucial constraint on their origin.

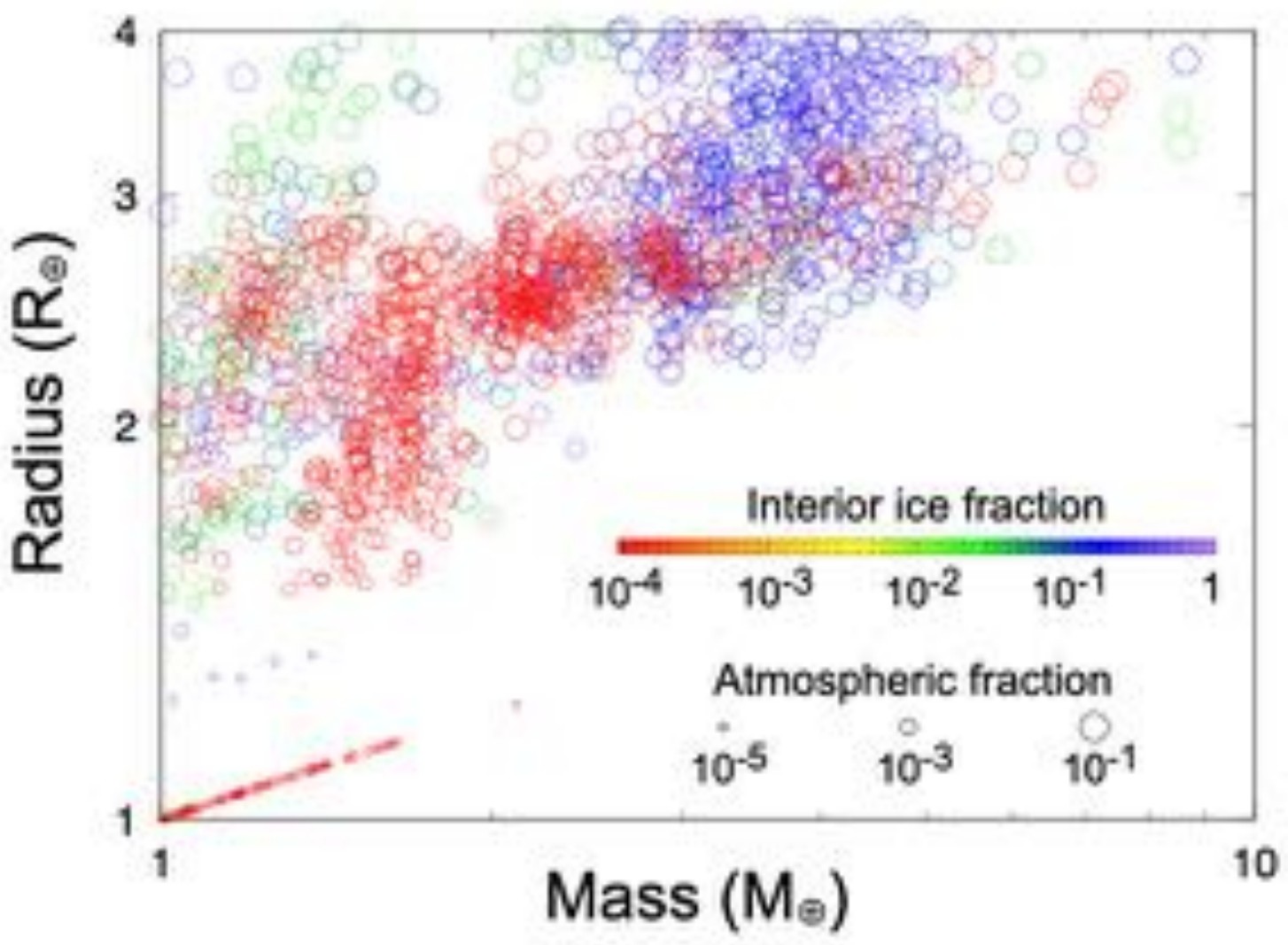

*Figure 2-21: Mass vs. radius relationships for low-gravity planets orbiting 0.3 $M_\odot$ stars predicted by population synthesis models (Tian & Ida 2015) with the effects of water production in the primary atmosphere (Kimura & Ikoma 2020). The color-coding indicates the ice content of the planetary interior, which is related to the location at which the planet starts to grow. The symbol size indicates the atmospheric mass relative to the planetary total mass.*

The most reliable way to investigate transitional planets is provided by the characterization of their atmospheric compositions in term of the main molecular species. An alternative approach permitted by Ariel is that of constraining the mean atmospheric molecular weight of these planets, to promptly discriminate those retaining significant amount of primordial hydrogen and helium from those which do not (Tinetti et al. 2018; Turrini et al. 2018). This second approach provides less detailed information than the first, but offers the advantage of allowing for quickly surveying tens of planets in the transition size range (Turrini et al. 2018; Edwards et al. 2019). The estimation of the mean atmospheric molecular weight can be used to make a preliminary classification of target planets and identify the best candidates for in depth atmospheric characterization. Finally, the photochemical regime created by stellar radiation in the upper layers of exoplanetary atmospheres (see Sect. 2.2.2.4) has been recently proposed as a third window for discriminating primary from secondary atmospheres. By creating spectroscopically-detectable ionic species, such as $H_3^+$ (Bourgalais et al. 2020), the effects of stellar radiation can allow Ariel to directly detect and quantify the atmospheric hydrogen.

Ariel will be able to distinguish the nature of tens of transitional planets by estimating the mean atmospheric molecular weight and, for a representative sample of them, by performing an in-depth atmospheric characterization, also aimed at directly detecting the primordial hydrogen through observing features of H-rich ionic species.

As this approach will be applied to a significant number of objects under different conditions, we will gather a clearer view of how the two populations were shaped by competing physical phenomena that are dependent on different system parameters. Furthermore, thanks to the information on their atmospheric composition that Ariel will provide, we will be able to gather insight into the formation pathways of transitional planets, as different paths imply different elemental ratios. This will shed light on whether the divide among transitional planets revealed by the Kepler mission is mostly a reflection of different formation environments and timescales or if it is mainly dominated by evolutionary effects.

Another region of the parameter space believed to be sculpted mainly by planetary atmospheric photoevaporation and at reach by Ariel is the "sub-Jovian desert", which is a lack of Neptune-size planets subject to an incident stellar radiation higher than about 100 times that of the Earth (e.g. Davis & Wheatley 2009; Szabó & Kiss 2011; Mazeh et al. 2016). In a way, the sub-Jovian desert can be seen as an extension of the radius gap at even higher incident radiation. Ariel will be able to characterise the atmospheres of planets at the upper edge of the desert (i.e. hot-Jupiters), thus identifying whether there are trends in the atmospheric chemical abundances with respect to stellar irradiation and/or planetary mass/radius that may be connected to the effects of photoevaporation. Thanks to their larger size compared to transitional planets, Ariel will be also able to characterise in depth the atmospheres of the few planets known to lie inside the desert, thus providing key observational constraints to the physical processes responsible for its formation.

#### *2.2.4.3 The diversity and survival of secondary atmospheres*

Terrestrial planets' atmospheres form and evolve through a complex combination of processes that includes solid and gas accretion, orbital migration, collisions and giant impacts, mass loss and atmospheric erosion, geophysical evolution, and outgassing. The interplay between these processes and the environment in which they act can produce an extremely wide range of results that we are just starting to comprehend.

Among the processes listed above, orbital migration plays a pivotal role in increasing the diversity of terrestrial planets in the inner orbital regions that Ariel will observe, as it allows for the presence of volatile-rich planets close to the star where otherwise the local planetary building blocks would be volatile-poor due to the high temperatures. Despite the known limitations of density as a probe of composition (see Sect. 2.2.3), this diversity is already demonstrated by the range of planetary densities of close-in terrestrial planets, as illustrated by Figure 2-22. While terrestrial planets can form during the lifetime of protoplanetary discs and capture H-rich atmospheres from the surrounding environment, the same collisional processes responsible for their growth can strip significant fractions if not the majority of these early primary atmospheres (e.g. Biersteker & Schlichting 2019; Lammer et al. 2020 and references therein for a recent discussion), which will be replaced by or chemically integrated into secondary atmospheres produced by outgassing from their interiors (e.g. water production resulting from the interaction of captured nebular H and outgassed O, Kimura & Ikoma 2020) and/or delivered by impacts of smaller bodies in the final stages of the planetary formation process (Elkins-Tanton 2012; Ikoma et al. 2018). For these reasons, the atmospheric compositions and properties of terrestrial planets are very difficult to predict, though we can reliably expect them to reflect the large diversity of the planetary interiors (see Figure 2-12 and Section 2.2.3).

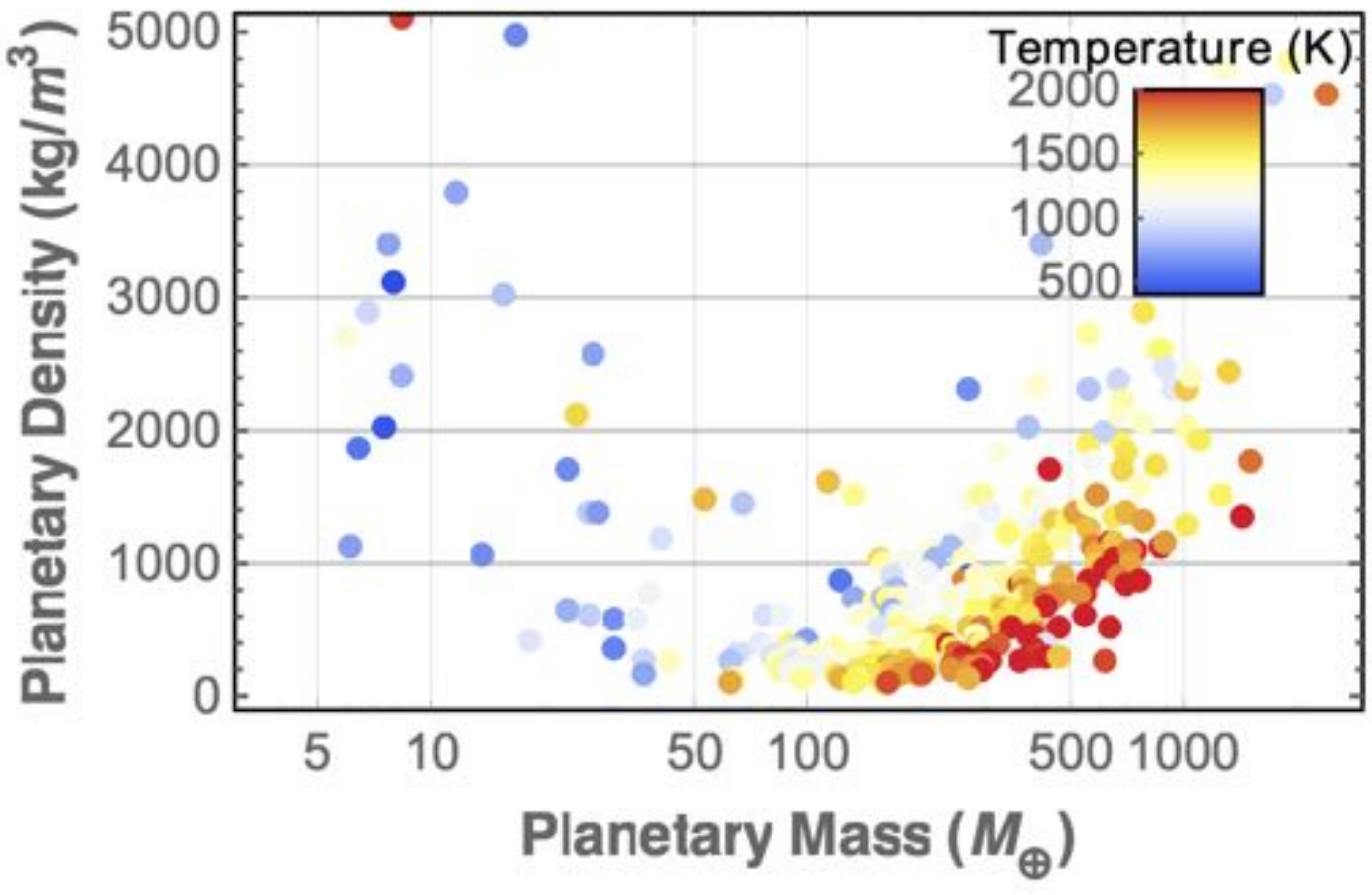

*Figure 2-22: Diversity in planetary densities of exoplanets as a function of planetary mass (the colour of the dots represents the equilibrium temperature of the planet). For low mass planets, density alone illustrates the variability in terms of content of volatile elements, yet it does not provide us insight into the nature of these volatile elements (e.g. water, multiple ices, organics, etc.) and, therefore, of the planets themselves. Data from exoplanets.org, retrieved March 2020.*

For small rocky worlds below 1.5 Earth radii, Ariel will be limited to the observation of hot objects. Nevertheless, their study will shed light on a number of outstanding questions related to the formation and nature of terrestrial planets. Interestingly, for such hot terrestrial planets *atmospheric escape* is expected to be an important process, that changes the mass and composition of an atmosphere with time, given selective escape of atmospheric molecules (e.g. preferred loss of $H_2$ and $N_2$ compared to $CO_2$, see Section 2.2.2.4).

By detecting the presence and composition of an atmosphere for terrestrial planets and super-Earths in various environments, Ariel will address an important scientific question: i.e. what are the conditions for a terrestrial planet to keep a significant secondary atmosphere? The characterization of secondary atmospheres will also provide important constraints to the interiors of these planets.

For some objects, this might be feasible directly in primary transit, although secondary eclipse is probably better suited for the hottest targets. For planets such as LHS 3844b ($T_{day}$ ~1000 K), Kreidberg et al. (2019b) have demonstrated using Spitzer that full phase curves in a single channel are attainable and yield powerful constraints. By repeating such observations in an extended spectral domain for numerous targets would therefore be transformational in this domain.

Even after the magma ocean solidifies, the amount and composition of atmospheric gas continue to be modified by gas supply from the interior (or outgassing) and loss to space (i.e., escape or erosion). Outgassing is controlled by mantle dynamics, tectonics, and volcanism, which depend on gravity and water content. Thus, the atmospheric mass and composition can be linked to planetary degassing history and interior, unless impacts of small bodies make the dominant contribution. Some studies predict inefficient mantle convection, which likely results in the mantle, and therefore magma, of super-Earths retaining volatile material. This material is later released in the atmosphere due to the slow interaction between atmosphere and interior mediated by such inefficient convection (Miyagoshi et al. 2018; Tackley et al. 2013). Meanwhile all the primordial hydrogen and water in the ancient atmosphere could have been evaporated away by the host star (Kurosaki et al. 2014, Lopez 2017, see also Sect. 2.2.2.4).

The redox state and water content of planetary mantles are controlling factors for the composition of outgassing gases. Some theoretical studies predict the presence of coreless rocky exoplanets (Elkins-Tanton & Seager 2008) carbon-rich exoplanets (Madhusudhan et al. 2012; Miozzi et al. 2018) and highly refractory planets (Dorn et al. 2018), in addition to silicate planets like the solar-system terrestrial planets. A coreless planet has a mantle that is highly oxidized with abundant iron-bearing oxides (Elkins-Tanton & Seager 2008), while a carbon-rich planet has a mantle that is highly reduced with abundant SiC, pure carbon and Si (Hakim et al. 2018; Allen-Sutter et al. 2020). Highly refractory rocky interiors are depleted in Fe and enriched in Al and Ca which are otherwise minor elements in solar-system terrestrial planets. We expect that their atmospheres reflect the differences in the interior composition, consisting of oxidizing gases such as $CO_2$, $SO_2$, and $H_2O$, similarly to the current Earth and Venus, in one case, and of reducing gases, such as CO, $H_2S$, and $CH_4$, for the carbon-rich planets (Gaillard & Scaillet 2009). Metal-rich atmospheres are predicted for hot and highly refractory planets. To remove such degeneracy in the interior structure and composition of rocky exoplanets, atmospheric observations are therefore essential to provide additional constraints.

Among the most extreme examples of the hot terrestrial planets that Ariel will allow us to study are the so-called *lava planets*, such as Corot-7b or 55-Cnc e. These planets are so close to their host star that the temperatures reached on the day side are sufficient to melt the surface itself and maintain them in a "permanent magma ocean" stage (e.g. Nikolaou et al. 2019). As a result, some elements, usually referred to as "refractory", become more volatile and can form a thin silicate atmosphere (Léger et al. 2011; Schaefer & Fegley 2009, Miguel et al. 2011; Kite et al. 2016). As it has been shown that the composition and thickness of the atmosphere is highly dependent on the composition of the crust and on the efficiency of the atmospheric escape processes, the direct observation of such targets, especially in secondary eclipse, will place strong constraints on these parameters (Ito et al. 2015). Lava planets would be the best targets for atmospheric observations to constrain their interiors because their secondary atmospheres are likely to be composed of materials directly vaporized from their magma ocean due to the rapid vaporization/condensation, i.e, gas-melt equilibrium condition. If lava planets are dry, they are likely to have atmospheres composed of rocky materials such as Na, K, $O_2$ and SiO (Schaefer & Fegley 2009; Miguel et al. 2011; Ito et al. 2015). On the other hand, if lava planets have volatile elements such as H, C, N, S and Cl, they will probably have atmospheres composed mainly of $H_2O$ and/or $CO_2$ with rocky vapours such as Na and SiO (Schaefer et al. 2012). This second sub-class of lava planets would enable us to explore runaway greenhouse effects, a paradigm widely discussed for the early history of Venus (Ingersoll 1969; Way et al. 2020). The exhaustion of $H_2O$ from the interior provides an additional thermal blanketing that can maintain planets in a "long-term magma ocean" state (Hamano et al. 2013) even if they are not preconditioned to be molten by their equilibrium temperature.

Thus, detection of rocky vapour in the atmospheres would provide a definitive piece of evidence for rocky planets. Especially, for silicate atmospheres, the detectability of the SiO feature around 4 μm through eclipse observations with Ariel has been investigated for a wide range of substellar-point equilibrium temperatures and distances of the observed exoplanetary system (Ito et al. 2020). Furthermore, identifying the atmospheric constituents could give constraints on the bulk composition and formation processes of the lava planets. Note that so far there are no studies on the vaporised atmospheres of molten coreless or carbon-rich planets, but they should be addressed in future studies.

At the opposite end of the atmospheric composition spectrum compared to the mineral atmospheres of the lava planets, Ariel will also allow us to address a fundamental question: *how widespread is the presence of water and volatiles in terrestrial planets*? Even for the terrestrial planets of the Solar System, it is still a matter of debate how water and the other volatile elements, most importantly carbon, were incorporated into the forming planets. In particular, it is still unclear whether the presence of massive planets orbiting beyond the water snow line is a required ingredient or an obstacle. Several studies have shown how Jupiter played a major role in the delivery of water in the inner Solar System and, consequently, for the formation of Earth's habitable environment we know today (e.g. Turrini & Svetsov 2014; Raymond & Izidoro 2017; Pirani et al. 2019).

Theoretical studies investigating the formation of terrestrial extrasolar planets, however, highlighted how such massive planets can also act as barriers to the inward migration of volatile-rich planetary building blocks (dust, pebbles and/or planetesimals), effectively hindering a transport process that otherwise would be significantly more efficient (e.g. Quintana & Lissauer 2014). Interestingly, higher amounts of water in the building block of terrestrial planets could favour the onset of the "long-term magma ocean" state (Hamano et al. 2013), linking the apparently disconnected studies of lava planets and water, as these planets are expected to be bright water-saturated atmospheric targets with characteristic albedos (Marcq et al. 2017; Pluriel et al. 2019). Ariel's observations will supply a new insight into this conundrum.

By studying the atmospheres of terrestrial planets and super-Earths inhabiting both single-planet and multi-planet systems with different architectures, i.e. with and without more massive outer planets and showing marks of chaotic or orderly dynamical evolution (Turrini et al. 2018, 2020b), Ariel will reveal whether or not the architectures of the planetary system result in systematic differences in the atmospheric composition of the terrestrial planets and in their budget of volatile elements. Planets significantly enriched by the delivery of cold material from the outer regions of the system are expected to have higher C/O values compared to planets for which, like the terrestrial planets in the Solar System, this delivery was less efficient. Depending on the equilibrium temperature of these planets and their C/O, their atmospheres might be dominated by molecules like HCN, CN, CO, $CH_4$, $NH_3$, $C_2H_4$ (C/O >1, the exact molecule combination depending on the atmospheric temperature) or CO, $CO_2$ and $H_2O$ (C/O < 1) (Zilinskas et al. 2020, see Atmospheric Chemistry WG report 2020 for additional discussion). All these molecules have features in the spectral range covered by Ariel. In particular, the overall spectral signatures of water-rich atmospheres are easily distinguishable from those of mineral atmospheres (see Atmospheric Chemistry WG report 2020). The systematic comparison between the atmospheric signatures of tens of terrestrial planets will provide an unprecedented insight into whether the incorporation of volatile material is a built-in mechanism in the planetary formation process or on whether special circumstances are required.

# 3 Strategy to Achieve the Science Objectives and Requirements

Ariel will employ five complementary methods to probe exoplanet atmospheric properties. This objective is facilitated by its exceptionally broad simultaneous wavelength coverage. A 4-tier observing plan will build from a large population study to a more detailed characterisation of selected high-interest objects. The target candidates cover a wide, representative range of density, size, effective temperature, as well as stellar host type, metallicity, and activity level

## 3.1 Strategy to Achieve the Science Objectives

### 3.1.1 How do we observe exo-atmospheres?

For transiting planets, we have *five complementary methods* to probe their atmospheric composition and thermal structure, which are described briefly in the following paragraphs. Ariel will use them all.

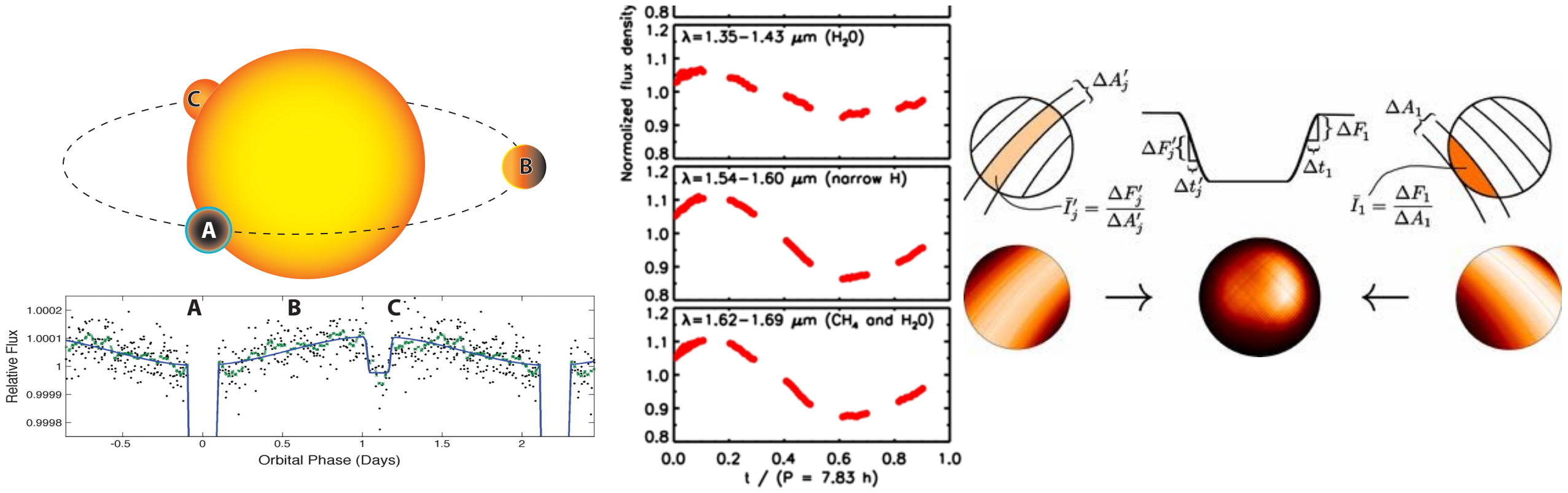

*Figure 3-1: Methods adopted by Ariel to probe the exoplanet composition and structure. Left: phase-curve of the transiting exoplanet HAT-P-7b as observed by Kepler (Borucki et al. 2009). The transit and eclipse are visible. Centre: time series of brown-dwarf narrowband light curves observed with HST-WFC3 (Apai et al. 2013). The spectral bands have been selected to probe specific atmospheric depths and inhomogeneities in the cloud decks. Right: slice mapping with ingress and egress maps as well as a combined map of HD189733b at 8 μm. These were achieved with Spitzer (Majeau et al. 2012; De Witt et al. 2012).*

1. When a planet passes in front of its host star (*transit*), the star flux is reduced by a few percent, corresponding to the planet/star projected area ratio (transit depth). The planetary radius can be inferred from this measurement. If atomic or molecular species are present in the exoplanet's atmosphere, the inferred radius is larger at some specific (absorption) wavelengths corresponding to the spectral signatures of these species (Seager & Sasselov 2000; Brown 2001; Tinetti et al. 2007a).

   The transit depth $\delta(\lambda)$ as a function of wavelength ($\lambda$) is given by:

$$\delta(\lambda) = \frac{2\int_0^{z_{max}}(R_p + z)(1 - e^{-\tau(z,\lambda)})dz}{R_*^2} \quad (1)$$

   where $z$ is the altitude above $R_p$ and $\tau$ the optical depth. Eq. (1) has a unique solution provided we know $R_p$ accurately. $R_p$ is the radius at which the planet becomes opaque at all $\lambda$. For a terrestrial planet, $R_p$ usually coincides with the radius at the surface. For a gaseous planet, $R_p$ may correspond to a pressure $p_0 \sim$ 1-10 bar.

2. A direct measurement of the planet's emission/reflection can be obtained through the observation of the planetary *eclipse*, by recording the difference between the combined star+planet signal, measured just before and after the eclipse, and the stellar flux alone, measured during the eclipse, Figure 3-1. Observations provide measurements of the flux emitted/reflected by the planet in units of the stellar flux (Charbonneau et al. 2005; Deming et al. 2005). The planet/star flux ratio is defined as:

$$\phi(\lambda) = (R_p/R_*)^2 \; F_p(\lambda)/F_*(\lambda) \quad (2)$$

3. In addition to transit and eclipse observations, monitoring the flux of the star+planet system over the orbital period (*phase curve*) allows the retrieval of information on the planet emission at different phase angles (Figure 3-1). Such observations can only be performed from space, as they typically span a time interval of more than a day (e.g. Cowan et al. 2007; Borucki et al. 2009; Stevenson et al. 2014; Demory et al. 2016). So far, phase-curve observations recorded with Spitzer, Hubble, Kepler, and TESS have been published for 28 planets.

The combination of these three prime observational techniques utilized by Ariel will provide us with information from different parts of the planet atmosphere; from the terminator region via transit spectroscopy, from the day-side hemisphere via eclipse spectroscopy, and from the unilluminated night-side hemisphere using phase variations.

4. In addition, eclipses can be used to spatially resolve the day-side hemisphere (*eclipse mapping*). During ingress and egress, the partial occultation effectively maps the photospheric emission region of the planet (Rauscher et al. 2007). Figure 3-1 illustrates eclipse mapping observations obtained with Spitzer (Majeau et al. 2012; De Witt et al. 2012). For a review of all aspects of planet mapping, see Cowan & Fujii (2018).
5. Finally, an important aspect of Ariel is the repeated observations of a number of key planets in both transit and eclipse mode (*time series of narrow spectral bands*). This will allow the monitoring of global meteorological variations in the planetary atmospheres, and to probe cloud distribution and patchiness (see e.g. Apai et al. (2013) for similar work on brown dwarfs, Figure 3-1).

### 3.1.2 Ariel observational strategy: a 4-tier approach

The primary science objectives summarised in Chapter 2 call for atmospheric spectra or photometric light-curves of a large and diverse sample of known exoplanets covering a wide range of masses, densities, equilibrium temperatures, orbital properties and host-stars. Other science objectives require, by contrast, the very deep knowledge of a select sub-sample of objects. To maximize the science return of Ariel and take full advantage of its unique characteristics, a four-tiered approach has been formulated, in which observations are analysed binning the raw spectra –taken always at the focal plane native resolution– in bins of different widths, such that the desired SNR appropriate for each tier is reached. A summary of the survey tiers is given in Table 3-1. In the following subsections we report the expected performances of the Ariel mission following the 4-tier strategy.

*Table 3-1: Summary of the survey tiers and the detailed science objectives they will address. The corresponding science requirements are listed in Table 3-2.*

| Tier name | Observational strategy | Science case |
|---|---|---|
| **Tier 1**<br>**Reconnaissance Survey** | Low spectral resolution observations of ~1000 planets in the VIS & IR, with SNR ~ 7 | - What fraction of planets are covered by clouds?<br>- What fraction of small planets have still retained H/He?<br>- Classification through colour-colour diagrams?<br>- Constraining/removing degeneracies in the interpretation of mass-radius diagrams<br>- Albedo, bulk temperature and energy balance for a subsample. |
| **Tier 2**<br>**Deep Survey** | Higher spectral resolution observations of a sub-sample in the VIS-IR | - Main atmospheric component for small planets<br>- Chemical abundances of trace gases<br>- Atmospheric thermal structure (vertical/horizontal)<br>- Cloud characterization<br>- Elemental composition (gaseous planets) |
| **Tier 3**<br>**Benchmark planets** | High SNR observations in 1-2 events, re-observed over time | - Weather and temporal variability<br>- Very detailed knowledge of the planetary chemistry and dynamics |
| **Tier 4**<br>**Phase-curves & bespoke observations** | Phase-curves and bespoke observations | - Targets of special interests<br>- Spatial variability<br>- Very detailed knowledge of the planetary chemistry and dynamics |

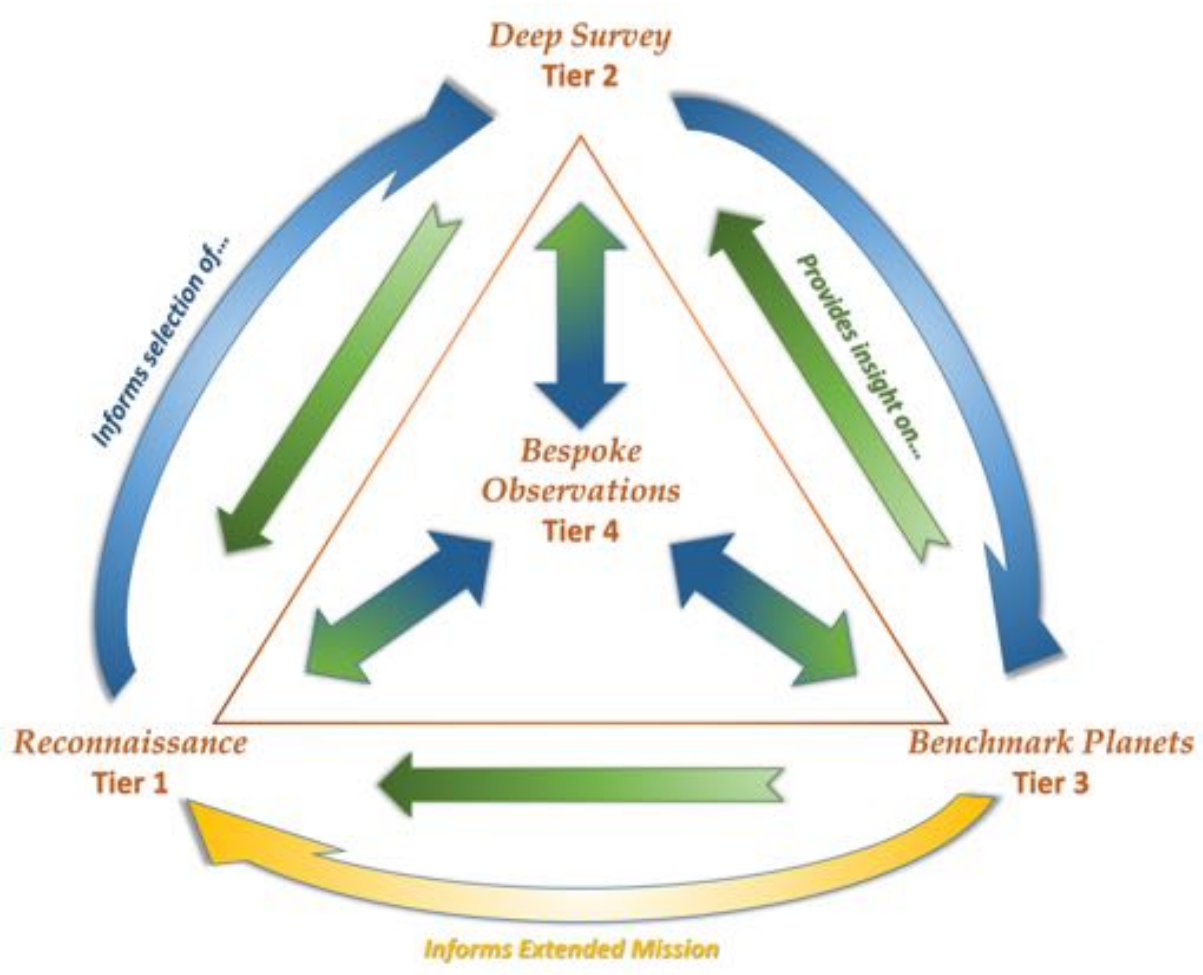

*Figure 3-2: Schematic diagram of Ariel's 4-Tier ecosystem. Tier 1 will provide the base for selecting Tier 2 planets, which in turn will inform the selection of Tier 3 targets. The planets observed in each Tier will give us deeper insight into the nature of the planets in preceding Tiers. This interdependence among Tiers means that the scientific value of the data collected by Ariel will grow over time. Tier 4 observations will benefit from the insight gained by the planetary populations of the other Tiers and will in turn provide a better understanding of their atmospheric behaviour. At the end of the nominal mission, the trove of information supplied by the 4-Tier approach will provide the basis for the selection of the new observational sample.*

## 3.1.3 Ariel Tier 1 Reconnaissance Survey: exoplanet population analysis

Ariel Tier 1 will analyse ~ 1000 exoplanets to address science questions for which a large population of objects needs to be observed. As pointed out in Section 2.1, a large number of objects needs to be analysed to fully appreciate the underlying properties of the planetary population, a requirement which cannot be fulfilled today. Ariel will allow to increase by two orders of magnitude the number of planets characterised and by an order of magnitude the number of colour-filters available in the 0.5-8 µm wavelength range compared to current observations.

The Ariel Tier 1 survey mode will also allow rapid and broad characterisation of planets so that decisions can be made about priorities for future observations with Tier 2 and Tier 3. For the majority of the targets observed by Ariel, the necessary performance can be reached in 1 or 2 transits/eclipses (Figure 3-3 left).

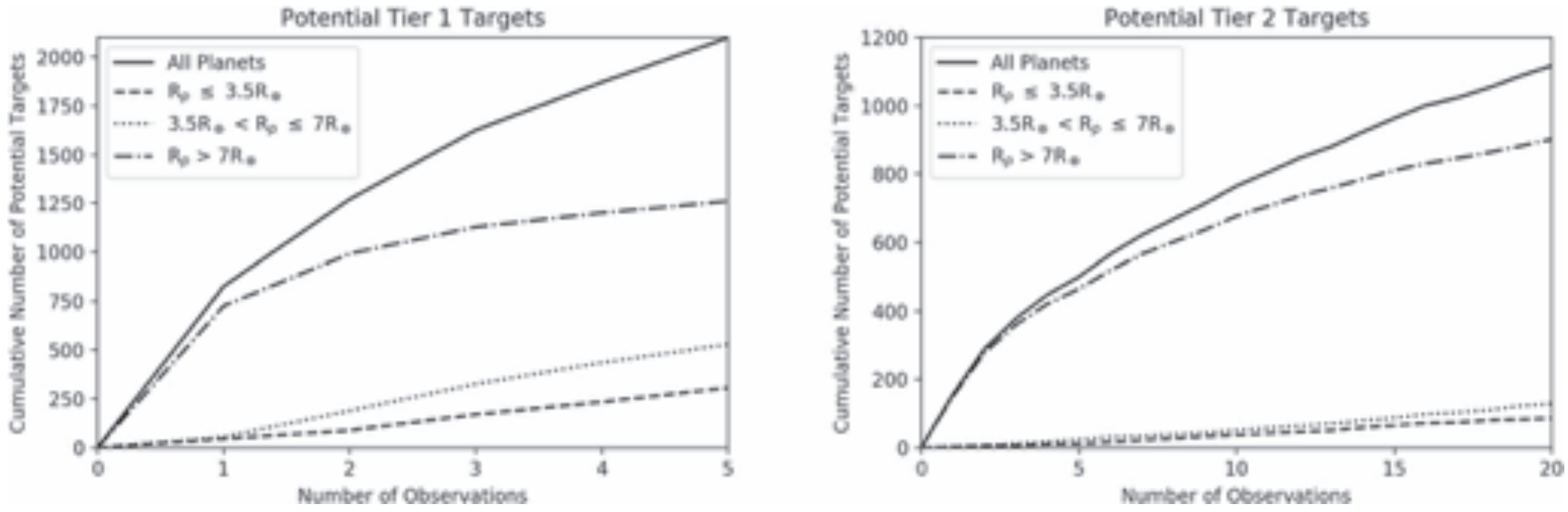

*Figure 3-3: Planet candidates observable by Ariel in the Tier 1 (left) and Tier 2 (right) modality, from Edwards et al. (2019). About ~1000 planets of various sizes and temperatures can be observed in Tier 1 in ~40% of the mission life-time. About half of them can be observed in Tier 2 in ~25% of the mission life-time, see Chapter 7.*

The science questions addressable by Ariel Tier 1 observations include:

- <u>*What fraction of planets are covered by clouds?*</u>

  Tier 1 mode is useful for discriminating between planets that are likely to have clear atmospheres, versus those that are so cloudy that no molecular absorption features are visible in transmission. Extremely cloudy planets may be identified simply from low-resolution observations over a broad wavelength range. This preliminary information will therefore allow us to take an informed decision about whether to continue the spectral characterization of the planet at higher spectral resolution, and therefore include or not the planet in the Tier 2 sample. In addition, it may be possible to discriminate between broad cloud types simply using photometric indices and optimised metrics. This information will also be useful for planning follow-up observations. Examples of how such metrics could work are given in Figure 3-4.

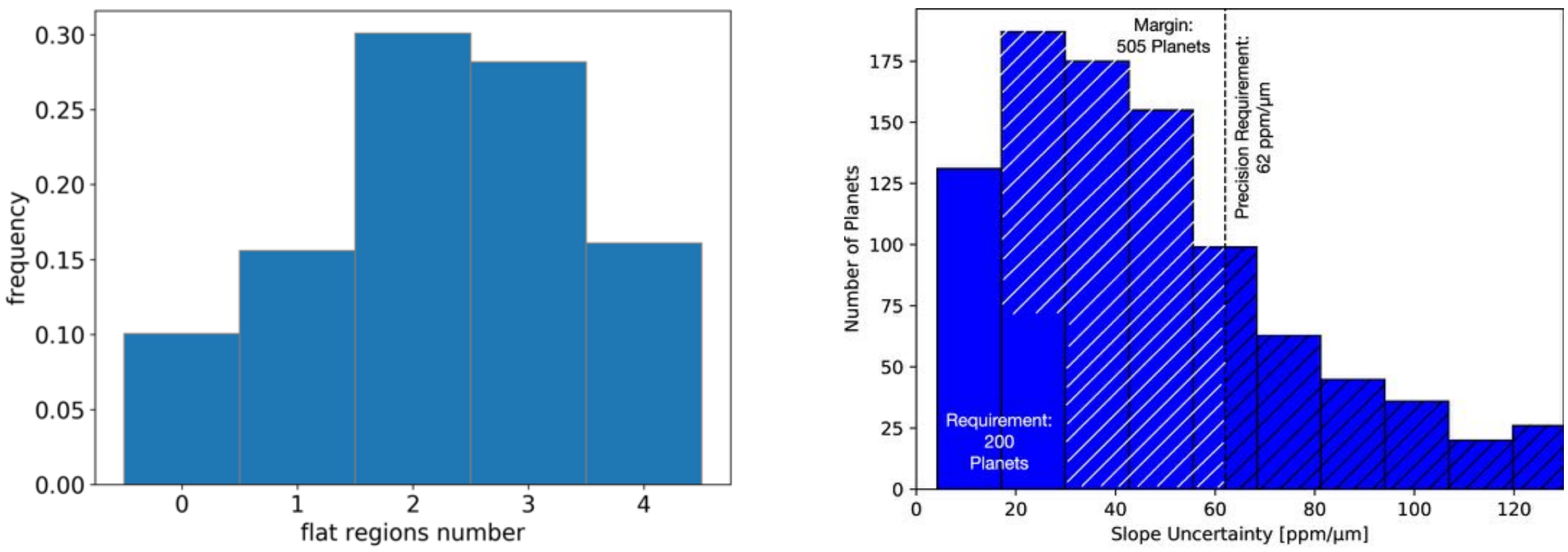

*Figure 3-4: Left: Simulations of Ariel Tier 1 observations with the Alfnoor tool, run for 1000 planets from the Ariel planetary candidates, from Mugnai et al. (2021). Here four broad bands are considered: one for the photometers (VisPhot, FGS1, FGS2) and one for each spectrometer (NIRSpec, AIRS CH0 and AIRS CH1). Each band is compared to a constant value by the $\chi^2$ metric to determine its flatness. In the spectral sample the presence of clouds was simulated by adding a cloud deck at random pressure between 1 bar (no clouds) and $10^{-2}$ bar (very opaque clouds). The histograms show the frequency of planets in the population that have a certain number of flat bands. The results are consistent with the ground truth. Right: Simulations from Zellem et al. (2019) showing how Ariel could measure exo-atmospheric aerosol and cloud spectral slopes to high precision. In these simulations only one visit of transit measurement for each planet was assumed, with an uncertainty of 15 ppm/µm, enabling high-precision measurements to determine if a planet's atmosphere is clear, cloudy, or hazy.*

- *What fraction of small planets have still retained molecular hydrogen?*

As described in Section 2.2, small planets, particularly those which could be rocky, are an intriguing population of bodies, especially since the discovery of the radius valley at ~1.8 $R_\oplus$ by the California- Kepler Survey (CKS; Fulton & Petigura 2018). Characterising the atmospheres of planets with radii smaller than 3.5 $R_\oplus$, and in particular, those within the transition region from rocky to gaseous, is pivotal to uncover the nature of this population and would be very informative for planetary formation and evolution theories. More specifically, understanding whether the atmosphere is still primordial (i.e., H/He-rich, possibly thick) or more evolved (i.e., richer in heavier elements, thin or completely absent) may constrain formation (formed in situ or remnants of more massive bodies which have migrated to closer orbits) and evolution scenarios (e.g. hydrogen escaped, a secondary atmosphere which might hint at the interior composition).

- *Preliminary classification through colour-colour diagrams or other metrics*

  Colour-colour or colour-magnitude diagrams are a traditional way of comparing and categorising luminous objects in astronomy. While these diagrams will never replace a full spectrum, they can give valuable constraints to select interesting targets for Tier 2 and Tier 3 surveys or even uncover unknown properties/trends in the population (Triaud et al. 2014; Dransfield et al. 2020; Figure 3-5).

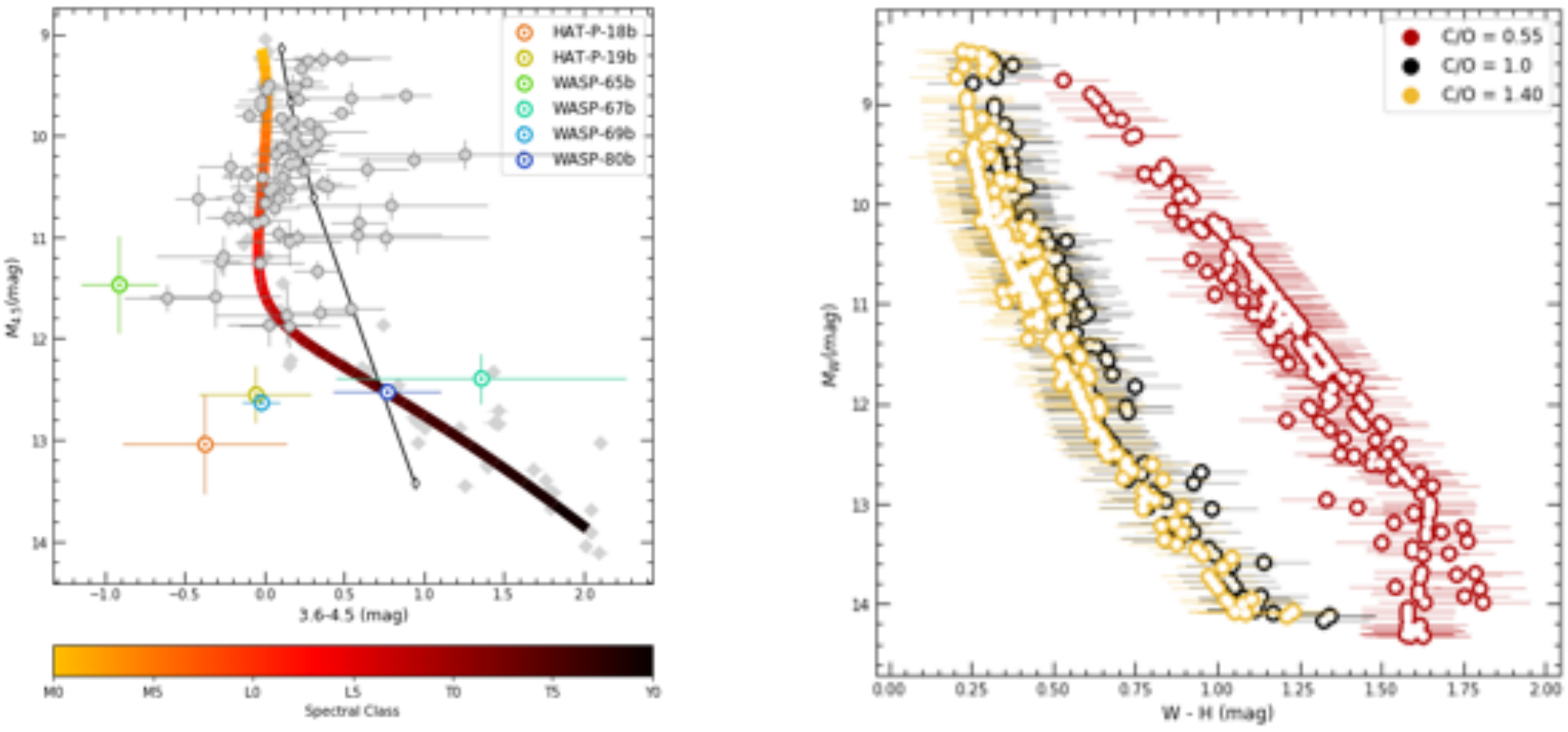

*Figure 3-5: Figures from Dransfield et al. (2020). Left: Mid-infrared colour-magnitude diagrams, using Spitzer's IRAC photometric bands at 3.6µm and 4.5µm measured during eclipse. The six planets highlighted have equilibrium*

*temperatures of between 800-1000K, with HAT-P-18b and WASP-80b being the coolest of the set. Planetary magnitudes have been scaled to a 0.9$R_J$ object for better comparison with the brown dwarfs. The black line shows the position of a 0.9$R_J$ blackbody with the white-filled diamonds showing the position of the blackbody at temperatures of 750K, 1750K, 2750K, 3750K and 4750K. The polynomial showing the mean position of the brown dwarfs sequence is coloured according to spectral type. Right: simulated colour-magnitude diagram featuring Mollière et al. 2015 model atmospheres plotted on a $M_W$ vs [$W_{JH}$ -H] colour-magnitude diagram, assigning C/O ratio values of 0.55, 1 and 1.40. According to these simulations, two populations with different C/O ratios can be identified with Ariel by simply placing all planets in a colour-magnitude diagram.*

To assess Ariel performances for a large sample of planets we built a new algorithm: Alfnoor, the thousand lights simulator, which combines TauREx3 (Al-Refaie et al. 2020) and ArielRad (Mugnai et al. 2020). Alfnoor returns a simulation of the planet spectrum as observed from each of the Ariel mission's Tiers. Iterating this procedure for different planets or compositions, Alfnoor automates the process of building entire planetary populations and therefore a data set that is representative of the one Ariel will provide.

In Mugnai et al. (2021) a large sample of H/He-rich planets were simulated with an atmosphere with randomised quantities of $H_2O$, $CH_4$, $CO_2$ , $NH_3$ and clouds to assess the information content of planets observed in Tier 1 mode, by defining an optimal metric able to constrain their atmospheric composition. Figure 3-6 illustrates how deep learning techniques applied to Tier 1 observations can provide an initial disentanglement of the contribution of different molecules and clouds.

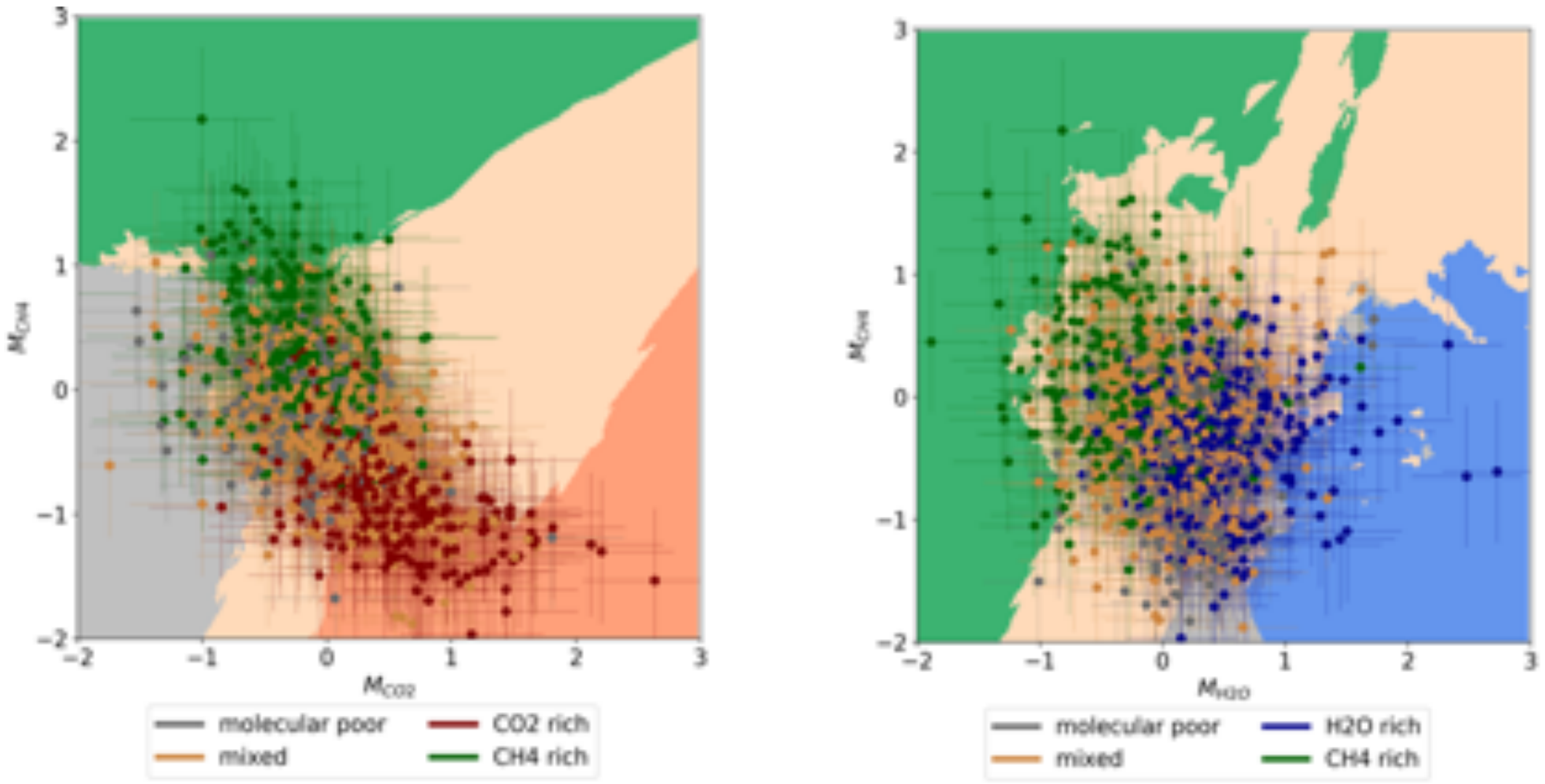

*Figure 3-6: Simulations of Ariel Tier 1 observations with the Alfnoor tool, run for 1000 planets from the Ariel planetary candidates, from Mugnai et al. (2021). KN neighbours analysis results with k = 20 for $CH_4$-$CO_2$ (left) and $CH_4$-$H_2O$ (right) cases. Superimposed dots are from the population observed data and error bars represents the metric uncertainties. Grey dots: planets that contain less than $10^{-5}$ in mixing ratio for the considered molecules; green points: planets that contain 10 times more $CH_4$ than the other molecule; red points: planets that hold 10 times more $CO_2$ than $CH_4$; blue points: planets with 10 times more $H_2O$ than $CH_4$; yellow dots: all the other possible solutions. The same colour scheme applies to the painted region of the diagram. Grey area: planets with low quantities of water and methane; green area: methane-rich planets, blue: water-rich planets; yellow: mixed atmospheres.*

In the diagrams of Figure 3-6 we can distinguish four regions: two regions where atmospheres are rich in a single molecule; an area where atmospheric spectra are flat or have no feature emerging from the clouds for the two molecules considered and an area where planets have features from both the molecules.

*Albedo, bulk temperature & energy balance*

Eclipse measurements in Ariel's broad optical and IR bands may provide the bulk temperature and albedo of the planet, thereby allowing the estimation of the planetary energy balance and whether the planet has an internal heat source or not.

The albedo of an exoplanet is related to the optical properties and cover of clouds/haze. In addition, we expect transitions in the values of the albedo due to the disappearance of some clouds with increasing temperature, as for the L-T transition for brown dwarfs (Parmentier et al. 2016). Measuring the albedo would help to understand the nature of these aerosols. Ariel could measure exoplanetary geometric albedos with high precision for a fraction of planets, see Figure 3-7.

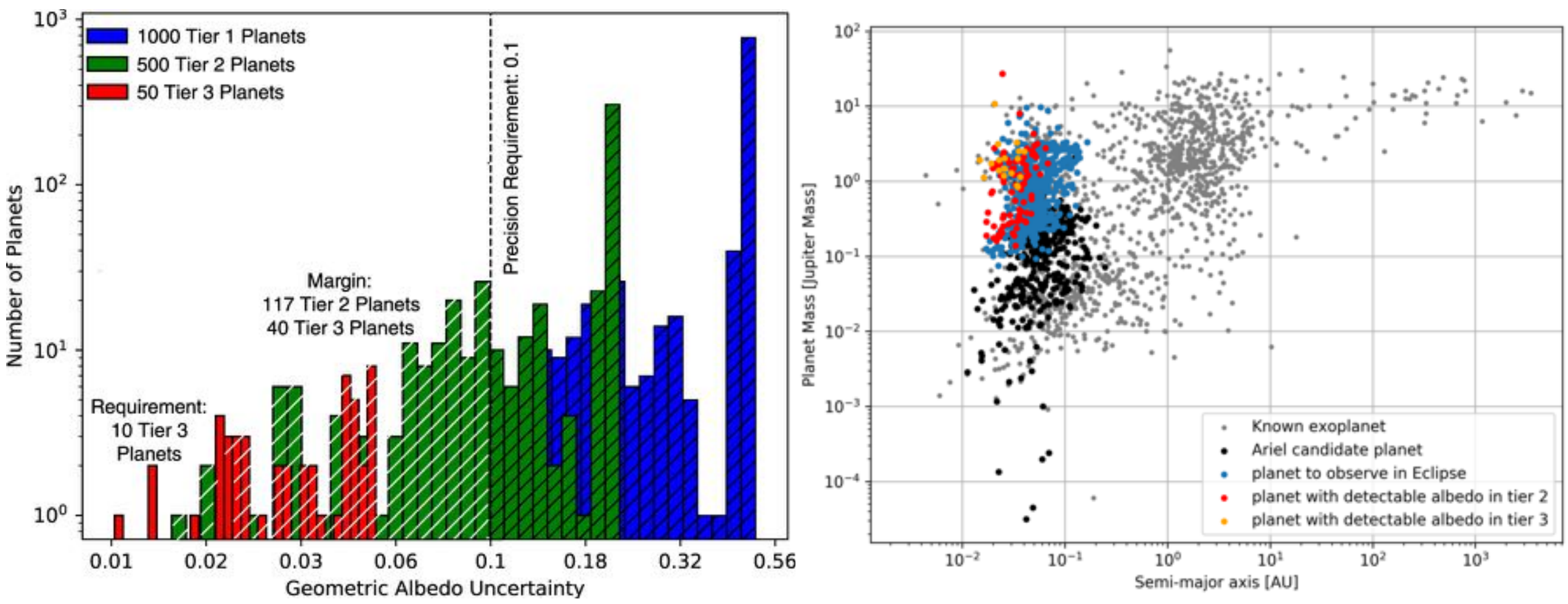

*Figure 3-7: Left: Simulations from Zellem et al. (2019): Ariel could measure exoplanetary geometric albedos with high precision. Right: simulations done with Ariel Rad (Mugnai et al 2020) showing the fraction of planets observed in Tier 2 and Tier 3 mode for which the albedo is detectable very precisely, assuming the albedo is = 0.5. This fraction reduces to one half and one quarter if the albedo is =0.3 and 0.1 respectively.*

### 3.1.4 Ariel Tier 2 Deep survey: single planets & population analysis

Tier 1 was created to deliver a reconnaissance survey, in which all planets are first observed at low spectral resolution, and only a subset of Tier 1 planets will be further observed to reach higher spectral resolution and SNR (Tier 2, Tier 3). Tier 2 spectroscopic observations will be essential for uncovering the atmospheric structure and composition, as well as to search for potential correlations between atmospheric chemistry and basic parameters such as planetary radius, density, temperature, stellar type and metallicity.

We report in Figure 3-8 an example extracted from the Ariel *Spectral Retrieval Data Challenge* https://arielmission.space/data-challenges/ for a fully blind retrieval of a cloudy sub-Neptune. As showcased in the figure, different retrieval codes provided impressively consistent results for the retrieved molecular abundances, atmospheric temperature and cloud parameters.

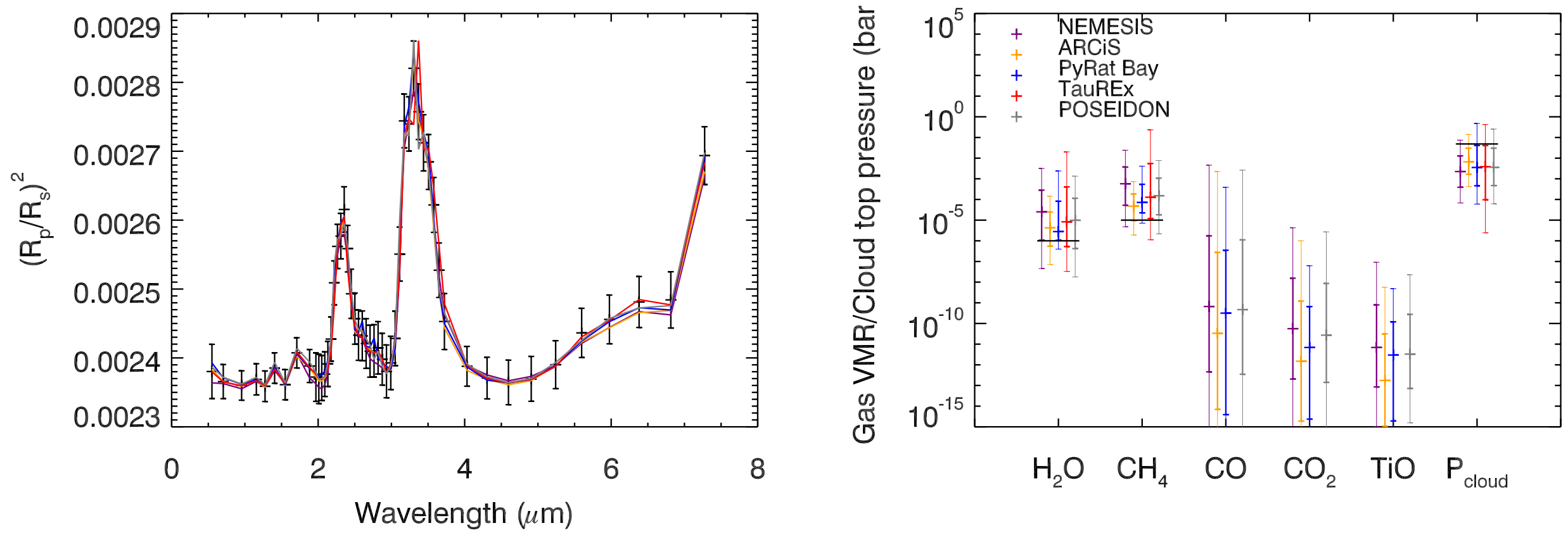

*Figure 3-8: Example from the Ariel Spectral Retrieval Data Challenge for a fully blind retrieval of a cloudy sub-Neptune (input parameters: $R_p$=0.15 $R_J$, $R_*$=0.45 $R_\odot$; $M_p$=0.01 $M_J$; $H_2O=1.0\times10^{-6}$; $CH_4=1.0\times10^{-5}$; CO=0; $T_p$ =700 K; $P_{cloud}=4.52\times10^{-2}$). The results submitted by the participants with different codes are indicated (Barstow et al. 2020).*

We studied the ability of Ariel to detect $H_3^+$ and $H_3O^+$ in the observational spectra of small planets based on realistic mixing ratio assumptions. According to Bourgalais et al. (2020), these ionic species may help to distinguish super-Earths with a thin atmosphere from $H_2$-dominated sub-Neptunes, to address the critical question whether a low-gravity planet is able to retain its volatile components (Figure 3-9 left).

Ito et al (2020) have shown that 10 secondary eclipse observations for 55 Cnc e with Ariel would suffice to distinguish a mineral atmosphere from a cloud-free, hydrogen-rich or water-rich atmosphere (see Figure 3-9 right).

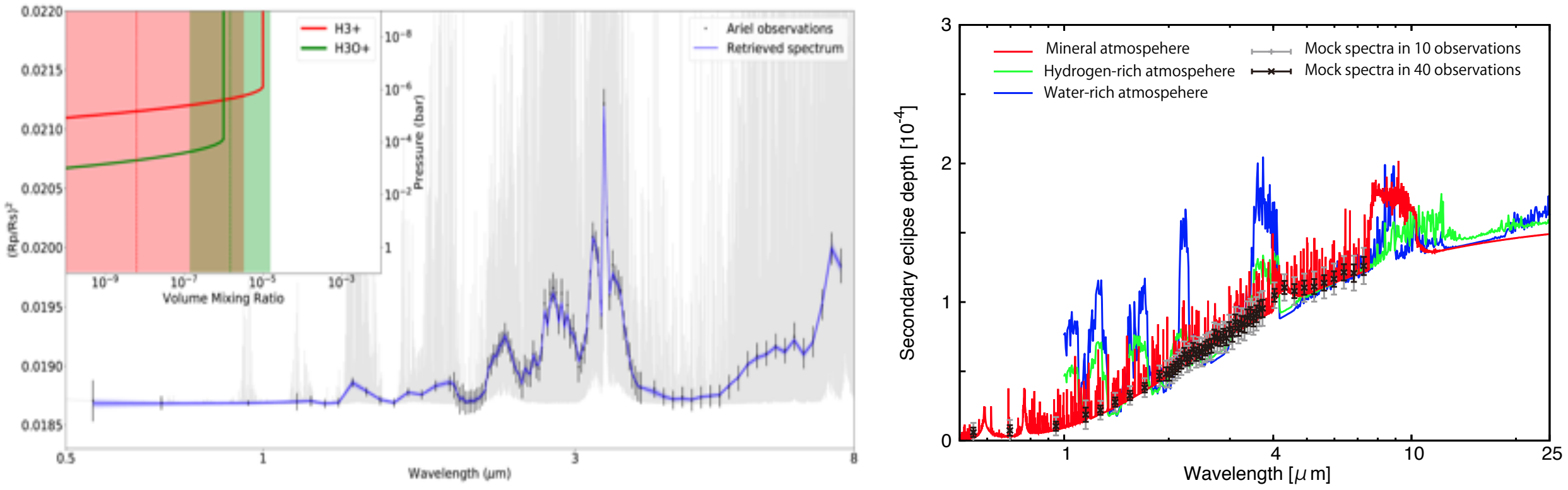

*Figure 3-9: Left: Figure from Bourgalais et al. (2020): Ariel simulations of a GJ1214b-like planet (i.e. a cloudy sub-Neptune with the planetary, stellar and orbital parameters of GJ1214b) with $H_3O^+$ and $H_3^+$ ions. For the abundances the solid line is the true value, the dashed line the mean retrieved value and the shaded area represents the 1σ retrieved value. Right: figure from Ito et al. (2020): theoretically predicted spectrum of secondary eclipse depths and the mock observational spectrum for a 55 Cnc e-like transiting small exoplanet with a mineral atmosphere, which are compared to those with hydrogen-rich and water-rich atmospheres from Fig. 15 in (Hu & Seager 2014). The solid lines show theoretical spectra for the mineral atmosphere (red), the hydrogen-rich atmosphere (green) and the water-rich atmosphere (blue). The bars in the mock spectra represent cumulative errors estimated for with Ariel for cumulated observations in 10 times (grey) and 40 times (black) eclipse observations with Ariel for 55 Cnc e.*

### 3.1.5 Ariel Tier 3: atmospheric variability and benchmark planets

Ariel Tier 3 observations will focus in particular on the study of the variability through time of the exoplanet atmospheres. "Weather planets" are selected from the very best targets, i.e. planets for which the maximum Ariel spectral resolving power and SNR > 7 can be reached in one or two observations. Repetition through time of the same observations will cast light on the temporal variability of the exo-atmospheres due to variations in the cloud coverage or patterns in the global circulation (e.g. Parmentier et al. 2013; Cho et al. 2015; Fromang 2016; Rogers 2017; Komacek & Showman 2019).

Ariel Tier 3 observations will identify variations in the thermal vertical and horizontal structure through time and provide critical insight into the complex circulation patterns of these exotic atmospheres. These results will be used also to quantify for the first time the uncertainty introduced when we obtain only disc and time integrated spectra.

By combining all the observations obtained through time for a Tier 3 planet, unprecedented SNR will be achieved, enabling an extremely detailed study of the atmospheric chemistry and dynamics.

| Molecule | Tier 2 | Tier 3 |
|---|---|---|
| $H_2O$ | $10^{-6}$ | $< 10^{-7}$ |
| $CH_4$ | $10^{-7}$ | $< 10^{-7}$ |
| CO | $10^{-4}$ | $10^{-6}$ |
| $CO_2$ | $10^{-7}$ | $< 10^{-7}$ |
| $NH_3$ | $10^{-6}$ | $10^{-7}$ |

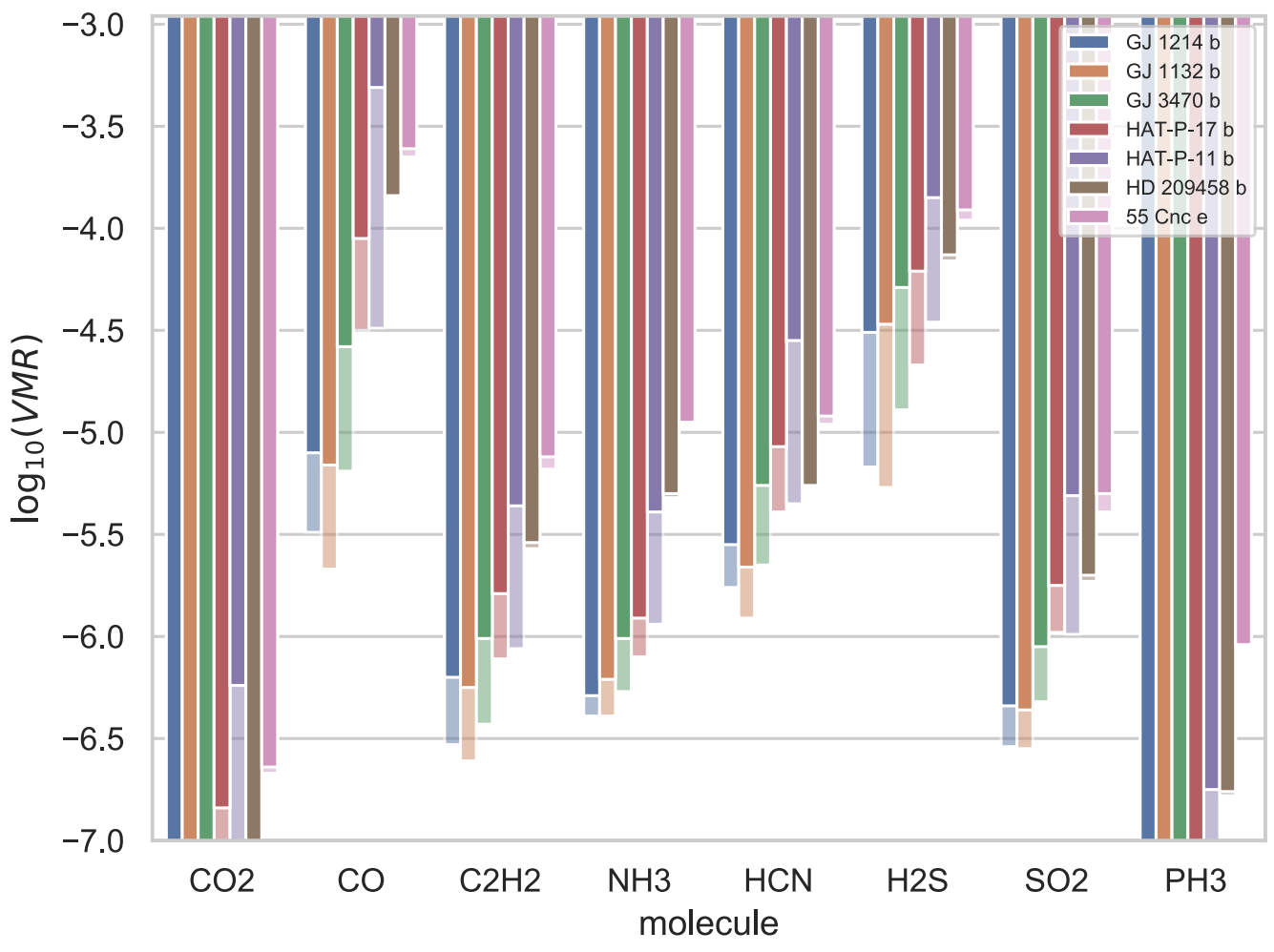

*Figure 3-10: Left: Detection limits (expressed as volume mixing ratios) for key molecules in Ariel Tier-2 and Tier-3 samples using the Alfnoor tool to analyse a population of H/He-rich planets (Changeat et al. 2020a). The simulations include clouds. Right: Sensitivity study for several exoplanets in Tier2 and Tier3 (more transparent bars) observation mode obtained with ARCiS (Min et al. 2020). We show the range over which a certain molecule can be significantly detected against a background of 100 ppmv $H_2O$ and 10 ppmv $CH_4$.*

### *3.1.5.1 Ariel sensitivity to molecular detection with Tier 2 and Tier 3 observations*

In Figure 3-10 left we show the detection limits for key trace gases when gaseous planets are observed with Ariel Tier 2 and Tier 3 modes (Changeat et al. 2020a). These detection limits were obtained by simulating a population of giant planets and Neptunes with the Alfnoor code.

To analyse the sensitivity of Ariel for certain molecular species in the atmosphere a threshold test was performed using the retrieval code ARCiS (Min et al. 2020). For this test illustrated in Figure 3-10 right we selected seven targets representative of the spread in temperature and planet mass. For these seven targets computations were performed to estimate the expected noise levels for Tier 2 and Tier 3 observations. The concept of the threshold study is to define a standard atmosphere and add to this an increasing amount of a certain molecule. We then perform a retrieval on the simulated spectrum and check at which level the retrieval can detect the molecule. We define a detection when the difference in Bayesian evidence between a retrieval with and without the molecule in question is larger than 5. The standard atmosphere contains by default $H_2O$ at a 100 ppmv and $CH_4$ at 10 ppmv. The temperature of the atmosphere is isothermal and computed from the distance to its host star assuming zero albedo and perfect redistribution of the irradiation. For the threshold tests we include 8 molecules other than $H_2O$ and $CH_4$. This is a very clean test and does not take into account possible confusion between different molecules. As an example, we also performed some tests where we included both CO and $CO_2$ and we conclude from these tests that at low molecular abundances there is a confusion between these two molecules. However, the numbers from the clean test do provide a good order of magnitude estimate of the sensitivity of Ariel spectra in these two different Tiers. Figure 3-10 right shows the results of these tests. The bars show which volume mixing ratios can be detected for each molecule and each target in our small sample of eight planets. We see that $CO_2$ and $PH_3$ can be detected already at an abundance around 0.1 ppmv. The most difficult to detect is CO, since this molecule has a few, weak bands overlapping with other species in the Ariel spectral range. Interesting to note here is that there is a temperature dependence of the detectability of especially $C_2H_2$, $NH_3$ and $SO_2$. These species are more easily detected in cooler atmospheres.

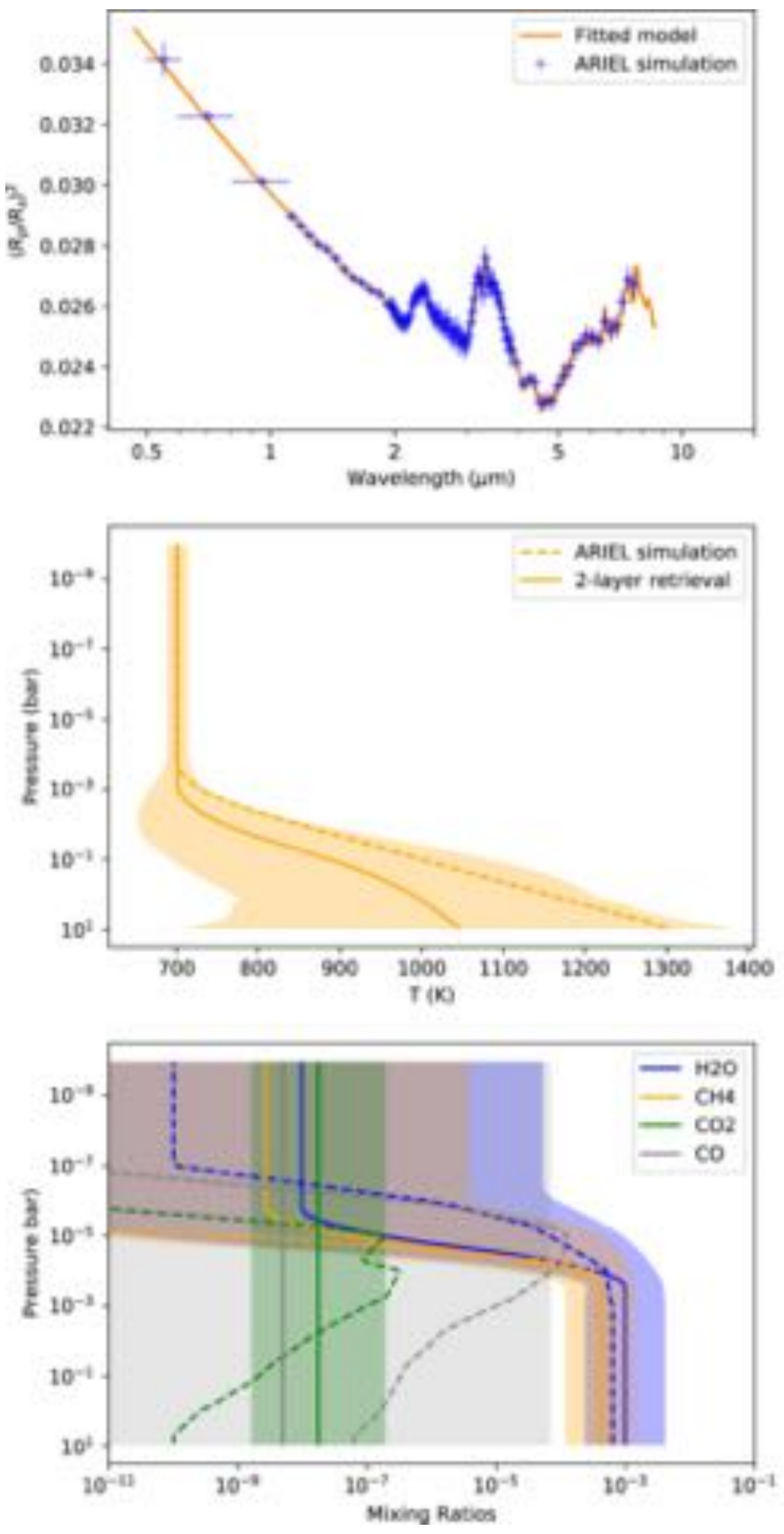

*Figure 3-11: Simulated spectrum (top) and retrieved chemical and thermal profiles (central and bottom) from Ariel Tier 2 transit observations of a hazy sub-Neptune, with planetary parameters similar to GJ1214b, from Changeat et al. (2019). Dotted lines correspond to the input values, solid lines show the retrieved profiles. Coloured areas show the uncertainties in the retrieved posteriors. Top: temperature profile as a function of pressure. Bottom: chemical profiles of $H_2O$, $CH_4$, $CO_2$, and CO. These plots demonstrate that Ariel performance enables an accurate and precise characterisation of both the atmospheric thermal and chemical profiles.*

We show in Figure 3-11 an example of Ariel's performances for an individual planet: here chemical and thermal profiles as a function of atmospheric pressure are retrieved from Ariel Tier 2 transit observations of a hazy sub-Neptune. This kind of information, which is essential to inform chemistry and dynamical models, cannot be extracted from current observed spectra of exoplanet atmospheres, because they do not cover a broad enough spectral window. Additional examples are given in Changeat et al. (2019). These simulations confirm that Tier 2 Ariel data will provide very accurate results and will fulfil the science objectives as stated in Section 2.2. For the majority of the targets observed by Ariel, these performance levels can be reached between 1 and 10 transits/eclipses (Figure 3-3).

### *3.1.5.2 Ariel sensitivity to chemical trends in exoplanet populations*

Changeat et al. (2020a) studied the ability of Ariel to reveal chemical trends in exoplanet populations with Tier 2 observations. To assess that, they built biased samples in which artificial chemical trends were introduced to test whether we were able to recover the trends from Ariel observations. For the example presented in Figure 3-12 left, they imposed a linear relationship between the logarithmic abundance of water and the temperature.

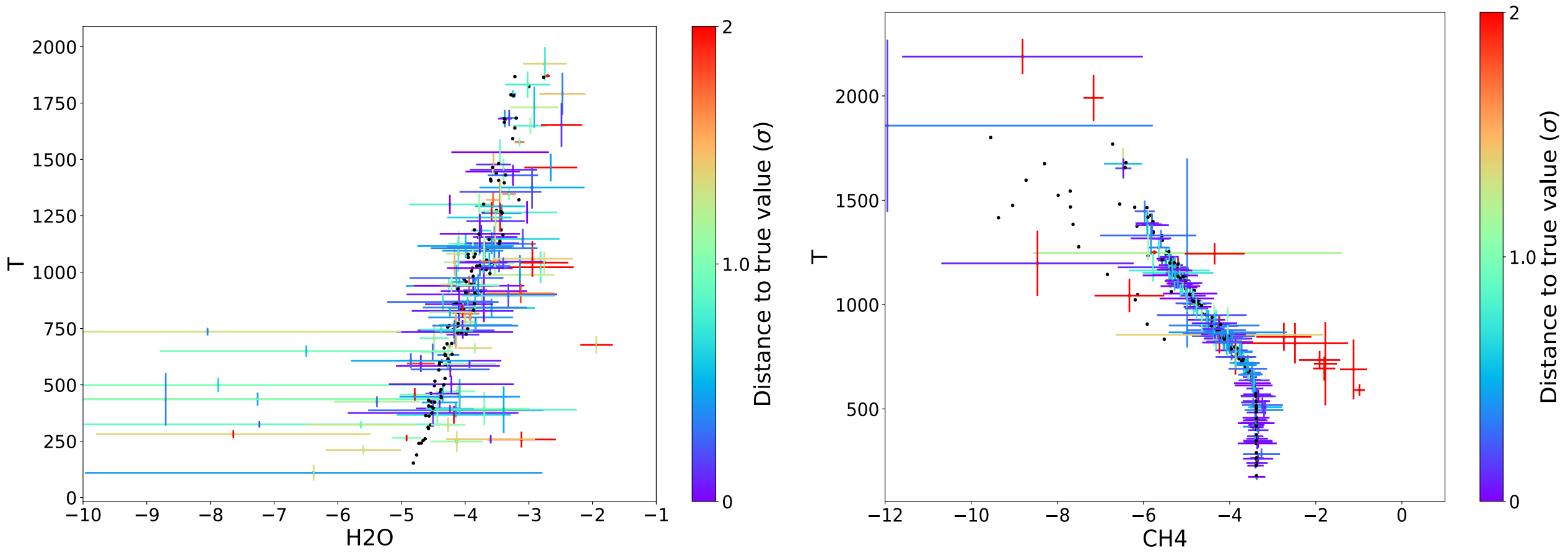

*Figure 3-12: Correlation maps of the retrieved abundance of water (left) or methane (right) and the temperature for a sample of 150 H/He-rich transiting planets. These tests obtained with Alfnoor model show that Ariel Tier 2 performance enables retrieval of chemical trends in planetary populations if these are present. The colours represent the distance to the true value (black dots) normalised to the corresponding retrieved 1-sigma error bars. From Changeat et al. (2020a). Left: an arbitrary linear trend between the water abundance and the effective temperature was imposed. Right: a trend dictated by chemical equilibrium between the methane abundance and the effective temperature was imposed.*

They enacted this trend by requiring a mixing ratio of $10^{-4}$ for an effective temperature of 1000 K and $10^{-3}$ for an effective temperature of 2000 K. Figure 3-12 (right) illustrates Ariel's ability to recover a non-linear relationship, caused by equilibrium chemistry, between the logarithmic abundance of methane and the temperature. Figure 3-12 clearly shows that the imposed trends are easily recovered from Ariel Tier 2 spectra.

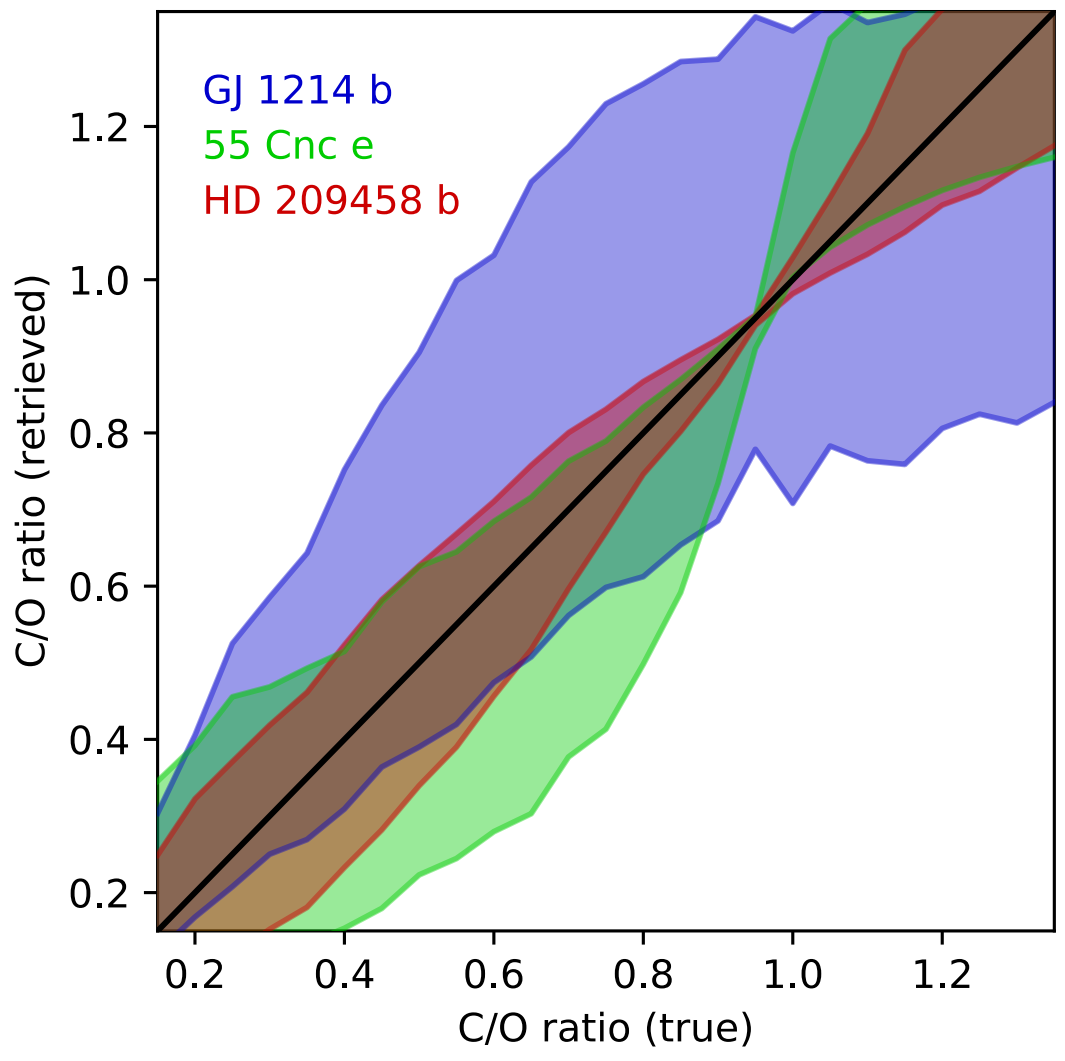

*Figure 3-13: Simulations generated with ARCiS exploring the ability of Ariel to retrieve the C/O from Ariel Tier 3 spectra (Min et al. 2020). Three representative exoplanets are considered: a hot Jupiter (HD209458b), a warm (GJ1214b) and a very hot (55 Cnc-e) low mass planet.*

### *3.1.5.3 Ariel sensitivity to constrain elemental ratios with Tier 2 and Tier 3 observations*

As explained in detail in Chapter 2, from planet formation theory we can predict the elemental abundance ratios in the atmosphere. To test the ability of Ariel to differentiate various formation scenarios, Min et al. (2020) used a range of elemental ratios as input to chemical models to simulate the molecular abundances in the atmosphere and the corresponding Ariel Tier 3 spectra. The simulated spectra contain equilibrium chemistry and a temperature structure computed from radiative-convective equilibrium. The ARCiS spectral retrieval tool was then used to assess the information content of the Ariel spectra generated with the different elemental ratios. Figure 3-13 shows the retrieved C/O, with uncertainties, in the case of three representative exoplanets: a hot Jupiter (HD 209458b), a warm small planet (GJ 1214b) and a very hot (55 Cnc-e) small planet.

### 3.1.6 Ariel Tier 4: phase-curves and bespoke observations

Missions flying earlier than Ariel or ground-based surveys, as well as Ariel in its Tier 1, are expected to identify "oddballs" that can have an impact on our general perspective of exoplanets. As a dedicated mission, Ariel has the capability and the time for a detailed study of these identified interesting objects which might require bespoke observations rather than the standard Tier-based observational strategy.

Thermal phase curves of a short-period planet constrain its atmospheric circulation—specifically the transport of energy from the day hemisphere to the night hemisphere of a synchronously-rotating planet (for a recent review, see Parmentier & Crossfield 2018). As described in Section 2.2 and in the Phase-Curve WG report 2020, Ariel phase-curve spectra will provide key observational constraints on the range of transport processes shaping the composition and vertical/horizontal structure of these atmospheres, e.g. Figure 2-9, Figure 3-14.

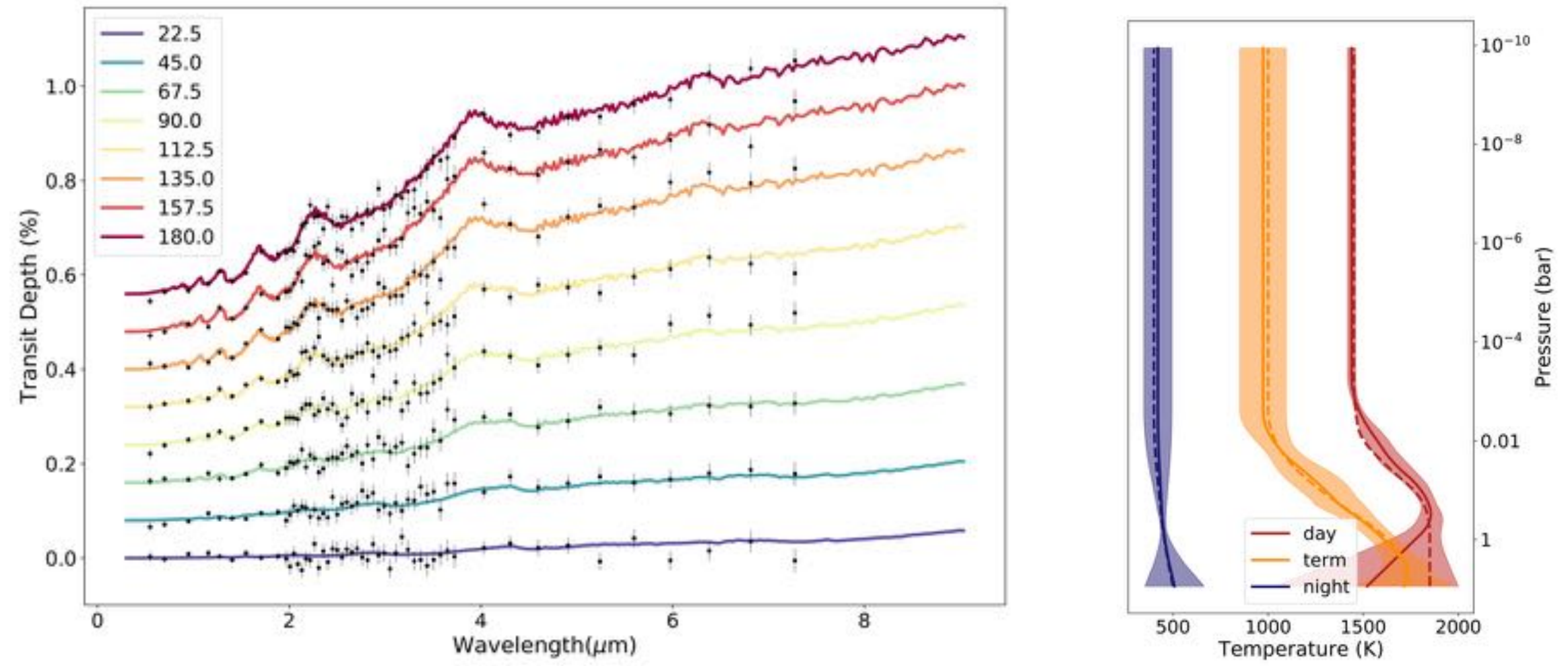

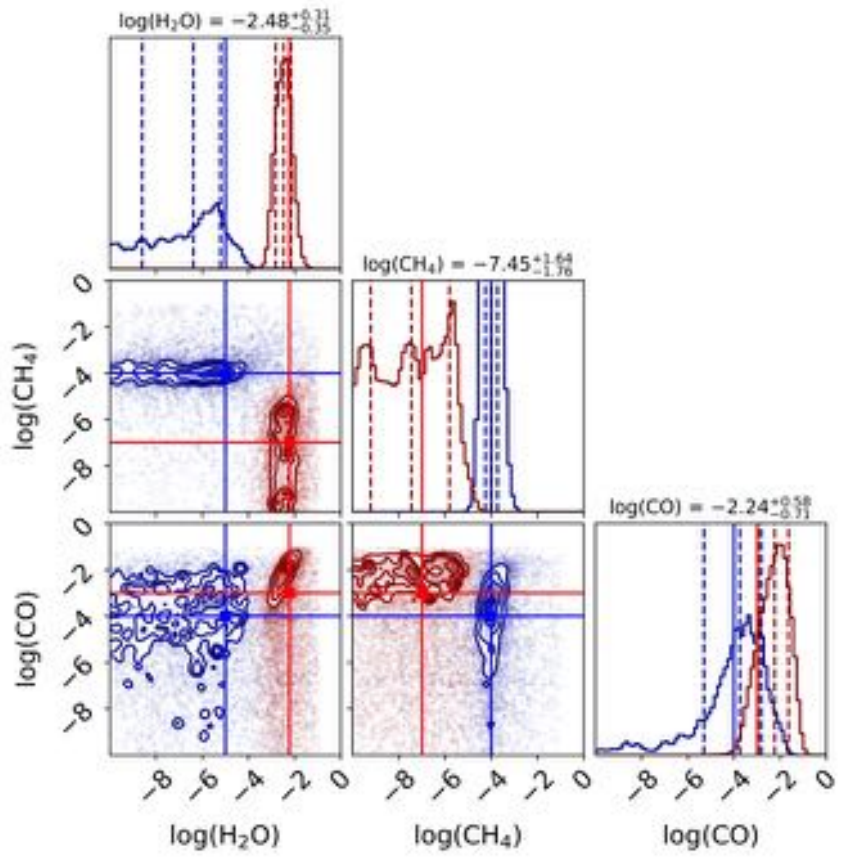

*Figure 3-14: Top left: Ariel Tier 2 spectra obtained at different phases for the hot-Jupiter WASP-43b. Top right: retrieved T-p profiles at the terminator, day-side and night-side. Left: retrieved trace gases. From Changeat et al. (2020c).*

Additionally, 3D climate models predict that the atmospheric heat redistribution decreases with atmospheric metallicity (Tremblin et al. 2020). Consequently, low-mass planets are expected to have high-amplitude phase curves (Charnay et al. 2015b), see Figure 3-15. The estimations of metallicity from Ariel phase-curves will be compared to measurements from transit spectroscopy. The evolution of atmospheric metallicity with planetary mass is an important constraint for planetary formation mechanisms.

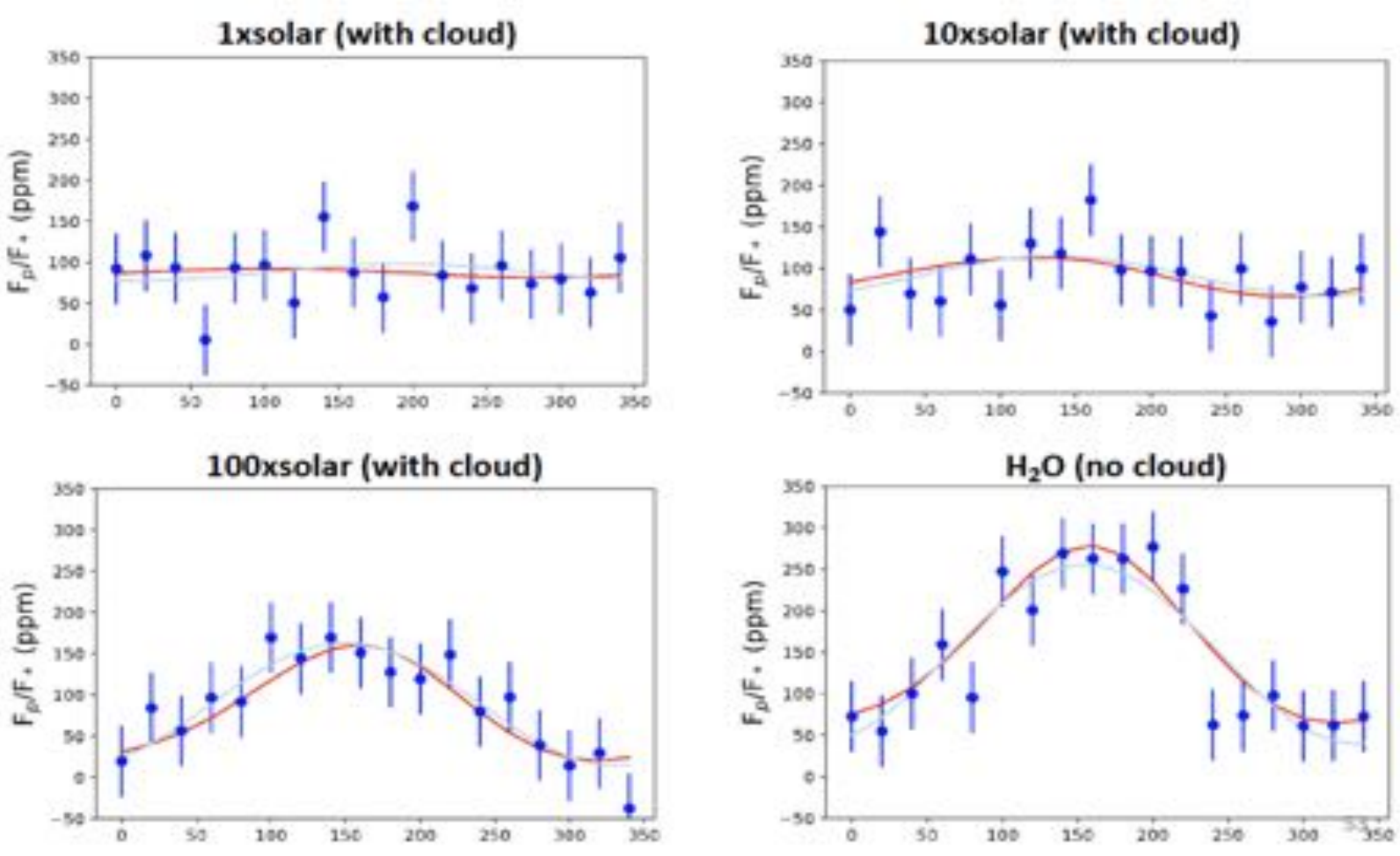

Phase-curves can also reveal the presence of an atmosphere on a rocky planet from the heat redistribution (Kreidberg et al. 2019b). Detecting atmospheres on hot super-Earths will provide a test for models of atmospheric escape.

*Figure 3-15: Simulated photometric phase-curve of GJ 1214b (from Charnay et al. 2015) for one orbit with AIRS-CH1 and for different atmospheric compositions. The red line is the GCM flux variations, blue points are simulated Ariel observations and the light blue curves show fits with simple sine functions.*

# 3.2 Science Requirements

## 3.2.1 Wavelength Coverage and Spectral Resolving Power

To fulfil the science requirements, Ariel has been specifically designed to have a stable payload and satellite platform optimised to provide a broad, instantaneous wavelength coverage to detect many molecular species, probe the thermal structure, identify/characterize clouds and monitor the stellar activity. The chosen wavelength range covers all the expected major atmospheric gases from, e.g. $H_2O$, $CO_2$, $CH_4$, $NH_3$, HCN, $H_2S$, through to the more exotic metallic compounds, such as TiO, VO, and condensed species (Figure 3-16).

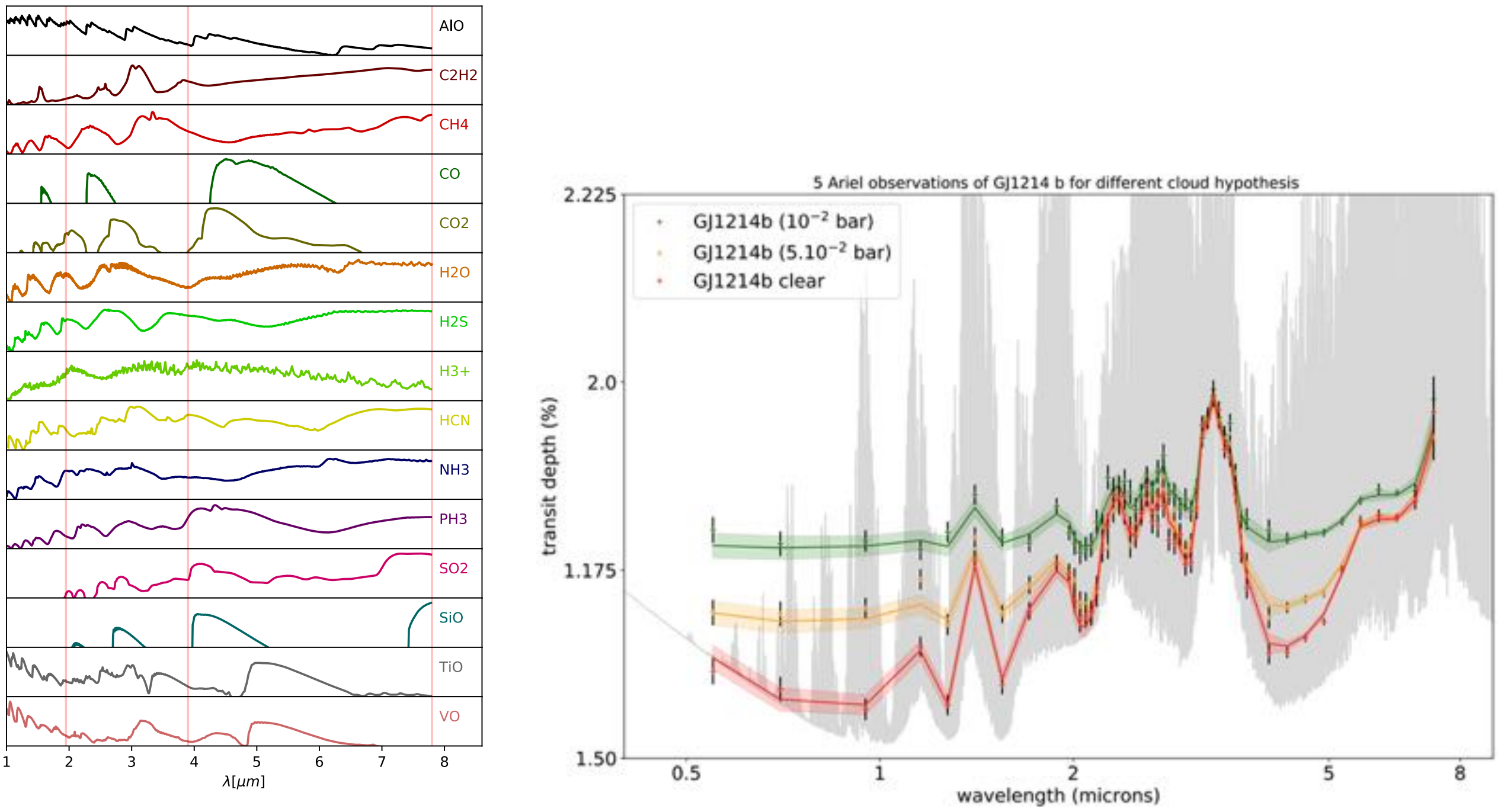

*Figure 3-16: Left: Molecular signatures in the 1-8 μm range at a constant resolving power = 300 (Chubb et al. 2020b). We notice that the molecular features are very broad in the IR and do not require a resolving power of 300 to be captured (see also Tinetti et al. 2013). The opacity strengths have been normalised to help the comparison. Right: simulated Ariel transit observations of the sub-Neptune GJ1214b in the 0.5-7.8 μm range. The atmosphere is assumed to be either cloud-free (red plot) or with clouds at different pressures (yellow and green plots). Grey area: cloud-free spectrum at the native spectral resolution of the model. Through transit, eclipse and phase-curve measurements Ariel will detect the presence of clouds/hazes, and constrain cloud parameters such as altitude, thickness, particle-size.*

Before and during Phase A, very thorough scientific and performance analyses were carried out to optimise the wavelength coverage and spectral resolving power of Ariel (e.g. Tinetti et al. 2013; Tessenyi et al. 2013; Pascale et al. 2015; Barstow et al. 2015; Encrenaz et al. 2015; Sarkar et al. 2016; Tinetti et al. 2018; see Table 3-2 for a summary of the scientific motivations). During Phase B, a suite of spectral retrieval simulations performed in parallel by different teams, were used to assess Ariel capability to observe individual planets and populations. The results of said analyses were used to inform an optimal design.

In Section 3.1.5.1 we show examples of how the simultaneous broad wavelength coverage provided by Ariel will enable detection of the key molecular species at high accuracy/precision (e.g. Changeat et al. 2019, 2020a; Mugnai et al. 2020; Min et al. 2020).

Compared to the initial proposal and the Phase A design, a few critical changes were introduced in Phase B: the FGS now provides 3 photometric bands and a low-resolution spectrometer from 1.1 µm to 1.95 µm. By increasing the number and optimising the positions of the photometric bands in the optical/NIR and adding a spectrometer in the NIR (Table 3-2), Ariel's capabilities have dramatically improved to characterise clouds and hazes in the atmospheres of exoplanets (Figure 3-17), as well as to monitor and mitigate the effects of stellar variability (see Chapter 7).

*Table 3-2: Summary of the Ariel required spectral coverage (left column) and resolving power (central column). The key scientific motivations are listed in the right column.*

| Wavelength range | Resolving power | Scientific motivation |
|---|---|---|
| **VISPhot**<br>0.50 – 0.60 µm | Integrated band | - Correction stellar activity (optimised early stars)<br>- Measurement of planetary albedo<br>- Detection of Rayleigh scattering/hazes |
| **FGS1**<br>0.60 – 0.80 µm | Integrated band | - Correction stellar activity (optimised late stars)<br>- Measurement of planetary albedo<br>- Detection/characterisation of clouds/hazes |
| **FGS2**<br>0.80 – 1.10 µm | Integrated band | - Correction stellar activity (optimised late stars)<br>- Detection/characterisation of clouds |
| **NIRSpec**<br>1.10 – 1.95 µm | R ≥ 15 | - Correction stellar activity (optimised late stars)<br>- Detection/characterisation of clouds/hazes<br>- Detection of molecules (e.g. $H_2O$,TiO,VO, metal hydrides)<br>- Measurement of planet temperature (optimised hot)<br>- Retrieval of molecular abundances<br>- Retrieval of vertical and horizontal thermal structure<br>- Detection temporal variability (weather/cloud distribution) |
| **IR spectrograph (AIRS)**<br>1.95 – 7.8 µm | R ≥ 100 (below 3.9 µm)<br><br>R ≥ 30 (above 3.9 µm) | - Detection of atmospheric chemical components<br>- Measurement of planet temperature (optimised warm-hot)<br>- Retrieval of molecular abundances<br>- Retrieval of vertical and horizontal thermal structure<br>- Detection of temporal variability (weather/cloud distribution) |

### 3.2.2 Science requirements for the four Ariel tiers.

In Section 3.1 we have described the four-tiered approach adopted to optimise Ariel scientific return. We report in Table 3-3 the summary of the science requirements for the survey tiers listed in Table 3-1.

*Table 3-3: Summary of the science requirements for the survey tiers given in.*

| Tier name | Science requirements |
|---|---|
| **Tier 1**<br>Reconnaissance Survey | - All planets in the sample<br>- 5+ spectral resolution elements covering the 1.10 – 7.80 µm range) measurements with average SNR ≥ 7<br>- Transit or eclipse |
| **Tier 2**<br>Deep Survey | - Spectroscopic measurements for a subsample (e.g. 50% of sample)<br>- R~10 for 1.10 < λ < 1.95 µm; R~50 for 1.95 < λ < 3.90 µm; R~15 for 3.90 < λ < 7.80 µm, with average SNR ≥ 7<br>- Transit and/or eclipse |
| **Tier 3**<br>Benchmark planets | - Spectroscopic measurements for a subsample (e.g. 10% of sample)<br>- R~15 for 1.10 < λ < 1.95 µm; R~100 for 1.95 < λ < 3.90 µm; R~30 for 3.90 < λ < 7.80 µm, with average SNR ≥ 7 achievable in 1-2 observations<br>- Transit and/or eclipse, repeated in time |
| **Tier 4**<br>Phase-curves and bespoke obs. | - Phase-curves, eclipse mapping, bespoke observations<br>- Multiple-band photometry/spectroscopy with SNR ≥ 10 |

## 3.3 Candidate Target List and Sample Diversity

To achieve the mission objectives described in Chapter 2, the Ariel target candidate sample is expected to be very diverse and include gaseous and rocky planets with a range of temperatures around stars of different spectral type, metallicity and activity level.

With this aim, we discuss in the following sections the diversity of the already known exoplanets (Section 3.3.1) and the "expected" ones yet to be discovered (Section 3.3.2). Additionally, we provide in Section 3.3.3 examples of exoplanets in extreme/unusual environments which could be included in the sample.

### 3.3.1 Ariel Candidate Target List today

Ariel will study a large population of planets already discovered by other facilities. In particular, it will focus on hundreds of warm/hot gaseous objects (Jupiters, Saturns, Neptunes) and of super-Earths/sub-Neptunes around bright stars of all types. The Ariel Radiometric model (ArielRad, Mugnai et al. 2020) has been developed to provide a comprehensive model of the instrument performance and ArielRad provides systematic noise on a case by case basis. A list of potential planet candidates to be observed by Ariel was created based on the expected performance predicted with ArielRad (Edwards et al. 2019). There are > 500 currently known planets complying with these requirements; see Figure 3-17. In the first two years of operations – the TESS extended mission was approved in 2019 – about 70 planets and 2400 TESS planetary candidates (TESS Objects of Interest; TOIs) have been identified. As expected, TOIs are significantly brighter than currently known Ariel planet candidates (see Figure 3-18).

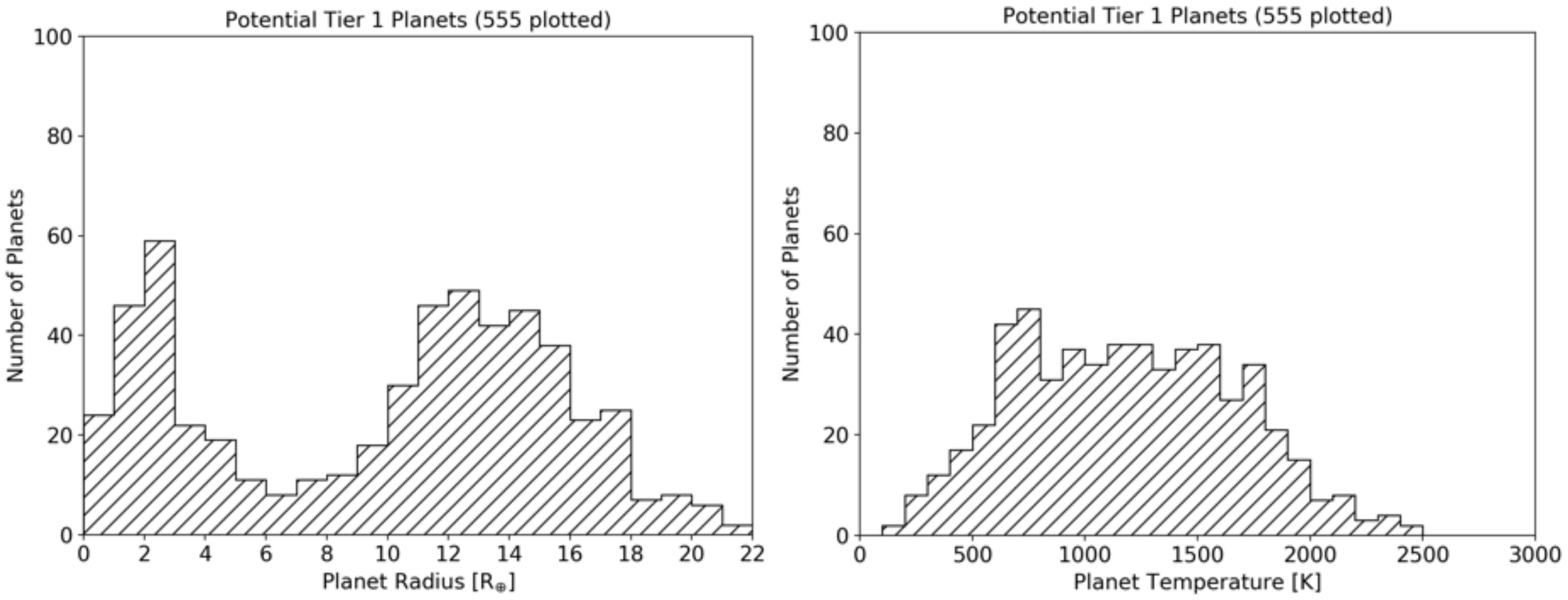

*Figure 3-17: Currently known Ariel planetary candidates plotted as a function of planetary radius (left) and planetary temperature (right).*

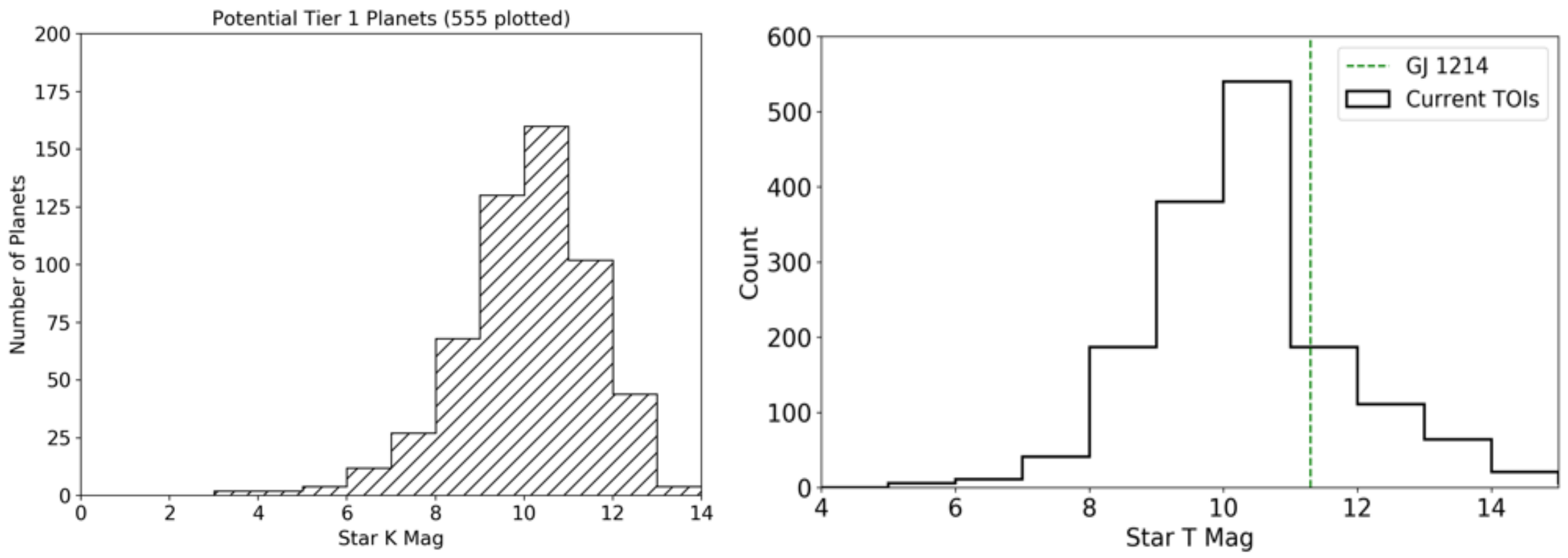

*Figure 3-18: Left: Brightness in K magnitude for currently known Ariel target candidates. Right: brightness of current TESS Objects of Interest shown in TESS magnitude (i.e. an integrated band over 0.6-1 μm). For comparison, we show the brightness of GJ 1214, which has K magnitude = 8.8. As expected, TOIs are brighter than currently known Ariel target candidates. Most of the 70 confirmed planets discovered by TESS orbit stars brighter than K mag = 8.*

Among the currently known planets, ~90 planets (including three small planets and three Neptunes) fulfil the requirements for phase-curve observations, described in Section 3.2 (Phase-curves WG report 2020; Charnay et al. 2020).

The current 500+ known targets have been discovered mainly close to the ecliptic plane because provided by ground-based surveys, as shown in Figure 3-19, illustrating also the sky visibility for Ariel.

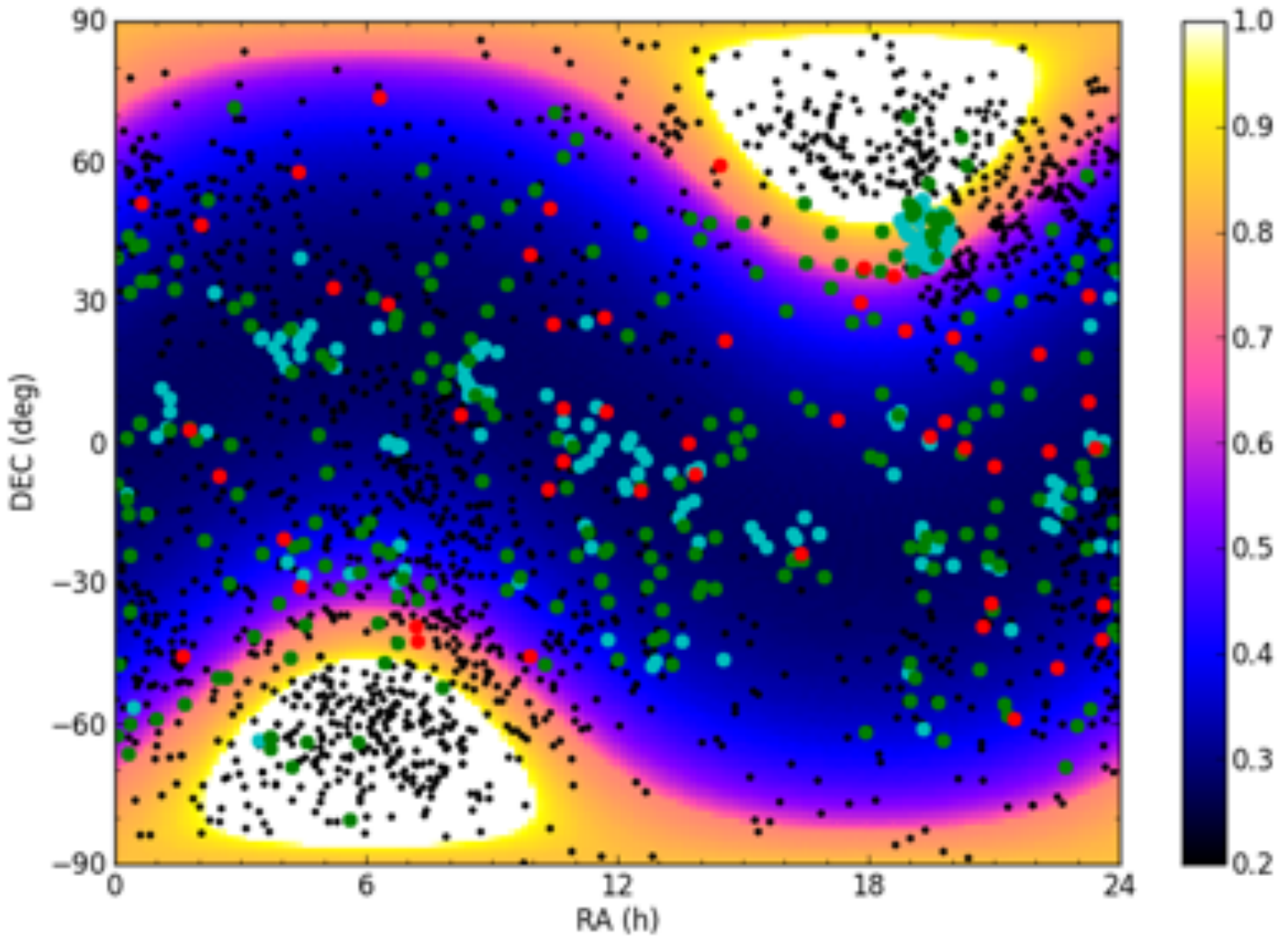

*Figure 3-19: A plot illustrating the fraction of the year for which a given location in the sky (in equatorial coordinates) is visible to Ariel, as seen from a representative operational orbit of Ariel around L2. Currently known transiting planets observable by Ariel are indicated as coloured dots, TOIs as black dots. Light blue dots: Tier 1 targets. Green dots: Tier 2 targets. Red dots: Tier 3 targets. Plots from Morales et al. (2020).*

K2, CHEOPS, Next Generation Transit Survey (NGTS) and other ground-based facilities (see Table 3-3-3) are expected to complete the search for planets around bright sources closer to the ecliptic plane. TESS and PLATO will extend the planet search closer to the ecliptic poles which are where Ariel has continuous coverage (see distribution of TOIs in Figure 3-20).

### 3.3.2 Ariel Candidate Target List at launch

Over 4300 extra-solar planets have been detected to date and TESS is anticipated to detect over 4500 planets around bright stars and more than 10,000 giant planets around fainter stars (Barclay et al. 2018).

Several surveys both from ground and from space will provide targets with the necessary characteristics to meet the objectives of the mission. Table 3-4 and Table 3-5 summarise the most important surveys from which we expect a significant contribution to the final core sample. The list is not exhaustive.

*Table 3-4: Summary of the main transit surveys that will provide targets for Ariel in the next ten years.*

| Transit from ground | Survey/Facility |
|---|---|
| Past / Ongoing | HATNet/HATSouth (Bakos et al. 2004)<br>WASP/SuperWASP (Pollacco et al. 2006)<br>MEarth (Nutzman et al. 2008)<br>TRAPPIST (Gillon et al. 2016)<br>APACHE (Sozzetti et al. 2013)<br>XO (McCullough et al. 2005)<br>TrES (Alonso et al. 2004)<br>NGTS (Chazelas et al. 2012) |
| **Space Transit & astrometry** | **Survey/Facility** |
| Past / Ongoing | CoRot (Auvergne et al. 2009; Fridlund et al. 2006)<br>Kepler (Borucki et al. 2010)<br>K2 (Howell et al. 2014)<br>Gaia (Perryman et al. 2014)<br>TESS (Ricker et al. 2014; Barclay et al. 2018)<br>CHEOPS (Benz et al. 2020) |
| Future | PLATO (Rauer & Heras 2018) |

*Table 3-5: Summary of the main Radial Velocity (RV) surveys/projects that will provide targets for Ariel in the next ten years (from Ariel RV WG report 2020). The instrument name, telescope name and size are indicated. Future instruments are indicated in orange. The wavelength domain is also indicated. † precise radial velocity (~1-2 m/s); ‡ extreme precision radial velocity (sub-m/s precision).*

| RV Instrument | Telescope | Size | RV Instrument | Telescope | Size |
|---|---|---|---|---|---|
| **Optical** | | | **Optical** | | |
| APF/Levy† | Lick 2.4m | 2.4 | OES | Perek telescope | 2.0 |
| CAFÉ | 2.2m | 2.2 | PFS† | Magellan II | 6.5 |
| CARMENES† | 3.5m | 3.5 | PLATOSPec† | 1.52-m ESO | 1.5 |
| CHIRON† | 1.5m | 1.5 | SALT HRS | SALT | 10.0 |
| CORALIE† | Euler | 1.2 | SONG | Hertzsprung | 1.0 |
| Dharma RVSurvey | DEFT | 1.3 | SONG | Chenese node | 1.0 |
| Espadons | CFHT | 3.6 | SONG | Australian node | 0.7 |
| ESPRESSO‡ | VLT | 8.2 | SOPHIE† | 1.93m | 1.9 |
| EXPRES‡ | DCT | 4.3 | TOU | AST | 2.0 |
| FIES | NOT | 2.6 | TRES | Tillinghast | 1.5 |
| FOCES | Frauenhofer | 2.0 | Tull Echelle | Harlan J. Smith | 2.7 |
| G-CLEF | GMT | 25.0 | **NIR** | | |
| GRACES | Gemini-N | 8.0 | VELOCE | AAT | 3.9 |
| HARPS† | ESO 3.6 | 3.6 | APOGEE | 2.5m Sloan | 2.5 |
| HARPS-N† | TNG | 3.6 | CARMENES† | 3.5m | 3.5 |
| HARPS-Terra† | INT | 2.5 | CRIRES+ | VLT | 8.2 |
| HERMES | Mercator | 1.2 | CSHELL /iSHELL† | IRTF | 3.2 |
| HIRES† | Keck | 10.0 | GIANO-B | TNG | 3.6 |
| HRS† | HET | 10.0 | HPF‡ | HET | 10.0 |
| KPF‡ | Keck | 10.0 | iGRINS | Gemini South | 8.1 |
| MAROON-X‡ | Gemini North | 8.1 | iGRINS | DCT | 4.3 |
| MARVEL† | MARVEL | 0.8 | iGRINS | Harlan J. Smith | 2.7 |
| MINERVA | MINERVA | 0.7 | iLocater | LBT | 11.9 |
| MINERVA-Australis | MINERVA-Australis | 0.7 | IRD | Subaru | 8.2 |
| NEID‡ | WIYN | 3.5 | NIRPS† | ESO 3.6 | 3.6 |
| NRES | LCO | 1.0 | NIRSPEC2 | Keck | 10.0 |
| NRES | LCO | 1.0 | PARAS/PARAS-2 | MIRO | 1.2 |
| NRES | LCO | 1.0 | PARVI | Hale | 5.1 |
| NRES | LCO | 1.0 | SPIRou | CFHT | 4.2 |

The predicted TESS discoveries from Barclay et al. (2018) were combined with the currently known exoplanets to create a list of potential planets for Ariel. Potential discoveries by other surveys (PLATO, NGTS etc.) have not been included yet in this analysis but are expected to provide thousands of planets which could be suitable for study with Ariel, enhancing the population of planets from which the final target list will be selected. A large number of planets were found to be suitable for observing with Ariel in each tier. For Tier 1, there are around 2000 potential planets for which the science requirements can be reached with five observations or less (Edwards et al. 2019). Being over-saturated in the number of possible targets is useful as it allows for redundancy in the scheduling of observations and it means there is a large catalogue of planets to draw from to allow for a diverse sample to be observed.

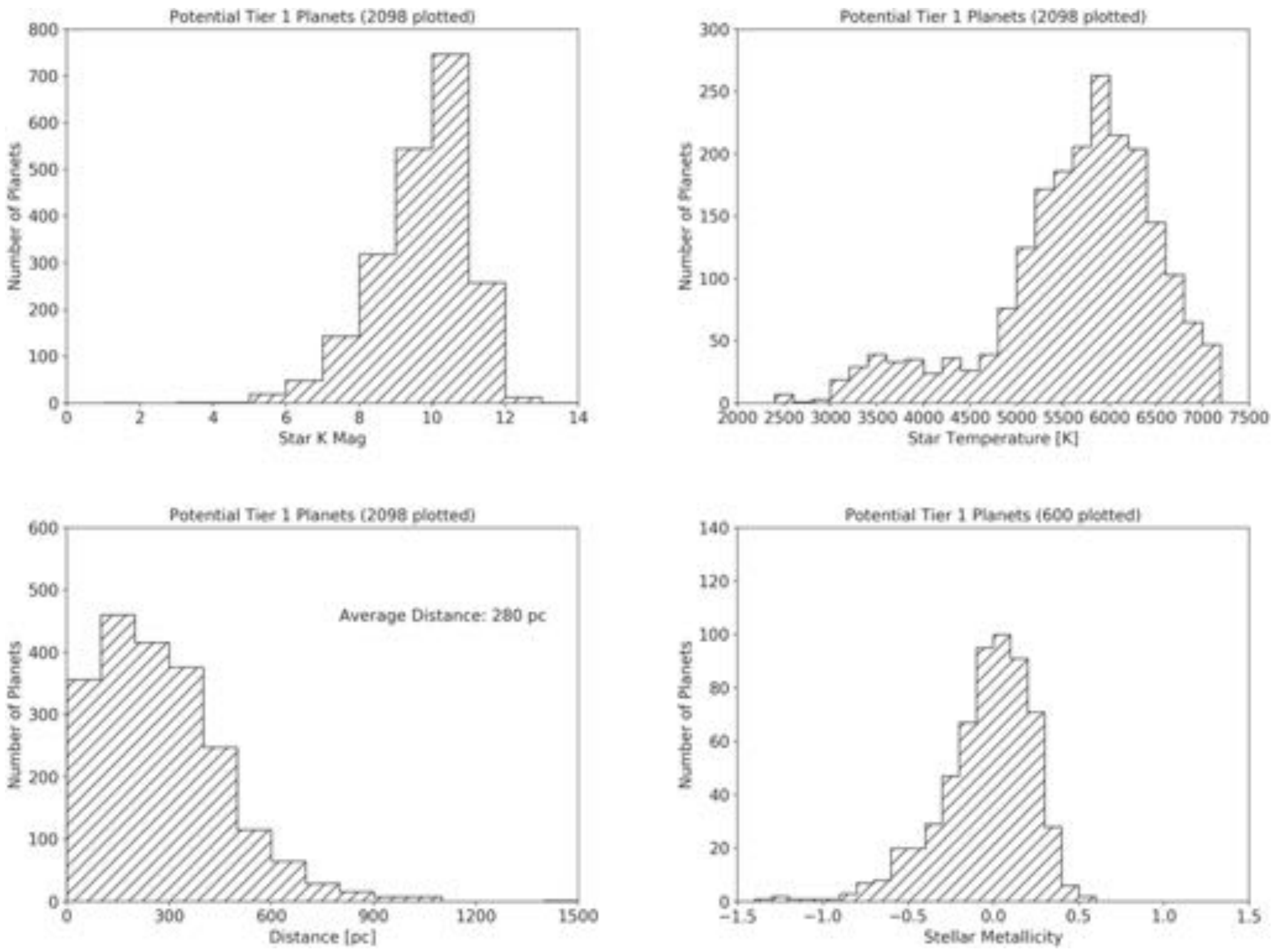

*Figure 3-20: Histograms of the properties of the stellar hosts within the potential Ariel Tier 1 catalogue. Metallicities were not available for all host stars (Edwards et al. 2019).*

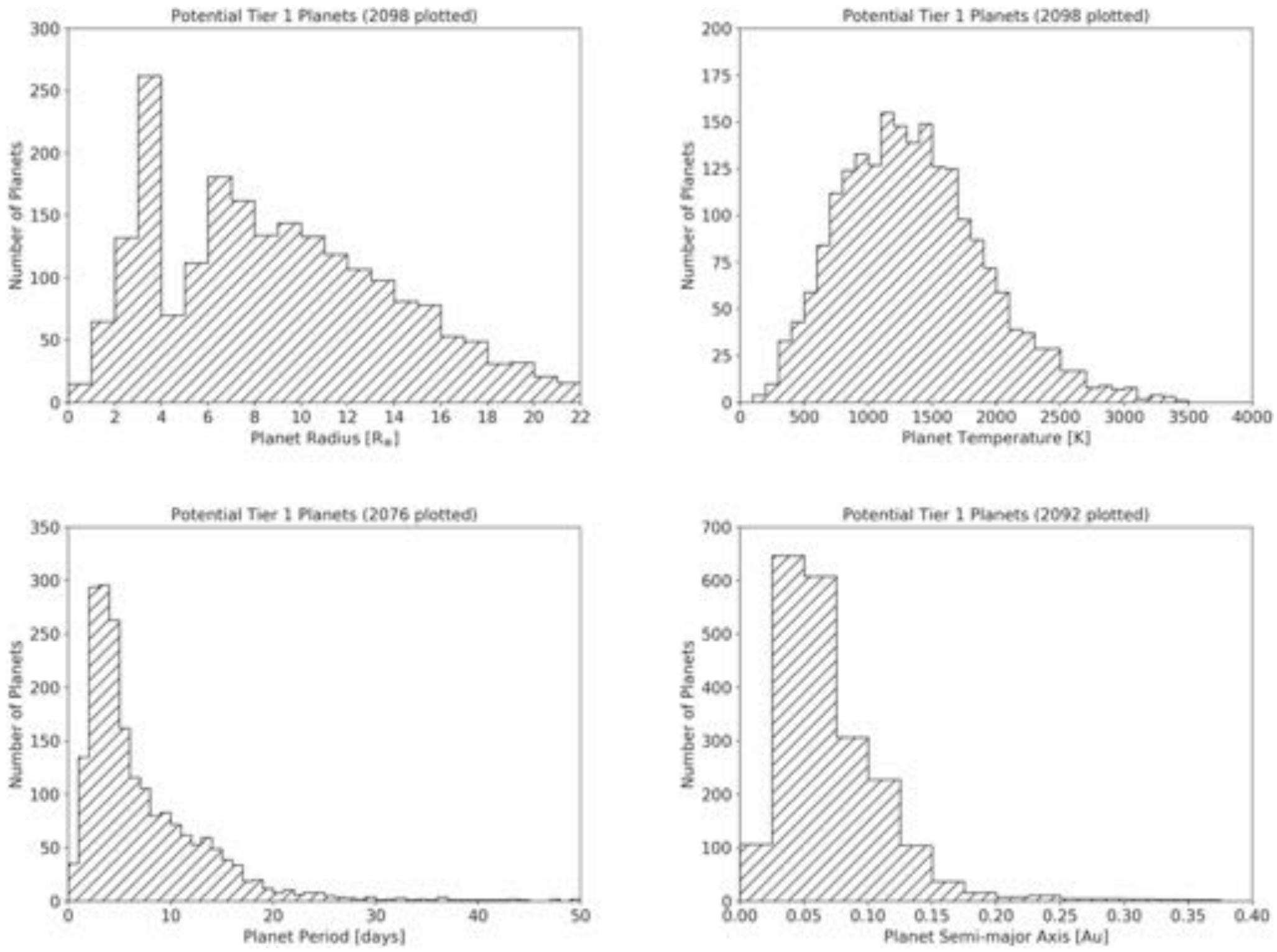

*Figure 3-21: Histograms of the planetary properties within the potential Ariel Tier 1 catalogue. In some cases, not all planets are plotted for aesthetic reasons (Edwards et al. 2019).*

The distribution of various stellar and planetary parameters for these potential Tier 1 targets is shown in Figure 3-20 and Figure 3-21. These show that: (i) to achieve a sample of ~1000 planets, Ariel does not need to observe faint stars (except for special targets of interest); (ii) there is a large diversity in planetary temperatures and radii; (iii) the spectral type of the planet-hosting stars is varied although FG stars are more dominant; (iv) the majority of potential targets are located within a few hundred parsecs; (v) most potential targets are close to their stars and have orbits of under 20 days; (vi) although the metallicities of many of the host stars is unknown, there is a wide range of values included in the sample.

Ariel will have uninterrupted visibility of the ecliptic poles with a partial visibility of the whole sky at lower latitudes. The sky locations of possible planets for study in each tier with Ariel are shown in Figure 3-22 left and they are found to be well distributed across the sky but with a noticeable gap close to the ecliptic due to a lack of TESS coverage in its primary mission.

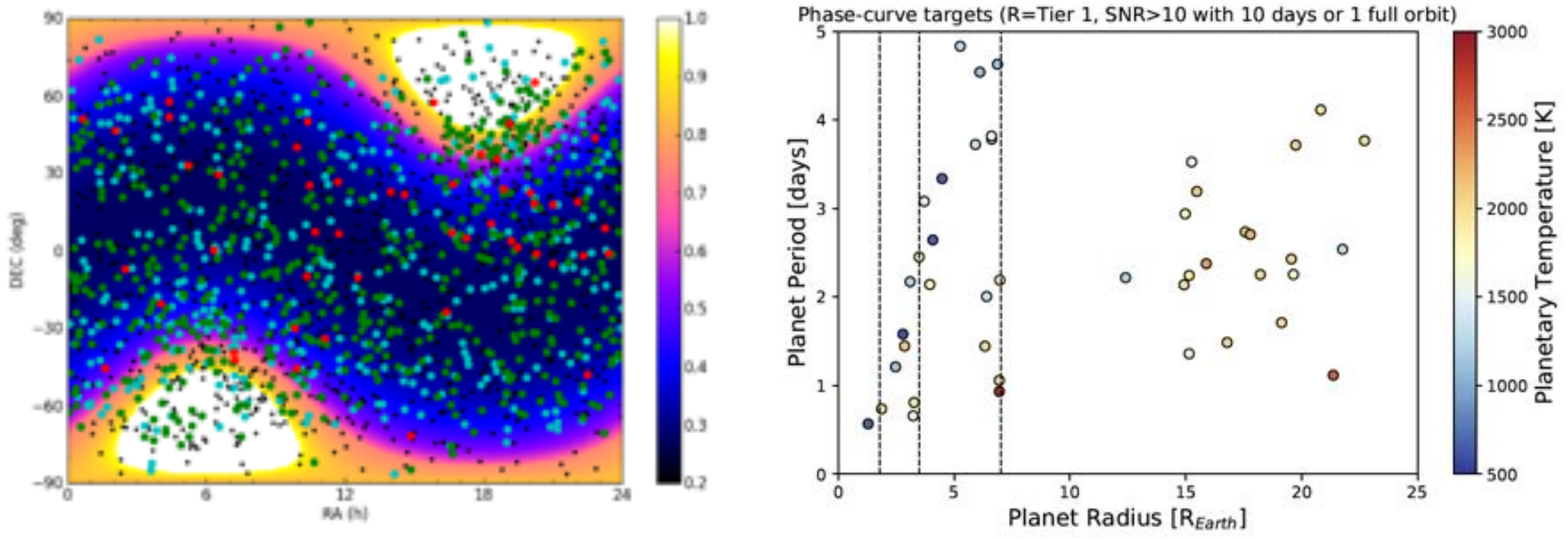

*Figure 3-22: Left: A plot illustrating the fraction of the year (colour bar) for which a given location in the sky (in equatorial coordinates) is visible to Ariel, as seen from a representative operational orbit of Ariel around L2. Currently known and expected transiting planets observable by Ariel are indicated. Light blue dots: Tier 1 targets. Green dots: Tier 2 targets. Red dots: Tier 3 targets. Black dots: back-up targets. Plot from Morales et al. (2020). Right: Optimal targets for phase-curve observations as a function of planetary radius, planet period and temperature (colour bar) as identified by the Phase-Curve WG. Vertical dashed lines represent the radius limits.*

The Ariel Phase-Curve Working Group has estimated that the general requirement to address the key science questions is to reach a SNR ≥ 10 (i.e. detection at 10 sigma) for photometric phase curves, assuming no heat redistribution and no offset. For spectroscopic phase-curves, they recommended the requirement to reach SNR≥ 10 at Tier 2 resolution to get measurements for ten different phases. If one includes in the simulations TESS expected targets, a list of 44 optimal planets has been identified (Figure 3-22 right). About 17% of the mission life-time (13% if we remove the time needed for the transit and eclipse measurements) would be required to complete the observations for all 44 planets.

### 3.3.3 Study of planets in rare and/or extreme conditions with Ariel

As well as the population studies described in the previous paragraphs, Ariel will observe a number of planets in extreme/odd conditions to test the physics in those extreme/unusual environments and get a glimpse of the characteristics of those exotic objects. We indicate here a few key examples:

- **Planets in highly eccentric orbits –** In contrast to Solar System planets, a large fraction of exoplanets discovered today have eccentric orbits around their parent stars. In some cases the eccentricity is extreme, e.g. 0.98 for HD 80606b. These planets are important from the perspective of planetary formation and evolution as their large orbital eccentricity is a strong indication that they underwent some violent dynamical event in their past (e,g. chaotic evolution or planet-planet scattering). This makes them prime candidates to study the different compositional signatures produced in planets by different dynamical histories (e.g. disc-driven migration vs. planet-planet scattering; see Turrini et al. 2018 and Ariel Planetary Formation WG report 2020 for a discussion). Also from a climate/chemistry perspective these planets represent a very interesting challenge (e.g. Williams & Pollard 2002): for instance Laughlin et al. (2009) measured with Spitzer the thermal properties of HD 80608b at periastron, finding that the planet temperature increased from 800 K to ~1500 K over a six-hour period. Maggio et al. (2015) observed with XMM the highly eccentric HD 17156b: its parent star showed enhanced chromospheric and coronal emission a few hours after the passage of the planet at the periastron suggesting a complex planet-star interaction. Ariel will observe these and other similar planets to study the climate, chemistry, and composition of highly eccentric planets.

- **Circumbinary planets –** Over twenty circumbinary planets, i.e. planets orbiting binary stars, including both S-orbits (planets orbiting just one of the 2 stars) and P-orbits (planets orbiting both stars at large distance) have been discovered. As in the case of eccentric planets, from a climate/chemistry perspective these circumbinary planets represent a very interesting challenge. Ariel will observe transiting circumbinary planets to study the climate, and chemistry of these exotic bodies.
- **Transiting multi-planet systems –** Among the 4300+ planets discovered so far, about 700 planets are part of a multi-planet system. Special examples are Kepler 90 (eight transiting planets, Lissauer et al. 2011; Shallue & Vanderburg 2017), and TRAPPIST1 (seven Earth-size planets transiting in the temperate zone of an ultra-cool dwarf, Gillon et al. 2017). Transiting multiplanet systems give us a unique opportunity to study not just a single object, but to enable comparative exoplanetology for planets within the same extrasolar system and to gain a deeper insight on the compositional signatures created by different dynamical evolution paths of planetary systems as a whole. As discussed in Sect. 2.2.4, population studies revealed how widespread chaos and violent dynamical events are in the life of multi-planet systems (Limbach & Turner, 2015; Zinzi & Turrini, 2017; Laskar & Petit, 2017; Turrini et al. 2020b). Part of the efforts of the Ariel Mission Consortium focused on developing metrics linked to the angular momentum deficit of planetary systems (Turrini et al. 2020b) to extract information from their architectures and constrain the violence of their past history. Ariel observations of those and other transiting multi-planet systems will reveal the intra-planetary-system diversity outside our solar system and frame it within the context of the dynamical histories of the systems seen as whole entities.
- **Planets around flaring stars** – Venot et al. (2016) investigated how the activity of a star can influence the chemical composition and resulting spectra of typical exoplanets. They focused on the effect of stellar flares (see section 8.1.2 for a discussion about flaring stars), and found significant changes in the chemistry of the atmospheres of two typical planets around an active M star. These changes are visible in the transit spectra of these planets, and the resulting differences are observable with Ariel. Ariel's unique ability to measure a broad wavelength spectrum in one shot will enable the study of the atmospheres of planets around flaring stars with great accuracy and the testing/validation of theoretical predictions about planetary atmospheres in these extreme environments.
- **Highly irradiated brown dwarfs** – There are currently only 23 brown dwarfs known to transit a main sequence star (Carmichel et al. 2020 and references therein), and a further 10 known to orbit white dwarfs (e.g. Casewell et al. 2020, 2018). The list of transiting brown dwarfs, despite being rare objects among substellar companions, is increasing after the analysis of the precise TESS light curves combined with radial velocity follow-up (e.g. Pallé et al. 2020, in preparation). These brown dwarfs are highly irradiated and as such provide an opportunity to explore the effects of irradiation on a brown dwarf atmosphere. As brown dwarfs have higher internal heat than most hot-Jupiters (e.g. Showman 2016), these rare objects allow to test irradiated model atmospheres not only in a higher gravity environment, but also with a different temperature-pressure profile than is usually assumed for exoplanets. Performing transit spectroscopy for these brown dwarfs is extremely challenging due to the small scale height of the brown dwarf atmosphere, but emission spectroscopy with Ariel will be possible for some of these objects. Beatty et al. (2020) have recently observed Kelt-1b in the near-IR with HST. The broader wavelength range covered by Ariel, will make this technique particularly powerful for these irradiated brown dwarfs, as well as for the highly irradiated objects orbiting white dwarfs where the brown dwarf spectral energy distribution can be directly detected. These observations will offer an excellent opportunity for the comparison of the Ariel data with the observations available for free-floating and widely separated brown dwarfs of similar mass and temperature. Through the comparative analysis of the spectral properties, we will discover whether the atmospheric properties are equal or deviating, which may be a signpost for different formation scenarios.

# 4 Payload

## 4.1 Payload Architecture

The baseline architecture for the Ariel payload is illustrated in Diagram 4-1. This diagram also shows the nationalities of the members of the Ariel Mission Consortium (AMC) who are taking responsibility for each element.

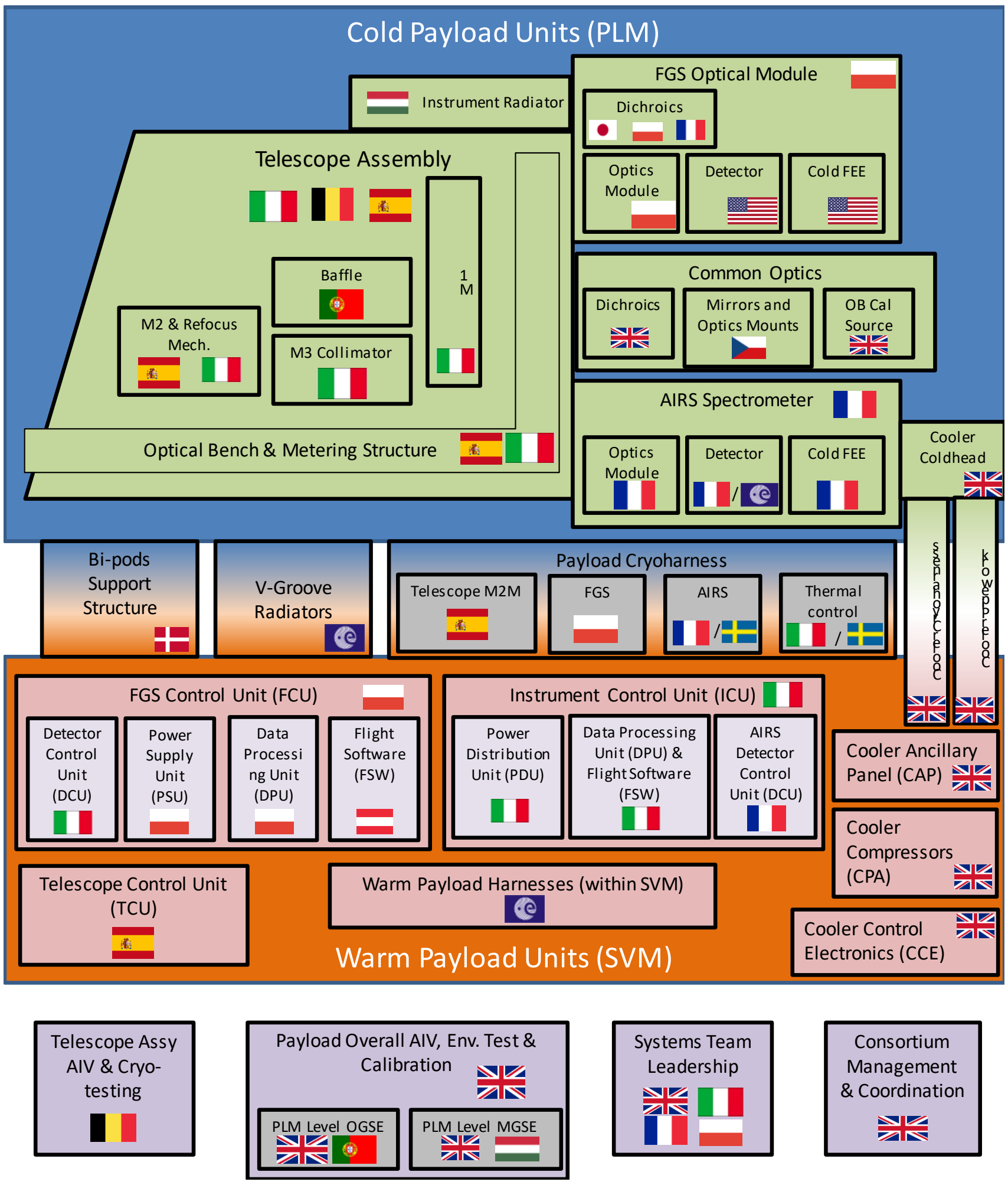

*Diagram 4-1: Ariel Payload architecture and responsibilities*

The baseline architecture splits the payload into two major sections, the cold payload module (PLM) and the items of the payload that mount within the spacecraft service module (SVM). The major items are:

- Cold PLM:
  - Telescope system, incorporating M1, M2 & M3 mirrors, a re-focusing mechanism on the M2 mirror (M2M), the telescope structure and baffles.
  - An optical bench / telescope metering structure onto which the telescope and instruments are mounted.
  - A set of common optics including fold mirrors, dichroics, formatting optics, and a common calibration source.
  - The Ariel IR Spectrometer (AIRS) instrument optical module and cold front end electronics module.
  - The Fine Guidance Sensor / Visible Photometer / Near-IR Spectrometer (FGS / VISPhot / NIRSpec) instrument optical module and cold front end electronics module.
  - Thermal hardware including: active cooler coldhead, passive radiators for cooling of the PLM & the V-grooves and support structures (bi-pods) to isolate the PLM thermally from the warmer SVM.
- Warm SVM mounted units:
  - Instrument Control Unit (ICU) housing the AIRS detector control unit (DCU), a central data processing unit (DPU), and a power supply unit (PSU);
  - the Telescope Control Unit (TCU) including the drivers for the M2M mechanism and the on-board calibration source plus the readout and control for most of the payload temperature sensors;
  - FGS Control Unit (FCU) electronics incorporating the FGS / VISPhot / NIRSpec wFEE and the processing electronics and SW for determining the pointing from the FGS data;
  - Active Cooler System (ACS) consisting of the Cooler Drive Electronics (CDE), the Cooler Compressors and the Cooler gas handling panel (for further details see Section 4.4).

Extensive details of the design of the Ariel payload are contained in Ariel Payload Design Description (2020). The overall layout of the payload and the baseline design are shown in Figure 4-1.

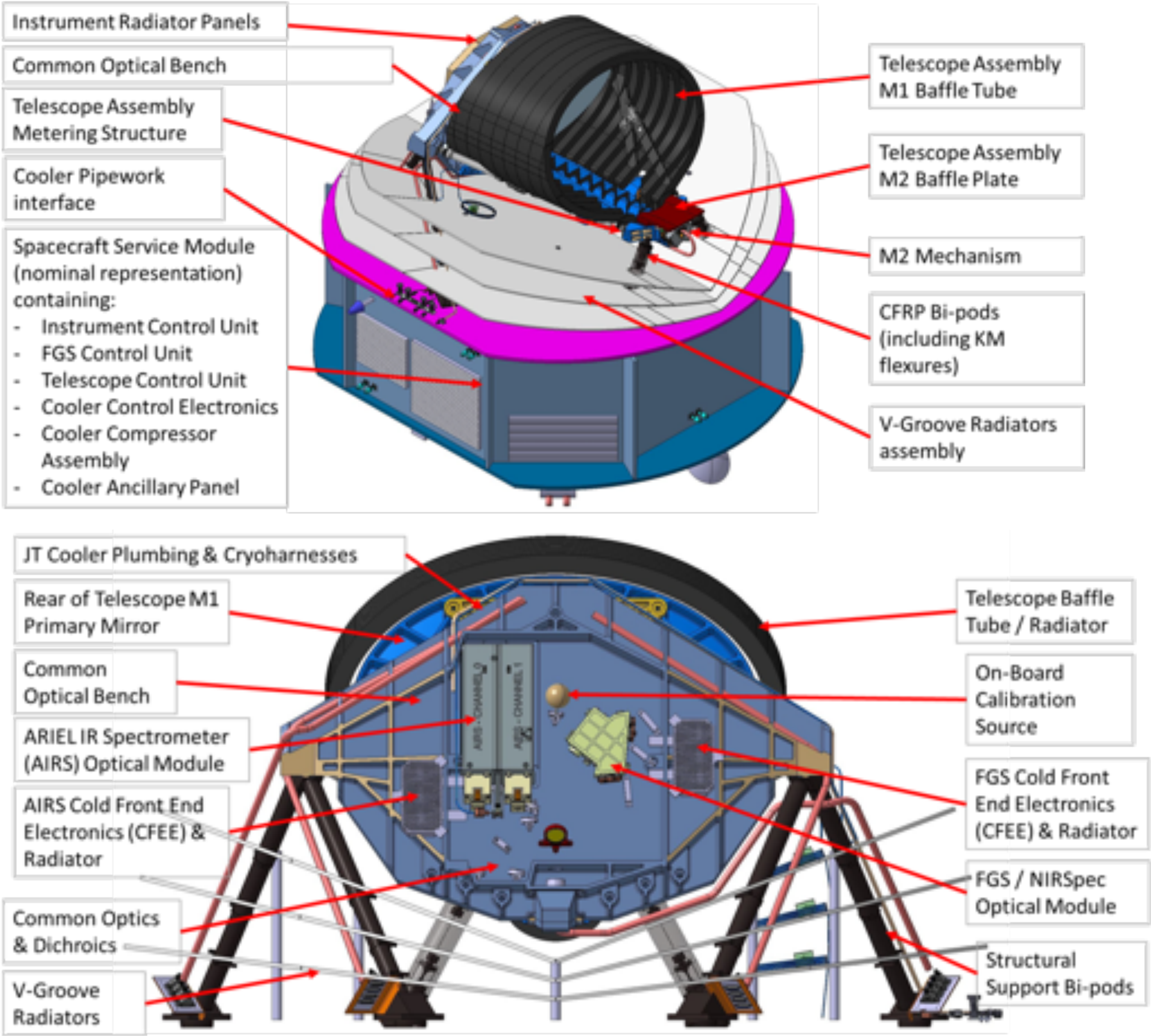

*Figure 4-1: Ariel PLM overall mechanical configuration and layout.*

### 4.1.1 Power and Data Rate Budgets

The power and data rate budgets for the Ariel payload are reported in detail in the Ariel Payload Budgets Report (2020). This shows that the estimated total payload power required during observations (including uncertainty and margin) is 173.6 W.

The data rate budget from the payload for all science data and payload housekeeping is estimated to be 236.0 Gbits / week, made up of 54.1 Gbits/week from the FGS, 126.8 Gbits/week from AIRS, 3.1 Gbits/week for housekeeping and the ACS monitoring and an ~22% data volume margin.

### 4.1.2 Throughput / Figure of Merit Budget

At system level, a budget of the Figure of Merit of the payload (product of collecting area, throughput and detector QE) is maintained to as a top level metric of the overall payload efficiency and to inform the performance modelling (reported in Section 4.10 below). The throughput budget accounts for efficiency losses due to: surface reflections, internal transmission, ageing, pupil shear vignetting, contamination effects and filter efficiencies. The current best estimate for the end-to-end figure of merit is shown in Figure 4-2 and compared to the performance requirements derived by the payload team which are needed to meet the top level science objectives of the mission.

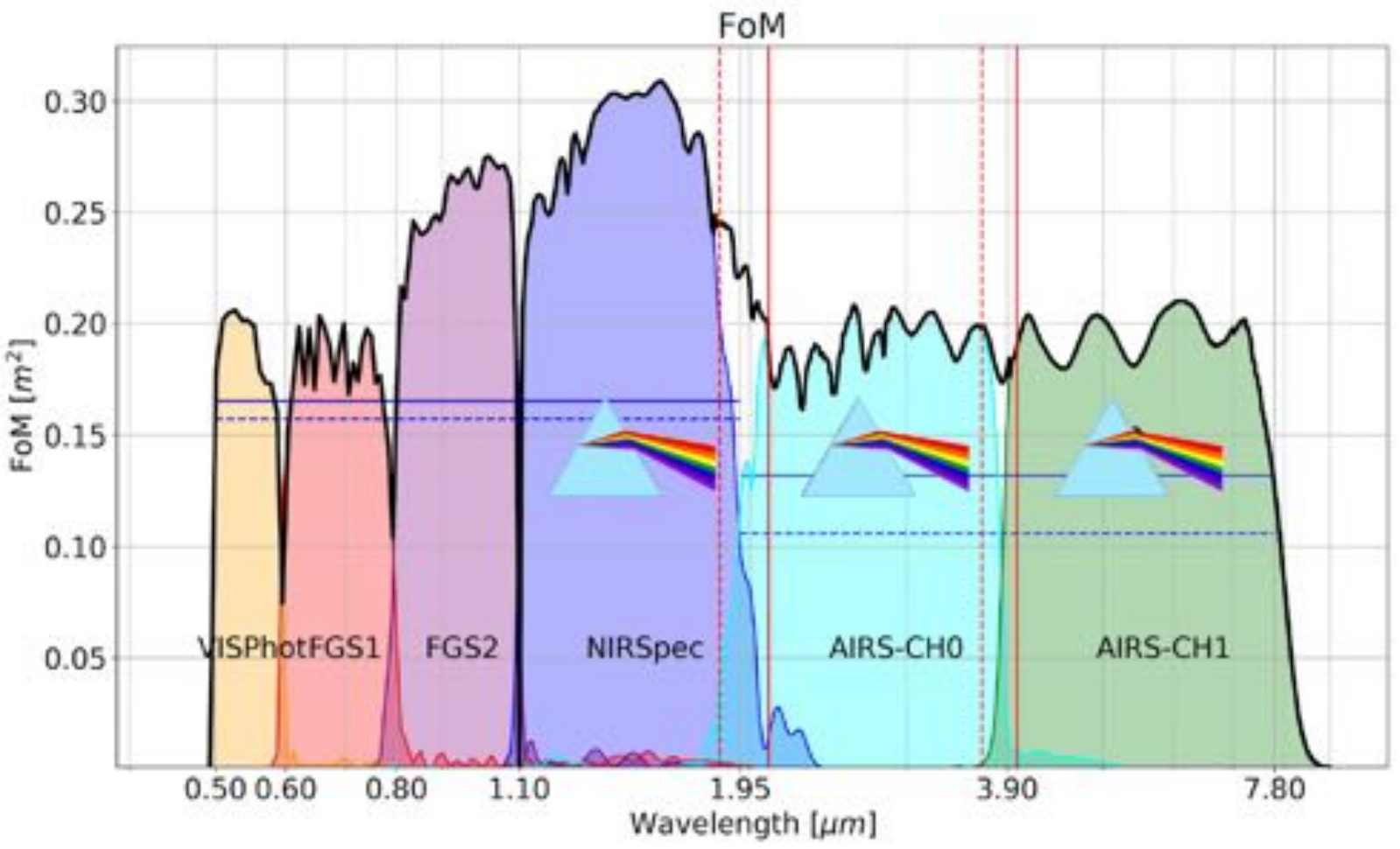

*Figure 4-2: End-to-end Figure of Merit for the payload channels. The derived requirement from the PRD is also shown, the solid line is the required band average and the dotted line is the minimum allowable at any point in the band..*

## 4.2 The Ariel Telescope Design

### 4.2.1 Optical Design

The baseline telescope design is an afocal unobscured off-axis Cassegrain telescope (M1 and M2) with a recollimating off-axis parabolic tertiary mirror (M3). The system aperture stop is located at the M1 surface. The M1 aperture is an ellipse with major axis dimensions of 1100 mm × 730 mm. M4 is a folding mirror. See Figure 4-3 left for the telescope layout. The nominal optical design is unchanged since the Mission Selection Review. Figure 4-3 right shows the telescope optical layout also including the first common optics mirror, M5.

The entrance baffle, a cylinder extending the length of the optical bench, limits M1's view of the sky as best practice to limit straylight and to act as a large passive radiator. In combination with placing the stop at the first optical surface (M1), this provides the first line of defence to block out-of-field light. An additional baffle is positioned over the 'top' of M2 (as viewed in Figure 4-3 left) to block any direct view of the sky from M2 past the end of the entrance baffle. The Cassegrain focus after M2 provides the location to insert a field stop to aid stray-light rejection. After the Cassegrain focus, the beam is recollimated by M3. The baseline design results in a recollimated beam of size 20 mm × 13.3 mm, which is directed onto flat mirror M4, used to direct the beam onto the optical bench. The telescope is required to be diffraction limited at approximately 3 µm, which equates to an rms Wave Front Error (WFE) of about 200 nm.

M2 has a focus adjustment mechanism with three degrees of freedom as a baseline (axial movement in focus and tip/tilt). The purpose is to correct for one-off movements due to launch loads and cool-down and potentially to make occasional adjustments (for example to compensate for any long term drifts in structural stability). The optimum setting is determined by ground analysis of FGS image vignettes and AIRS spectral data during commissioning.

A detailed trade-off of the material to be used for the telescope mirrors and structures has been carried out during the assessment phase. The conclusion is that for the consortium provision of the telescope the optimum solution is a telescope with all mirrors and structures made from Aluminium 6061 T651 alloy.

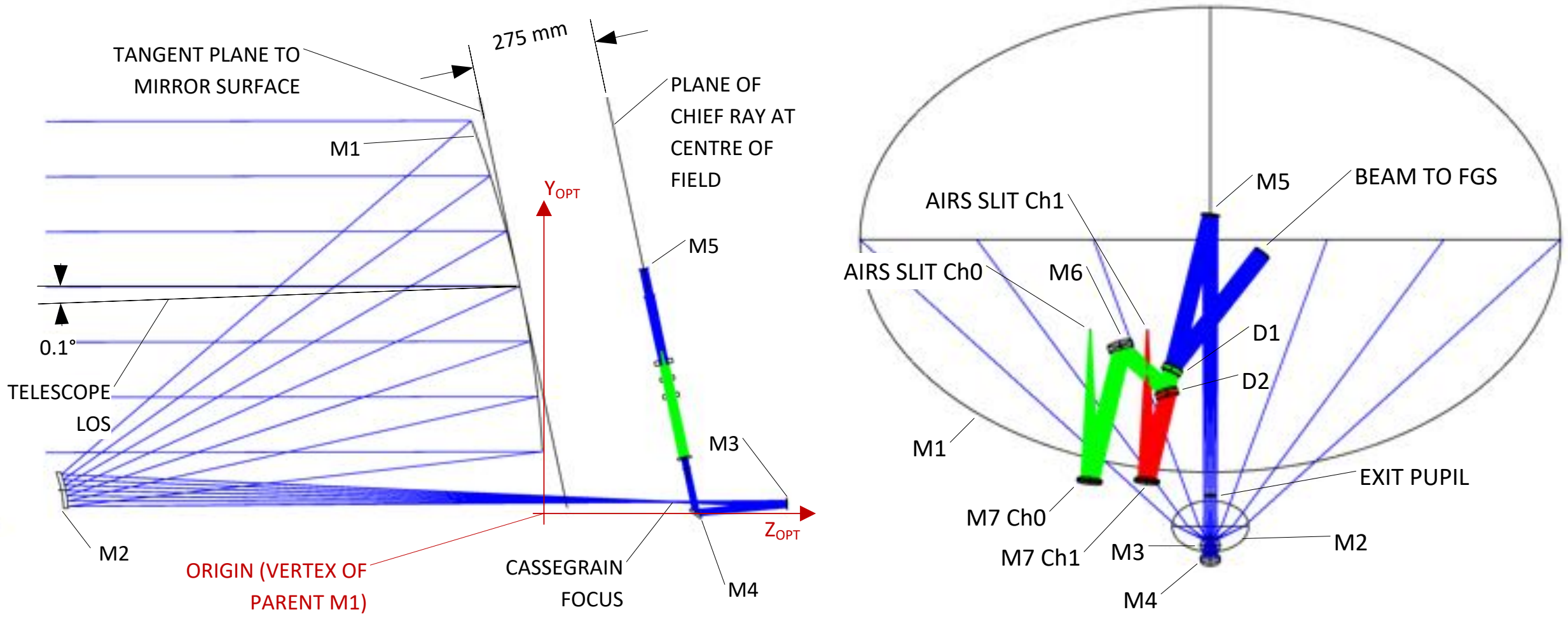

*Figure 4-3: Scale drawings of the telescope and common optics. (Left): view in $Y_{OPT}$-$Z_{OPT}$ plane. The 0.1° offset is exaggerated for clarity. (Right) view in $X_{OPT}$-$Y_{OPT}$ plane.*

### 4.2.2 M2 Mirror (M2M) Mechanism

The mechanism which is necessary to ensure that the telescope is in best focus and meets its WFE requirements when in operations is located on the M2 mirror. This mechanism builds on the heritage from other similar M2 mechanisms which has been developed at Sener, ES. Specifically, it builds on the design heritage from the Gaia and Euclid M2 mechanisms. The design allows for a range of motion of ±300 µm in the focus direction and rotations of ± 2000 µrad in tip & tilt.

During phase B1, a Technology Development Activity (TDA) was initiated with Sener to progress the design and carry out the necessary delta qualification for the actuators to lower temperature. The TDA includes all the activities required to bring M2M to TRL6 by the time of mission adoption at the end of phase B1. In particular, the activity will test the capability of the M2M actuators (upper) to operate at the Ariel required temperatures; the activity passed a TRR in February 2020 and tests are in progress at the time of writing. A preliminary M2M assembly design was also provided as part of the activity (see lower).

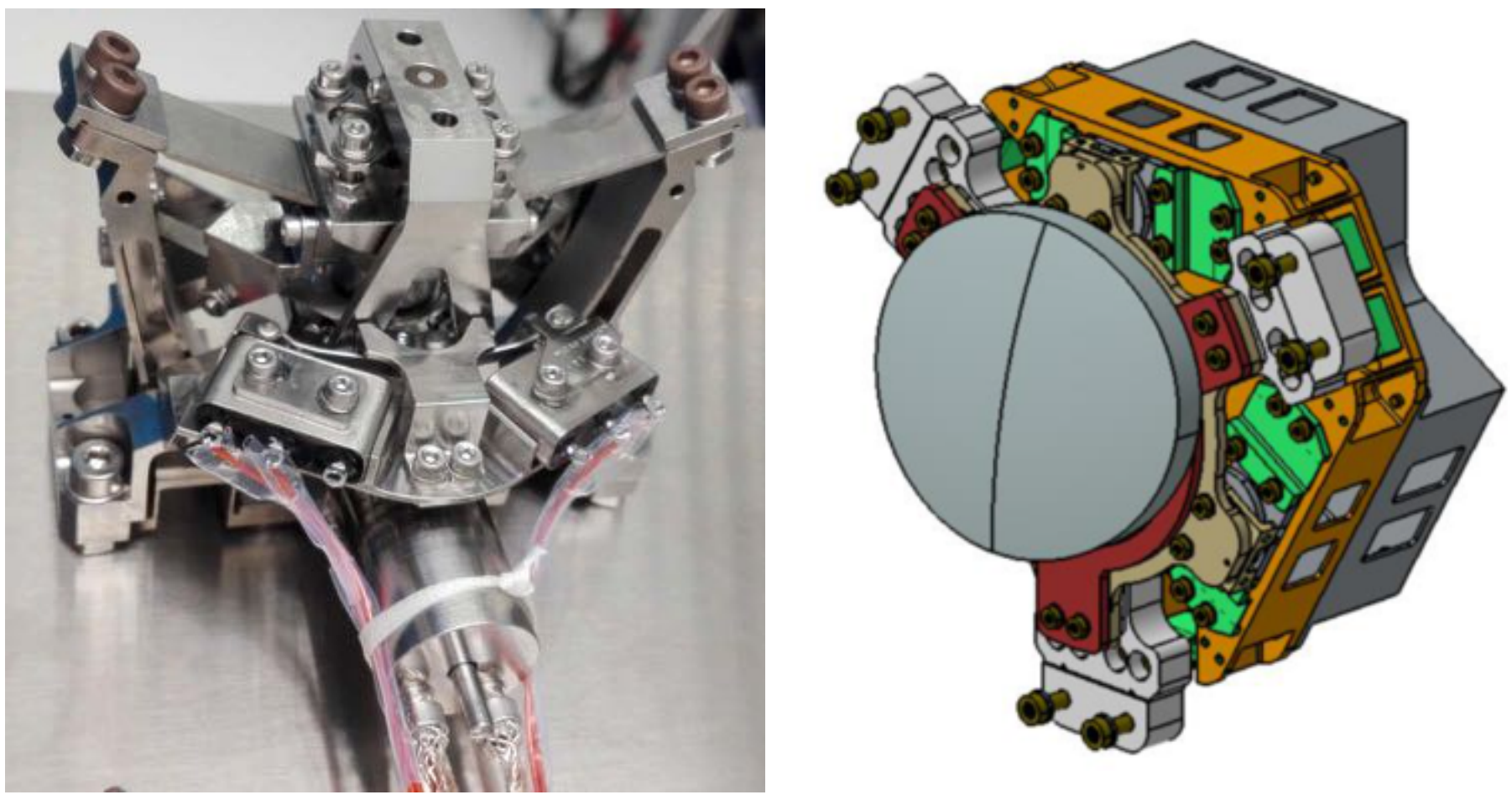

*Figure 4-4: (Left) M2M actuator developed for Ariel TDA. (Right) Ariel M2M preliminary design from phase B1.*

## 4.3 The Ariel InfraRed Spectrometer (AIRS) Design

### 4.3.1 Functional Design and Architecture

AIRS is the Ariel scientific instrument providing low-resolution spectroscopy in two IR channels (called Channel 0, CH0, for the [1.95-3.90] µm band and Channel 1, CH1, for the [3.90-7.80] µm band). The input to the instrument located at the intermediate focal plane of the telescope and common optical system. The AIRS instrument has two main functional chains.

- The **Optical Chain** function which is converting the entrance optical object into a spectrum in the image focal plane of the system.
- The **Acquisition Chain** function which convert the incoming optical spectrum into digital science data packets that will be sent to the ground through the ICU and the SVM.

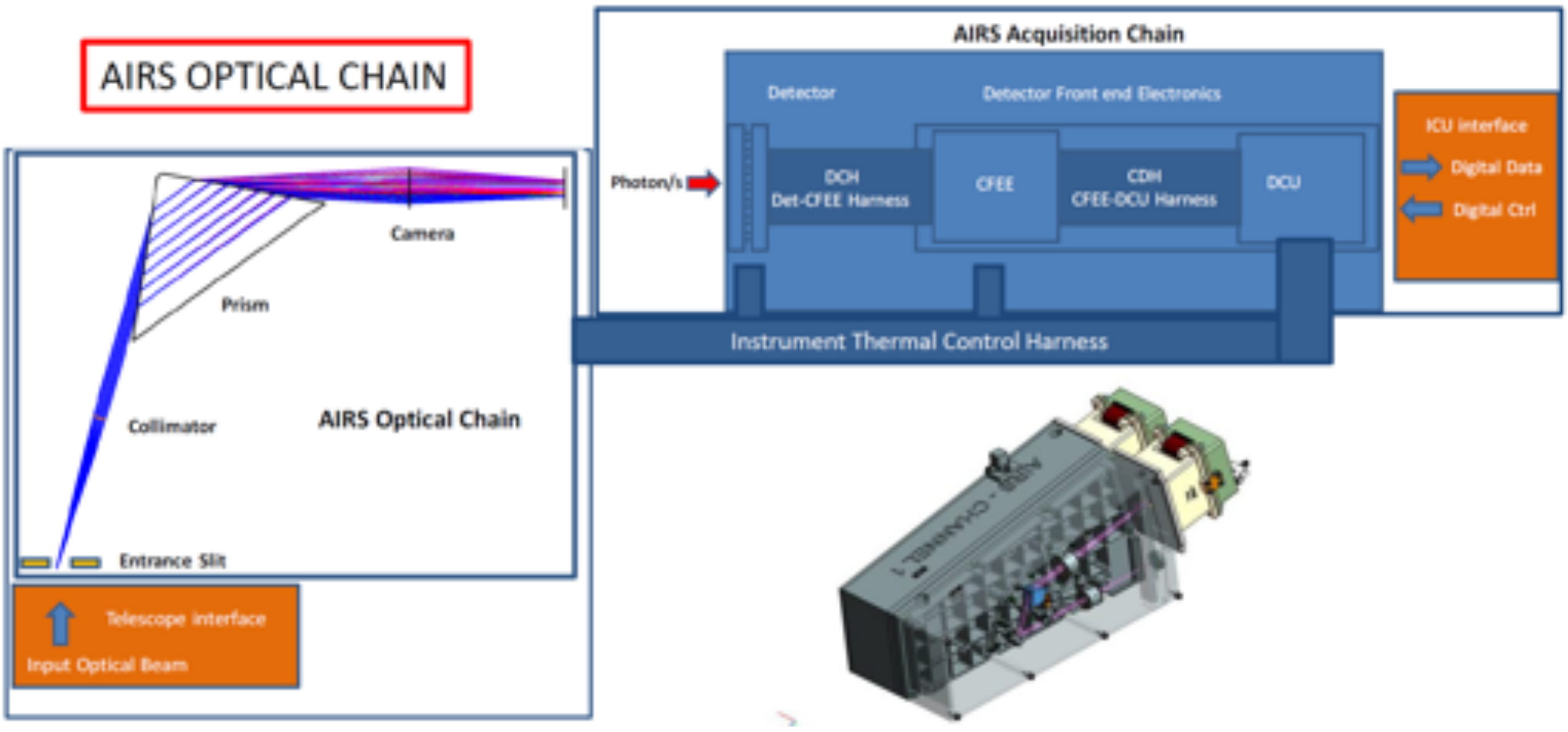

*Figure 4-5: Architectural block diagram of the Ariel Infra-Red Spectrometer.*

AIRS-OB, SCA and CFEE are located on the cold section of the PLM, while the AIRS-DCU is located in the warm part of the SVM. More details of the AIRS design can be found in Amiaux et al. (2020).

### 4.3.2 Optical Design

Following a detailed trade-off analysis in phase A, a prism system was selected as the baseline design for AIRS. The implementation is made at operational temperature of 55 K and the indices of refraction are defined at this temperature. The selected material for the prisms is CaF2, which has heritage from previous IR space missions.

Detailed Zemax modelling showed the need to introduce doublet systems for the Camera (CaF2/Sapphire) and the Collimator (CaF2/ZnSe) in order to control the chromatic aberrations. With this correction the system is diffraction limited over the operational wavelength ranges. A fold mirror is inserted in the optical path between the collimator and the prism in order to allow having both channel entrance planes and exit planes (detector plane) collocated and to optimize the location of the exit focal plane above the entrance slit. This solution optimises the volume implementation of the overall instrument.

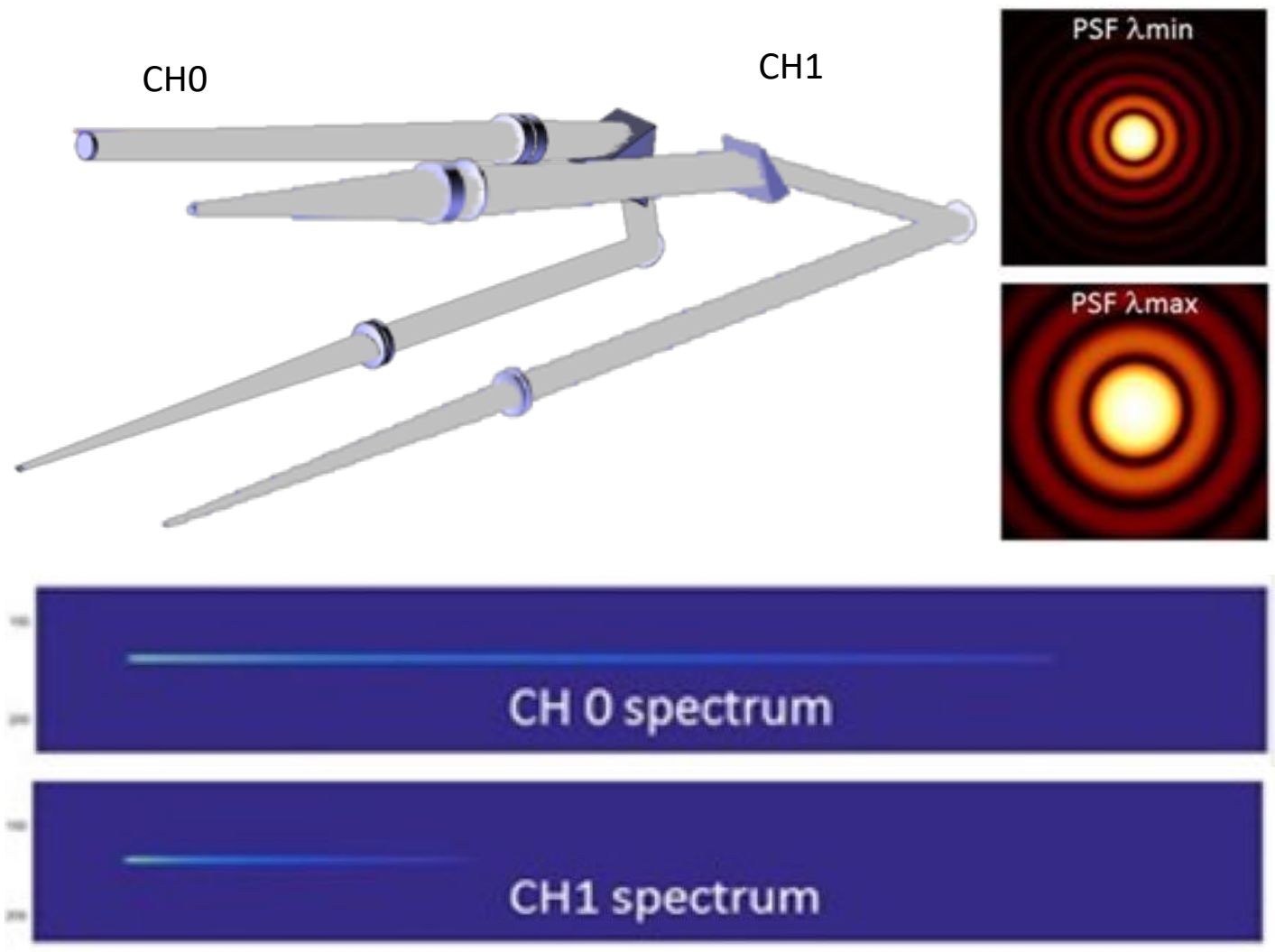

*Figure 4-6: Detailed Optimized folded Zemax implementation of Channel 0 and Channel 1 spectrometer. The windows on the detector needed for the spectra are 325 x 64 pixels in Ch0 and 120 x 64 pixels in Ch1 where the spatial direct size is set by the need to monitor the background and dark current of the detector. The PSFs are shown for Ch0 for a region of 10x10 pixels.*

### 4.3.3 Optics Mounting

Both AIRS prisms are in CaF2 which is known to be brittle. Furthermore, CaF2 is birefringent, so it is important to avoid any stress in the used volume. However, CaF2 prisms and lenses have some heritage in space programs at cryogenic temperatures (JWST-NIRSpec low resolution prism, with resolution similar to AIRS, JWST MIRIm filters). In phase B1 a breadboard of the optical prism has been manufactured, assembled and successfully tested to vibration levels expected within AIRS, see Figure 4-29(d).

### 4.3.4 Mechanical and Thermal Design

The approach for the mechanical design is to have two independent spectrometer half-boxes, each containing one channel, that are ultimately assembled together. The folding of the two channels in the volume is described in Figure 4-8. The overall volume is set by the CH0 longest arm. As the beam is collimated for both channels between the collimator and the camera, the design can easily be adapted for folding and matching length and alignment of both arms to a single plane for the entrance focal plane and for the detector focal plane with a single folding mirror.

The AIRS instrument has an estimated mass of 13 kg (including maturity margin). The thermal interfaces are through the mounting plane for the passively cooled optical bench and optics, and on the rear of the detector modules for the attachment by thermal straps to the Cooler Coldhead heat exchanger which extracts heat from the two actively cooled detectors.

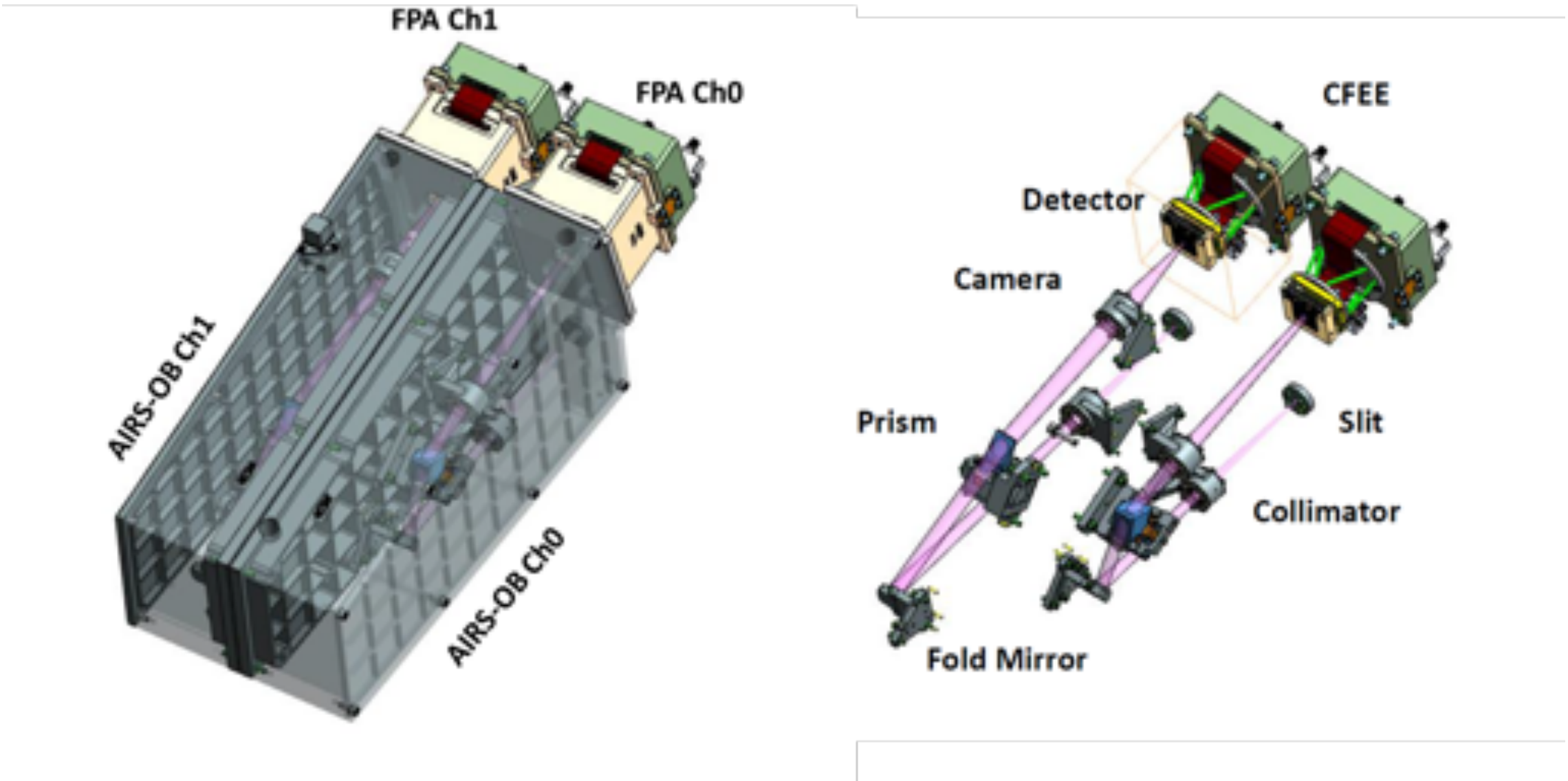

*Figure 4-7: Implementation of AIRS optical design within the allocated volume. The size of the AIRS OB volume is 482 × 206 × 184 mm; the optical interface is at the slits which are 40 mm above the mounting plane and separated by ~108 mm for the two channels.*

### 4.3.5 Detector and Signal Chain System

The AIRS detection chain is defined as the functional sub-assembly of the AIRS that is necessary to detect the AIRS spectrometer images and to pre-process scientific data to fit with the telemetry allocation. The AIRS detection chain electrical system encompasses the detector array assemblies, the Cold Front End Electronics (CFEE) and the Detector Control Unit (DCU) which is the warm front end electronics. The electrical architecture is shown in Figure 4-9; note that the AIRS DCUs are a subassembly of the ICU unit. Besides the interfaces to the cold AIRS subsystems the DCU shares unit-internal interfaces for power supply and control / command functions.

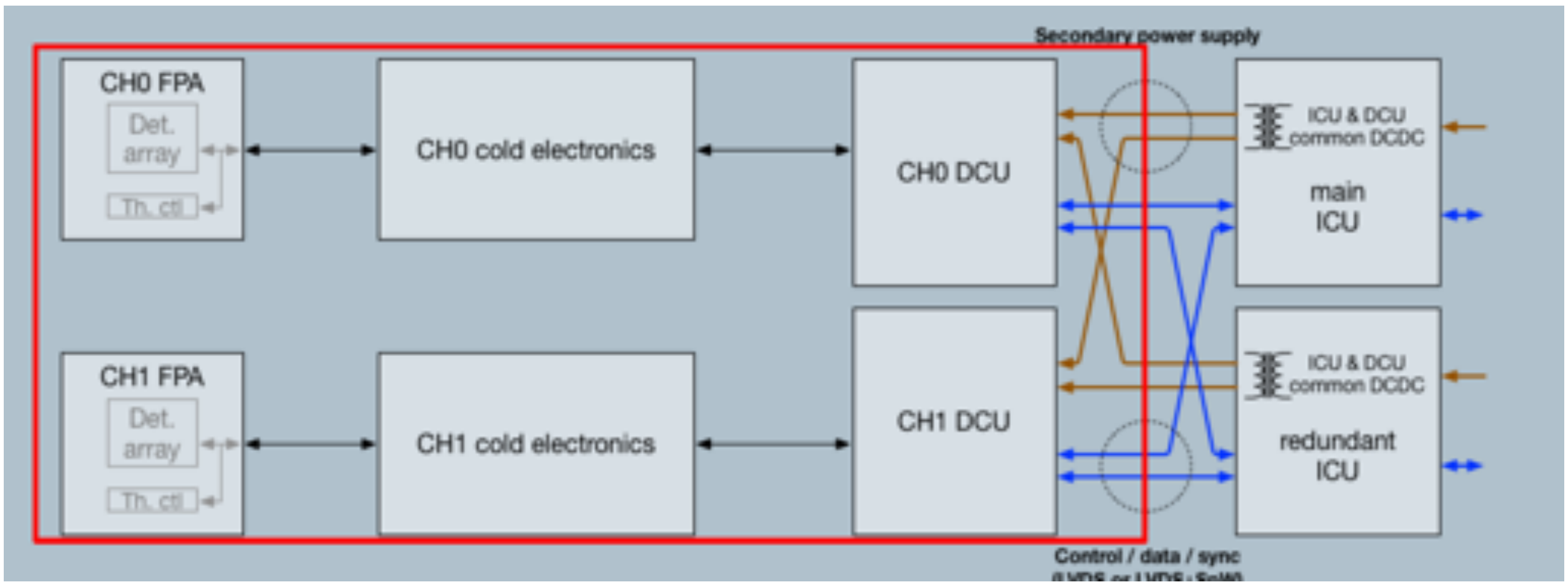

*Figure 4-8: Overview of AIRS electrical system. Red line delineates AIRS instrument boundaries.*

### 4.3.5.1 Detectors

The two AIRS Focal Plane Assemblies (FPA) are thermal mechanical boxes containing the following for each channel (CH0 and CH1):

- Sensor Chip Assembly (SCA) with the detector and associated Read Out Integrated Circuit (ROIC), and the thermal control (two temperature probes + two heaters per SCA). The SCA is thermally decoupled from the mechanical structure and regulated through a thermal control loop at 42 K+/-0.05 K.
- The Cold Front End Electronics (CFEE), linked to the Detector ROIC with a dedicated harness, and conductively cooled to 55 K through the AIRS-Optical Bench in the case of a bespoke analogue CFEE consisting of discrete components. There is an open trade as to whether this will be implemented or whether a CFEE consisting of a SIDECAR ASIC running at ~130K on a separate temperature stage (similar to the FGS architecture) will be used – this trade will be closed out at least six months in advance of the payload PDR.

The baseline detector for CH1 is Teledyne Imaging Sensors (TIS) H1RG array with a cut-off wavelength at approximately 8.0 µm. The heritage of this device builds on that in the frame of the NEOCam program (McMurty et al. 2013) but tailored to the Ariel needs. Two engineering model devices have been manufactured and tested during phase B1, and further calibration of these devices is ongoing at CEA and ESTEC. The baseline CH0 detector is a standard H1RG product with cut-off at 5.3 µm.

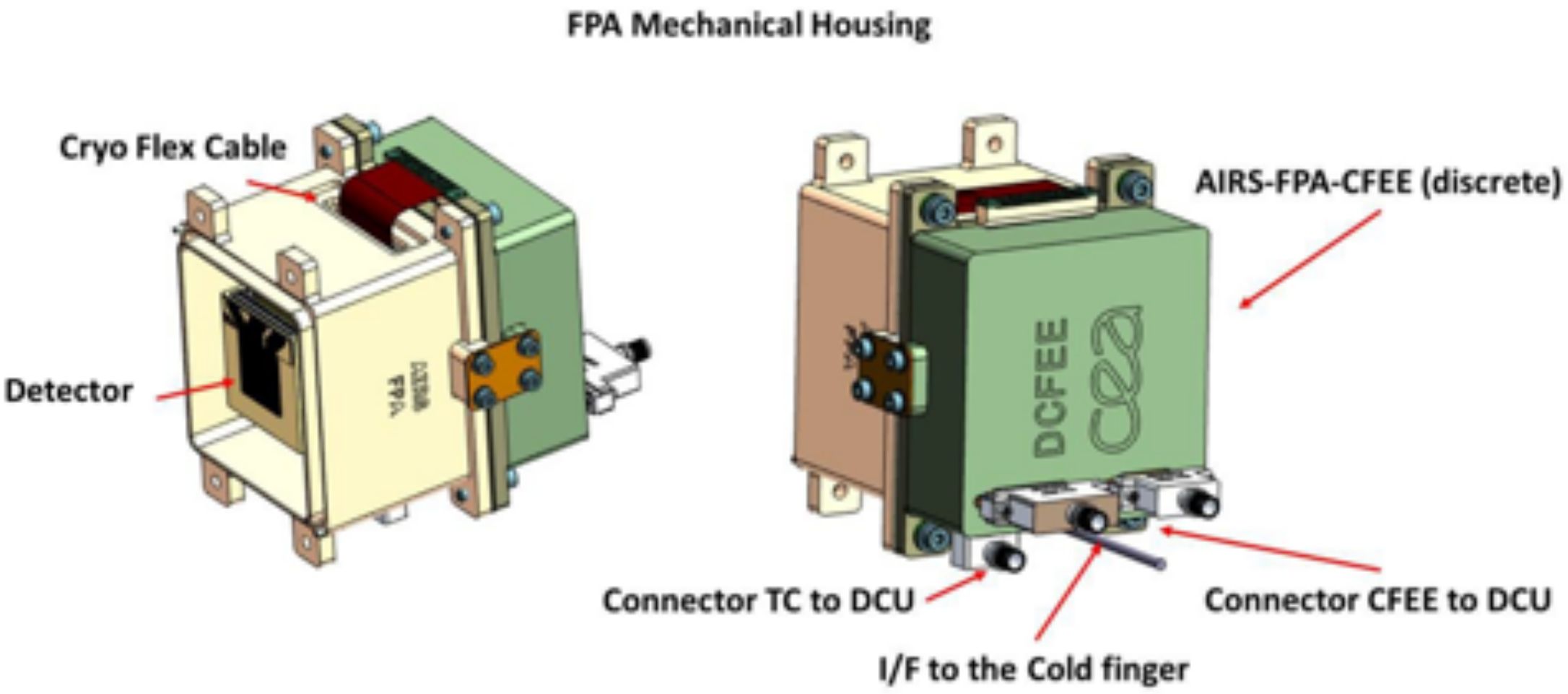

*Figure 4-9: Design of the AIRS Focal Plane Assembly with the detailed units (one FPA per channel). Note that the green box housing the CFEE is only installed in the case of the Discrete CFEE solution; in case of the SIDECAR ASIC option these are housed in a separate enclosure outside the payload instrument cavity (as shown in Figure 4-20).*

# 4.4 The Fine Guidance System (FGS), Visible Photometer (VISPhot) and Near-IR Spectrometer (NIRSpec) Design

## 4.4.1 Functional Design and Architecture

The main task of the Fine Guidance System (FGS) is to ensure the centering, focusing and guiding of the satellite, but it will also provide high precision photometry of the target in the visible and additionally a low resolution near-IR spectrometer. The term "FGS instrument" will be used to refer to this combined functionality of both guidance and science channels. The FGS function uses light from the exoplanet target host star coming through the optical path of the telescope to determine the changes in the line of sight of the Ariel payload (i.e. pointing fluctuations). The attitude measurement is then fused (by the S/C AOCS) with the rate information from the star tracker, and used as input for the control loop stabilizing the spacecraft.

To meet the goals for guiding and science, four spectral bands are defined: FGS–1 (0.6-0.8 μm), FGS–2 (0.8-1.1 μm), VISPhot (0.50-0.60 μm) and NIRSpec (1.10-1.95 μm spectrometer with R ≥ 15). The two FGS channels (FGS-1 and FGS-2) deliver redundant guiding information to the S/C by providing the centroid position of the target star at 10 Hz. Data from FGS-1 and FGS-2 will also be used for photometric analysis. The spectral bands are selected using a series of dichroic mirrors internal to the FGS Optical Module as shown in Figure 4-12 (left).

## 4.4.2 Optical Design and Performance

The FGS optical module interfaces to the collimated beam fed by the telescope and common optics. It then includes an off-axis Gregorian mirror telescope before the input to the dichroics system to divide the field of field into the four separate channels. Figure 4-12 (right) presents the baseline optical design solution.

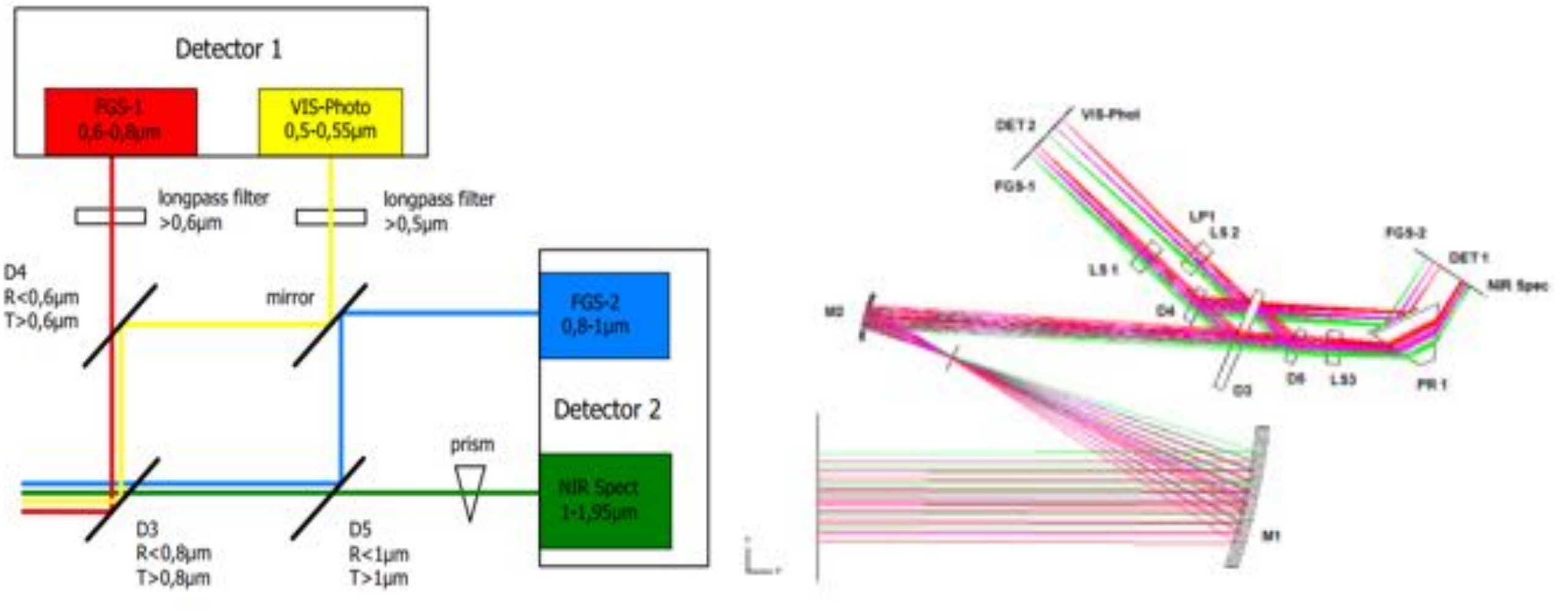

*Figure 4-10: Left: Dichroic schematic showing how channels are split. Right: Baseline optical layout for FGS.*

To obtain a prediction of the optical performance of the FGS the two main sources of aberration have to be taken into account: the optical system of the FGS and (substantially more influential) the main Ariel Telescope assembly. The Telescope Assembly WFE was implemented in the Zemax model using Zernike polynomials based on the current expected distribution of error terms to match the allowable WFE for the Ariel M1. Results of the optical analysis are presented in Figure 4-13 with the PSFs as seen on a 6 x 6 pixel region of the detector for three wavelengths 0.55 μm (VISPhot), 1.0 μm (FGS-1 / FGS-2 boundary) and 1.95 μm (NIRSpec longest wavelength). As can be seen from these plots, the PSF remains coherent in all cases and will allow centroiding of the position of the star (see the central plot for the wavelength of FGS-1/2). These simulations are also used in the evaluation of the pointing system performance shown in Section 4.4.6.

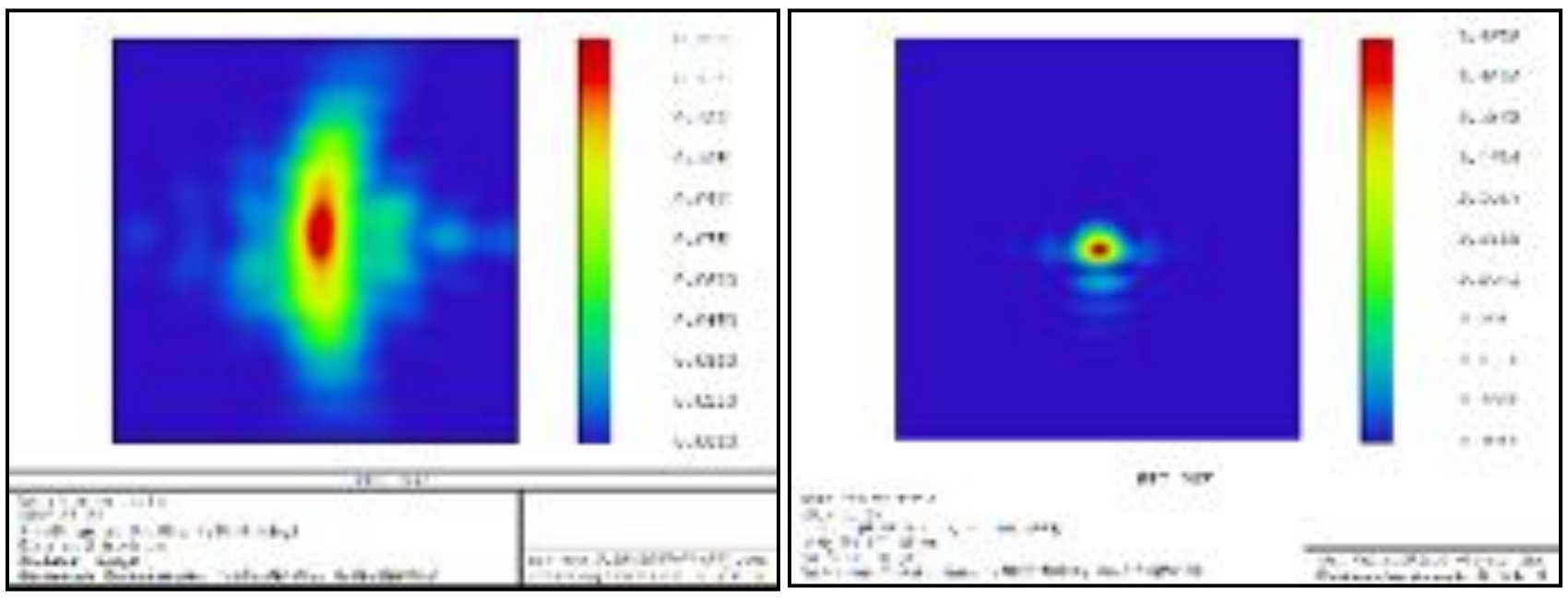

*Figure 4-11: PSF plots for a 6x6 pixel region for the end-to-end optical performance of VISPhot (right) at λ=0.55 µm, FGS-1/2 (centre) at λ=1.0 µm and NIRSpec (right) at λ=1.95 µm including telescope assembly WFE.*

### 4.4.3 Mechanical Design

Based on the optical design and the mechanical constraints of the holders for the optical elements and detectors, the design of the FGS cold optical unit is shown in Figure 4-14.

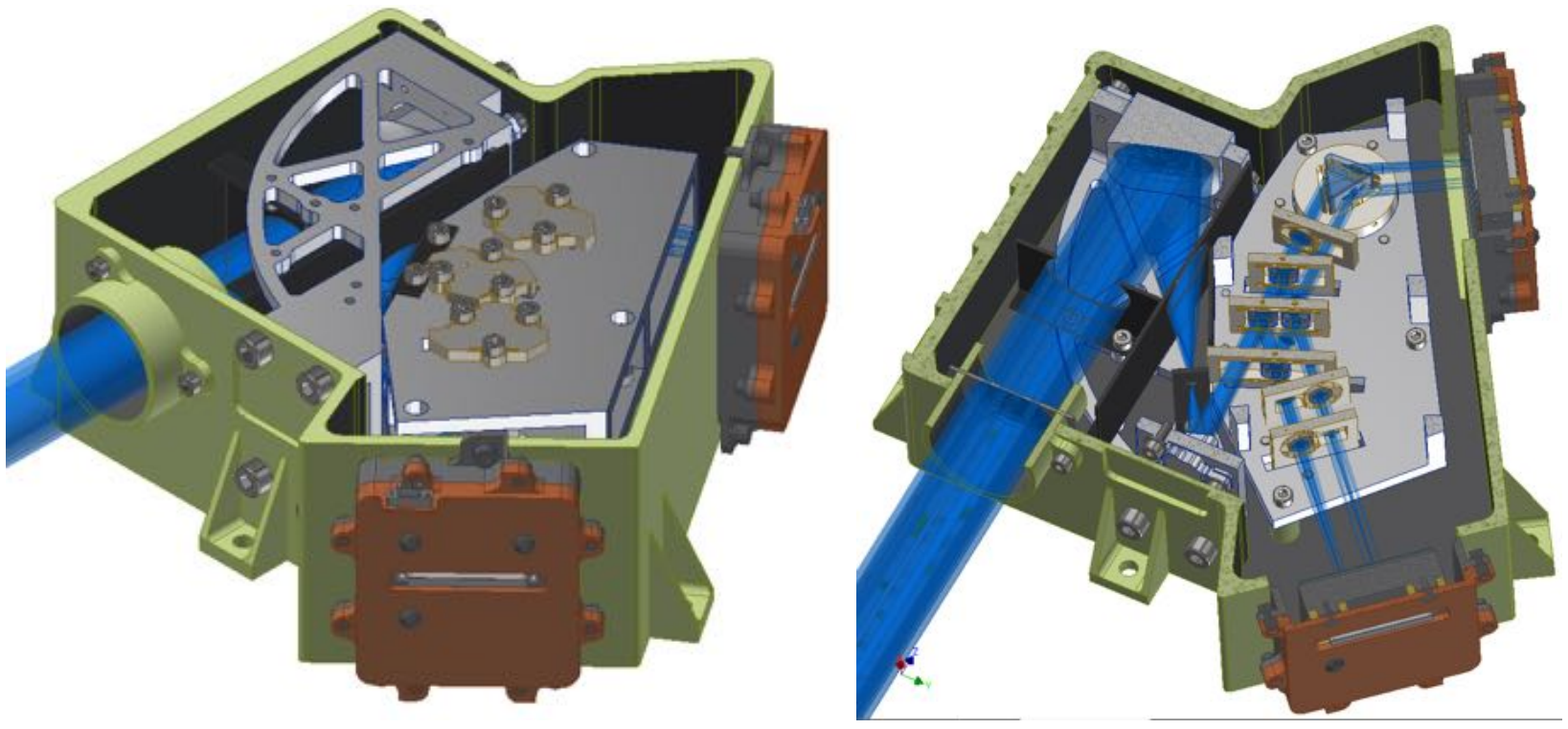

*Figure 4-12: FGS Optical Module mechanical design with lid removed (left) and in cross section view (right)*

The mirrors of the Gregorian off-axis telescope are fixed in the Telescope Cage, which gives the possibility to align an optical axis of the FGS relay telescope to the main payload Telescope Assembly. A set of baffles is placed at the entrance to the FGS. After the FGS relay telescope, the set of dichroic mirrors is placed, integrated in one Beam Splitter Assembly. All three dichroics, three lenses, and a prism are mounted in one Beam Splitter Cage. The optical beams are then focused by the final optical elements onto the two H2RG MCT detectors. The two independent cage systems allow relatively easy assembly and alignment of the optical subsystems inside the housing.

The FGS optical unit envelope size is 230 × 230 × 75 mm, with a mass (excluding the focal plane modules housing the detectors) of 2 kg.

As well as the main FGS optical unit, the FGS FPE housing the cold Front End Electronics (cFEE) based on the SIDECAR ASICs are also housed in the cold payload module. These are housed in an independently thermally isolated and temperature controlled box in order to ensure a temperature of > 130 K throughout the mission. The FPE mechanical design is shown in Figure 4-15.

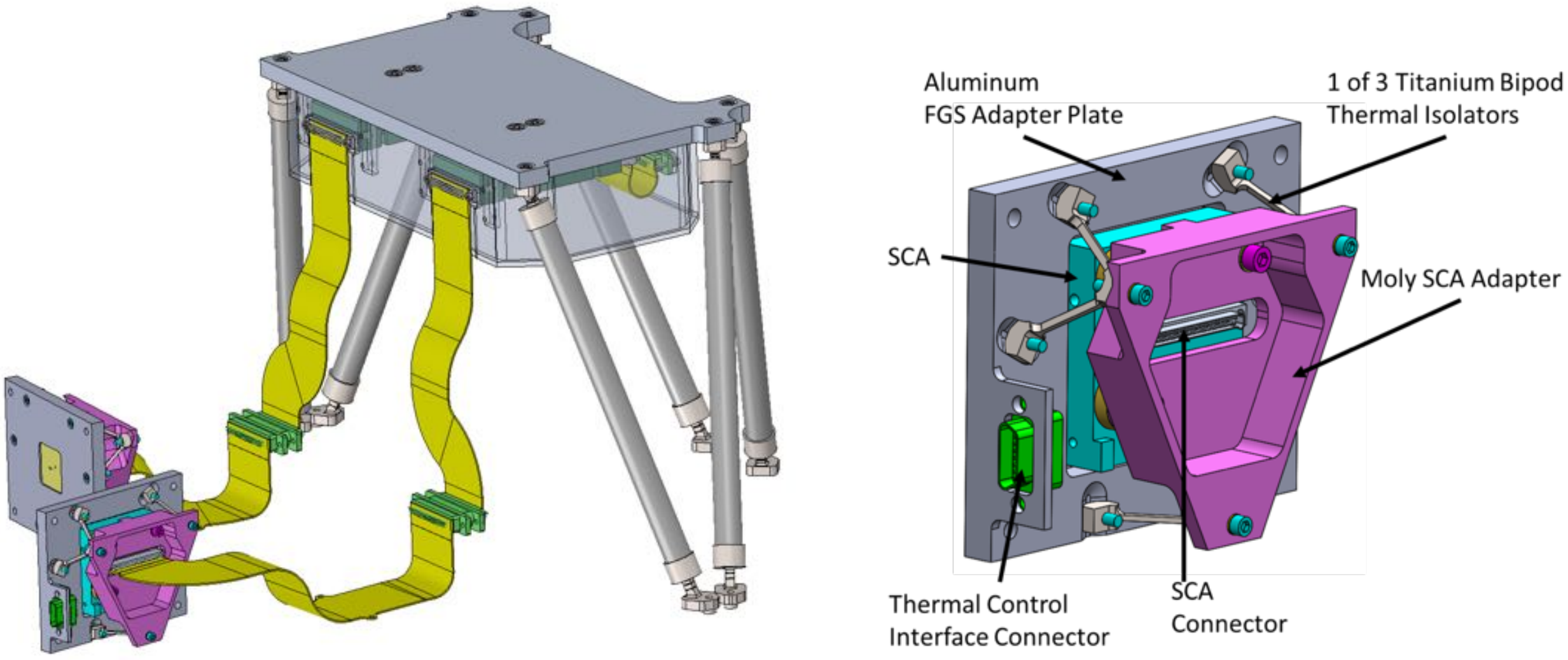

*Figure 4-13: Conceptual design of the FGS detector system. Right – FPE (housing CFEE SIDECAR ASICs operating at ~130K, with cryo-flex harnesses running to detector modules. Right – Focal Plane Module (FPM) design*

### 4.4.4 Detector System

The detector system (FPM, CFC & FPE) for the FGS is being provided by NASA / JPL as part of the CASE (Contribution to Ariel Spectroscopy of Exoplanets) participation in Ariel. The detectors to be used are from the flight production run for the Euclid program, so are already available and well characterised; they have performance which meets or exceeds all the key performance requirements in terms of wavelength coverage, QE, read noise, dark current and linear well capacity, while also meeting the necessary engineering constraints in terms of operating temperature, power consumption and packaging for Ariel. They are H2RG 2k x 2k Mercury-Cadmium-Teluride arrays from Teledyne, and are mounted within a Focal Plane Module (see Figure 4-15). These detectors use the SIDECAR ASIC as the cFEE, with the interface and control signals for the cold electronics coming from the FGS Control Unit (FCU). The detector modules are connected to the cFEE via a low thermal conductance cryogenic flex cable (CFC).

### 4.4.5 FGS Control Unit (FCU)

The FGS has its own control electronics (the FCU) in the SVM to carry out all necessary communication, control and data processing tasks. It will drive and read the FGS detectors through a Detector Control Unit (DCU) based on the Euclid design, establish a control loop with the spacecraft, and deliver scientific data products from the FGS channels to the S/C mass memory. The FCU will consist of a mechanical chassis holding a number of electronics boards providing the functions as shown in Figure 4-17.

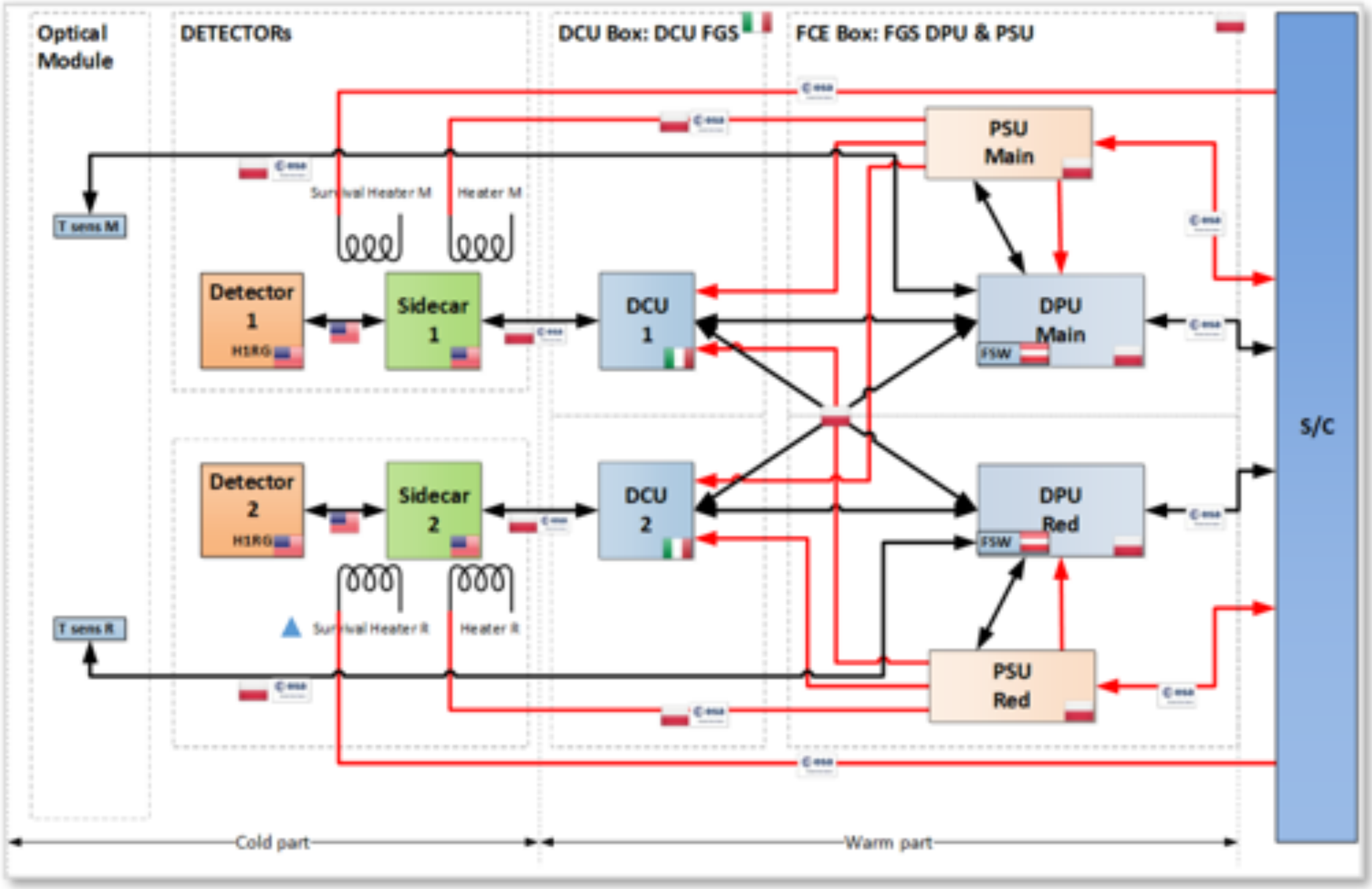

*Figure 4-14: Scheme of the FGS control unit.*

Based on the heritage designs from Solar Orbiter STIX and the JUICE SWI instruments from which the FCU draws heritage, the mass including maturity margin is estimated to be <9.0 kg and the maximum operating power consumption is 26 W.

### 4.4.6 Centroiding and Guidance Software

The FGS will have to carry out and support a number of different tasks through its application software. There will be functions to control the FGS subsystems, process the detector data and communicate with the spacecraft. The FGS application software will offer its functionality in the form of ECSS service commands and reports. The software includes some low level FDIR functionality and offers several modes of operation to support maintenance and calibration activities

The main requirement of the FGS is the centroiding performance of 20 milli-arcsec (mas) accuracy for bright and 150 mas for faint targets at 10 Hz with 99.7% confidence levels. For the best support of the operating modes, several centroiding and data extraction algorithms will be implemented, fully configurable by parameter and command. In the warm FGS control electronics the data will be processed in real-time. Output data products will be reformatted images, centroid coordinates, dimensions and errors in both axes, photometry, glitch count and housekeeping.

Figure 4-18 presents the results from a centroiding study demonstrating that the precision requirement can be met for the whole range of targets with the anticipated worst-case optical performance of the telescope and FGS. A sample of 864 simulated images was used as input for the algorithm under test. The dots represents the computed centroid estimation errors. The coloured solid line represent the mean true Centroiding Estimation Errors (CEEs). The vertical dashed lines indicate the expected electron fluxes on the detector for faint and bright targets.

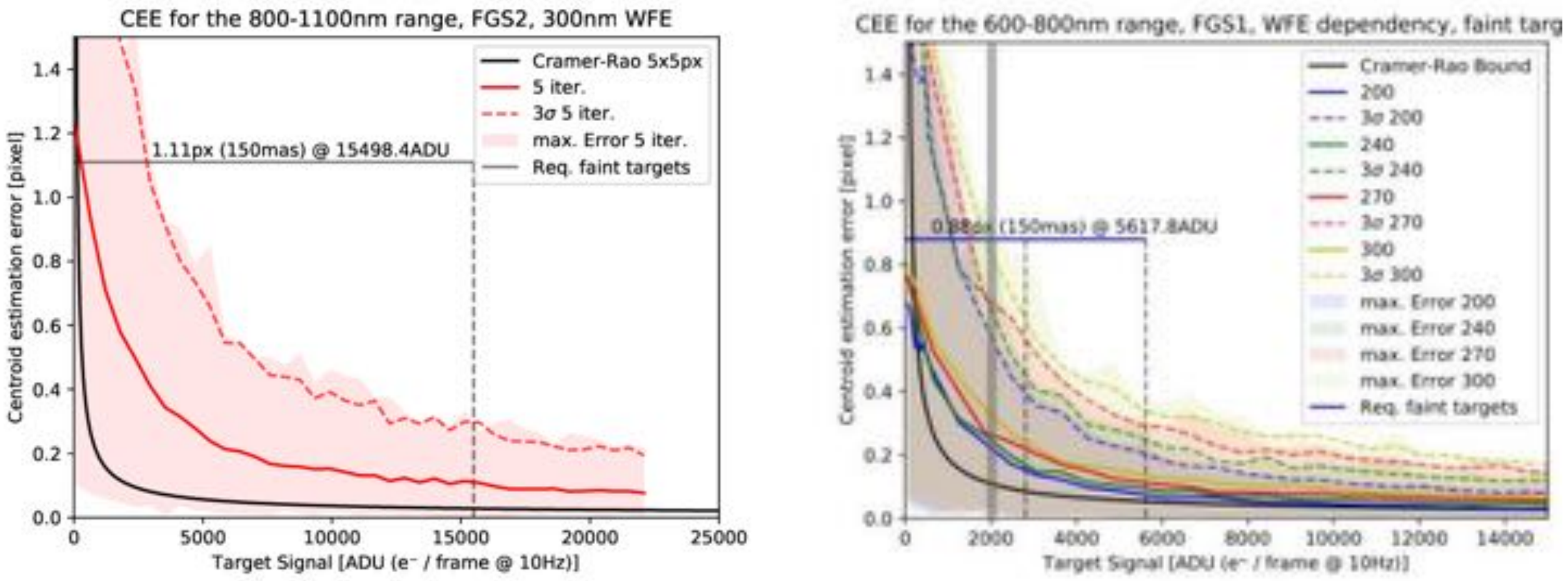

*Figure 4-15: Centroiding accuracy estimates for faintest target in nominal FGS2 channel (left) and back-up FGS1 channel (right). Note the simulations shown in different colours for the FGS1 case are for different WFE levels showing that even with 300nm WFE (vs derived requirement of 270nm on the combined TA, CO & FGS systems) the centroiding requirement is met.*

## 4.5 Common Optics, Dichroics and In-Flight Calibration Unit

The function of the common optics is to split the light from the telescope into three channels of science instrument input (FGS; AIRS Ch0 and AIRS Ch1), deliver the collimated beam to the FGS, deliver focussed beams, of the correct f/number, at the AIRS input slits, and to provide an input position for the On-Board Calibration Unit (OBCU). The common optics positions can be seen in Figure 4-3.

The splitting of the instrument channels by wavelength is done by dichroic mirrors. Details of the dichroic designs obtained from vendors (IML at University of Oxford (previously Reading) and CILAS) are used in calculation of the overall throughput performance as shown in Figure 4-2.

The OBCU is placed in the common optics and can provide a quasi-uniform illumination on all focal planes. It will be used during commissioning and in-flight calibration to monitor variations in QE (detector array flat fielding), as an alternative to diffuse astronomical sources, and to transfer over time the calibration obtained on the ground.

The OBCU makes use of a tungsten-filament thermal source, serving AIRS and NIRSpec, and light emitting diodes (LEDs), serving the FGS photometer channels. The light from the emitters is fed to a 1 mm diameter hole in the M5 mirror, which in the collimated beam results in less than 0.4% loss of throughput. JWST-MIRI and NIRSpec use tungsten filament sources for the same purposes in a temperature range from 500 K (on JWST MIRI (Wright et al. 2015)) to 1600 K (JWST NIRSpec (Bagnasco et al. 2007)). LED emitters with characteristics suitable for the ARIEL photometric channels have been qualified for operation at 150 K for the Euclid VIS instrument (Carron et al. 2017) and are currently undergoing evaluation for operation in the ARIEL optical bench operating temperature range. The brightness range of the OBCU emission will be carefully optimised to provide the high detector SNR needed to achieve the required 0.5% accuracy in flat-field monitoring while avoiding saturating the detectors.

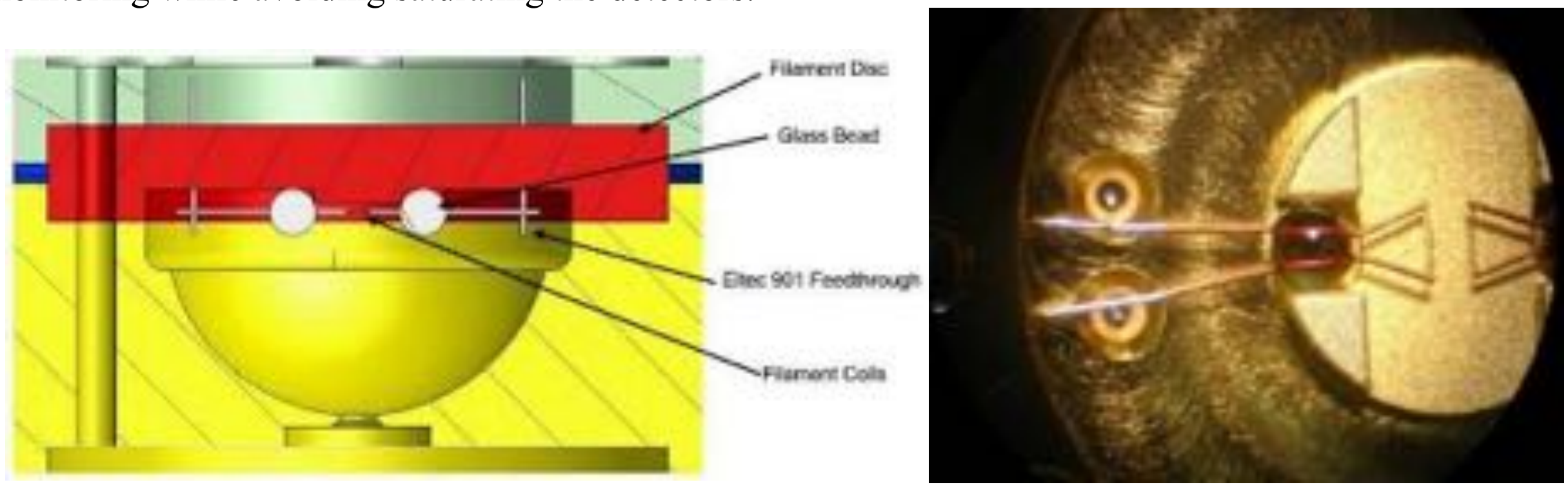

*Figure 4-16: (Left) A schematic of the JWST MIRI calibration source. The two filaments (prime and redundant) are shown at the centre of the assembly. The glass beads which achieve mechanical bonding of the filaments are also visible. (Right) A microscope picture of the JWST MIRI filaments.*

# 4.6 Payload Module Structural Design and Analysis

The main mechanical units composing the PLM are the supporting bipods, the Telescope Assembly (optical bench, mirrors, struts and baffle) and the Instrument Enclosure (AIRS & FGS optical modules, common optics, radiator). The overall PLM mechanical design is shown in Figure 4-2. The payload mass budget is shown in Table 4-1, the values shown are the Nominal Mass consisting of the current Best Estimate (CBE) value, plus a design maturity margin set depending on the level of design detail and heritage. Note that this mass budget has been updated following the recommendations of the Payload SRR.

The finite element analysis shows that the first modes of the payload (when fixed to a rigid interface) comply with the required frequencies, and the loads from this model are used in verification of positive structural margins of safety in detailed lower level models for the critical items (such as the bi-pod struts and flexures, instruments and internal optics etc.).

## 4.6.1 Common Optical Bench & Metering Structure Design

The Optical Bench (Figure 4-2, right) is a key mechanical component as it forms the main structure of the PLM: the telescope M1 is mounted to its front surface and telescope support beam projects from its bottom edge forward to support the M2 near the front bi-pod. On the other side the Instrument enclosure has been incorporated into the bench design, increasing the bench thickness and thus stiffening the structure. The OB is directly supported on either side from the top of the rear bi-pods; the front bi-pods mount to the front of the telescope support beam. The design of this platform, and the units mounted on it, will be as isothermal as possible except for the components that need thermal decoupling. The bench, the telescope mirrors, the modules and the common optics supports are to be manufactured from the same aluminium alloy (Al6061-T6) in order to minimize possible optical effects that could occur during thermo-elastic contractions.

## 4.6.2 Support Bi-pods Design and Sizing

The PLM is supported by three bipods mounted onto the Payload Interface Plate (PIP) as shown in Figure 4-21. The bipod thermo-mechanical configuration is based on the Planck and JWST MIRI design heritage, with optimisation for the Ariel interfaces and requirements. They are hollow cylinders, made of CFRP (a trade-off with GFRP and other materials has been conducted in phase B1 and concluded that these offered the best combination of heritage and thermo-mechanical performance). To prevent radiative transfer up the inside of the struts, and to increase their mechanical stiffness, the inner volume of the cylinders is filled with low thermally conductive rigid foam. The bi-pods incorporate Titanium flexures in the end fittings which are within the tubes in order to maximise the thermal length of the bi-pod; the flexures will be inspectable via dedicated endoscope access points.

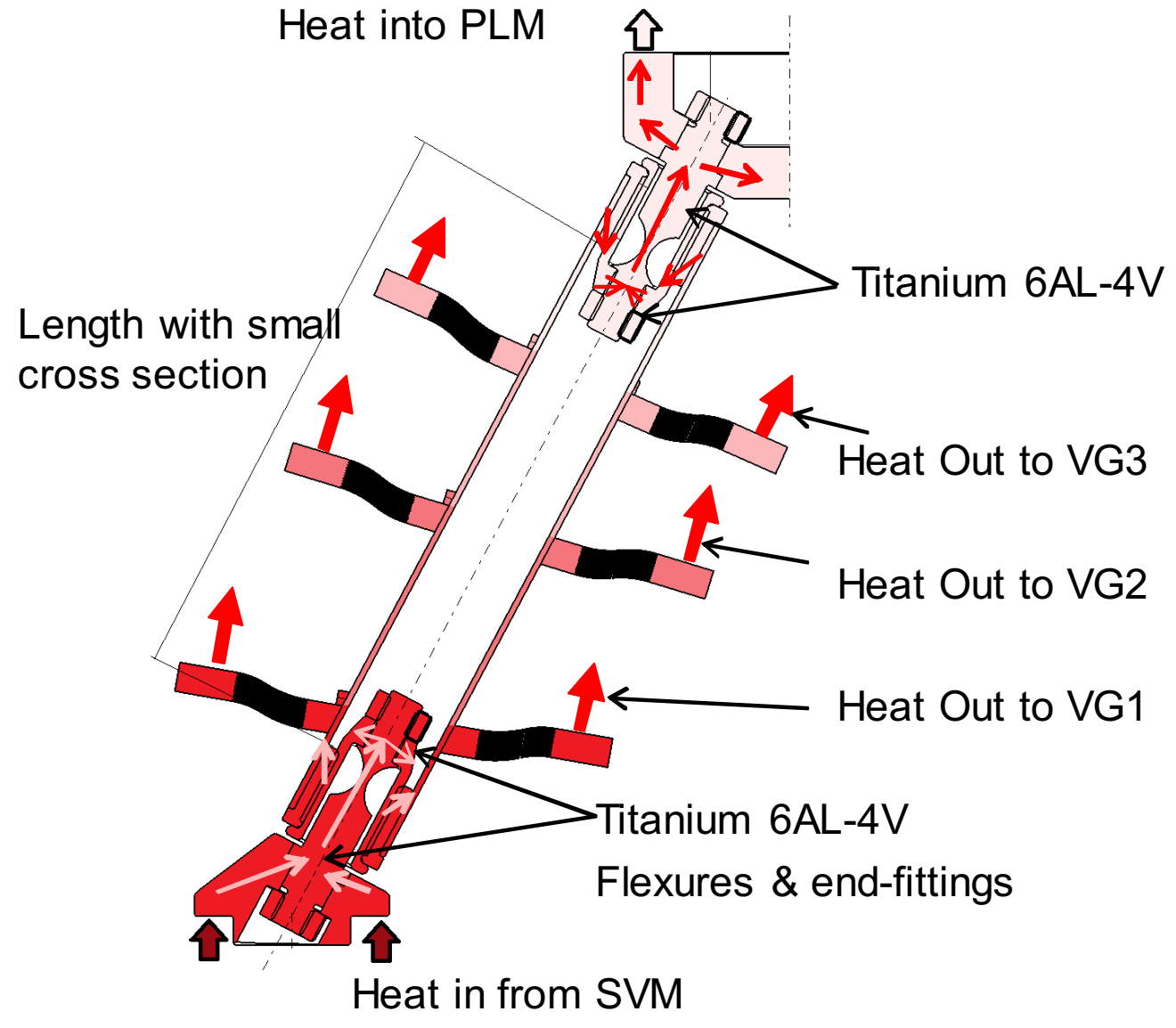

*Figure 4-17: Cross-section view of individual bi-pod tube showing flexures and interfaces to V-Grooves*

| Item | Nominal Mass (kg) |
|---|---|
| **Telescope Assembly** | |
| Optical Bench | 99.1 |
| Metering Structure & Stiffeners | 48.8 |
| Telescope Baffles B1 & B2 | 21.5 |
| TA Heaters, Harnesses | 3.0 |
| M1 Mirror | 123.9 |
| M1 Brackets & blades | 25.8 |
| M2 Mirror | 1.8 |
| M2 Refocus Mechanism | 6.0 |
| M3 & M4 mirrors and mounts | 1.7 |
| Fasteners | 2.4 |
| **Instrument Subsystems** | |
| FGS (Including OM, FPMs, CFEE & harness) | 10.4 |
| AIRS (Including OM, detectors, CFEE & harness) | 17.4 |
| Common Optics (inc cal. source & harness) | 3.8 |
| Instrument Radiator | 7.7 |
| JT Cooler Cold Head and disconnect plate | 3.5 |
| Decontam. & Survival heaters & harnesses + purge HW | 3.9 |
| Instrument interface fixings & harness connector brackets | 5.4 |
| **Thermal Shield Assy** | |
| PLM Bi-pods and Interfaces | 19.2 |
| V-Groove Assembly | 64.0 |
| ***PLM Total:*** | ***470.5*** |
| **SVM Components** | |
| Instrument Control Unit (ICU) | 8.0 |
| FGS Control Unit (FCU) | 8.9 |
| Telescope Control Unit (TCU) | 8.4 |
| Cooler Compressor & Anc. panel | 11.6 |
| Cooler Drive Electronics | 8.0 |
| ***SVM Total:*** | ***45.0*** |
| **Payload Nominal Total:** | **515.4** |

*Table 4-1: Ariel Payload Mass Budget Estimates*

# 4.7 Payload Module Thermal Design and Analysis

## 4.7.1 Thermal Architecture

The spacecraft thermal design (Figure 4-22) is based on a cold Payload Module (PLM) sitting on the top of a warm Service Module (SVM). The Ariel thermal control is accomplished by a combination of passive and active cooling systems. The SVM is thermally controlled in the 253K-323K range for nominal operations of all the S/C subsystem units. The function of the cold PLM is to shield the scientific instrumentation (the Ariel Instruments and the Telescope Assembly) from the warm section of the S/C and to provide it with the required cooling and thermal stability at temperatures < 60K.

Passive cooling is achieved by a high efficiency thermal shielding system (the V-Grooves) based on a multiple radiators configuration that, in the L2 environment, can provide stable temperature stages down to the 50 K range. At 1.5 million km from the Earth in the anti-Sun direction, the L2 orbit allows the spacecraft to maintain the same attitude relative to the Sun-Earth system, while having access to the whole sky during the each six-month period. Limiting the allowed Solar Aspect Angle (SAA) range enables Ariel to operate in a very stable thermal environment keeping the coldest section of the PLM always shaded from the Sun/Earth/Moon illumination. For this reason, the SAA allowed during nominal observations is limited to ±1.5° around the spacecraft Y-axis and to ±25° around the X-axis, see Figure 5-1.

There will be active cooling provided for the AIRS detectors using a Neon JT cooler – see section 4.8. All of the detectors will be located on dedicated isolated thermal control stages to enable active stabilisation of their temperatures to accuracies of ±10 mK for FGS and ±5 mK for the AIRS detectors.

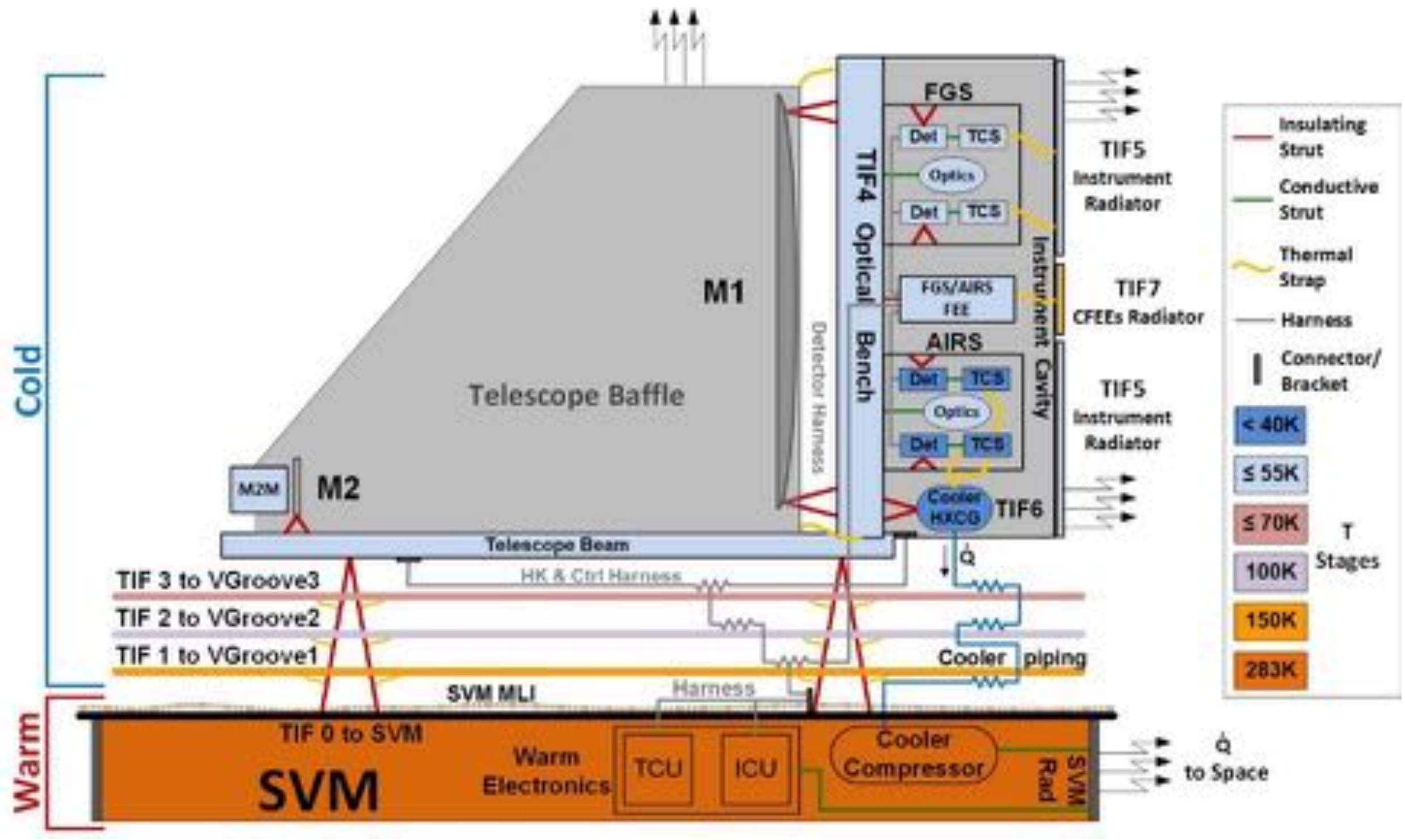

*Figure 4-18: PLM Thermal Architecture Scheme*

### 4.7.2 V-Groove Radiation Shields

The V-Grooves (VGs) are high efficiency, passive radiant coolers, providing the first stage of the PLM cooling system. The Planck mission has demonstrated their efficiency as passive cooling systems. Parasitic heat from warmer sections of the S/C is intercepted by the VGs and radiated to space after multiple reflections between the adjacent shields. To achieve this, VGs surfaces have a very low emissivity coating. Only the upper surface of the last VG (VG3), exposed to the sky, is coated with a high emissivity material to maximize the radiative coupling, and so heat rejection, to deep space. The VG's are a simple honeycomb structure of Aluminium alloy, thermally attached to the three bipods (to intercept the conducted parasitics through the mounting bi-pods) and are separately mechanically supported from the PIP by GFRP struts.

### 4.7.3 Thermal Analysis Results and Temperature Margin Analysis

Figure 4-23 shows the calculated steady-state temperatures of the cold and hot case for the nominal conditions, with the S/C orbiting at the Earth-Sun L2 point. The cold and hot case are different for the fixed boundary temperatures applied at the SVM/PLM I/F due to the S/C attitude and SVM internal conditions assumed. The temperatures of all passively and actively cooled units are fully compliant with the requirements including margins. The temperature gradients in the PLM (including those on the M1 mirror) for the hot case are shown in Figure 4-23. Full details are contained in Morgante et al. 2020.

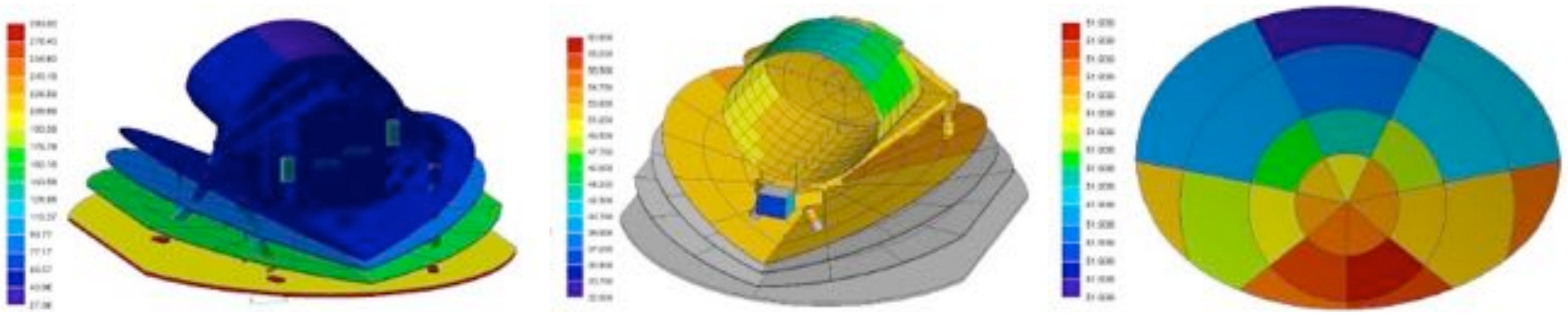

*Figure 4-19: Thermal map of hot case temperatures of (from left to right) the whole PLM, the coldest part of the telescope assembly, the M1 mirror surface – note that the colour scales are different in each view.*

### 4.7.4 Transient Thermal Analyses

In order to size the active control systems on the detectors and M1 thermal control stages, transient analysis has been carried out whereby the SVM interface temperature is changed by 10K (instantaneously) and the effect on the temperatures on the PLM is modelled. The results show a maximum temperature change of the detectors of 9mK over a 10 hour observation period, with the M1 mirror temperature changing by only 2mK. Additional transient thermal analysis for the cooldown cases and for the sizing of the decontamination heaters has also been conducted in this phase and the control capability of the thermal hardware has been sized appropriately.

## 4.8 The Ariel Active Cooler System

The baseline active cooling system for Ariel is a closed cycle Joule-Thomson (JT) system with Neon as the working fluid, this is shown schematically in Figure 4-24. It is used to cool the detector for the AIRS instrument to an operational temperature of ≤ 42K. It is based on the successful 4K-JT cooler for the Planck-HFI instrument and also draws from developments for the 2K cooler for the Athena mission. Key to success is the high reliability and lifetime of the cooler mechanisms, which stems from the philosophy of non-lubricated, non-contacting moving parts, enabled by a flexure bearing suspension system operating in the infinite fatigue strength regime; there have been no failures of this design of cooler over many years operation in space.

The JT cooler system comprises two reciprocating linear motor compression stages to perform an expansion of the gas across a JT orifice to produce the cooling. Neon is selected as the working fluid because its boiling point (27.05K at 1atm) is well matched to the temperature requirements for Ariel. The system makes use of the pre-cooling stages that are available from the V-Groove radiators. Counter-flow heat exchangers are also used between these stages to reduce the heat rejected to the radiators. An ancillary panel carries gas handling and measuring equipment as well as filters and a reactive getter to ensure gas cleanliness. The cooler is controlled by a set of drive electronics which provide the electrical input power for the compressors, perform all controlling functions such as active vibration cancellation and return of cooler housekeeping data.

*Table 4-2: Cooler system parameter sizing*

| **Ariel two-stage Ne JT cooler performance estimates (@ 28K)** | |
|---|---|
| **Parameter** | **Baseline Est.** |
| cooling power at 28K | 88 mW + margin |
| heat rejected at VGs at 190K / 120K / 60K | 101 / 45 / 99 mW |
| total input power | 88 W |
| working pressure (high / low) | 20.0 / 3.5 bar |
| Mass flow | 20.0 mg/s |
| **Masses** | |
| Compressors | 9.1 kg |
| Ancillary panel | 2.5 kg |
| Heat exchangers etc | 3.5 kg |
| Drive electronics | 8.0 kg |
| **Total mass** | 23.1 kg |

The baseline solution for Ariel is to use a two stage compressor pair, based on the RAL 2K-JT cooler mechanism. Although oversized for this application, they offer significant margin on cooling power, while still satisfying the available mass and power budgets for Ariel, and give maximum flexibility in future phases. The cooler system baseline performance predictions and mass and power resource needs are summarised in Table 4-2.

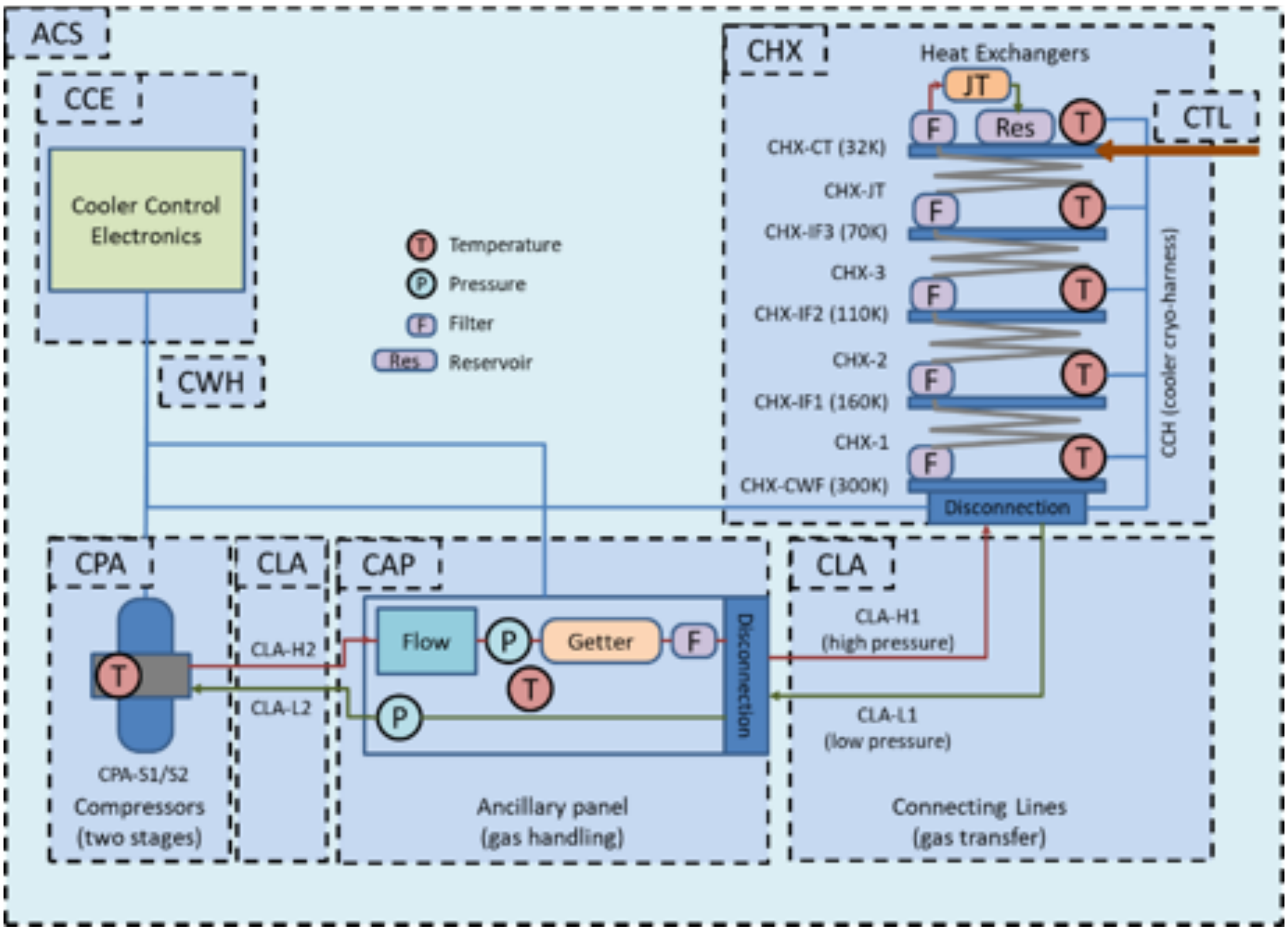

*Figure 4-20: JT Cooler Schematic*

## 4.9 Payload Warm Electronics Units

A number of warm electronics units for the payload are housed within the warm service module. These include the Cooler compressor, electronics and ancillary panel and the FGS control electronics which are all discussed elsewhere in this chapter. There are two additional warm units which provide control, commanding and read-out of the telescope, AIRS spectrometer and other payload items. These items are all connected by warm harnessing (supplied by the S/C Prime) to interface connector panels on the PIP.

### 4.9.1 Instrument Control Unit

The Ariel ICU (for more details refer to Focardi et al. 2020) implements the commanding and control of the AIRS Spectrometer and is interfaced on one side with the instrument and on the other side with the S/C SVM Data Management System (DMS) and Power Conditioning and Distribution Unit (PCDU). The baseline ICU architecture includes four (active or switched-on at the same time) units as shown in Figure 4-25:

- 1 PSU - Power Supply Unit (PSU)
- 1 DPU – Command and Data Processing Unit (CDPU)
- 2 DCU - Detector Control Unit (DCUs) for AIRS

For P/L operation and control, the Telescope Control Unit (TCU) management shall be in charge of the ICU processor; this is the only ICU&TCU Integrated Circuit (IC) running a high-level software (SW), controlling both the AIRS DCUs and the TCU as well. The TCU shall host, as baseline, only an FPGA with embedded firmware designed to receive commands from the ICU processor and provide it with TCU and P/L housekeeping.

The main functions of the DCU are:

- providing housekeeping for AIRS-OB (optical bench) and AIRS-FPA (detectors and CFEEs);
- controlling the data acquisition at detector level through the CFEE from users entry;
- pre-processing digital data from the detector (on-board its FPGA);
- controlling the thermal stabilisation of the detectors through temperature probes and heaters;
- interfacing the ICU CDPU for TC reception, HK and Science Data transmission.

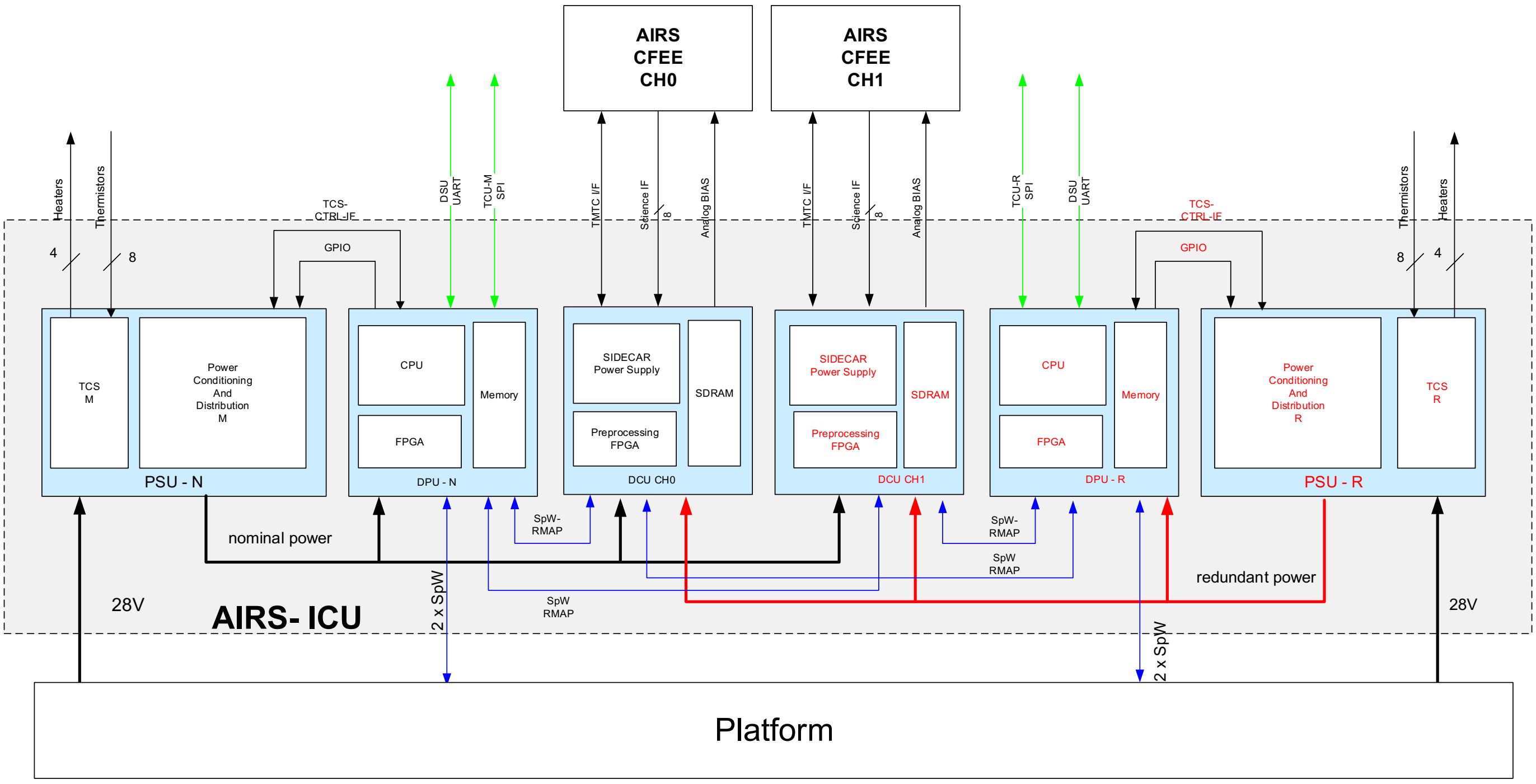

*Figure 4-21: Ariel ICU baseline solution block diagram and electrical I/F*

It is desirable to minimise the level of on-board data processing – data-processing on the ground gives the maximum flexibility in the algorithms and the chance to improve the processing during the mission to produce the best possible science data. The baseline design has the capability to implement all the potentially necessary processing such as flat-fielding, ramp fitting, glitch detection etc. if needed with flexibility to select the algorithm. However this is not the baseline plan.

### 4.9.2 Telescope Control Unit

The Telescope Control Unit (TCU) acts as an ICU slave subsystem and accomplishes the following tasks: drives the M2 refocusing mechanism, drives the OBCU sources, monitors the thermal state of several PLM elements and controls the operational heaters on the telescope assembly.

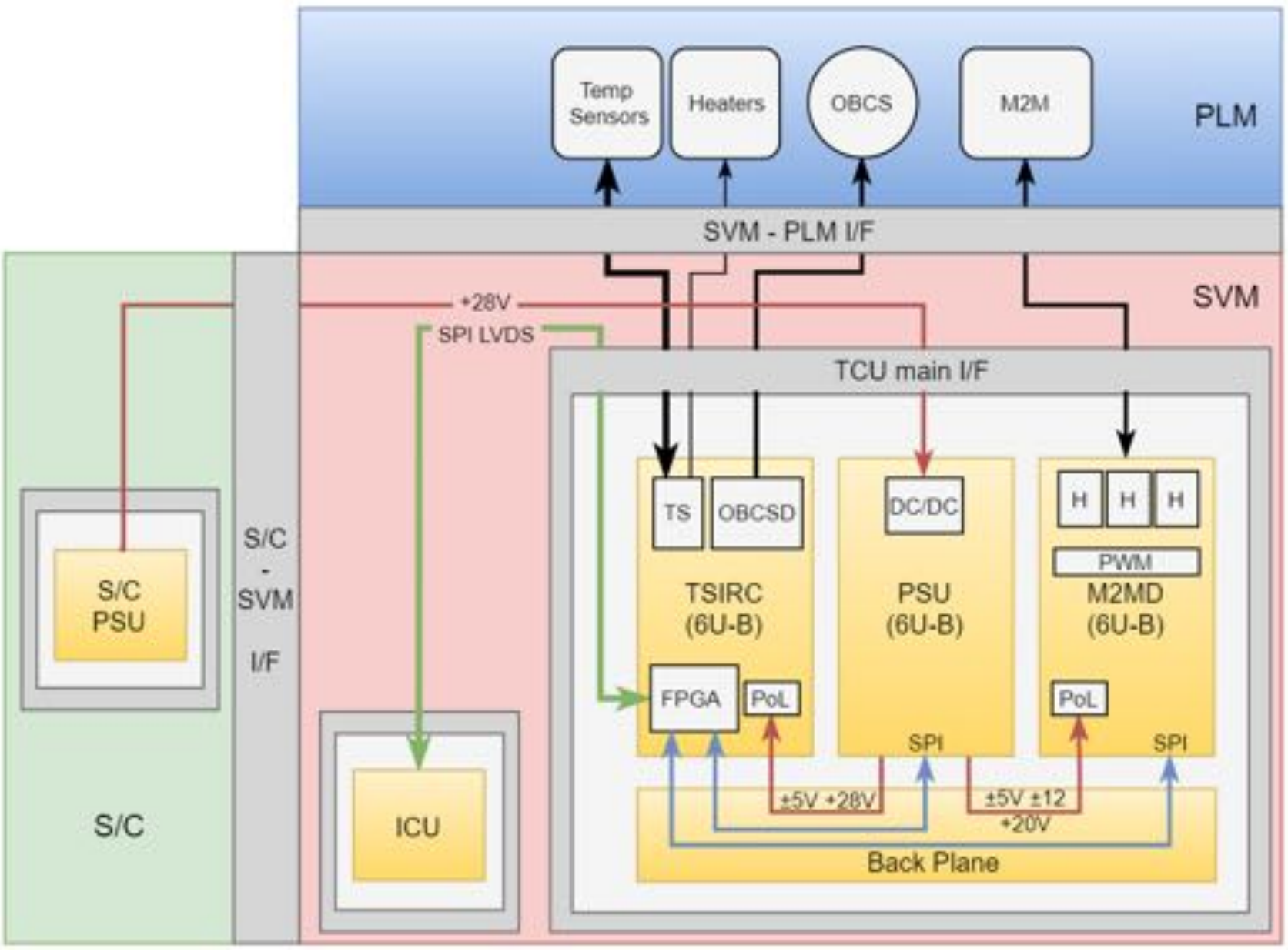

*Figure 4-22: TCU block diagram and electrical I/F*

The ICU will control TCU and share information with it by means of a LVDS SPI bus. The system will be managed by a rad-hard FPGA with a FSM to switch between states that perform specific tasks. A 6U-B PCB will host the PLM thermal monitoring and its multiplexing stages, M1 heaters driver, OBCU driver and the digital system in charge of managing the whole TCU. For the driver electronics of the M2 mechanism, an upgraded version of Euclid's M2MM (which was manufactured by the company Sener) is foreseen: the driver will require a separate 6U board for both nominal and redundant systems.

## 4.10 Payload Performance Simulation

The performance of the Ariel payload is evaluated through numerical simulations implementing a detailed description of the payload design. We have developed two simulators: ExoSim and ArielRad. ExoSim (see

Sarkar et al. 2020) is a time-domain, end-to-end simulator, see Figure 4-27. ExoSim starts from a representation of the science target to be observed (spectral light curve + emission and absorption processes of the star and the planet) and simulates the detection providing as output spectral images vs time similar to Ariel Level-1 data (see Section 6.4.3 for description of the data levels). A representative data reduction pipeline has been developed to reduce ExoSim data and to provide noise performance metrics (see Ariel Data Processing Pipeline Design Description 2020). ArielRad (Mugnai et al. 2020) is a radiometric model which implements a physically motivated description of the uncertainties relevant for the detection. Both ArielRad and ExoSim make use of a parametric description of the payload design using the same parameters captured from a central data repository.

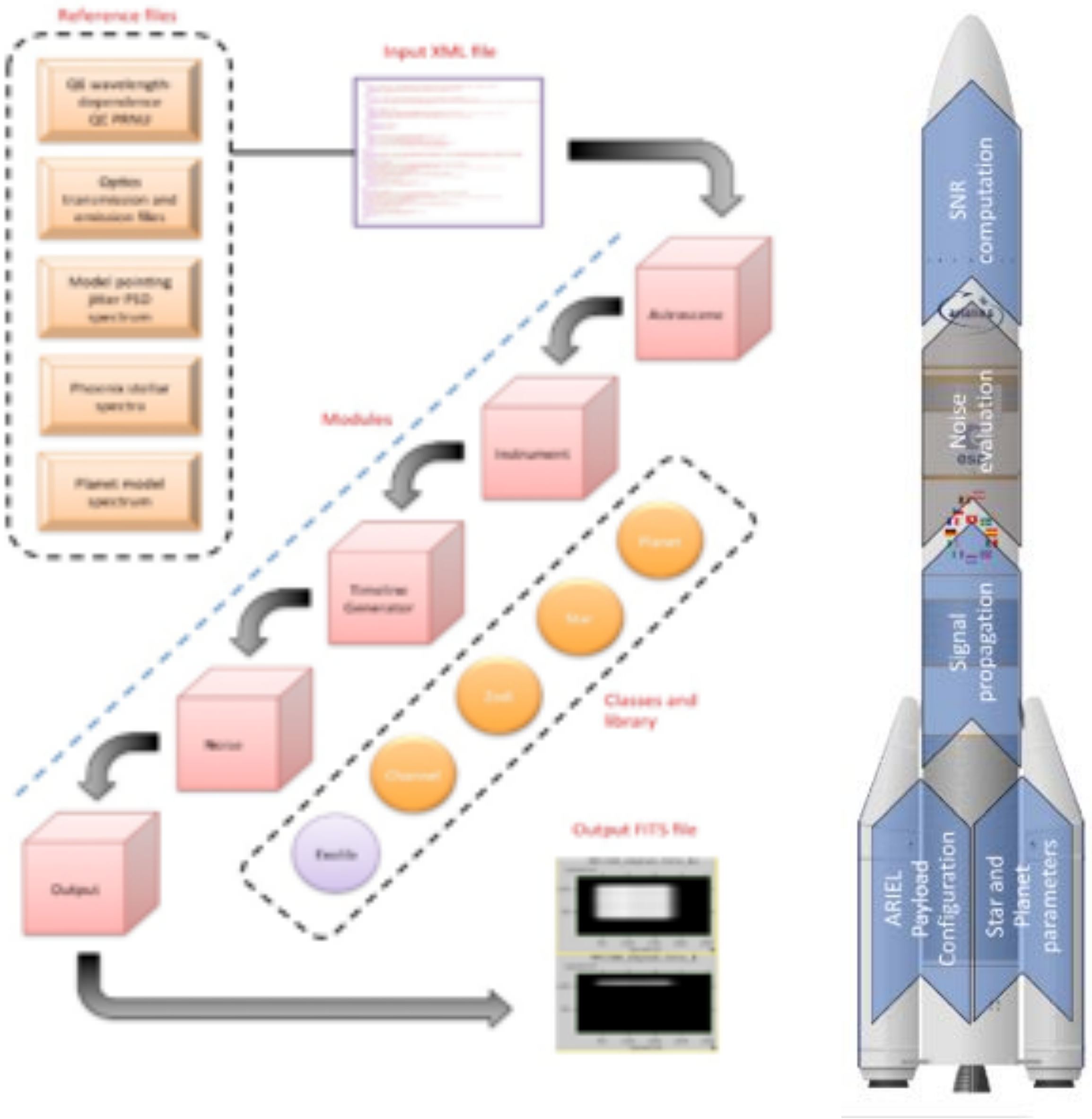

*Figure 4-23: Left: Logical description of the ExoSim simulator time-domain: The payload is defined in a parametric form which also contains all quantities required to represent how Ariel converts the light arriving at the telescope to the data timelines and how these are processed to produce spectra. Right: the ArielRad radiometric simulator workflow: the simulation starts with two input files: a payload configuration file and a candidate planet list. ArielRad propagates the target host star signal thought the payload, then evaluates the noise based on outputs from the time domain simulations. Finally, the simulator estimates the transit or eclipse observation and the resultant SNR.*

## 4.11 Payload AIV and Ground Calibration Plans

The Ariel payload will follow a pFM approach to overall qualification. Some major design aspects will be de-risked earlier in the program using the development models where possible within the programmatic constraints. Two models of the Payload module are now foreseen: a Structural Model (SM) used to de-risk the structural design of the overall PLM, which is then upgraded with functional mirrors and instruments to become an Engineering Model (EM) and then proto-Flight Model (pFM). There will also be Avionics Models (AVM) for the electronics units which interface to the spacecraft (the ICU, FCU, TCU and CCE) to verify the electrical and electronics interfaces. The SM, AVM and pFM will be delivered to ESA for inclusion in S/C level AIV activities; the SM is returned to the AMC after the S/C SM environmental test to be upgraded and become the payload EM.

The PLM performance verification and ground calibration will be carried out in a dedicated test facility at RAL Space, in the UK. This will provide a low temperature environment allowing simulation of the thermal environment of the PLM to verify the thermal design and correlation of the thermal model. A dedicated set of OGSE with an ultra-high stability (and monitored) source will be used to verify the payload system overall photometric stability. Further details can be found in the Ariel Calibration OGSE Design Definition (2020).

## 4.12 Payload Technology Development Status and Plans

The development status of all baseline technologies for the payload is being assessed as part of the payload SRR review process, with a number of the key technologies (identified during the phase A study) having been subject to prototyping and testing recently. Some of the key technologies that are being developed and demonstrated are:

- the dichroic D1 (cutting at 1.95µm), a prototype of which is being manufactured by the InfraRed Multilayer Laboratory (IML), now at University of Oxford;
- the 1m class primary mirror made in Aluminium 6061-T651. The manufacturing capability, stability of the mirror and the coating at cryogenic temperature have been the subject of initially the manufacture of the PTM mirror in phase A by the consortium, and now an ESA funded TDA activity. During phase B1 work has continued in demonstrating the feasibility, through the additional processing of the PTM through heat-treatment (to stabilise the material), polishing and coating of the material. It has undergone cryogenic testing during summer 2020. In addition during this phase the qualification of the key aspects of the mirror manufacturing processes have been on-going on sub-scale models of the mirror.
- the M2M refocus mechanism has included a development activity by Sener (Spain) to delta-qualify the design of Gaia and Euclid M2M mechanism actuators into one fully suitable for the Ariel temperature requirement, see Figure 4-4;
- the Visible / NIR LED sources for in-flight calibration are undergoing evaluation for use in the necessary environment at Cardiff University;
- the discrete electronics components for a bespoke cold front end electronics of the AIRS detector chain has been undergoing testing at operational temperature at CEA (France)
- two Engineering grade MCT detectors (with a tailored cut-off wavelength designed specifically for AIRS Ch1) have been procured by ESA and are undergoing environmental testing (including radiation exposure) and performance testing at both ESTEC and CEA;
- the TCU team have developed breadboard for both the driver circuits for the On-Board Calibration Source (to verify that the required accuracy and stability can be achieved) and for the temperature control circuits on the TSIRC board;
- the FGS team have breadboarded the manufacturing of the Aluminium mirrors for the FGS optical module, and the mounting of the FGS dichroics and lenses and tested these at cryogenic temperatures
- the Ne-JT cooler has been subject to a TDA funded by ESA for both the demonstration of the performance of Neon as the working fluid, and for the demonstration of a complete system including the next generation compressor design.

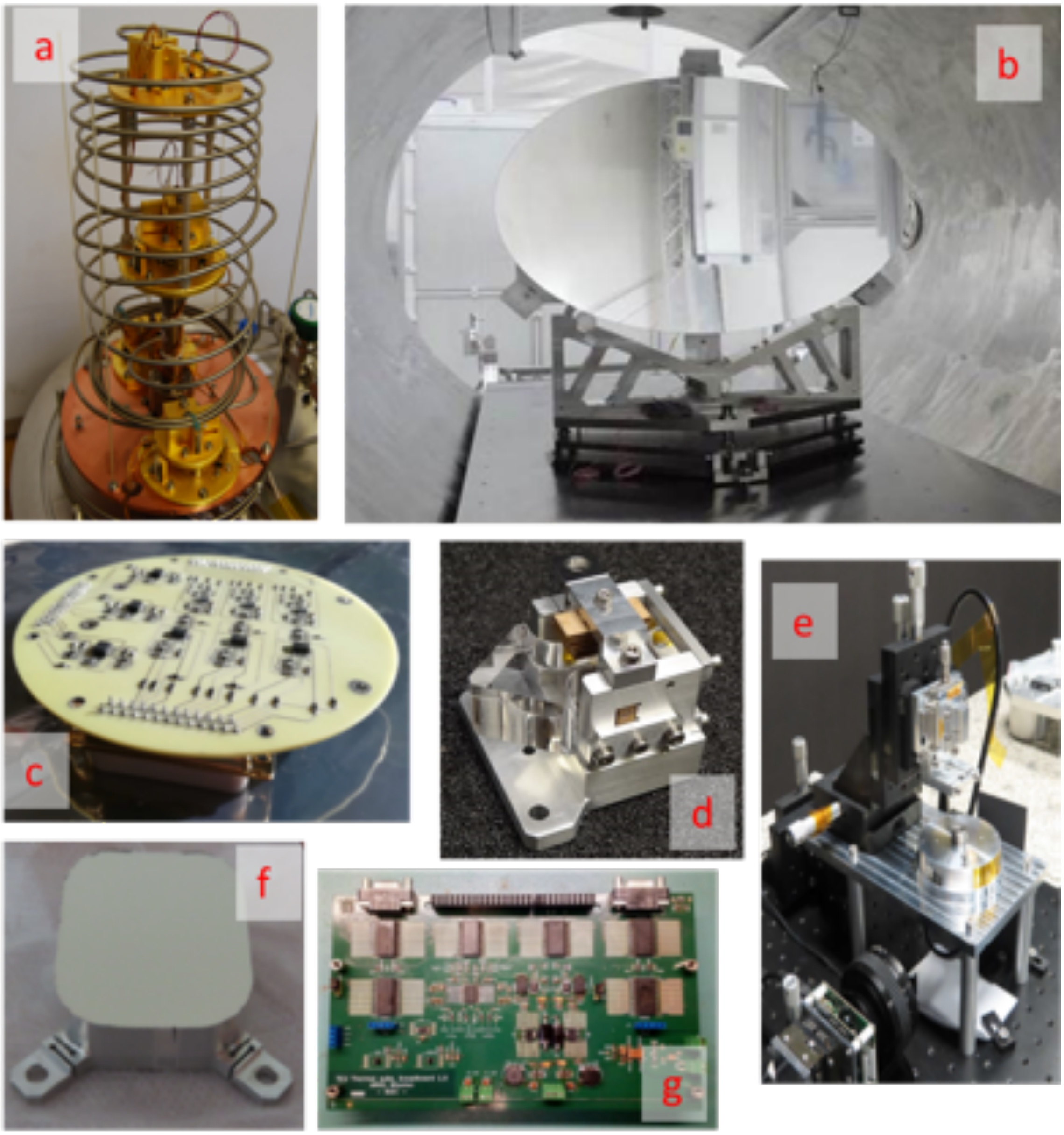

*Figure 4-24: A selection of the TDA hardware developed during phase A and B1. (a) demonstration hardware for Neon JT HX at RAL, (b) Pathfinder Telescope Mirror (PTM) from Italy in TVAC chamber at CSL; (c) AIRS CFEE candidate discrete components on test board at CEA; (d) breadboard optically polished prism for AIRS in holder; (f) breadboard FGS M1 mirror;(e) mounting stand to verify bonding of small FGS optical components; (g) TCU OBCS driver breadboard*

# 5 Mission and Spacecraft Design

This Chapter presents the Ariel mission design (Section 5.1) a brief summary of the S/C design coming from the parallel industrial studies (Section 5.2), followed by a description of the S/C development plan (Section 5.4) and an evaluation of the technology readiness (Section 5.5).

## 5.1 Mission Design

The Ariel nominal science operations orbit is an eclipse-free (Earth and Moon) large amplitude halo orbit around the Sun-Earth L2 point (see Figure 5-3). This orbit is key to meeting two of the most important science requirements: it offers a very stable thermal environment (constant view to the Sun on one side and to cold space on the other side, and no eclipses), combined with a very large instantaneous field of regard. In addition this orbit also enables a simple design of the communications and power subsystems (stable distances to the Earth and Sun and stable Sun-S/C-Earth angle), along with a benign radiation environment compared to Earth orbits that cross the radiation belts.

The baseline launch strategy consists of an Ariane 6.2 launch from CSG, Kourou, in a dual launch configuration with the Comet-Interceptor (F1) mission. Ariel sits in the bottom position under the Dual Launch Structure (DLS), and Comet-I on top. Both spacecraft will be injected into similar direct transfer trajectories towards L2.

The separation sequence will consist of the ejection of Comet-I first, followed by the ejection of the DLS, and then Ariel. The ejection ΔVs will be adjusted to minimise the risk of collision between both spacecraft, and with the DLS. The ejection modes will be optimised separately for both missions (e.g. 3-axis stabilised or spin-stabilised) taking into account their specific constraints (e.g. need to minimise the risk of direct Sun illumination of the cryogenic PLM for Ariel).

The PLM thermal design is optimised for the nominal cryogenic conditions and therefore any PLM Sun illumination should be minimised to avoid any risk of damage (e.g. through over-heating of structural/thermal elements or payload detectors etc.) and to ensure that all elements and materials stay within their qualified temperature range, in all mission phases. There are three separate phases to consider:

- Launch: during this phase, Ariel is fully protected by the DLS. After ejection of Comet-I and then the DLS, the launcher upper stage attitude will be controlled within the Ariel allowed Sun angles to completely avoid any direct Sun illumination on the PLM.
- Launcher separation: until booting, Sun acquisition and slew to a safe attitude, the spacecraft is uncontrolled and entirely dependent on the initial separation conditions provided by A62 (this is a short transient phase expected to last < 10 minutes). The standard tip-off rates provided in the A6 User Manual are insufficient to ensure no PLM Sun illumination. Work has been initiated on two fronts to address this:
    - Arianespace has been contracted to analyse the launch trajectory and perform a separation analysis. Optimised separation conditions are expected to be identified (i.e. improvements compared to the conservative values provided in the generic User Manual) as a result of this activity. This will also identify whether the 3-axis or spin stabilised separation mode is best suited to Ariel. Kick-off of this activity occurred in Q2/2020.
    - The payload Consortium are performing transient thermal analyses representative of this phase, to determine whether the conditions provided by Arianespace (resulting Sun angles) and the industrial contractor (time to boot and slew Ariel to a Sun-safe attitude) are acceptable.
- Rest of the mission till End of Life (EoL): the Ariel spacecraft is strictly designed to ensure that the Sun can never illuminate the PLM in all operational conditions (nominal and failure cases). This is achieved with cut-angles defining the overall S/C architecture and geometry (see Figure 5-1), with the SVM fully shading the PLM at all times in all possible attitudes, with margins and associated Failure Detection Isolation and Recovery (FDIR) procedures to ensure that this remains true at all times.

Given the status of the A6 launcher (new development, still awaiting first flight), several uncertainties remain, including the actual performance. Adequate mass allocations have been given to both missions, margins are kept above both projects, and the possibility to make use of a bespoke/lighter DLS, as opposed to the standard "medium" DLS sized for large ≥3 t telecommunication satellites (significantly oversized for Comet-I), is under investigation. Launcher evolutions are also planned between the first launch and 2029, which may result in further improvements in the expected performance.

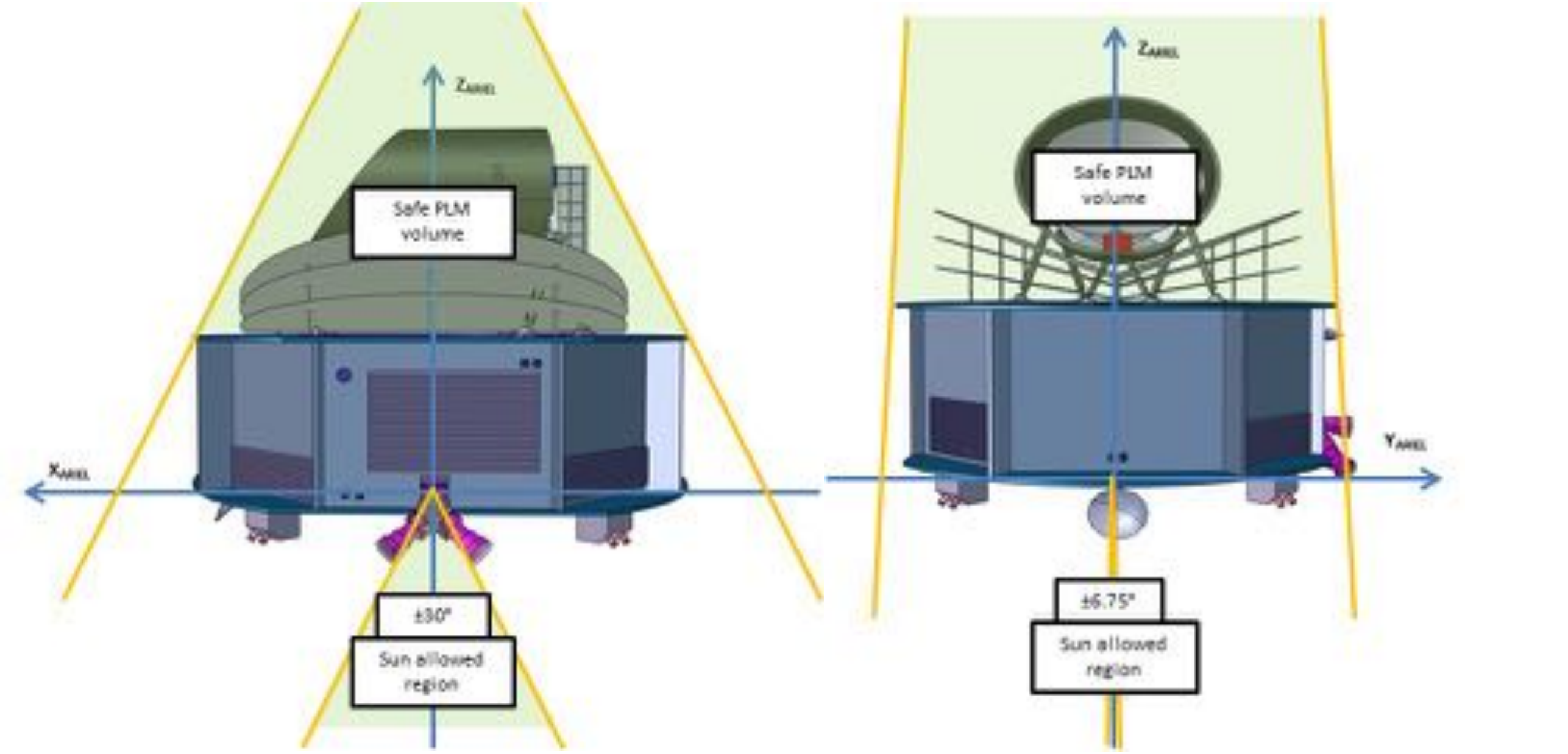

*Figure 5-1: Ariel nominal cut angles and PLM shading (illustrative, not to scale). Left: ±30° around Y. Right: ±6.75° around X. In addition to the 360° slew capability around Z, this provides access to over 40% of the sky at any given time (requirement ≥ 30%). Note that these cut angles include margin to accommodate for FDIR aspects. The resulting angular range that can be used for science observations is therefore a bit smaller.*

Near daily launch windows exist with a duration of ~1 to 2 hrs, as depicted in Figure 5-2.

The transfer and insertion into L2 is designed to be propellant free, however three opportunities at day 2, 5 and 10 after launch are available for Transfer Correction Manoeuvres (TCM) to correct for launcher dispersion and perigee velocity errors.

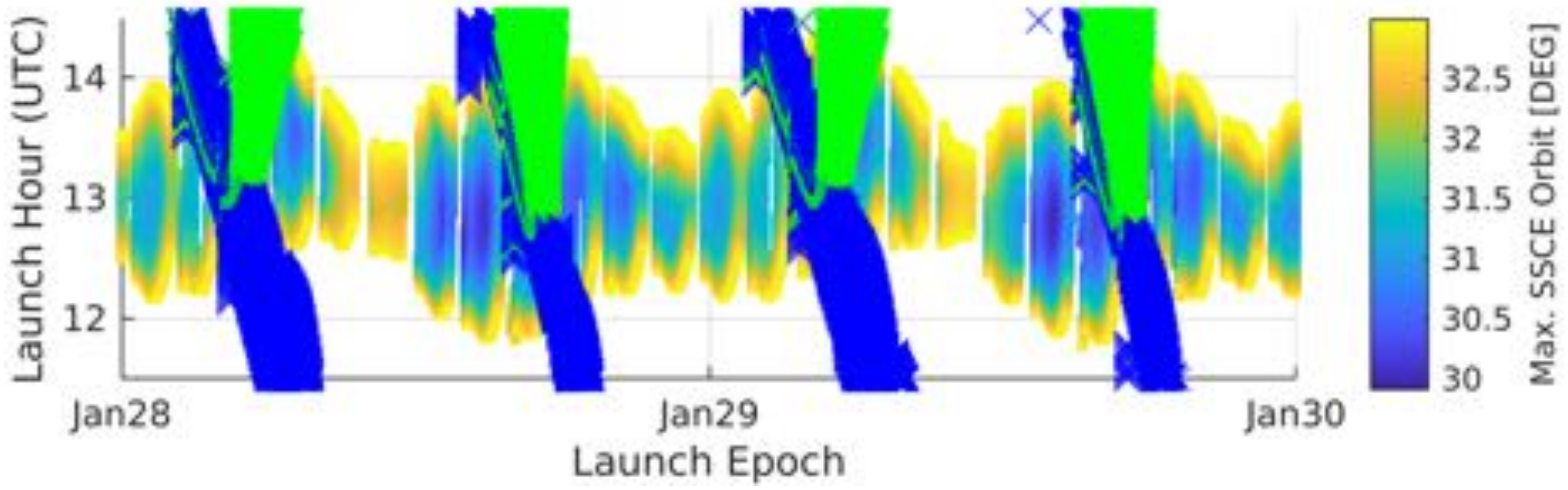

*Figure 5-2: Example launch window for a direct transfer to L2 provided by ESOC (to be confirmed by Arianespace). Similar results are valid for any launch year. The colour scale indicates the resulting Sun-S/C-Earth (SSCE) angle in L2 as a function of the launch time. Launches resulting in eclipses during the transfer to L2 and during the nominal operations phase are crossed-out.*

After launch, three months are allocated to the S/C commissioning phase, followed by another three months for the payload verification and science demonstration phases. The remaining 3.5 years are dedicated to the nominal science operations phase, while the mission is sized for an additional two years goal extension, with small impacts on units degradation over lifetime, radiation dose and consumables (mainly solar cells and propellant).

During the nominal science operations phase, the S/C is designed to provide an observation efficiency ≥ 85%. Very few events impact this efficiency (eventual safe modes, monthly station keeping manoeuvres, reaction wheels off-loading and slews between targets), while the rest of the time the S/C is available for science observations (including calibration observations). Ground contacts do not degrade the observation efficiency as parallel telemetry downlink and science observations are possible thanks to the accommodation of the antenna on a mechanism with two degrees of freedom (DoF). This simplifies the ground contact schedule, while at the same time avoiding any interference with the time critical observations of exoplanet transits/eclipses.

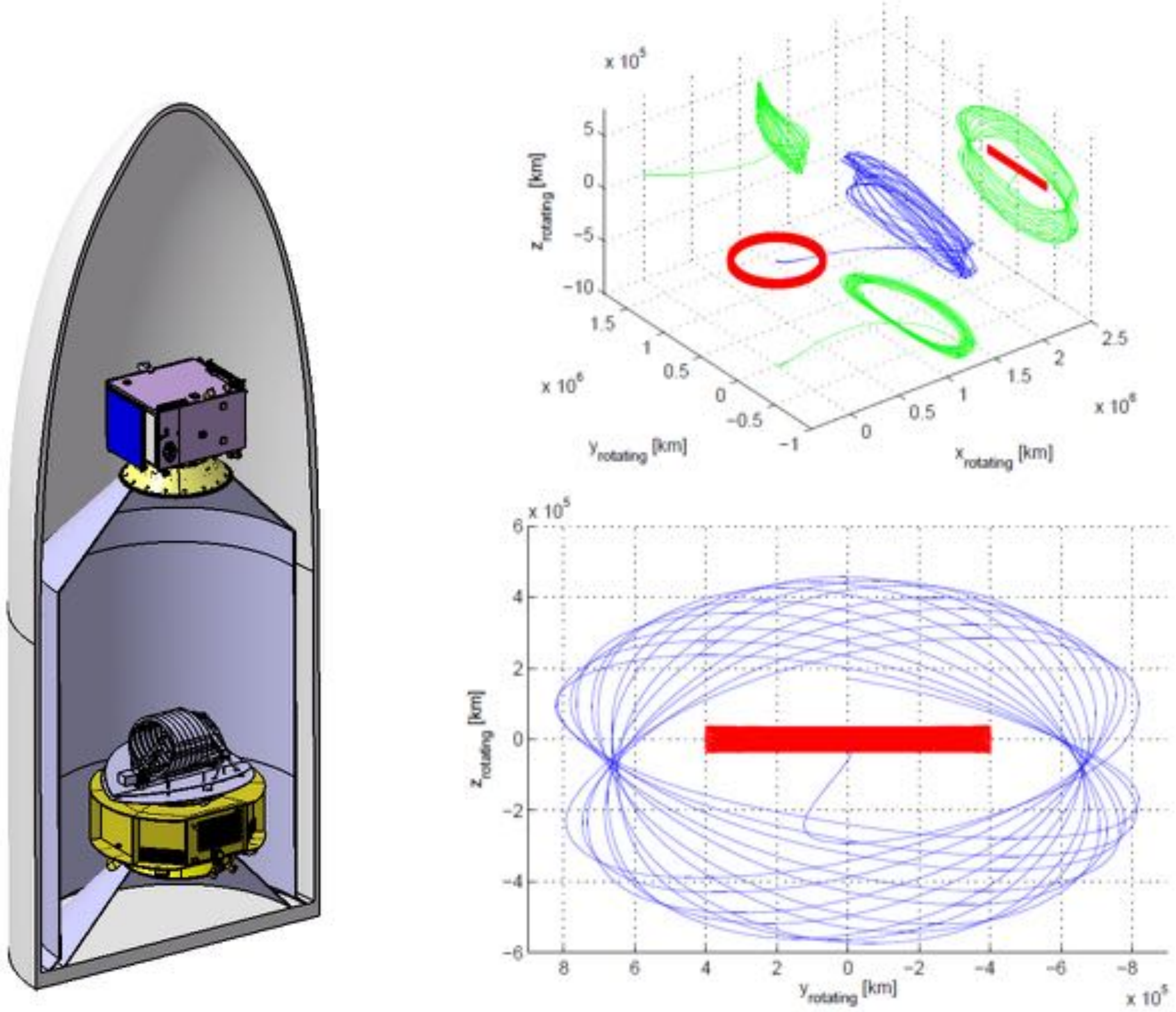

*Figure 5-3: Left: illustration of the Ariel S/C in the lower position inside the A62 dual launch fairing (Comet-Interceptor sits in the top position). Right: example of the Ariel quasi-halo orbit around L2 and how eclipses are avoided. Blue shows the 3D trajectory, while green shows the projections in the rotating frame. Red indicates the Moon and its projection in the YZ plane of the rotating frame.*

At end of life, de-commissioning will be performed with a de-orbiting manoeuvre to ensure that the S/C does not return to Earth and pollute the protected LEO and GEO regions. This is a probabilistic analysis resulting in a likelihood > 90% that the S/C does not return to Earth within the next 100 years. Higher probabilities can be achieved, simply by restricting the amount of days every year during which the disposal manoeuvre can be performed. Based on this, a re-entry casualty analysis has been performed that demonstrates that both S/C designs comply with the Agency's requirement (casualty risk $\leq 10^{-4}$).

## 5.2 Spacecraft Design

An illustration of the Ariel S/C is given in Figure 5-4.

Note that although the two S/C designs from the industrial contractors have many similarities, the design specifics cannot be elaborated upon in detail here due to the competitive nature of their work. As Phase B1 ends, both primes are initiating preparations for their proposal to the upcoming B2/C/D ITT and do not wish to describe their design in this document. Instead, this chapter will simply present the S/C driving requirements and general design features deriving from those requirements.

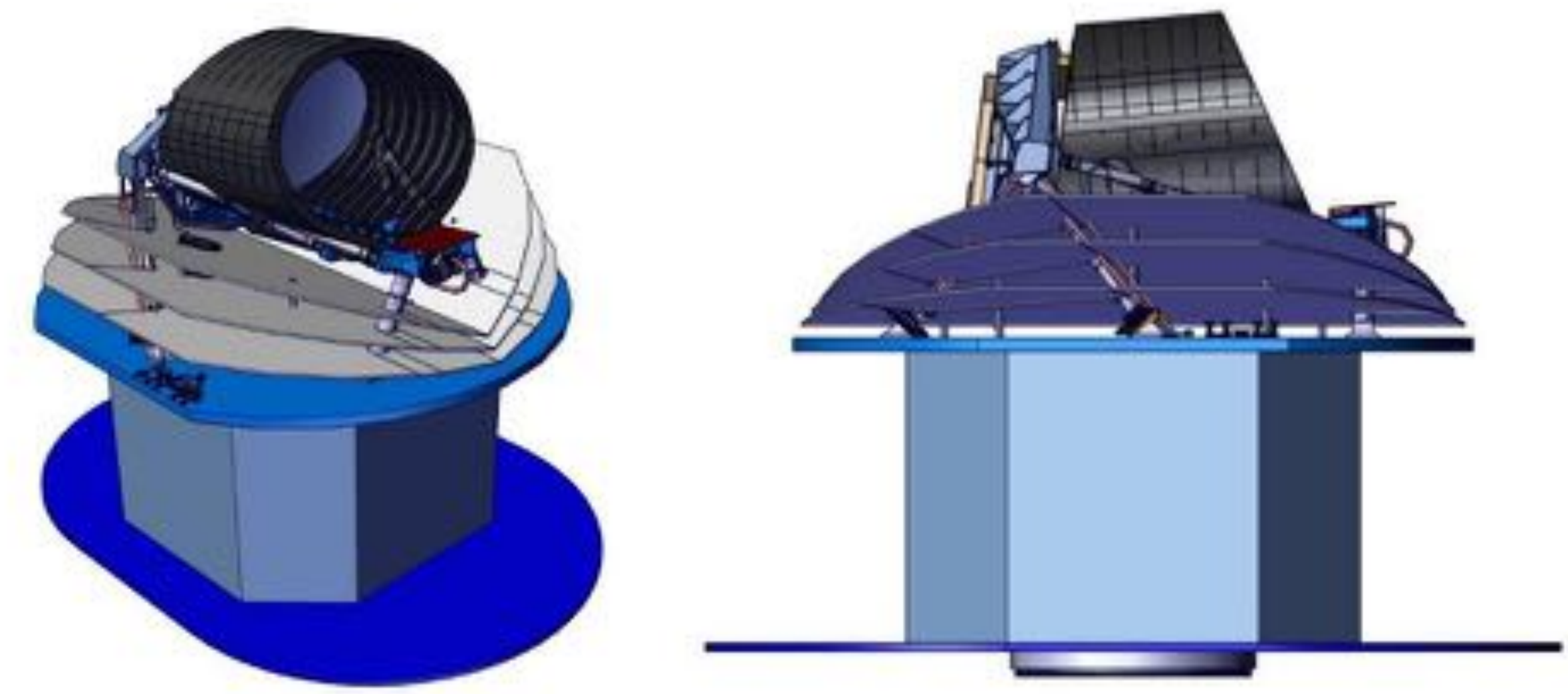

*Figure 5-4: Illustrative representation of the Ariel S/C from two angles, composed of the Consortium provided PLM, and a generic dummy SVM. An upwards tilt (5°) can be seen on the right picture between the telescope pointing axis and the SVM "horizontal" XY plane. This means the accessible sky described in Figure 3-19 is a 50° wide annulus region near orthogonal to the ecliptic plane (-20° / + 30°)*

### 5.2.1 Design drivers

The overall design architecture is driven by the PLM accommodation, having to be in constant shade from the Sun, behind the SVM, and covering the cut angles detailed in Figure 5-1.

The M4 programmatic constraints have resulted in the need to follow a strict design-to-cost approach. The S/C design is therefore optimised to simply meet requirements, as opposed to provide the best possible performance at all cost. The resulting implications are listed in Section 5.2.2.

Following the selection of Comet-Interceptor, this design-to-cost philosophy is now topped by a design-to-mass philosophy, to ensure that both missions can be implemented and launched with an A62. There has so far been no adverse effect on the Ariel spacecraft design due to this change. Nonetheless it means the mass evolution will need to be carefully monitored and controlled during the development phase. Any potential structural issue will not be simply resolvable by stiffening or strengthening the design with more mass, but detailed design optimisation features will need to be investigated instead.

*Table 5-1: Summary of the main Ariel pointing stability requirements. "c" pointing metrics refer to the coarse pointing mode, "f" to the fine pointing mode. The various pointing metrics relevant to Ariel are the Absolute Performance Error (APE), Relative Performance Error (RPE), Mean Performance Error (MPE) and Performance Drift Error (PDE) around the Line of Sight (LoS) of the various instrument channels. The most stringent requirement is the high frequency (≥ 10 Hz) fine RPE across LoS (180 milli-arcsec), which imposes strict control of all micro-vibration sources (reaction wheels and cryo-cooler). "Around LoS" pointing errors are omitted in this table, as they are typically relaxed by ~1 order of magnitude compared to the "across LoS" given the nature of the Ariel observations (single target viewed as a point source in the centre of the FoV).*

| Mode | Submode | Pointing metric | Bright targets | Faint targets | Time scale |
|---|---|---|---|---|---|
| Coarse Pointing Mode | Acquisition | cAPE across LoS | 10 " | | |
| | Tracking | cAPE across LoS | 3.5 " | | |
| Fine Pointing Mode | | fAPE across LoS | 1 " | | |
| | | fRPE across LoS | 180 mas | 280 mas | 0.1 s |
| | | fRPE MPE across LoS | 130 mas | 280 mas | 0.1 s to 90 s / 300 s |
| | | fPDE across LoS | 70 mas | 300 mas | 90 s / 300 s to 10 hrs |

Beyond the PLM accommodation and the programmatic constraints, the main S/C design drivers reside in the interface requirements to the payload. In particular, all interface stability aspects that have a direct impact on the measurement noise or photometric stability are of prime importance, and result in rather stringent requirements. This is particularly true for the pointing stability requirements (see Table 5-1 below), but also for the SVM to PLM thermal interface (both conductive heat exchange through the PLM bipods, and the radiative heat exchange between the SVM top panel and the V-grooves) and the thermo-elastic stability, and to a lesser extent the power stability and the EMC environment inside the SVM for the PL warm units.

### 5.2.2 Generic design considerations

The SVM is composed of a main body, housing all the S/C equipment and warm PL units, between a bottom panel and a top panel. The bottom panel holds the function to host the body mounted Solar cells, and any other equipment that needs to be pointed towards the Sun or the Earth (e.g. antenna), as well as propulsion thrusters. Together with the SVM body, this bottom panel also acts as a 1$^{st}$ partial Sun shade for the PLM. The top panel, also referred to as the Payload Interface Panel (PIP), is the mechanical and thermal interface to the PLM, and acts as a complete Sun shade (“V-groove #0”) for the PLM in all possible attitudes.

The design-to-cost philosophy has resulted into enforcing the following design constraints:

- the AOCS subsystem uses Reaction Wheels as the sole actuators to achieve the fine pointing requirements (i.e. no micro-propulsion system).
- given the low amount of data produced (no imaging, only a few photometric channels and coarse resolution spectroscopy), the communication subsystem makes use of X-band (i.e. no Ka band).
- the amount of ground contact passes needed for data downlink is limited to only 15 hrs per week, split in 3 passes (no daily passes, no week end passes).
- the power budget is constrained (~1kW) to enable body mounted Solar cells only (i.e. no deployable Solar array).
- the mass budget is constrained (~1.5t) to enable the dual launch with Comet-I (i.e. Ariel cannot increase in capability and complexity any further with new add-ons to the design).
- the lifetime is limited to 4 years, while the design and consumables are sized for a goal of 6 years.

The results of the Phase A/B1 studies, combining the conclusions of both industrial primes, indicate that the SVM design is feasible, can make use of high TRL equipment only (i.e. no technology developments are needed inside the SVM), and can meet the science and mission requirements.

## 5.3 Spacecraft and Payload AIV and Development Plans

The spacecraft development plan is based on a standard Proto Flight Model (PF) approach, although with some adaptations due to the specifics of the Ariel mission: optics in cryogenic conditions, and a clear SVM / PLM separation (both physically and in terms of responsibility between ESA / industry and the payload Consortium / Member States). This implies that all final qualification tests are conducted on a S/C PFM. Such an approach carries an acceptable amount of risk and avoids having to produce and pay for two full S/C models in parallel (one for testing / qualification and one for launch). To consider environmental tests only at final S/C FM level would further decrease the cost but would increase the risks to a level that is no longer acceptable (i.e. large cost impact if a test fails so late in the development phase and results in the need for re-design and re-qualification).The PFM will be built assembling the SVM and PLM PFM models together. The FM units of all payload elements (telescope, instruments, cryo-cooler etc.) will be included, but the PLM internal alignment, and the overall payload performance verification in operational conditions will be done on the PL PFM only, prior to delivery to the S/C prime for integration on the SVM. Therefore, no “deep” cryogenic (beyond liquid Nitrogen) or optical tests are needed at full S/C PFM level. Only functional verification of the PL will be done on the S/C PFM. The S/C PFM model is typically supported by additional models earlier in the development plan. The following models are baselined:

At S/C level:

- an early Structural Model (SM) to de-risk the structural design, and confirm / lower the mechanical loads specified to the PLM. This S/C SM will be developed with a simplified PLM SM, procured by ESA / industry, so as to de-couple the S/C development from the payload development under the responsibility of the payload Consortium.
- an Avionics Model (AVM): this model typically enables to check all the functional and electrical (power, data and communications) interfaces between all units, to verify the functionality of the avionics including the on-board software and the AOCS control loop, and potentially EMC and RF tests etc.

At PL level:

- A payload S(T)M/EM, for functional and performance testing of the PL in operational conditions. This payload model test campaign should be completed by pCDR, to give the go-ahead for production of the PL FM model.
- a PL AVM, composed of the instrument EMs. Only the warm parts of the PL elements will be sent to prime for integration on the S/C level AVM, while the cold instrument EMs will be replaced by a cold PLM simulator.

A representative text matrix is shown in Table 5-2. Beyond these hardware models, additional software models and simulators are planned, and dedicated EGSE as well. Of particular importance:

- The S/C prime will provide a Spacecraft Interface Simulator (SIS) to the payload Consortium, to enable control and testing of the PL AVM, EM and pFM models.
- The payload Consortium will provide two early FGS models (a Functional Simulator and a High Fidelity Simulator), for inclusion and testing within the AOCS closed control loop tests at prime level.

Some major payload design aspects will also be de-risked earlier (some having already started in Phase B1) in the program using development models where possible within the programmatic constraints (see Section 4.12). Details on the planned models and the test philosophy of the PLM are outlined in Section 4.11 above.

*Table 5-2: PL and S/C test matrix. Tests at lower levels (e.g. subsystem or instruments) are not shown here. The general principles are agreed, further refinements may still take place.*

| **Model** | **SM** | | **AVM** | | **PFM** | |
|---|---|---|---|---|---|---|
| **Level** | **PL** | **S/C with simplified PLM** | **PL** | **S/C** | **PL** | **S/C** |
| **General** | | | | | | |
| Optical alignment | X | | | | X | (X) |
| Functional | (X) | (X) | X | X | X | X |
| Performances | | | | | X | |
| Mission | | | | X | | X |
| Polarity | | | | | | X |
| Mechanical Interfaces | X | X | | | | X |
| **Mechanical** | | | | | | |
| Physical properties | X | X | | | X | X |
| Modal survey | X | X | | | X | X |
| Static | X | X | | | | |
| Acoustic | X | X | | | X | X |
| Random vibration | | | | | | |
| Sinusoidal vibration | X | X | | | X | X |
| Shock | | X | | | | X |
| Micro-vibration susceptibility | | (X) | | | | (X) |
| Mechanisms | (X) | (X) | | | X | (X) |
| **Structural integrity** | | | | | | |
| Proof pressure | | | | | | X |
| Pressure cycling | | | | | | |
| Design burst pressure | | | | | | |
| Leak | | | | | | X |
| **Thermal** | | | | | | |
| Thermal vacuum | | | | | X | X |
| Thermal balance | | | | | X | X |
| **Electrical / RF** | | | | | | |
| EMC (conducted & radiated) | | | | X | X | X |
| ESD | | | | X | | |
| Electromagnetic auto-compatibility | | | | | | X |

X: test to be performed
(X): test that may be performed, depending on the AIV programme.

# 6 Ariel mission and science operations

## 6.1 Ground Segment Overview

### 6.1.1 Overview of the operational centres

The Ariel Ground Segment (GS) provides the means and resources with which to manage and control the mission via telecommands, to receive and process the telemetry from the satellite, and to produce, disseminate and archive the generated products (for more details Pearson et al. 2020).

Responsibility for and provision of the Ariel GS is split between ESA and the Ariel Mission Consortium (AMC); the Ariel GS will be comprised of the following elements:

- The Operational Ground Segment, comprising:
  - The Mission Operations Centre (MOC) – provided by ESA
  - The ESA tracking station network (ground stations) – provided by ESA
- The Science Ground Segment, comprising:
  - The Science Operations Centre (SOC) – provided by ESA
  - The Instrument Operations and Science Data Centre (IOSDC) – provided by the AMC

A schematic view of the operational interfaces for the Ariel mission is presented in Figure 6-1. Full details of the operational interfaces and plans are contained in Ariel Science Operations Concept Document (2020). The responsibility for the SGS tasks and activities are distributed and shared between SOC and IOSDC.

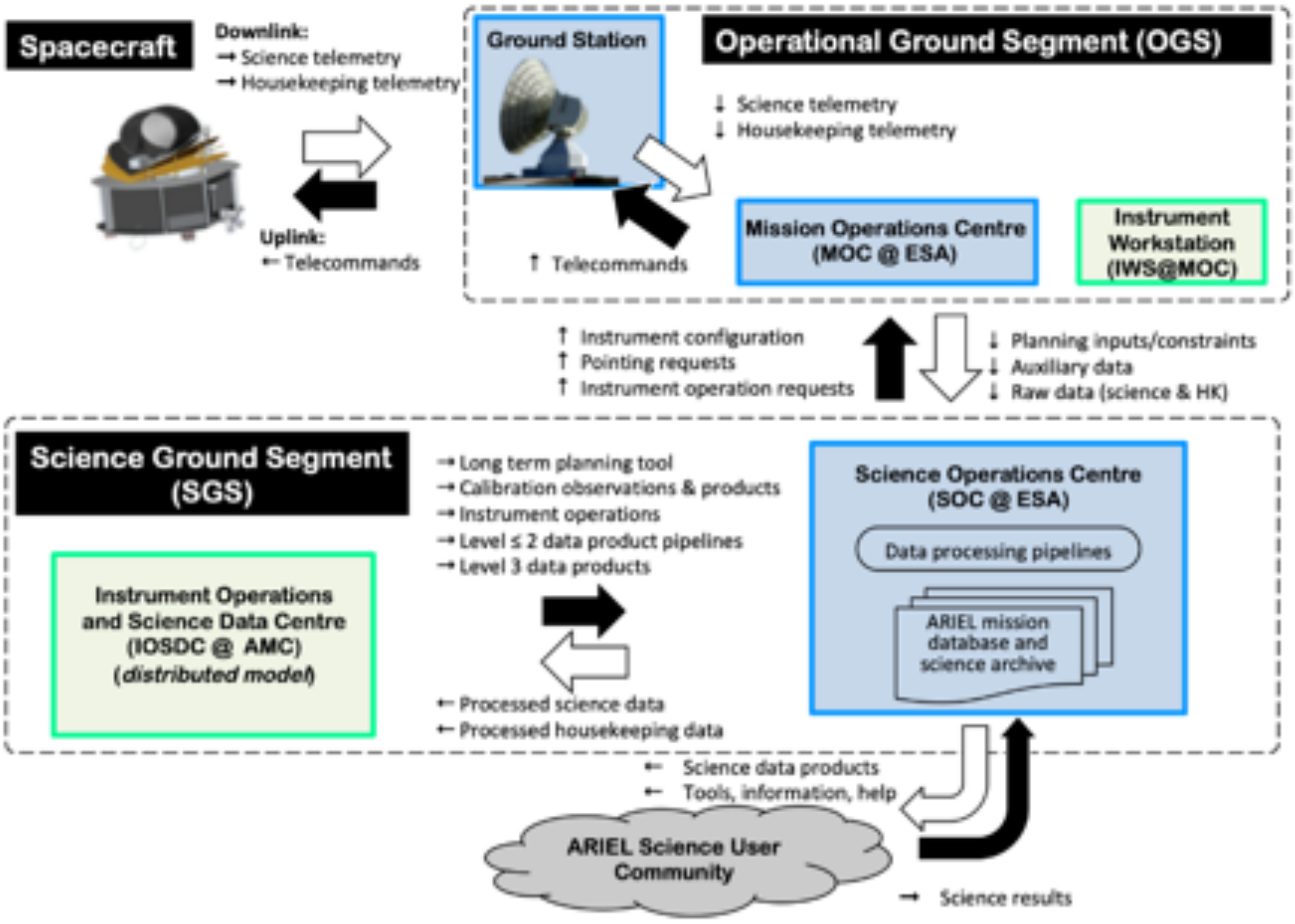

*Figure 6-1: Schematic view of the ground segment and data flow in routine science operations phases, with the main responsibilities of the various actors indicated; blue boxes are ESA and green boxes are AMC responsibilities.*

The following is a list of the key tasks that will be performed during Ariel operations:

- Science planning (target selection and long term planning)
- Mission operations planning
- Production and execution of operations requests (including spacecraft and instrument sequences)
- Ground control and monitoring at MOC (including spacecraft health and orbit maintenance)

- Contingency isolation, management and recovery
- Instrument/payload operations and calibration
- Mission (including science) data processing, population of archive, provision of data products
- Ground infrastructure operations and management
- Spacecraft and ground segment performance analysis and tuning
- Science data archive and community support including community calls for core and complementary science targets

### 6.1.2 Mission Operations Centre and Ground Stations

The ground segment and operations infrastructure for the Mission Operations Centre (MOC) of the Ariel mission will be set up by ESA at the European Space Operations Centre (ESOC) in Darmstadt, Germany. The role of the Ariel MOC is fairly ‘standard’, and the MOC will be based on extensions of the existing ground segment infrastructure, customised to meet the mission specific requirements thereby maximising the sharing and reuse of facilities and tools made available from other ESA Science missions. ESOC will prepare the ground segment including all facilities, hardware, software, documentation, the respective testing, validation and training of staff required to conduct the mission operations.

The MOC will be responsible for all operations of the Ariel spacecraft and instruments during all mission phases covering both nominal and contingency operations. In particular the MOC is responsible for the following tasks:

- Planning, scheduling and execution of the ground station contacts;
- Performing uplink of the satellite and payload telecommands and receiving telemetry through the ground stations and communications network;
- Monitoring the spacecraft health and safety;
- Monitoring the payload safety and reacting to contingencies and anomalies according to procedures provided by the AMC;
- Performing mission planning of spacecraft activities;
- Alerting the SOC of all significant anomalies or deviations from nominal behaviour of the satellite and payload for onward transmission to the AMC as relevant;
- Executing procedures to safeguard the spacecraft and payload and preserve data integrity;
- Performing maintenance of the satellite's on-board software;
- Performing uplinks of payload on-board software updates as generated, validated and delivered by the AMC via the SOC;
- Providing flight dynamics support, including determination and control of the orbit and attitude of the satellite;
- Handling provision of the science and housekeeping telemetry to the SOC;
- Producing and providing ancillary data to the SOC (orbit files, pointing information, housekeeping telemetry, etc.);
- Supporting the SGS on all aspects concerning spacecraft operations.

### 6.1.3 Science Operations Centre

The Ariel Science Ground Segment (SGS) will be provided jointly by ESA and AMC, and will consist of the ESA Science Operations Centre (SOC) and the Instrument Operations and Science Data Centre (IOSDC) provided by the AMC.

The Ariel SOC will be set up by ESA at the European Space Astronomy Centre (ESAC) near Madrid, Spain, which also hosts the archives of all ESA science missions. The SOC will take the lead in the overall design and engineering of the SGS, and will organise and manage the end-to-end tests that are needed to validate the SGS uplink and downlink systems, interfaces, and operational processes.

The SOC will design, develop and operate the ESA-funded part of the Ariel science ground segment throughout all mission phases. It is the only interface to the MOC during routine operations, and is the point of contact for the community throughout all phases. In particular the SOC is responsible for the following tasks:

- Scientific mission planning in operations using the IOSDC provided long-term mission planning tool, and involvement in mission planning activities;
- Centralised scheduling system as interface to MOC to produce instrument commanding and pointing requests;
- Reception of science, housekeeping and auxiliary data from the satellite received via the MOC;
- Operation of the automatic data processing pipeline provided by the IOSDC to process science data products from Level 0 up to Level 2, housekeeping and other ancillary data, and data distribution to the IOSDC;
- Operation of a quality control system dedicated to instrument/payload data quality analysis and mission planning feedback;
- Operational support, i.e. for instrument operations, performance checks, health monitoring, trend analysis and survey execution tracking;
- Development, operation and maintenance of the Ariel archive comprising mission data base and science data archive, and data dissemination (data product Levels 1 to 3 and ancillary data) to the scientific community;
- User Support (in collaboration with the IOSDC), including ESA-led community calls, proposal handling system, documentation, helpdesk, and workshops.

### 6.1.4 Instrument Operations and Science Data Centre

The Instrument Operations and Science Data Centre (IOSDC) will be provided by the AMC. The IOSDC architecture will follow a distributed model, across the participating consortium countries, but will provide a single-point interface to the SOC (and under special circumstances to the MOC). The IOSDC will design, develop and operate the AMC-supported element of the Ariel science ground segment throughout all mission phases. In particular the IOSDC is responsible for the following tasks:

- Design, production, and delivery to the SOC of the long term mission planning tool;
- Design, production, and delivery to the SOC of the Level 0 to Level 2 Data processing pipeline;
- Provision of science instrument calibration requirements and plan;
- Calibration products production, delivery and maintenance;
- Definition of operation observing modes to observe bright, normal and faint targets;
- Maintenance of the payload onboard software and procedures;
- Health Monitoring and trend analysis System (HMS) and Quick Look Analysis (QLA) tools implementation and operation;
- Level 3 data product production and delivery to the SOC;
- Maintenance of all software provided to the SOC;
- Planning and execution of the payload Commissioning Phase activities;
- IWS provision and maintenance;
- Participating in pre-flight SGS integration and testing;
- Supporting the SOC user support on payload specific matters.

Functionally, IOSDC will be organised into teams. The tasks of each functional Team are described in the IOSDC Management Plan 2020. The IOSDC Manager will be the single point interface to SOC for the IOSDC.

## 6.2 Mission Operations

Mission operations is under the responsibility of MOC throughout all phases of the mission, from the start of the mission definition phase through preparation, implementation, and in-flight operations to the very final command beyond the Routine Science Phase (RSP).

The Ariel mission is planned with a short 2-day LEOP, included in the 3-month Commissioning Phase (CoP). While underway towards the operational orbit around L2 the transfer activities run in parallel to the early Commissioning Phase. Flight operations teams will be optimised throughout this period to ensure full MOC team and ground station coverage during the critical early stages of the mission reducing to smaller teams and day passes during the later commissioning phases. After the In-Orbit Commissioning Review (IOCR), the mission transitions into the Performance Verification Phase (PVP), in which ground station pass durations and shift team sizes are gradually further reduced to achieve those as planned for routine science operations at the beginning of the Science Demonstration Phase (SDP)

For Routine Science Phase (RSP) operations the Ariel mission is foreseen to be operated in a highly automated way. The Ariel mission on board capability is foreseen to handle commanding for execution of slews between observations based on requested target position. Reaction wheel momentum management will be controlled on board and all operations executed on board from preloaded mission timelines. This approach reduces the need for ground based definition of the slew profiles and momentum management. The mission timelines are uplinked on a periodic basis (e.g. weekly) and the shortened pass durations (~14 hours per week) are primarily used for downlink of the science data and collection of tracking data needed for orbit determination. The orbit control is anticipated to be of a predictive nature allowing the mission to apply rule- based planning for engineering windows for platform operations, allowing the remaining science planning to be done on a long term basis by the SOC. It has been assumed that based on the moveable antenna the ground station passes can be scheduled independently of the science observations.

In the event of ground anomalies Ariel is foreseen to continue nominal operations without the loss of Science data for a period of 5 consecutive days and in all post LEOP mission phases the spacecraft will be able to survive without ground contact for 7 consecutive days.

## 6.3 Science Operations, Data Handling and Archiving

### 6.3.1 Mission planning

The scientific mission planning consists of generating a long-term observation plan of the Ariel 'core survey' based on the list of potential targets (Mission Candidates Sample (MCS)), the prioritisation scheme provided by the Ariel Science Team (AST), calibration observations (on sky and internal), complementary science targets and mission constraints (spacecraft attitude constraints, data rates, straylight from bright out-of-field sources such as planets, etc.). The targets to be observed are detailed in the long-term observation plan and form the Mission Reference Sample (MRS), a subset of the MCS; see further Section 7.2.

The IOSDC is responsible of performing the scientific mission planning activity before launch and delivery of the long-term planning tool including training to SOC. SOC will then carry out the long-term planning exercise during operations under the guidance of the AST, chaired by the ESA Project Scientist. The long-term observation plan will be produced on a monthly basis (to account for missed observations or new targets being added to the target list) always covering until end of mission and within a mission planning working group under participation of IOSDC, SOC and MOC. It is foreseen to produce several long-term plans based on different scientific scenarios. After final approval of the selected long-term plan by the Project Scientist, it serves as input to the observation scheduling and product generation under responsibility of SOC. These pointing and instrument operations requests are sent from SOC to MOC for uplink to the spacecraft.

### 6.3.2 Instrument operations and calibration

Due to the survey nature of the Ariel mission and the concise suite of science instruments, the complexity of instrument operations is inherently low with operational modes mainly conditioned by the target brightness.

Most calibration observations will be on-sky and use the same instrument operation modes as used for the science target observations. Some internal calibration operations may require dedicated modes.

The generation of the instrument configuration files and operations requests for target observations and calibration is under the responsibility of the IOSDC and goes along with the long-term mission planning exercise. The IOSDC will also maintain the payload On-board Software (OBSW), which will be stored and version controlled with any updates verified and validated by the SOC before the images are sent to the MOC for uplink to the spacecraft.

To support the commissioning phase and contingency cases, an Instrument Workstation (IWS) will be developed and operated by the IOSDC as a direct interface to the MOC data distribution system. The IWS is based on the Electrical Ground Support Equipment (EGSE) already used during the pre-launch on-ground test campaign and together with the training and continuity of expertise of instrument teams, support the concept of a "smooth transition" from development to operations. The IWS will host a Quick-Look Analysis (QLA) system to quickly visualise and analyse the science and housekeeping data received from the satellite.

From the start of the PV phase, the SOC will utilise the data interface to the MOC and the received data will be disseminated through the Ariel archive.

To provide feedback on the quality of the received science data and the payload/instrument health, an operational Health Monitoring System (HMS) will be used to analyse the data received after each ground station pass and post pass health monitoring reports will be issued. The HMS tools are developed and operated by the IOSDC. SOC will operate a trend analysis system and analyse meta-data information and the health monitoring and quality control reports for feedback on the observation success and survey execution. Another aspect is the pointing information, also provided by SOC based on information and products from MOC, which is further used for calibration in the data processing. Currently only the FGS centroid and spacecraft roll angle timelines are foreseen for this purpose.

#### *6.3.2.1 Ground calibration*

Calibration activities begin with the on-ground testing to obtain instrument calibration data for the boot-strapping of the calibration products into the data processing. A central and critical part in this context is the detector system performance.

The ground test and calibration campaign will be the responsibility of the AMC and is described in Pearson et al. 2020 It is desirable to implement as many tests as possible at the lower testing levels (Component and Unit Level) to avoid complication further down the signal chain at instrument (Subsystem) or Payload Module level. The test plan will follow the methodology given below and will follow the development schedule outlined in Section 6.3:

- **Component Level Test:** Low level testing of components such as Detector SCA, CFEE, Dichroics, mechanism actuators, etc.
- **Unit Level (parts of a subsystem) Test:** Unit level tests are those carried out on individual components (i.e. parts of a subsystem) of the system without inter-dependencies / influences from other components. This would include the FGS and AIRS detector systems.
- **Subsystem Level Test:** These are tests at the instrument level and would include the entire AIRS instrument, FGS instrument (including FCE), the Telescope Assembly (including the Bipods and the V-grooves), Instrument Control Unit (ICU) and the Cooler unit. Instrument level tests are those carried out on the entire instrument including the optical paths and inter-dependencies / influences.
- **Payload Module Level Test:** This level comprises the integrated instruments and telescope payload and includes Payload Functional Tests, Test Facility Functional Tests, EGSE integration testing and Ground Segment End-to-end testing. The following payload module models are foreseen: structural model, avionics model, and proto-flight model.

#### *6.3.2.2 Flight Calibration*

Regular in-orbit calibration is required to establish, track and update calibration parameters, either if the required accuracy can only be obtained in orbit or due to environment changes, e.g. temperature and radiation effects. The on-sky and internal calibration observations are absorbed into the long-term observation plan.

The internal calibrator is described in Section 4.6 and will be used for monitoring trends in the detector systems, particularly concerning the flat-fielding performance. On-sky calibration as described in detail in Pearson et al. (2020) will be made using pre-selected highly stable target stars (for first selection of possible stars see Sky Calibrators for ARIEL 2017) with similar flux to the bright targets of Ariel.

### 6.3.3 Data flow and data level products

Data acquired in-flight by Ariel are provided to the SOC. The data flow of the signals produced by the Ariel detectors is described in Figure 6-2 for a single transit observation (star + planet). Charge is accumulated up the detector ramps until a detector reset is performed, forming a single exposure. The detector pixel clock drives the cadence at which the detectors are read. The signal up each ramp can be read out as a Non-Destructive Reads at a given cadence (NDR, see Figure 6-2).

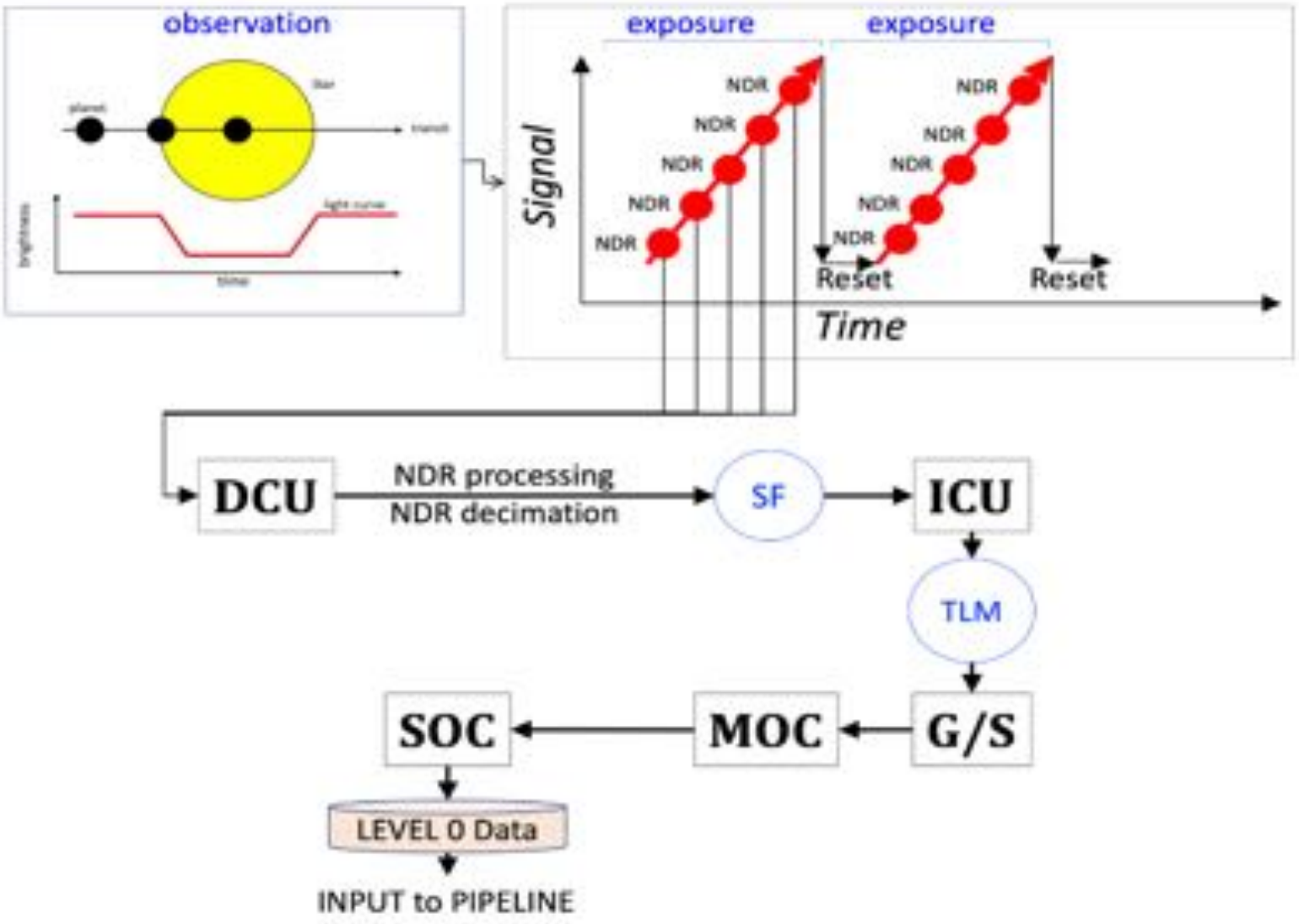

*Figure 6-2: Simplified Ariel detector data flow: (a) Detector NDRs (Non-Destructive Reads) read by CFEE (Cold Front End Electronics) at fixed cadence are sent to the DCU (Data Control Unit); (b) DCU performs decimation/processing (see text) of the NDRs to create Science Frames (SFs), and send SFs to the ICU (Instrument Control Unit); (c) ICU performs compression and storage and sends SFs as telemetry (TLM) to the Mission Operations Centre (MOC) via ground stations. Data is then transferred from MOC to the ESA Science Operations Centre (SOC) and ingested into the data archive ready for processing by the Ariel automatic pipeline.*

The science requirements of Ariel define the amount of required signal data (number of packaged NDRs) downloaded within the allocated telemetry budget. In principle, due to the correlated nature of the astronomical signal, not all NDRs are required, therefore the on-board Data Control Unit (DCU) will provide a level of rudimentary processing (decimation, averaging, etc) to remain within the telemetry budget limits. The output from the DCU is referred to as a Science Frame (SF) and forms the basic data packet. The time stamp assigned by the DCU is the time from the on-board clock corresponding to the reset of the first pixel in the array. The science frames are then passed to the Instrument Control Unit (ICU) which sends the telemetry (TLM) data to the ground station. These are received by the MOC and passed on to SOC. These data forms the basis of the Level 0 products, the starting point for the Ariel data processing pipeline as shown in Figure 6-3).

The Ariel Data Reduction Pipeline (ADaRP), will be produced and maintained by the IOSDC and delivered to ESA SOC. ADaRP will then be run automatically at SOC, starting with the Level 0 Data Products and producing Data Products up to Level 2, which are subsequently ingested into the Ariel science archive. Level 2 Data Products will then be provided back to the AMC/IOSDC for the production and delivery of the Level 3 Data Products to SOC, and ingestion into the Ariel Science Archive.

Data Levels define break points along the pipeline processing, at which specific science data products are generated and made available to the scientific community through the Ariel science archive. These data products are summarized in Figure 6-3 and Figure 6-4.

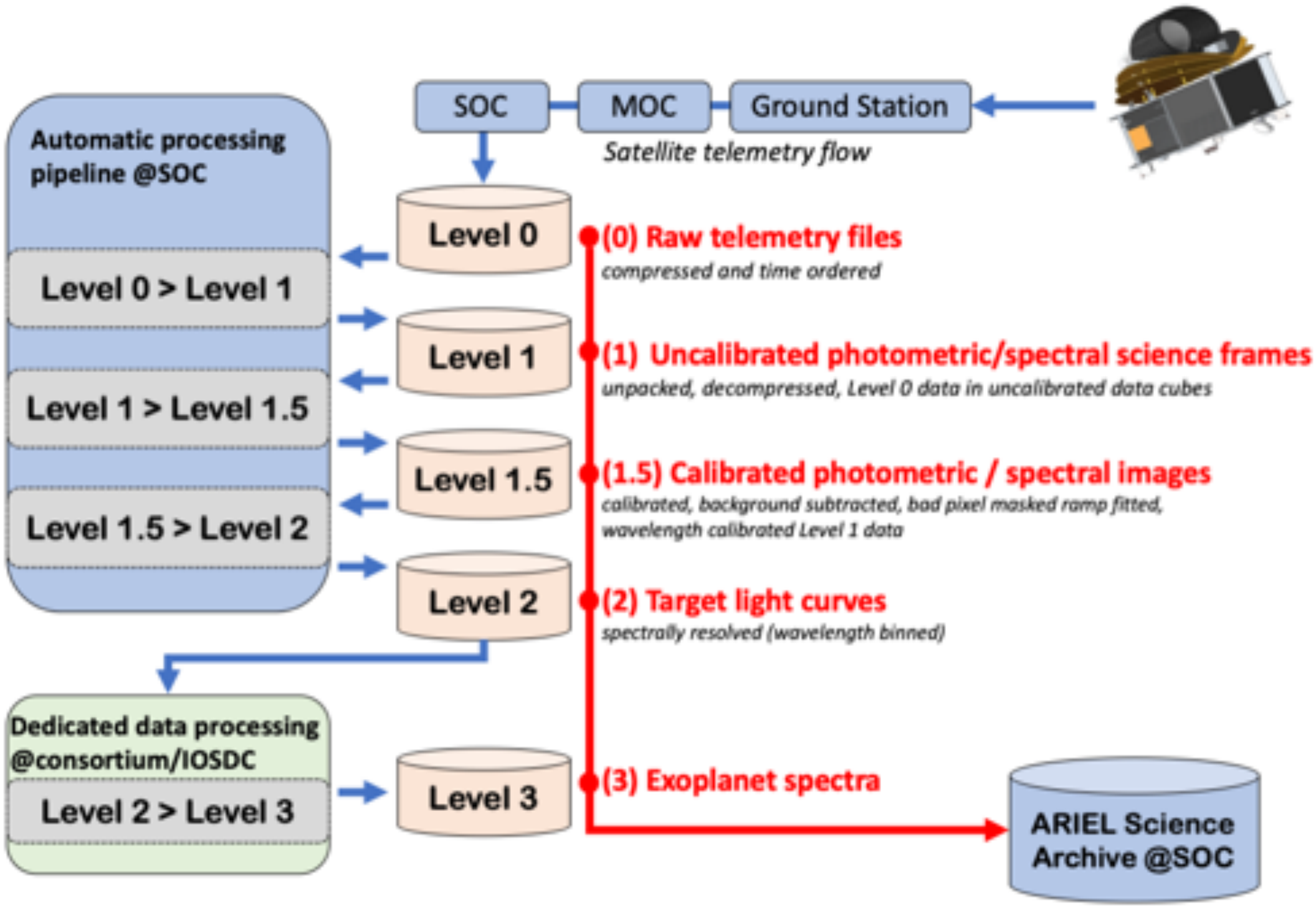

*Figure 6-3: Schematic of Ariel science and HK automatic (Level 0 to Level 2) and customised (Level 2 to Level 3) processing and data flow (left), together with a non-technical and science user oriented, description of Science Data Products contents (right).*

An overview of the Ariel data processing pipeline (ADaRP), including the automatic pipeline run at SOC and a visualization of the consortium Level 2 to Level 3 processing is provided in Figure 6-3. The current ADaRP processing steps are shown sequentially, along with any associated calibration products and auxiliary products (obtained from the satellite housekeeping data, FGS pointing information or meta data created by the pipeline itself). The current required calibration products are listed in Table 6-1.

*Table 6-1: Calibration products required by the Ariel pipeline. Initial versions of these products will be populated during the ground test campaign and then updated in flight where necessary. The pipeline module that uses each calibration product refers to the pipeline in Figure 6-4.*

| Cal Product | Description | Pipeline Module |
|---|---|---|
| INITPIX | Initial pixel data quality flags | DQ array initialisation |
| SATFILE | Pixel saturation thresholds | Flag saturated pixels |
| GAINFILE | ADU to electron gain per pixel | ADU to $e^-$ unit conversion |
| LINCORR | Non-linearity model coefficients per pixel | Non-linearity correction |
| CROSS | Crosstalk model coefficients per pixel | Pixel crosstalk correction |
| DARK | Dark image array | Dark subtraction<br>Error array initialisation<br>Image statistics (deglitching) |
| WAVSOL | Wavelength solution model | Flat fielding<br>Apply spectral mask<br>Extract 1D spectrum<br>Spectral binning |
| WFLAT | Wavelength-dependent QE per pixel | Flat fielding |
| PERSIST | Persistence model coefficients per pixel | Persistence correction |
| READNOISE | Read noise per pixel | Error array initialisation<br>Image statistics (deglitching) |
| POINTING | Converts FGS pointing timeline to channel-specific x and y offsets | Flat fielding<br>Jitter correction<br>Apply photometric mask |
| FLUX | Flux to $e^-$/s conversion factor | $e^-$/s to flux conversion |

### 6.3.4 Science data processing

All data levels are processed sequentially, not concurrently accessing the same data sets: Level 0 to Level 1, Level 1 to Level 2 and Level 2 to Level 3. An overview of the data processing pipeline and levels is shown in Figure 6-4, which also shows representative data products from Level 0 to Level 2.

Up to Level 2 is processed at SOC using the data processing pipelines and calibration products provided by the IOSDC, and automatically stored in the Ariel archive. The Level 2 science data products are the final output of the automatic pipeline. The pipelines are delivered in form of Virtual Machines (VMs) or containers for easy integration at SOC and are configured to run autonomously without the need for human interaction. Foreseen processing steps up to Level 2 are considered standard and of low complexity for the baseline detector systems and expected science image data: de-compression, bias correction, flat fielding, non-linearity correction, de-glitching, etc. The quality control will be part of the science data processing: the IOSDC will be responsible for the definition of the metrics, i.e. the products and meta-data that shall be generated for quality checks as part of the processing pipelines. Based on the quality reports from the pipeline processing, SOC will conduct the quality control during operations.

The Level 2 to Level 3 processing fetches the Level 2 data from the archive and will require more sophisticated procedures and manual intervention in order to extract the planet spectra, this is considered a scientific rather than operational task. This task is fully under AMC/IOSDC responsibility. Once delivered from IOSDC to SOC, the final Level 3 data products are ingested into the archive and made public after the proprietary period (cf. Section 9.5). Although the AMC will provide the Level 3 products to the Ariel science archive to be made available to the community, everyone may not necessarily follow the AMC selected procedures and tools to extract the final planet spectra (the Level 3 products), but may opt to produce these for themselves based on the science and ancillary data made available.

Re-processing of the data is envisaged when new calibration data becomes available or new insights into processing steps require modifications to the pipeline modules, but in general re-processing is connected to the regular data releases, currently foreseen to be twice a year.

The average data rate (science, HK and ancillary) is about 4.2 GB/day and the total data volume including all (re-)processed data levels over the nominal mission lifetime is estimated to around 90 TB. Due to the definition of the Level 1 data product it is currently not foreseen to reprocess Level 0 data.

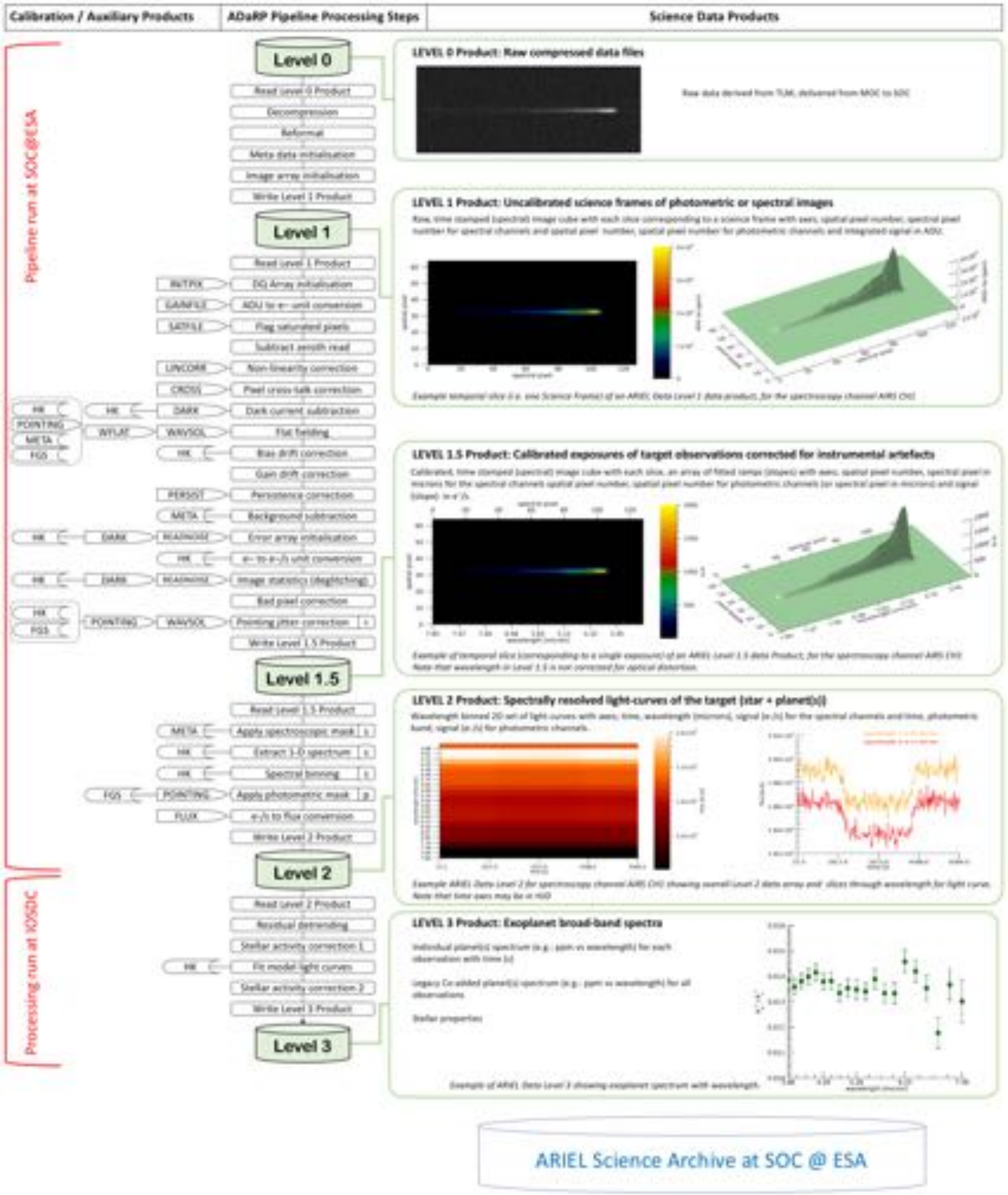

*Figure 6-4: Overview of the Ariel Data Processing Pipeline (ADaRP) and envisaged science products. Level 0 to Level 2 pipeline and data products at SOC@ESA. Level 3 processing and data products will be produced by IOSDC. All data products will be ingested into the Ariel Science Archive at SOC. The ADaRP processing steps are shown as sequential rectangles as steps between each Data Product Level. Note that processing steps used solely for the spectroscopic channels are marked with an 'S' and steps used solely for the photometric channels are marked with a 'P'. Also shown on the left of the pipeline are the required calibration/auxiliary products. Auxiliary products can be from the satellite housekeeping (HK) data, the FGS pointing timeline (FGS) or from meta data produced by the pipeline itself (META). The right-hand side of the figure shows example data products created from simulated data that has been processed through the pipeline.*

### 6.3.5 The Ariel archive and scientific community support

Ariel has a single archive serving the purpose of mission database and science data archive, which after completion of the post-operations phase becomes the legacy archive. SOC will also store for long-term preservation MOC database information such as telecommand (TC) history, housekeeping (HK), ancillary data, etc., in so far as the information is of relevance for the data processing or calibration. Together with the raw and processed science data the versioned data processing pipelines, their source code and the calibration data will be stored and based on the agreed data releases, made publicly available to the scientific community. Additionally all the uplink products including the MCS, generated long-term plans and MRS, selected complementary science targets, priority schemes, observation schedules, etc. will be stored in the archive.

The archive serves as the central interface and data repository for the data processing including calibration processing, science health monitoring, any off-line data analysis, and data dissemination of the final science data products to the general scientific community through the science data archive function.

The archive will be developed, operated and maintained by ESA at ESAC, which hosts all science data archives of ESA science missions. The archive interfaces are foreseen to support International Virtual Observatory Alliance (IVOA) protocols for meta-data access, queries and distributed data storage.

The community will be involved through ESA-led calls, at least one in the last year before launch and another one during the routine science operations phase. There will be two different type of calls, one for additional exoplanet targets to the MCS and another one for complementary science targets.

In preparation for and in support of the calls, a proposal handling system will be developed, operated and maintained by the SOC. The proposal handling system will provide the interface to the proposers, receive and store the proposals and provide them to the evaluation committee. Any additionally needed tools such as exposure time calculator will be provided by the IOSDC.

With help from the IOSDC, SOC will support the scientific community user in access and exploitation of the Ariel science data, activities that include:

- Helpdesk
- Operation of the proposal handling system and support to the ESA-led calls
- Documentation: Ariel users handbook, data processing guides, archive guide, etc.
- Science community support activities, including support to science conferences
- Ariel web portal, Ariel newsletter/announcements

Another planned activity will be the web publishing of the long-term observation plan and current mission status and target observation to aid follow-up and parallel observations of Ariel targets.

# 7 Core Science Implementation

## 7.1 Core Science Performance Simulation

Ariel's technical capabilities are evaluated with two complementary simulators: the ExoSim and ArielRad codes. A validated set of exoplanet spectral retrieval tools and phase-curve models, which will further improve by the launch date, will facilitate the data analysis. Line-lists, chemical networks and experimental data will be essential to guarantee the correct interpretation of Ariel data.

### 7.1.1 Ariel performance simulators

The performance of the Ariel payload is evaluated through numerical simulations implementing a detailed description of the payload design. We have developed two simulators: *ExoSim* and *ArielRad*. ExoSim (Sarkar et al. 2020) is a time-domain, end-to-end simulator that starts from a representation of the science target to be observed (spectral light curve + instrumental effects) and simulates the detection providing as output spectral images vs time similar to Ariel's Level 1 data. A data reduction pipeline has been developed to reduce ExoSim data and to provide noise performance metrics. ArielRad (Mugnai et al. 2020) is a radiometric model which implements a physically motivated description of the uncertainties relevant for the detection. Both ArielRad and ExoSim make use of a parametric description of the payload design. The output of the simulators are estimates of observed planet+star spectra, planet spectra, and their uncertainties. Spectral binning to the resolving power of the Ariel Tiers allows to estimate the amount of time and the number of observations required to achieve the desired SNR. In phase B1 we adopted a margin philosophy: i.e. for a given target and observation time, the SNR was assumed 30% lower than predicted from the current best knowledge of the payload performance.

### 7.1.2 Ariel spectral retrieval algorithms

Thanks to the broad wavelength coverage of Ariel, we will have access to many molecular species and to the most important molecular carriers of many atomic elements. This ability is critical for our understanding of planet formation and evolution and for the processes occurring in exoplanet atmospheres (see Chapter 2). Key tools to interpret correctly Ariel data are the spectral retrieval models developed to extract the information contained in the observed spectra and evaluate the likelihood that the solutions found are correct and unique. There are today several retrieval tools available in the community. The main details of the retrieval schemes currently used in the Ariel Mission Consortium are briefly summarised in the Ariel Spectral Retrieval WG Report (2020): NEMESIS (Irwin et al. 2008), ARCiS (Min et al. 2020), PyRat Bay (Cubillos & Blecic, in prep.) and TauREx (Al-Refaie et al. 2020; Waldmann et al. 2015). The retrieval code CHIMERA (Line et al. 2013) was thoroughly compared to NEMESIS and TauREx in Barstow et al. (2020). In addition, results from a further code, POSEIDON (MacDonald & Madhusudhan 2017a,b), were submitted from a researcher external to the Ariel team in response to the open retrieval Ariel Data Challenge 2019. As part of the retrieval Ariel Data Challenge 2019, simulated Ariel spectra were provided with known atmospheric properties. The codes used in the challenge performed their retrieval on these test spectra to see if the underlying atmospheric parameters could be retrieved. All codes showed excellent agreement (see e.g. Figure 3-9). The underlying parameters of the atmospheres could be retrieved and the uncertainties of these parameters due to spectral noise were also estimated consistently for all codes involved. This exercise provided strong confidence in the robustness of each individual retrieval code. For more details on the code comparison and detailed retrieval studies on a sample of Ariel targets we refer to the Ariel Spectral Retrieval WG Report (2020). Ariel will produce a wealth of phase curves which will allow for the global thermal and chemical structure of exoplanets to be investigated. To test the ability of the instrument to extract this information we used the NEMESIS code to perform retrievals of synthetic phase curves observations produced from the THOR Global Circulation Model (Mendonça et al. 2016) and TauREx3-PhaseCurve, a 1.5D model for phase-curve retrievals (Changeat & Al-Refaie 2020).

#### *7.1.2.1 Line-lists, chemical networks and experimental data in support of Ariel data interpretation*

- <u>*Molecular line-lists*</u>
  Molecular line lists will be provided by ExoMol, HITRAN, HITEMP and e-Phytheas. Table 7-1 summarises the main spectral features that Ariel plans to probe and indicates the current source of the opacity data. The spectroscopic properties of the atomic species, namely Na, K, H Hα, H Hβ, He, Ca, are well known and can be found in standard data sources such as NIST (see http://www.nist.gov/pml/data/asd.cfm). The sources of the collision-induced absorption (CIA) data are listed in Table 7-2.

*Table 7-1: Main spectral features between 2 and 8 μm. The asterisk indicates the molecular/atomic species already detected in the atmospheres of exoplanets. The main bands are illustrated in bold.*

| | 2-5 μm | 5-8 μm | Line list | Source |
|---|---|---|---|---|
| $H_2O$* | **2.69** | 6.2 | ExoMol | Polyansky et al. (2018) |
| $CO_2$* | 2.03, **4.25** | - | Ames/ExoMol | Huang et al. (2013), Yurchenko et al. (2020) |
| $C_2H_2$ | **3.0** | 7.53 | ExoMol | Chubb et al. (2020b) |
| HCN | **3.0** | - | ExoMol | Barber et al. (2014) |
| $O_3$ | 4.7 | - | HITRAN | Gordon et al. (2017) |
| HDO | 2.7, 3.67 | 7.13 | ExoMol | Voronin et al. (2010) |
| CO* | 2.35, **4.7** | - | HITEMP | Li et al. (2015) |
| $NH_3$ | 2, 2.25, 2.9, **3.0** | 6.1 | ExoMol | Coles et al. (2019) |
| $PH_3$ | 4.3 | - | ExoMol | Sousa-Silva et al. (2015) |
| $CH_4$* | 2.2, 2.31, 2.37, **3.3** | 6.5,7.7 | HITEMP/ExoMol/TheoReTS | Hargreaves et al. (2020), Yurchenko and Tennyson (2014), Rey et al. (2014a). |
| $CH_3D$ | 3.34, **4.5** | 6.8, 7.7 | TheoReTS | Rey et al. (2014b) |
| $C_2H_4$ | **3.22**, 3.34 | 6.9 | ExoMol/TheoReTS | Mant et al. (2018), Rey et al. (2016a) |
| $H_2S$ | 2.5, 3.8, ... | 7 | ExoMol | Azzam et al. (2016) |
| $SO_2$ | 4 | 7.3 | ExoMol | Underwood et al. (2016) |
| $N_2O$ | 2.8, 3.9, **4.5** | 7.7 | HITRAN | Gordon et al. (2017) |
| $NO_2$ | 3.4 | 6.2, 7.7 | HITRAN | Gordon et al. (2017) |
| $H_3^+$ | 2.0, 3-4.5 | - | ExoMol | Mizus et al. (2017) |
| TiO | 2-3.5 | - | ExoMol | McKemmish et al. (2019) |
| VO | 2-2.5 | - | ExoMol | McKemmish et al. (2016) |
| FeH | 2 | - | MoLLIST | Bernath (2020) |
| $C_2H_6$ | **3.35** | 6.79 | HITRAN | Gordon et al. (2017) |

*Table7-2: CIA data.*

| | 2-5 μm | 5-8 μm | Data | Source |
|---|---|---|---|---|
| $H_2$ | 1st overtone | Fundamental | HITRAN | Abel et al. (2011) |
| $O_2$ | None | Fundamental | HITRAN | Baranov et al. (2004) |

- ExoMol: The ExoMol project (Tennyson, Yurchenko et al. 2016) aims to produce high temperature line lists of spectroscopic transitions for key molecular species likely to be significant in the analysis of the atmospheres of extrasolar planets and cool stars (see complete list here: http://exomol.com/data/molecules/ ). The newly funded ExoMolHD ERC project will provide greatly improved pressure broadening parameters for hot exoplanets and particularly hot-Jupiters (Tennyson et al. 2020). This molecular data is crucial for accurate astrophysics models of the opacity and spectroscopy of these atmospheres and therefore it will be very useful to interpret Ariel data (see Chubb et al. 2020b). Molecular opacities for these and other molecules and atomic species have been computed by Chubb et al. (2020b) in the form of cross sections and k-tables for all molecules using data available at ExoMol, NIST, HITEMP, HITRAN and e-Phytheas. The data are formatted for use in various retrieval codes including Tau-REx, ARCiS, petitRADTRANS, and NEMESIS.
- e-Phytheas: 5 French laboratories and associated partners are working at improving the existing high-temperature spectroscopy data for several molecular species detected in exoplanets. The provision of infrared (IR) laboratory data of methane, acetylene, ethylene and ethane, between 500 and 2500 K will help to refine thermal profiles and provide information on the gaseous composition, the hazes and their temporal variability. More information is available on the project's website: http://e-pytheas.cnrs.fr

Accurate and complete line lists for about twenty isotopic species, including long lived greenhouse molecules are available from TheoReTS database (Rey et al. 2016b, 2017, 2018b) updated in https://theorets.univ-reims.fr/, http://theorets.tsu.ru. The TheoReTS line lists are validated by laboratory measurements both within the e-Pytheas project (Ghysels et al. 2018; Amyay et al. 2018; Georges et al. 2019; Dudas et al. 2020), and by comparison with experimental absorption cross-section in the near infrared (Wong et al. 2019; Hargreaves et al. 2020). The TheoReTS line list has been successfully used previously in planetology for an advanced modelling of methane absorption in Titan's atmosphere.

- *Thermal dependency of VUV absorption cross sections*
  One important source of uncertainty in current models of hot exoplanet atmospheres are related to the physico-chemical data included in those models. Indeed, most of these data are not known at the high temperatures prevailing in these exotic atmospheres. Consequently, by lack of better data, room temperature data are often use to model physico-chemical processes occurring at high temperature, leading to an undefined uncertainty. Venot et al. (2013, 2018) have started to address the question of the temperature dependency of Vacuum UltraViolet (VUV) absorption cross sections. Thanks to several campaigns of measurements in the synchrotron radiation facility BESSY and at the LISA, they have been able to measure the absorption cross section of $CO_2$ in the range 115-230 nm at different temperatures. An important thermal dependency of the absorption has been observed for this species. The study of other molecules of atmospheric interest (e.g. $NH_3$, $C_2H_2$) is currently in progress. A key aim of the newly funded ExoMolHD ERC project is the calculation of UV photodissociation and other cross sections for key exoplanetary species. This theoretical work will be performed in collaboration with and in parallel to the experimental work of Venot and co-workers.

- *Ionic chemistry in exoplanet atmosphere*
  Simplified exoplanetary thermospheres are reproduced in laboratory by irradiating with UV photons, relevant gas mixtures in a stain- less steel reactor chamber at low pressure ($10^{-2}$ - 1 mbar). UV photons are produced with a gas discharge lamp generating EUV wavelength at 73.6 nm (16.8 eV). The ionic photoproducts are monitored in the reactor by using mass spectrometry analysis. To interpret the experimental results, a 0D-photochemical model is used to reproduce the chemistry occurring in the reactor under similar conditions (geometry of the reactor, wavelength, pressure), see Bourgalais et al. (2020). Experimental results suggest that relatively cold sub-Neptunes are good candidates for the detection of $H_3^+$ and should be combined with the observation of other ions, such as $H_3O^+$ and $HCO^+$. These three compounds could provide a new parametric set for classifying bodies in the transition between super-Earths and sub-Neptunes. This work has stimulated the ExoMol project to produce molecular line lists for key ionic species such $H_3O^+$ and $HCO^+$ to facilitate their detection.

- *Laboratory simulation of atmospheric hazes and particulates*
  Solar system planetary atmospheres have photochemically generated hazes, including the organic ones of Titan and Pluto. At INAF Observatory of Palermo we aim to simulate haze formation in a wide composition range including hydrogen, nitrogen, water, methane, carbon monoxide, and carbon dioxide gases. The samples are studied with high energy irradiation sources (UV and electric discharge), we monitor the destruction of the initial gas components and the formation of new species with mass and ultraviolet spectrometers. Photochemical hazes form over a range of temperatures, pressures, and atmospheric compositions. A database of spectral features of interest for Ariel observations will be built before Ariel launch.

- *Chemical Networks*
  The chemical scheme is one of the most important ingredients of kinetic models. Thus, the reliability of chemical networks is crucial to trust the atmospheric composition predicted by these models and the interpretation of Ariel data. Several chemical schemes for H-dominated atmospheres have been published so far (e.g. Moses et al. 2011; Kopparapu et al. 2012; Miller-Ricci Kempton et al. 2012; Venot et al. 2012, 2015, 2020a; Tsai et al. 2017; Wang et al. 2017). Most of these networks involve H, C, O and N-bearing species and ignore minor species for simplicity. Nevertheless, other species such as P- and S-bearing species might also be relevant for exoplanets (Visscher et al. 2006; Zahnle et al. 2009b). Studying the relevance of other species in the chemistry is important because they might take some of the H inventory in the atmosphere and therefore we might be miscalculating the abundances of other, more abundant species. In a work by Wang et al. (2017), the authors included these species and found that both P- but mostly S-bearing species are extremely relevant for chemistry and spectra of warm and hot exoplanets. Finally, these networks involve hundreds of species in thousands of reactions, which is necessary to understand the chemistry and the most relevant species present in exoplanet atmospheres, but is not practical when using them to model 3D atmospheres, because of computational timescales. Therefore, the development of reduced schemes is also necessary (Venot et al. 2019, 2020a) to develop the first 3D kinetic model of exoplanet atmospheres (Drummond et al. 2020).

All this is work is done by the chemistry working group of Ariel and some future work within this group include the development of a reaction network specifically designed for hot rocky exoplanets and also to continue the work on S-bearing species with a more complete scheme. The results obtained in the chemistry working group (Ariel Chemistry WG Report 2020) highlight the importance of using reliable chemical schemes when modelling future Ariel targets and the relevant work done in support of future Ariel observations.

- *Application of large lasers*
  During the past two decades, the team at Heyrovsky Institute of Physical Chemistry, Czech Academy of Sciences, has conducted several experiments with the large terawatt laser facility PALS (Prague Asterix Laser System), focusing mostly on prebiotic molecule synthesis and their transformation in planetary atmospheres (Babankova et al. 2006). Most studies successfully demonstrated with a single laser shot the synthesis of important prebiotic substances (Šponer et al. 2016), for instance, canonical nucleobases (Ferus et al. 2012, 2014, 2015a,b, 2017) sugars (Civiš et al. 2016b) and amino acids (Civiš et al. 2004). They also demonstrated the transformation of simple molecules occurring on early terrestrial planets (Civiš et al. 2008), such as formamide (Ferus et al. 2011), isocyanic acid (Ferus et al. 2018a), hydrogen cyanide (Ferus et al. 2017), acetylene (Civiš et al. 2016a), methane (Civiš et al. 2017), or carbon monoxide (Ferus et al. 2009). In recent pilot studies, the team has also focused on applications of lasers to simulate meteoritic spectra and estimate the elemental composition of meteoroids (Ferus et al. 2017, 2018b, 2019, 2020a; Křivková et al. 2020). This experimental set-up simulates the high-density energy plasma created in the gas phase by an asteroid impact and also collisions of hot expanding plasma with solid or liquid surfaces. The method has been demonstrated feasible, however long computations are necessary.

  Possible applications of these experiments to the Ariel mission were explored by the Ariel Working Group of Prebiotic chemistry and Astrobiology and described in the Ariel Prebiotic chemistry and Astrobiology report 2020 and Ferus et al. (2020b).

# 7.2 Target Selection, the Mission Reference Sample & Scheduling

From the list of Ariel potential targets presented in Chapter 3, we can simulate mission scenarios where a subset of the planet sample is observed in different Tiers during the mission life-time. The list of Ariel targets will continually evolve with new exoplanet discoveries replacing predicted detections. Trade-offs between studying more planets, observing fewer targets but in greater detail, and/ or choosing interesting planets that require more observational time, will form a key part in the selection of the final mission scenario.

The approach adopted during Phase A for the so called Mission Reference Sample (MRS) consisted of choosing a very diverse, and as complete as possible, combination of star/planet parameters whilst minimising the number of repeated observations (Zingales et al. 2018; The parameter space of planetary systems explored by ARIEL 2017). The same strategy has again been adopted in Phase B and we obtained a distribution of planets by radius and temperature as displayed in Figure 7-3 and Figure 7-4.

## 7.2.1 Current and future MRS

ArielRad simulations of known exoplanets show that Ariel will be able to observe 500+ planets within the primary mission (Figure 7-1). The simulations of known exoplanets combined with the TESS yield predict that Ariel will be able to observe 1000 planets within the primary mission (Edwards et al. 2019, Figure 7-2). The number of planets within this updated version of MRS is similar to that of the Phase A study although we found an increase in the number of Tier 2 planets compared to the results of Zingales et al. (2018) on top of the 10% mission lifetime dedicated to other science observations (Tier 4).

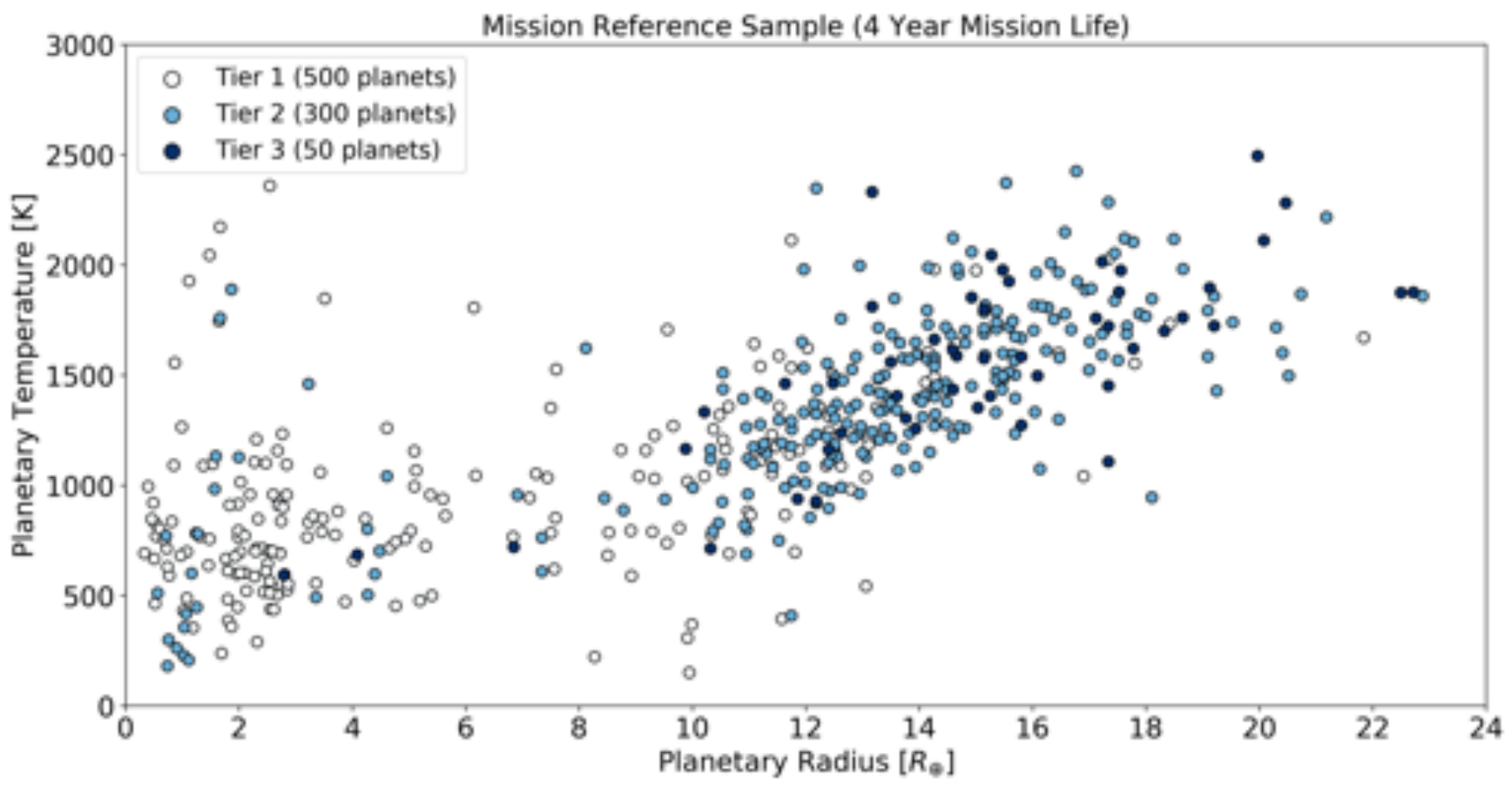

*Figure 7-1: Planetary radius and temperature distribution of a potential Ariel MRS including only known planets. In this simulation, 10% of the mission lifetime is allocated to Tier 4 observations.*

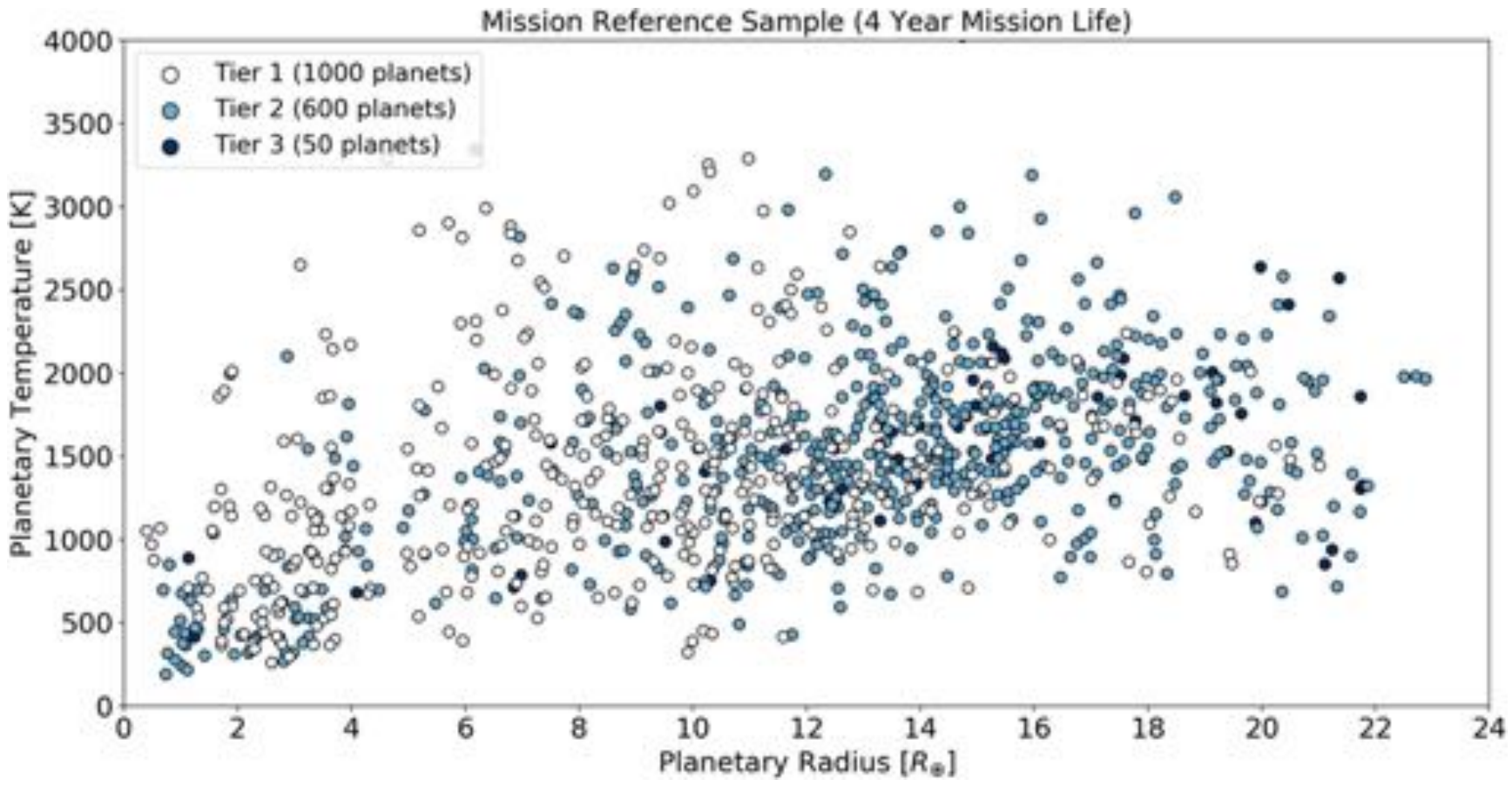

*Figure 7-2: Planetary radius and temperature distribution of a potential Ariel mission reference sample in 2029 (Edwards et al. 2019). In this simulation, 10% of the mission lifetime is allocated to Tier 4 observations.*

#### *7.2.1.1 Alternative MRS*

Different observing strategies have been discussed within the Ariel team including acquiring data in both transit and eclipse for some Tier 2 planets. Such observations would increase our ability to characterize the atmospheres of these targets but would reduce the total number of planets studied. Table 7-3 left highlights the science time (i.e., time on target) required to achieve different observations. These discussions are ongoing and further studies will be undertaken but it can be seen that acquiring data in the secondary method (i.e., the method which gives a lower S/N) for some of the best planets will not require significant mission time. However, the total number of Tier 2 planets may have to be sacrificed to achieve this. Here we explore a different option for the Ariel MRS, with more emphasis on the interpretation of the nature of smaller planets, by specifically devoting the mission lifetime to studying the dichotomy of small worlds as identified by Fulton et al. (2017). In the MRS shown in Figure 7-2, ~110 planets with a radius less than 3.5 Earth radii were selected for study over around 600 observations (~2100 hr of science time) in all three tiers. These planets are located on both sides of the radius valley (see Chapter 2 and Figure 2-1). A key goal of Tier 1 is to discover the fraction of small planets with hydrogen/helium envelopes. For this reason, the number of required observations to detect an atmosphere is estimated assuming a low mean molecular weight so that if a planetary atmosphere has a primordial composition, this atmosphere should be detected with high confidence. Additionally, the atmospheric trace gases will be accurately constrained if the planet is observed in Tier 2 or 3. If no detection is made, the planet either has (*i*) an atmosphere with a higher molecular weight, (*ii*) opaque clouds across all wavelengths, or (*iii*) no atmosphere at all. Phase-curves observations are helpful to address *ii* and *iii*. In all likelihood, some fraction of these planets will have far heavier atmospheres (higher mean molecular weight) and thus will be harder to characterize, requiring more observations to obtain the observational requirements in each tier. In particular, additionally to the H/He atmospheric content, the fraction of $H_2O$ present in an atmosphere is also very important to constrain formation/evolution scenarios and the delivery of volatiles to the inner part of the planetary system. Water worlds, i.e., planets with a significant amount of $H_2O$ on their surface or in the subsurface (e.g. Léger et al. 2004), or magma ocean planets with a steam atmosphere (e.g. Hamano et al. 2015), are expected to have atmospheres with a large fraction of $H_2O$. However, the characteristics of a planet's atmosphere (if present) cannot be known before observations are undertaken, unless these targets are observed previously with other facilities from space or the ground. To quantify the fraction of lifetime needed to characterize the atmospheric composition of small planets with an atmosphere heavier than H/He, we select the small planets ($R_p < 3.5\ R_\oplus$) from the example MRS for further study. The science time required to achieve Tier 1 resolutions (with S/N > 7) for different atmospheric compositions is determined and compared to the Tier 1 time assumed for the standard MRS.

*Table 7-3: from Edwards et al. 2019. The total science time over the 4 yr primary life is ~24,800 hr. Left: mission time required to achieve different observation goals. Note that for some bright targets (e.g. HD 209458b), Tier 2 or 3 resolutions would be reached in a single observation. Right: mission time required to achieve Tier 1 resolutions (at S/N>7) for the 113 Small Planets in the example MRS assuming different mean molecular weights. $t_0$ is the time spent observing small planets in Tier 1 of the standard MRS.*

Mission Time Required to Achieve Different Observation Goals

| Number of Planets | Observation Requirement | Required Science Time (hr) |
|---|---|---|
| 1000 | Achieve Tier 1 resolutions | ~10,600 |
| 400 | Increase resolution from Tier 1 to Tier 2 | ~3100 |
| 500 | | ~6000 |
| 600 | | ~10,500 |
| 200 | Achieve Tier 1 resolutions in the second method | ~1400 |
| 300 | | ~2500 |
| 400 | | ~4200 |
| 50 | Tier 3 (five repeated observations per planet) | ~1700 |
| … | Tier 4 (additional science time) | ~2300 |

Mission Time Required to Achieve Tier 1 Resolutions (at S/N > 7) for the 113 Small Planets in the Example MRS Assuming Different Mean Molecular Weights

| Atmospheric Mean Molecular Weight | Number of Planets | Required Science Time (hr) |
|---|---|---|
| 2.3 | All | ~1000 ($t_0$) |
| 5 | 50 | $t_0$ + ~360 |
| | All | $t_0$ + ~3000 |
| 8 | 50 | $t_0$ + ~1100 |
| | All | $t_0$ + ~9200 |
| 10 | 50 | $t_0$ + ~1900 |
| 15 | 50 | $t_0$ + ~4400 |
| 18 | 25 | $t_0$ + ~1700 |
| | 50 | $t_0$ + ~6400 |
| 28 | 25 | $t_0$ + ~4300 |
| | 50 | $t_0$ + ~15,600 |

#### *7.2.1.2 Ariel Mission Planning*

Exoplanet transit and eclipse events can only be observed at specific times defined by the orbital properties and ephemerides of the system. This imposes an important constraint on the mission planning. Two different approaches and tools have been presented in the Ariel Mission Planning (2020), one based on genetic algorithm optimisation techniques (team at IEEC) and the other on a hierarchical greedy heuristics-based principle (team at CNES). The two methods take as input the MRS and an agreed set of constraints derived from the mission requirements, but resolve certain specific problems in different ways, and produce schedules of the 3.5 years science mission of Ariel that, in a statistical sense, are very similar. This gives good confidence in the final results.

Long-term mission plans are obtained by allocating the requested number of events for each target indicated in the MRS, following the three Tiers approach. Phase-curves are also considered for several planetary systems. An inevitable outcome of the scheduling of time constrained events with a fixed duration is that some inaction slots will remain between observations of different targets. For this reason, mission planning tools are designed to maximize the number of surveyed targets and the time used for their observations.

The different simulations carried out demonstrate that, within the 3.5 years mission lifetime, the vast majority of targets can be observed as requested while meeting all mission and system requirements and constraints. In all cases, Tier 3 Benchmark planet observations are fully completed, as well as more than 90% of Tier 2 and Tier 3 observations, following the MRS planet types distribution.

Despite the fact that inaction slots are inevitable in the Ariel context, between 85%-90% of the mission time can be devoted to science as shown in Figure 7-3, including extra observations of targets in the list, observations of backup targets, or using time for ancillary science. In fact, by choosing cleverly the targets to re-observe, it is possible to approximately double the number of targets for which the Benchmark quality is reached. In the end, about 24000 hours (of 30720 hours available) are scheduled on observations of some 3500 transit or eclipse events and about 30 phase curves, which typically require more than a single planetary orbital period. The inactive time between observations is mainly distributed in time slots shorter than 1-2 hours, which will still be very useful to increase the baseline of transiting events or to schedule short calibration operations without interfering with the science goals.

As a further check, some alternative catalogues to give emphasis to small planets or taking into consideration actual known planets have also been considered, with similar results. In addition, the different tests performed adding targets from a backup list indicate that it is also possible to increase the number of surveyed planets above 1000 by re-distributing the observations in the different Tiers.

On a practical level, the current prototype tools produce a full schedule in just a few minutes, enabling the production of regular updates (in both design and operation phases). In particular, this will allow rapid feedback on new target lists, or to rapidly update the mission plan considering feedback from previous Ariel observations in order to include new targets, new priorities, or re-schedule corrupted observations.

The development of Ariel scheduling tools will continue in the future phases of the project with the goal of assessing mission performance with updated MRS's and new or updated constraints as well as user needs. Furthermore, we will investigate possible ways to optimize the efficiency of the Ariel survey adapting our scheduling algorithms, for instance, by adding observations of backup targets, optimizing the sequences for multi-planetary systems, or changing target priorities or optimization choices.

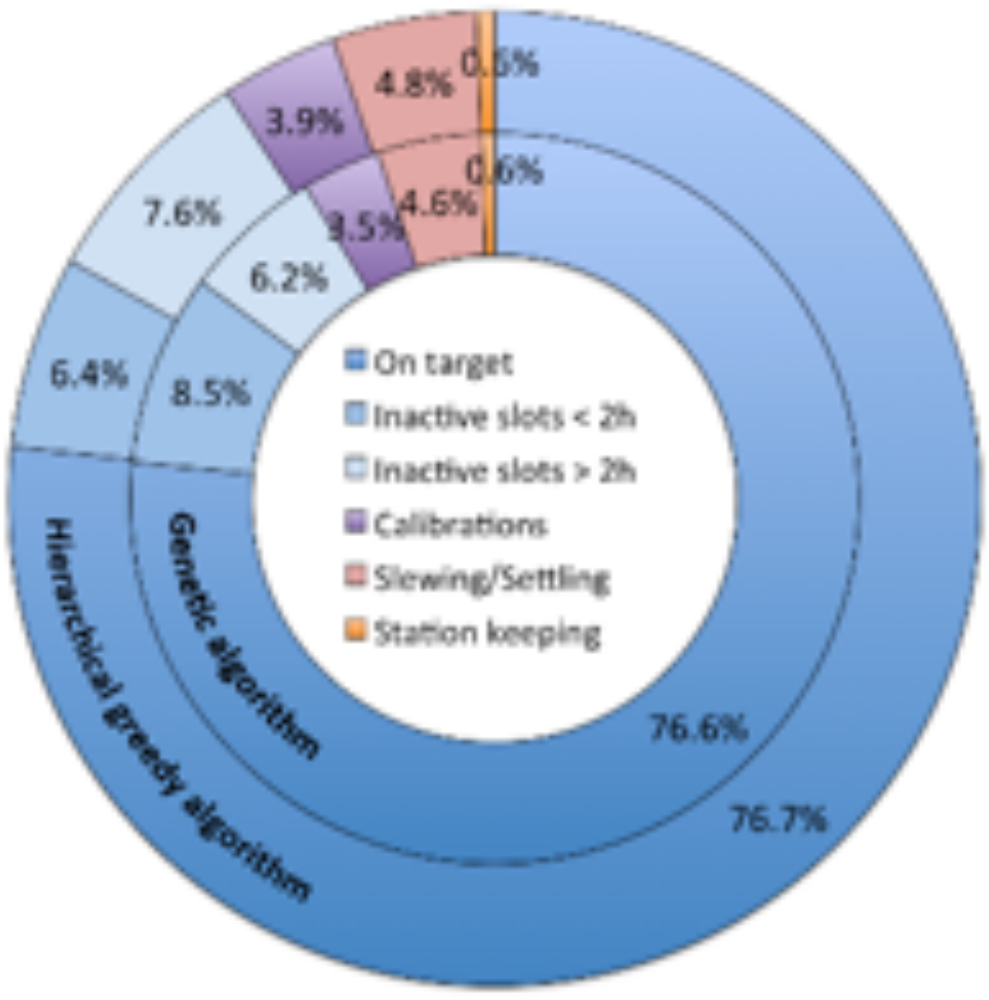

*Figure 7-3: Distribution of the mission lifetime between different operations considered in the Ariel mission planning of the MRS target list. Inner and outer ring charts correspond to the genetic and hierarchical greedy algorithms, respectively. More than 85-90% of the time can be used for MRS (on targets time), ancillary science targets (inactive slots), or science calibration observations.*

# 7.3 Stellar Characterisation

## 7.3.1 Homogeneous determination of host star parameters

Ariel's observational methodology is based on differential spectroscopy in and out of a transit, allowing us to measure atmospheric signals from the planet at levels of 10 - 50 part per million (ppm) relative to the star. Such a small signal requires an exact knowledge of the host star and its spectrum, which has to be known at least at the same level of the planetary signal. Any intrinsic variation of the stellar spectrum due to stellar variability (e.g. granulation, oscillations) needs to be mapped to avoid confusion with planetary features (e.g. Robertson et al. 2015, see Section 7.4).

The information on the composition of the host star, moreover, is the cipher key to decode the compositional signatures on planets left by their formation and migration histories (Tinetti et al. 2018; Turrini et al. 2018). Various papers in the literature point out that the comparison between the stellar metallicity and C/O ratio, with the analogous planetary quantities (e.g. the planetary C/O ratio being super-stellar or sub-stellar) provides information on the original formation region of the planet with respect to the $H_2O$, $CO_2$ and CO snow lines, and on the time when the planet migrated to its present orbit (e.g. Madhusudhan et al. 2016, and references therein). Further work in preparation of Ariel shows that similar considerations apply also to elemental ratios involving other elements (e.g. N, S, Ti, Al), and that their comparison with the stellar abundance ratios provides additional constraints on the planetary formation process (Ariel Planetary Formation WG report 2020; Turrini et al. 2018; Turrini et al. 2020a,c). The precise knowledge of the stellar abundances and their elemental ratios is therefore the founding stone for interpreting the composition and, hence, the origin and evolution of planets. Additionally, more and more observational evidence is being reported presenting correlations between the host stellar properties and their planetary companion properties. For instance correlations between planet radius versus host star metallicity (e.g. Buchhave et al. 2014; Schlaufman 2015), eccentricity versus metallicity (e.g. Adibekyan et al. 2013; Wilson et al. 2018), gas-giant planets occurrence rate versus metallicity (e.g. Santos et al. 2004; Sousa et al. 2008), low-mass planets occurrence rate versus stellar Mg/Si ratio (Adibekyan et al. 2015), or the fact that iron-poor planet-hosting stars are enhanced in α-elements (Adibekyan et al. 2012), are taking a clear shape aided by the constant and hefty increase of new planet discoveries. Such works, which present a robust investigation of the possible correlations and trends with the planetary properties rely upon homogeneously and precisely derived stellar parameters. Therefore, a refinement of the values of the stellar properties, by using both high-quality data and a uniform approach to the data analysis, is crucial (Santos et al. 2013; Biazzo et al. 2015; Andreasen et al. 2017; Sousa et al. 2018).

For this reason, the goal of the Stellar Characterisation Working Group was to develop a procedure to precisely characterise the stellar parameters of the Ariel Reference Sample (Edwards et al. 2019) in a homogeneous way, adopting both a model dependent and empirical approach when possible, in order to evaluate the robustness and the consistency of the parameters obtained with different methods. The concept of *homogeneity* is crucial for achieving Ariel goals: fundamental parameters of host stars are usually found in the literature as the result of a case-by-case analysis performed by different teams, resulting in an inhomogeneous census of stars with planets. An accurate, precise and uniform determination of the fundamental properties of host stars is a crucial step towards a comprehensive characterisation of a statistically significant number of planetary systems (Danielski et al. 2020, Ariel Stellar Characterisation WG 2020).

To achieve this goal the working group has chosen SWEET-Cat (Santos et al. 2013) as the stellar catalogue of reference. SWEET-Cat is a catalogue of parameters for stars with planets, some of which have been analysed using the same uniform methodology. The analyses have been performed adopting as inputs those stars of the Ariel Reference Sample (Edwards et al. 2019) that are included in the homogeneously determined part of SWEET-Cat, which accounted for 155 stars (based on 15/03/2019 Ariel Reference Sample).

A first check has been performed on the fundamental atmospheric parameters $T_{eff}$, log(g), and [Fe/H] through a benchmark study among the parameters detailed in SWEET-Cat and two more techniques (FAMA, Magrini et al. 2013; and FASMA, Tsantaki et al. 2018) to verify the consistency of the derived parameters. The first results show a good agreement for a bulk population around the solar values (Brucalassi et al. 2020); for values outside this range a more thorough analysis is planned for the next phase.

A similar analysis has been conducted to derive reliable and homogeneous stellar abundances (Danielski et al. 2020, Ariel Stellar Characterisation WG 2020). The abundances ratio [X/H] for sodium (Na, 91 stars), aluminium (Al, 86 stars), and magnesium (Mg, 92 stars) have been determined by using two different techniques (Equivalent widths and FAMA). The abundance values retrieved are consistent within 2 median absolute deviation. The analysis has been applied to stars with temperature $T_{eff} > 4500$ K: cooler stars present molecular bands that need a different approach. The determination of specific elements for M dwarfs in particular will need to await for high quality NIR data or improvement of current methodology. The determination of specific elements for M dwarfs in particular has proved a difficult task. Spectral synthesis has been used in several works. However, it has been tested usually on small number of stars, focusing on spectral windows known to be less affected by molecular lines or in NIR data. Furthermore, it is computationally expensive and requires a good knowledge of the atomic and molecular data. New methodologies based on the use of principal component analysis and sparse Bayesian's fitting methods applied on high-resolution optical spectra are under development (Maldonado et al. 2020). These methods are trained by using a set of M dwarfs in binary systems orbiting around an FGK primary. Carbon (C, 90 stars) and Nitrogen (N, 90 stars) abundances have also been determined through a method based on spectral synthesis, for stars with fluxes recorded below 4400 Å and $T_{eff} > 5000$ K. The method is currently being improved to also derive oxygen O abundances. We note that spectral synthesis is required to analyse molecular bands, especially in the case of cool stars where line blending is evident.

The stellar age is one of the parameters more difficult to derive. The age of the exoplanet-hosts in the Ariel Reference Sample has been estimated by using a model dependent approach: the technique uses Bayesian methods to estimate stellar fundamental properties (e.g. age, mass, and radius), by matching a set of observational constraints to a pre-computed grid of stellar evolutionary tracks (i.e. isochrones fitting technique). These values show a scatter with the literature values of approximately ~ 50% (Bossini et al. in prep.). Some tests will be performed in order to estimate the dependence of the obtained results from the models used.

Using the available high resolution spectra of the Ariel Reference Sample stars, is possible to extract useful activity indices (e.g. log $R'_{HK}$, S), relevant to study the impact of the stellar activity on the planetary atmospheric characterisation, deriving the average activity level of the exoplanets' hosts, as well as their time variation (Danielski et al. 2020). Detecting any variations would be an indication of stellar activity cycles, which have to be taken into account to correctly schedule and interpret exoplanet observations. The activity index S can be used also to derive an empirical age estimate to be compared to the ages obtained by isochrones fit method. Figure 7-4 shows the properties for the homogeneously analysed sample.

Finally, the Stellar Characterisation working group has begun a ground-based observational campaign to obtain high resolution spectra for those stars in the Ariel Reference Sample that have not yet been observed with the spectral quality needed to perform a precise characterisation.

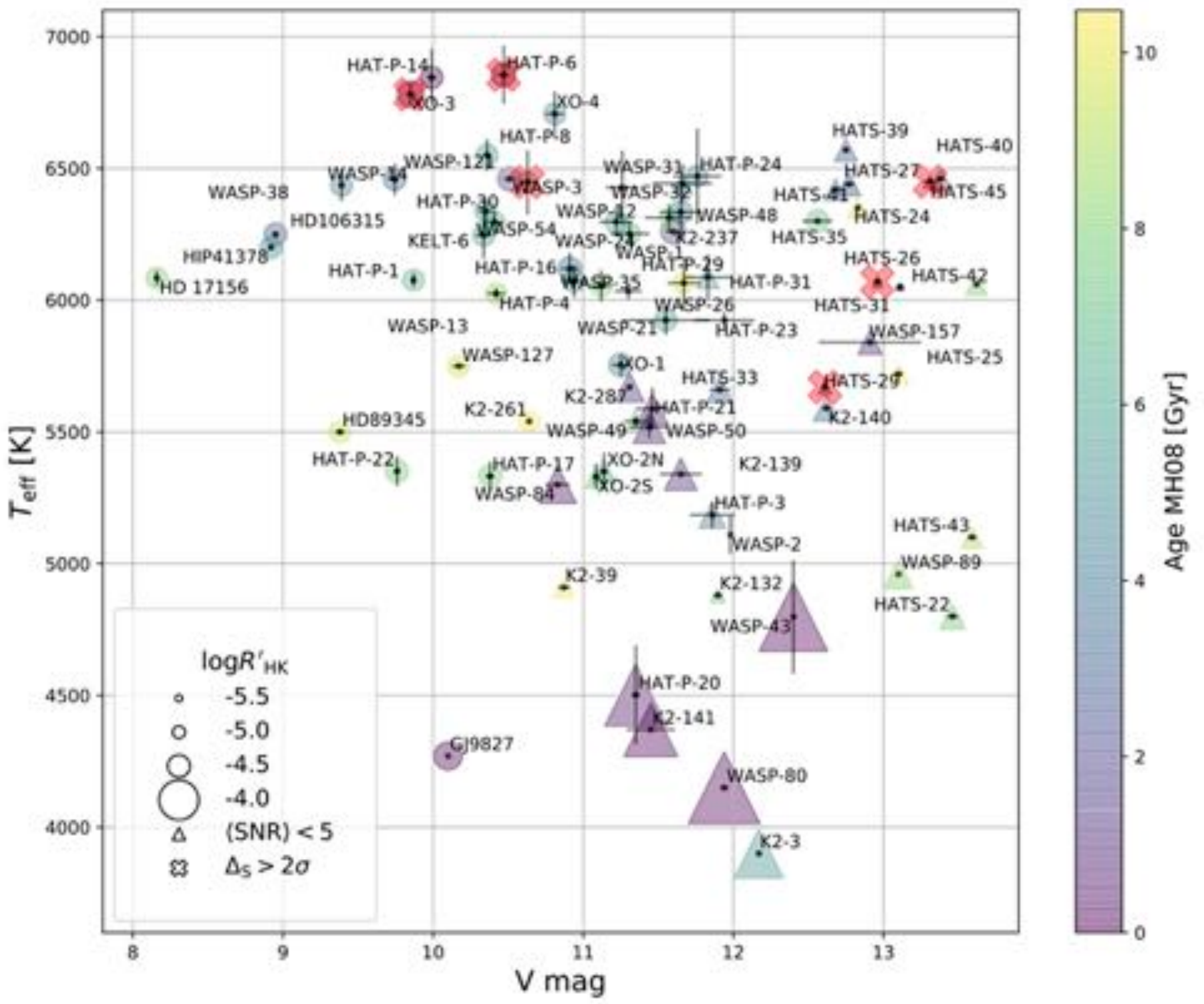

*Figure 7-4: Effective temperature vs. V magnitude of our sample. The average log R'HK target is indicated by the marker size. Triangles and crosses correspond to targets with average S/N< 5 spectra and to > 2σ observed variations in the S- index, respectively. Colours indicate derived stellar ages.*

## 7.3.2 The energetic environment of the exoplanets

Stellar activity evolves considerably during stellar evolution, with significant effects on the environment in which the planet is embedded, changing the nature and intensity of star-planet interaction. Stellar activity is governed by the intensity and structure of the magnetic fields which in turn depends on the rotation and internal structure of the star.

It is very difficult to determine directly the properties of the stellar magnetic field, so it is often necessary to resort to rather uncertain models. These uncertainties can be overcome with modelling efforts based on the Zeeman Doppler Imaging (ZDI) technique in combination with field extrapolation- and wind modelling methods. ZDI has become one of the most powerful astrophysical remote sensing techniques (e.g. Hussain et al. 2002; Jardine et al. 2002; Donati et al. 2008; Farès et al. 2009; Lüftinger et al. 2010a,b; Boro Saikia et al. 2018; Piskunov & Kochukhov 2002; Kochukhov & Piskunov 2002a). It has reached a state of maturity in which we can determine strength, distribution, polarity and even polarity reversals of stellar surface magnetic fields. At the same time temperature or brightness (with the most recent implementations of ZDI-codes), structures on the surface of stars can be reconstructed by inverting time series of high-resolution spectropolarimetric data. However, we still have only limited knowledge of statistical and evolutionary systematics of stellar magnetic fields.

We have already conducted several successful observing campaigns (and will continue to do so), applying for high-resolution spectropolarimetric data of selected stars of the Ariel Mission Reference Sample. Using this high resolution spectropolarimetric data we are able to determine magnetic field strengths, field orientations, field geometries, and the temperature distributions on the surface of well selected Ariel stars of the Ariel Mission Reference Sample. Figure 7-5 shows the analysis of the prototypical young Sun $\pi^1$ UMa, determining its magnetic field geometry via ZDI (Lüftinger et al. 2019). Based on maps like this, using field extrapolation methods and the freely available 3D MHD code BATS-R-US/AWSoM (Powell et al. 1999) it is even possible to determine the wind configurations and speed, wind temperature (Boro Saikia et al. 2019; Lüftinger et al. 2020) and mass loss values of the star (e.g. at the planetary orbit, which are crucially affecting the evolution and the erosion of planetary upper atmospheres and magnetospheres.

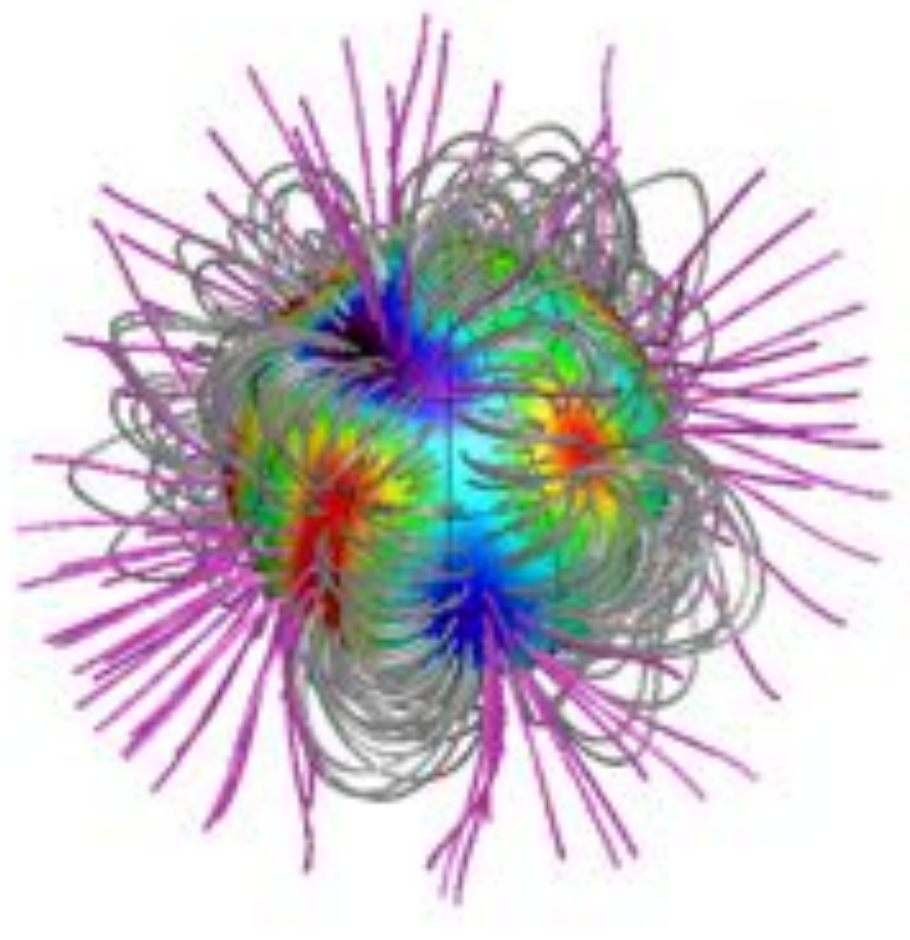

*Figure 7-5: Magnetic polarity map of $\pi^1$ UMa obtained via ZDI (red positive and blue negative components of the magnetic field.*

These same surface magnetic fields are at the root of diverse high-energy outputs of the star that fundamentally affect planetary atmospheres (see also Section 2.2.2.4): the magnetized winds lead to non-thermal interactions with the planetary upper atmospheres and magnetospheres, eroding atmospheres as seen in today's Venus, Earth, and Mars. Such processes can be amplified by coronal mass ejections associated with stellar flares. Energetic XUV radiation from magnetic interactions in the outer stellar layers (chromosphere to corona) can heat, ionize, and evaporate upper planetary atmospheres, selectively removing light gases. High-energy particles may drive, apart from ionization, also chemical reactions in upper atmospheres (Airapetian et al. 2016). Because XUV radiation systematically depends on stellar rotation, its initial condition, and stellar mass, any given observed planetary atmosphere is the result of the entire previous stellar evolution path (Penz et al. 2008; Sanz-Forcada, et al. 2011; Tu et al. 2015; Locci et al. 2019; Johnstone et al. 2020), thus in fact potentially also constraining the latter.

We need to select a sample of planets around stars spanning a range of activity levels to be characterized in detail in terms of magnetic fields, chromospheric and coronal emission, wind and any other specific activity indicators that will be used as a benchmark for analyzing the star-planet interaction phenomena.

Key targets that need to be monitored in detail to understand their magnetic configuration and predict, together with observations of activity indicators, the activity level during Ariel observations could include HD 189733, Corot 7, DS Tuc, Kelt-24, 55 Cnc, and HD 209458.

The response of atmospheric T-p profiles and chemical composition to the interaction with the various elements of stellar activity, including time-variable components, may be synthesized and convolved with the Ariel instrument response for planets around young stars. From Ariel observations we may have direct information on the physics and relevance of the star-planet interactions. Additional modelling and simulations will be performed to assess further this point.

## 7.4 Ephemerides Maintenance

To optimise the timing of individual Ariel observations, planetary orbital ephemerides need to be refined such that transit-time errors after launch are minimised. To achieve this objective, the consortium project ExoClock gathers professional and amateur photometric monitoring data. Additional data will be available from transit surveys (e.g. TESS, CHEOPS) and dedicated telescope facilities available to consortium members.

Many currently known planets have large ephemeris uncertainties. For instance an analysis by Dragomir et al. (2019) suggests that many TESS targets will have errors garter than 30 minutes less than a year after their discovery, due to the short baseline of TESS observations. By the time of the Ariel launch, many exoplanet transit ephemerides based on the initial discovery transit measurements will not be accurate enough to predict transits with the precision needed to efficiently schedule Ariel observations. Verifying and updating the ephemerides of these planets will require regular follow-up observations over the coming years from the ground and from space, to maximise the efficiency of Ariel scheduling and the outcome of every observation.

The *ExoClock Project* (https://www.exoclock.space, Kokori et al. 2020) is an open, integrated, and interactive platform, designed by the Ephemerides working group, to maintain the ephemerides accuracy by collecting all currently available data (literature observations, observations conducted for other purposes, both from ground and space) and by efficiently planning dedicated efforts to follow-up the riel targets. *ExoClock* is open to contributions from a variety of audiences — professional, amateur and industry partners — aiming to make the best use of all available resources towards delivering a verified list of ephemerides for the Ariel targets before the launch of the mission. The *ExoClock* target is to provide transit mid-time predictions (for the end of 2028) with an 1σ uncertainty equal to the 1/12 of the transit duration, so that the difference between the expected and the actual time of the transit will not exceed the 1⁄4 of the transit duration.

Given the brightness of the host stars and the large transit depths of many of these planets, modestly-sized telescopes can play a crucial role in maintaining the ephemerides of Ariel targets. Most of the planets already known today can be easily observed using relatively small telescopes, between 30-80 cm in aperture. Caines et al. (2020) have shown that a network of only two telescopes, one per hemisphere, of only 60 cm aperture are capable of tracking 600+ transiting exoplanets from the Ariel candidate target list over a period of ten years. Such telescopes are readily available and several networks already exist that can provide observing time on demand, for example *LCO* (Brown et al. 2013) or *Telescope Live* (https://telescope.live). Tartu Observatory in Estonia currently offers ephemerides tracking on its 30 cm and 60 cm telescopes (RMS~2 and 1mmag in 1 minute, respectively) on up to 50 clear nights annually. From 2023-2024, below 1mmag RMS will become possible with a new camera on the 1.5m telescope. As part of the Japanese and Spanish contributions to Ariel, there is the plan to conduct simultaneous photometric monitoring of Ariel target host stars using the three multi-colour cameras: MuSCAT in Okayama, Japan (Narita et al. 2015), MuSCAT2 in Tenerife, Spain (Narita et al. 2019), and MuSCAT3 in Maui, USA (under development and available after June 2020).

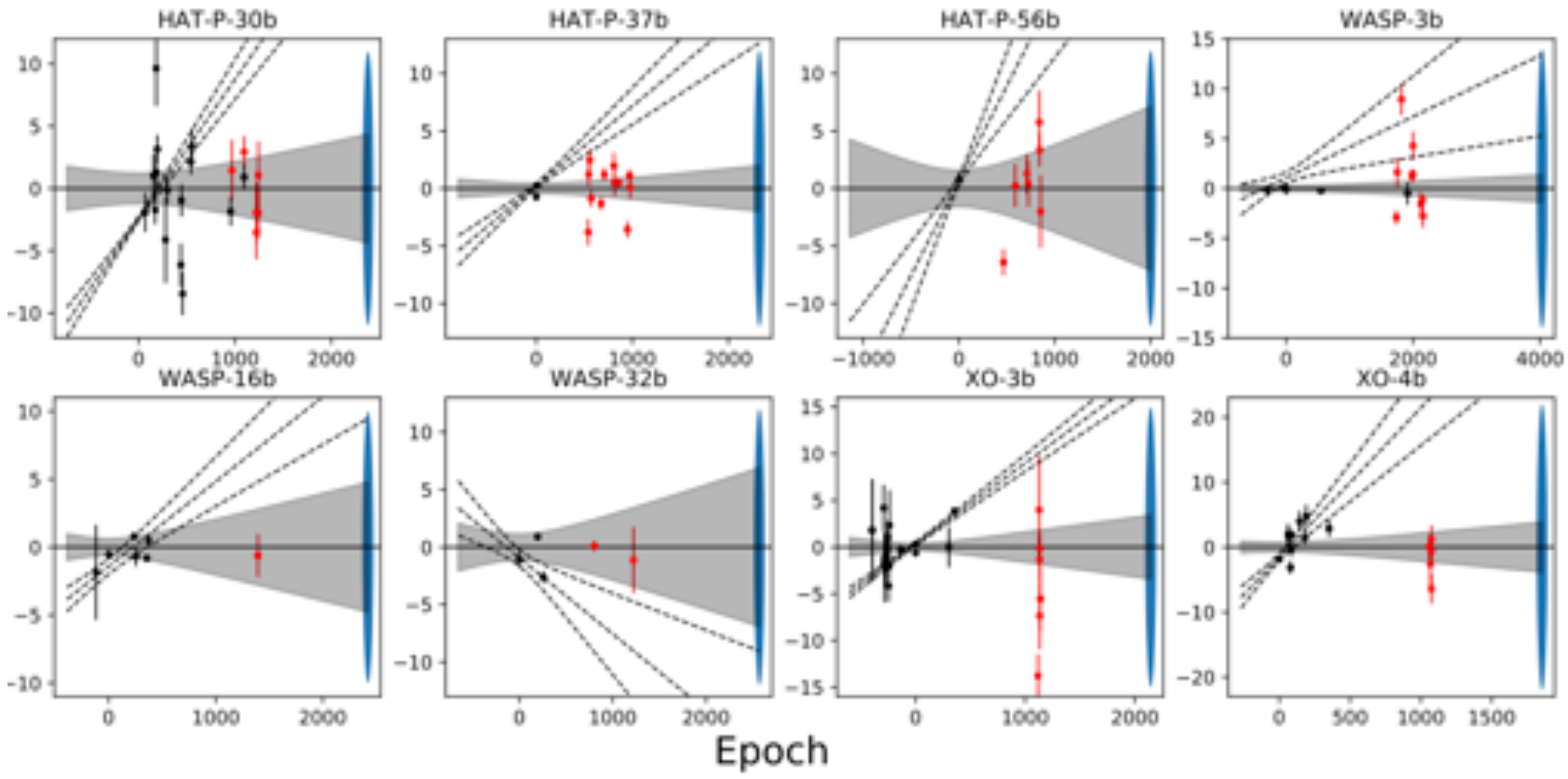

*Figure 7-6: Figure from Kokori et al. 2020. Comparison between the initial (dashed lines) and the updated ephemerides (solid lines) for the 28 planets that had their ephemerides refined. The black circles indicate literature values while the red ones indicate observations provided by telescopes registered in the ExoClock network. In blue, we indicate the target 1σ uncertainty for 2028 for each planet.*

An important part of the *ExoClock* project is to support and coordinate the efforts of a network of telescopes towards contribution to the ephemerides maintenance with dedicated observations (Figure 7-6). This network now includes 140 telescopes (from 6 to 40 inches) registered by 120 participants from 16 countries around the world, who are continuously providing new observations in a rate of 100 light-curves per month. The data provided are analysed in a homogenous way, and initial results are already being published (e.g. Edwards et al. 2020a; Kokori et al. 2020), with 120 planets observed so far. In this effort, amateur astronomers play a key role, as they provide 80% of the available telescopes. Other programmes with dozens of amateurs have been providing (*ETD*, http://var2.astro.cz/ETD/), or will provide (*Exoplanets Watch*, https://exoplanets.nasa.gov/exoplanet-watch/about-exoplanet-watch/), further follow-up observation. Such observations are another valuable resource which will be used towards maintaining the ephemerides of Ariel targets, though close collaborations with these other projects.

For shallower transit depth signals, ground-based measurements might be challenging. The updating of target ephemerides is part of the CHEOPS Guarantee Time Program and in its 3.5 year nominal mission, CHEOPS will point to many known transiting planets for precise measurements of transit depth and epochs. Additionally, 20% of the telescope time has been reserved for guest observers. The update of ephemerides for transits not observable from the ground is relevant for mission like Ariel that needs accurate knowledge of transit time at the time of observations. CHEOPS could provide follow-up observations for over half of the Ariel target sample, several hundred of which would be difficult to observe with ground-based photometry (Lendl et al. 2020, Figure 7-7).

Finally, as TESS is surveying almost the whole sky during its primary mission, and continuing to do so through mission extensions, it will collect data for many of the known planets. As of May 2020, TESS has observed nearly 200 planets within the Ariel target sample. The extended TESS mission will also re-observe many of the planet discovered by the mission, reducing the ephemeris uncertainties predicted by Dragomir et al. (2019).

The CUbesat for Refining Ephemerides (CURE) project aims at providing reliable ephemerides for Ariel targets, which have too shallow (~0.5%) or too long (~5 hours) transits to be followed by ground-based photometry and too high (> 60 deg) ecliptic latitudes to be followed by CHEOPS or TWINKLE. CURE will be able to provide ephemerides for 307-313 planets out of 366 such objects in the current Ariel target list (Coudé du Foresto and Aret, private comm.).

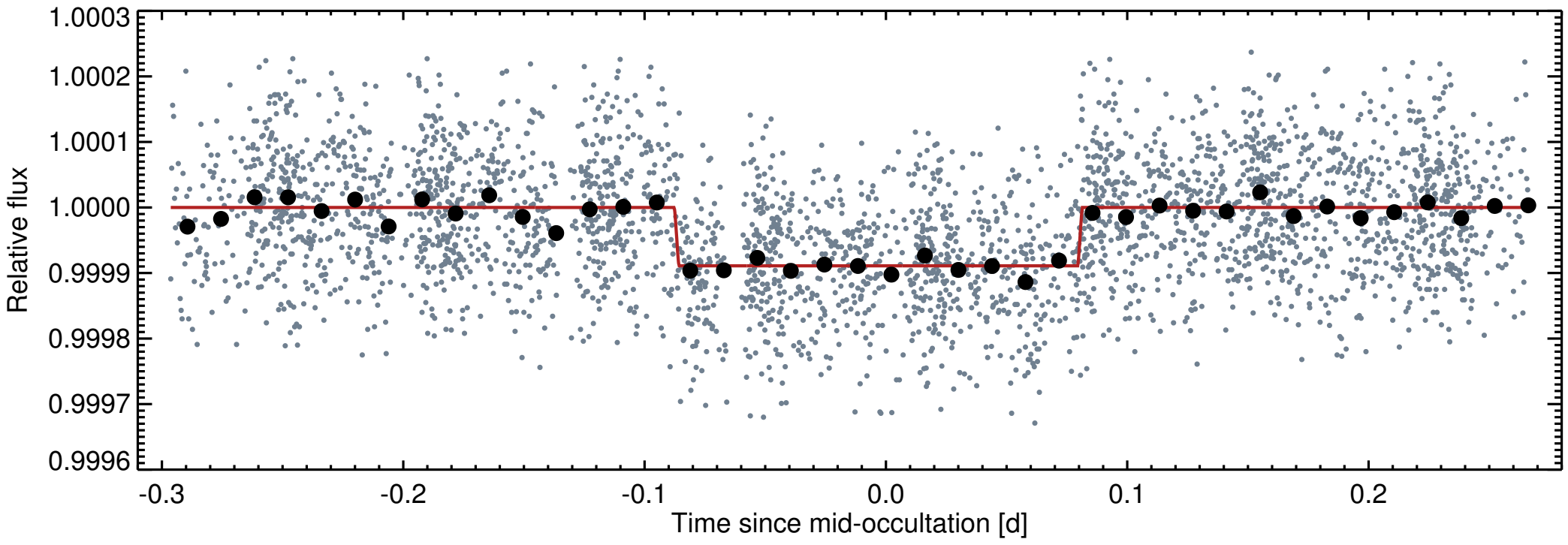

*Figure 7-7: Corrected and phase-folded CHEOPS occultation light curve of WASP-189 b. Black points show the light curve binned into 20-minute intervals and the red line shows the final occultation model (Lendl et al. 2020).*

Radial velocity measurements over a long baseline can also substantially improve the ephemerides accuracy. Additionally, orbital solutions based on RV measurements will also be able to find those cases of transiting exoplanets with small orbital inclination that will not be occulted by the star due to their highly eccentric orbit.

## 7.5 Planetary Mass Characterisation

Broad spectral range observations as delivered by Ariel permit to infer directly the planetary mass from the analysis of atmospheric observations; this option is not possible with current, narrow band observations as provided by HST/Spitzer (Changeat et al. 2020b; Batalha et al. 2019). Changeat et al. (2020b) shows, by analysis of the transit equations, that the main degeneracies in retrieving the mass are caused by the clouds (which are masking the reference planet radius) and the mean molecular weight (mmw). These results have also been confirmed in the case of Ariel observations by simulations and retrievals on various key cases. The findings confirm the theoretical analysis and show that it is possible to constrain accurately the planetary mass in the case of gas-rich planets or planets of known mean molecular weight (Changeat et al. 2020b).

By contrast, for the cloudy cases and the secondary atmosphere cases degeneracies are expected from the theoretical results and are confirmed by numerical simulations. In particular, when we deal with secondary atmospheres of unknown composition, the main chemical component ratio (e.g. $H_2/N_2$), shows a direct degeneracy with the planet mass. An important result of this work is that all the other parameters (temperature, trace gases and radius when clouds are not present) are unaffected by the mass uncertainties. All retrievals (both mass known and unknown) lead to the same posterior distributions for these parameters as they are totally independent in the radiative transfer equations. In Changeat et al. (2020b), this conclusion was highlighted for numerous cases, including a low signal to noise and limited wavelength coverage retrieval of the public HST data for HD 209458: the lack of preliminary mass information introduces a degeneracy in the planet radius and the cloud parameter but it does not affect the trace gases and the temperature.

Overall, this study indicates that the knowledge of planetary mass for cloudy and secondary atmospheres is important to fully exploit the information content of these spectra, even if most of the retrieved (trace gases and temperature) are unaffected by the planet mass uncertainties. These results, which are confirmed in Batalha (2019), suggest that a constraint with errors less than 50% on the mass should allow to guide retrieval analysis successfully, while a precision of 20% is recommended for an in-depth characterisation of the atmosphere. Mass characterization campaigns should therefore be focused on small planets for which their nature is ambiguous or for which a secondary atmosphere is suspected.

The Ariel Candidate Reference Sample is currently populated by 500+ known exoplanets, with the remaining being expected TESS discoveries. Mass measurements better than 50 and 30% for ~80 and 65% of the known targets in the Ariel sample are already available, with the fraction depending on planet characteristics. Many of the candidates discovered by TESS are currently observed with RV instruments by different teams from different telescopes. For the remaining Ariel targets that will need mass refinement, the Ariel Mission Consortium is organising dedicated radial velocity campaigns. Two parallel studies reported here indicate that the mass measurement of these targets can be obtained before launch and/or during the Ariel operations.

### 7.5.1 Current status of mass determination amongst Ariel targets

The Ariel target list will evolve over the decade that separates us from the launch of Ariel. We currently know more than 4300 exoplanets. and many more exoplanets are still to be discovered. Currently, the Ariel Candidate Reference Sample (CRS) is populated by 500+ known exoplanets, with the remaining being expected TESS discoveries. Although this CRS will rapidly evolve over the coming years, it is relevant to study the known targets as they provide a realistic estimate of the typical mass precision that the community achieves when announcing a new discovery. Mass measurements better than 50 and 30% for ~80 and 65% of the known targets in the Ariel targets are already available, with the fraction depending on planet characteristics. In particular, a large fraction of the giants and sub-giant targets in the Ariel CRS have their mass measured at the required precision, while for the known super-earths and mini-Neptunes this fraction decreases down to about 30 and 20%. In summary, the sub-giants and giants require only a small additional effort to achieve 50% of mass precision on all targets, while, the fraction of super-Earths and sub-Neptunes with known mass is still low. However, these planets are of particular importance to the community and they are and will be heavily targeted by radial velocity instruments. For the remaining Ariel targets that will need mass refinement, the Ariel Mission Consortium is organising dedicated radial velocity campaigns.

### 7.5.2 Assessing the additional radial velocity effort

It is possible to estimate statistically the amount of telescope time required to measure the mass of the exoplanets in the Ariel sample. This estimate is a function of several factors, including the spectral type of the host star, its magnitude, the level of stellar activity, the amplitude and timescale of the stellar activity with respect to the amplitude and timescale of the planetary signal, the presence of other planets in the systems. To assess the impact of these assumptions and the robustness of our estimates of the effort required to achieve the mass precision needed for Ariel, we have used two different approaches presented below. Both the approaches take advantage of the information known *a priori* of the planetary orbits and are based on CRS.

#### *7.5.1.1 Approach 1*

This approach is explained in Barnes et al. (2020). For each star with $V < 14$ in the CRS, the number of observations required to recover the mass of the orbiting planet or planets to within 30% and 50% was estimated. An assessment of spot noise, the dominant contribution to astrophysical radial velocity variability on solar rotation timescales (Meunier et al. 2010) was first made by simulating realistic stellar spot patterns. The spot models from Solanki (1999) were scaled from solar observations and models by Bogdan et al. (1988). These were used to obtain appropriate spot “contaminated” absorption line profiles using the Doppler imaging code, DoTS (Collier Cameron 2001). Spot induced radial velocity noise appropriate for solar min and max activity cases, with respective spot filling fractions of 0.02% and 0.3% were obtained. Spot temperature contrasts were estimated from Berdyugina (2005). The spot noise from the simulated solar min and solar max models was added to the planetary radial velocities. A stellar rotation period of $P_{\rm rot} = 25$ days was assumed to generate spot models, with $v \sin i = 2.0$, 1.5 and 1.0 km/s respectively for G2V, K2V or M2V models. HARPS was assumed as the reference instrument, achieving 1 ms$^{-1}$ precision for the brightest stars. The appropriate instrumental noise was added for each target according to its visual magnitude and adopted exposure times (at the $V = 14$ limit of our simulations, this is 4.1 m s$^{-1}$).

To optimise the observing strategy, radial velocity observations for each planet were simulated at quadrature points when the maximum RV semi-amplitude of each planet is expected. Each simulation trialled a minimum of 2 quadrature points (phases 0.25 and 0.75). Observations at phases 0.00 and 0.50 were then added, giving 4 observations. This helps to recover the correct injected period more quickly compared with only adding further quadrature observations. Further points were then similarly added in pairs such that a data set contained $N_{obs\text{-}trial}$ = 2, 4, 6, 8, 10, 16, 20...100 (in steps of 10), 120 - 240 (in steps of 20), 300 and 360. For two-planet systems, a minimum of $N_{obs\text{-}trial} = 4$ (i.e. 2 per planet) is required. The mass uncertainty for each sequence with $N_{obs\text{-}trial}$ observations was determined from posterior sampling via MCMC using the Radial Velocity Fitting Toolkit, *Radvel* (Fulton et al. 2018).

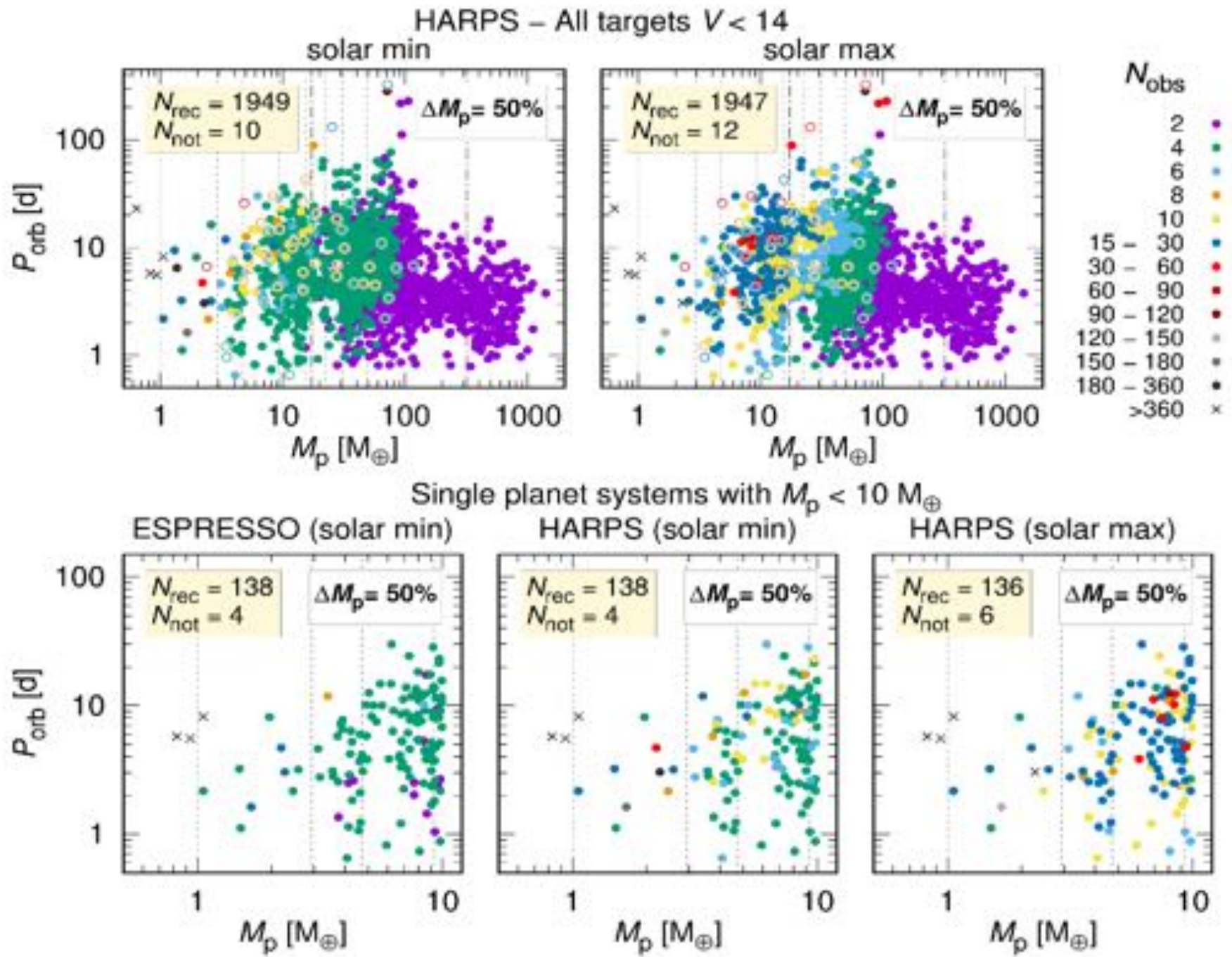

*Figure 7-8: The number of observations required for each simulated planet. The results are separated into two radial velocity r.m.s. bins (as indicated in the upper and lower sub-plot titles) and are plotted for $\Delta M_p$ < 50% (left) and $\Delta M_p$ < 30% (right). The number of observations required, Nobs, is indicated (see key for colour coding). Planets in two-planet systems are indicated by open circles. Solar minimum activity with Prot = 25 d is assumed. The total number of recovered planets, Nrec, and unrecovered planets, $N_{not}$ (i.e. >180 observations required - indicated by crosses), are indicated in each panel. The 7 planets with $M_p$ < 0.5 $M_\oplus$ are not shown.*

We obtained posterior mass estimates by imposing 10% planetary orbital period and transit time priors. In reality, these parameters will be somewhat better determined for Ariel targets. All simulated planets had circular orbits; we used priors for recovery, requiring $e < 0.05$. Assuming an uncertainty in stellar mass of 3%, the uncertainty in mass for each planet and each trial data set with $N_{obs\text{-}trial}$ observations was determined. Finally, the minimum number of observations, $N_{obs}$, to recover each planet with $\Delta M_p \leq 30\%$ or $\leq 50\%$ was estimated. Further details are in the report of WG on "Exoplanetary Mass Determination for the Ariel mission". The number of observations required for each target is presented in Figure 7-8 (see caption for details), revealing the majority of targets only require a minimum of 2 or 4 observations to obtain masses with $\Delta M_p \leq 50\%$ (and also for $\Delta M_p \leq 30\%$). The 23 two-planet systems in the Edwards et al. 2019 sample are simulated and recovered as two-planet systems (open circles in Figure 7-8). Only 10 - 12 very low mass planets are not recovered (i.e. $N_{obs} > 360$). Figure 7-8 shows a larger number of observations are required for the less massive planets (see lower panels for $M_p < 10$ $M_\oplus$), especially as stellar activity increases from solar min to solar max levels. A solar min comparison using ESPRESSO is also made. The results are summarised below.

*Summary of simulations (Approach 1)*

The times required to obtain planet masses with the optimised strategy described are summarised below. The estimates allow 5 mins for acquisition overheads, assume each night comprises 9 hrs of observing, and include 20% time lost due to bad weather. Drawing targets from the CRS, to recover the masses for **1000 Ariel targets** with a HARPS-like instrument on a 4m telescope requires:

**Solar min, $V < 14$ : 502 nights ($\Delta M_p \leq 50\%$) or 703 nights ($\Delta M_p \leq 30\%$)**

**Solar min, $V < 13$ : 427 nights ($\Delta M_p \leq 50\%$) or 579 nights ($\Delta M_p \leq 30\%$)**

**Solar max, $V < 14$ : 789 nights ($\Delta M_p \leq 50\%$) or 1500 nights ($\Delta M_p \leq 30\%$)**

**Solar max, $V < 13$ : 731 nights ($\Delta M_p \leq 50\%$) or 1443 nights ($\Delta M_p \leq 30\%$)**

Assuming that the 1000 planet Ariel sample will include the same fraction of low mass planets as in the CRS we foresee that we would recover ~70 single planet systems (depending on $V$ magnitudes considered) with $M_p < 10\ M_\oplus$. Recovery of masses for these systems requires:

**Solar min, $V < 14$ : 117 nights for $\Delta M_p \leq 50\%$ or 218 nights for $\Delta M_p \leq 30\%$**

**Solar max, $V < 14$ : 158 nights for $\Delta M_p \leq 50\%$ or 432 nights for $\Delta M_p \leq 30\%$**

Thus, up to 31% of the time commitment is spent on the ~7% of planets with $M_p < 10\ M_\oplus$. Increasing activity has a greater impact for recovery with $\Delta M_p \leq 30\%$ compared with $\Delta M_p \leq 50\%$**.** The actual number of nights required is likely to lie between the solar min and max estimates. Use of ESPRESSO at VLT would be a huge benefit for low activity systems - for the same $M_p < 10\ M_\oplus$ targets, we require

**Solar min, $V < 14$ : 28 nights for $\Delta M_p \leq 50\%$ or 45 nights for $\Delta M_p \leq 30\%$**

since instrumental noise is much lower than for HARPS. Thus for solar min stellar activity and recovery with $\Delta M_p \leq 50\%$, with it's greater precision and photon collecting ability, ESPRESSO recovers the masses ~4 × faster than HARPS. For $\Delta M_p \leq 30\%$, recovery is ~5 × more efficient.

#### *7.5.1.2 Approach 2*

The approach described in Benatti et al. (2020) simulates individual planets as realistic as possible. The HARPS-N instrument has been taken as reference, and a precision of 50% (20%) in the measured mass has been assumed. Results of simulations are validated through the comparison with published data obtained with HARPS-N. For the distribution of observing times, we adopted the cadence of 300 observations over 6 years for all the 1920 planets in the Ariel candidate list. This cadence is the same obtained for the target HD 164922 observed within the GAPS program with HARPS (Benatti et al. 2020). The RV noise is built by using a Gaussian distribution with a dispersion defined by different contributions, as described in Cloutier et al. 2018:

- from the instrument: in the case of HARPS-N at TNG, $\sigma_{instr} = 0.5$ m/s
- from the photon noise: by using the definition in Bouchy et al. (2001) and the stellar V magnitude provided in the CRS
- from additional planets in the system: kept fixed, $\sigma_{pl} = 2$ m/s

The measurement errors are derived from a sample observed within the GAPS program as a function of the stellar magnitude and exposure time (between 0.8 and 10 m/s). The use of real data also allows us to consider the impact on the RV uncertainties due to the stellar activity or the line broadening occurring for intermediate/faster rotators.

The contribution to the RV jitter due to stellar activity is injected in the time series as a sinusoid. To determine the amplitude and period of the sinusoid we generated a uniform set of ages between $10^8$ and $10^{10}$ years associated with each star of the sample. We apply the Gallet & Bouvier (2015) relation to each star to derive the rotation period and, from the period and spectral type, we determine the activity level (Mamajek & Hillembrand 2008). Finally, we can estimate the amplitude of activity contribution to the radial velocity from Santos et al. (2000) for FGK stars, while for M-dwarfs, the RV dispersion is derived from a set of data of the GAPS program. Planetary signals are injected in the time series as sinusoidal functions, assuming circular orbits, a good approximation for close orbits. The RV semi-amplitude is derived from the Mass-Radius relation by Chen & Kipping (2017).

For each target, we simulated the RV values corresponding to the 300 measurements, as previously described. To quantify the RV data needed to fully recover the injected planetary signal, we calculated the GLS periodogram (Zechmeister & Kürster 2009) for the first 10 data points of the series, adding recursively five RVs until the constraints are reached. For "fully recovered" we intend that the orbital period must be identified within the uncertainties of the fit (obtained with the Levenberg-Marquardt MPFIT IDL routine) and impose that the detected signal has a five (two) σ recovery of the RV semi-amplitude. With these simulations, we successfully recovered 1860 planetary masses (1862 at 50% of precision) out of 1920. Among them, 377 (379 at 50%) have radii smaller than 4 $R_E$ and masses lower than 15.4 $M_E$. Considering the number of targets with specific magnitude and corresponding exposure time, adding five min of overheads for each observation and nine hours per night, the entire sample (1860 planets) require ~4361 (4280) nights, while the "small" planets (377 planets) need ~1397 (1307) nights . For the majority of the sample the number of required RVs is small, with only about 98 targets of the small planet sample requiring more than 60 observations. Figure 7-9 reports the number of planets with measured mass as a function of the observation time.

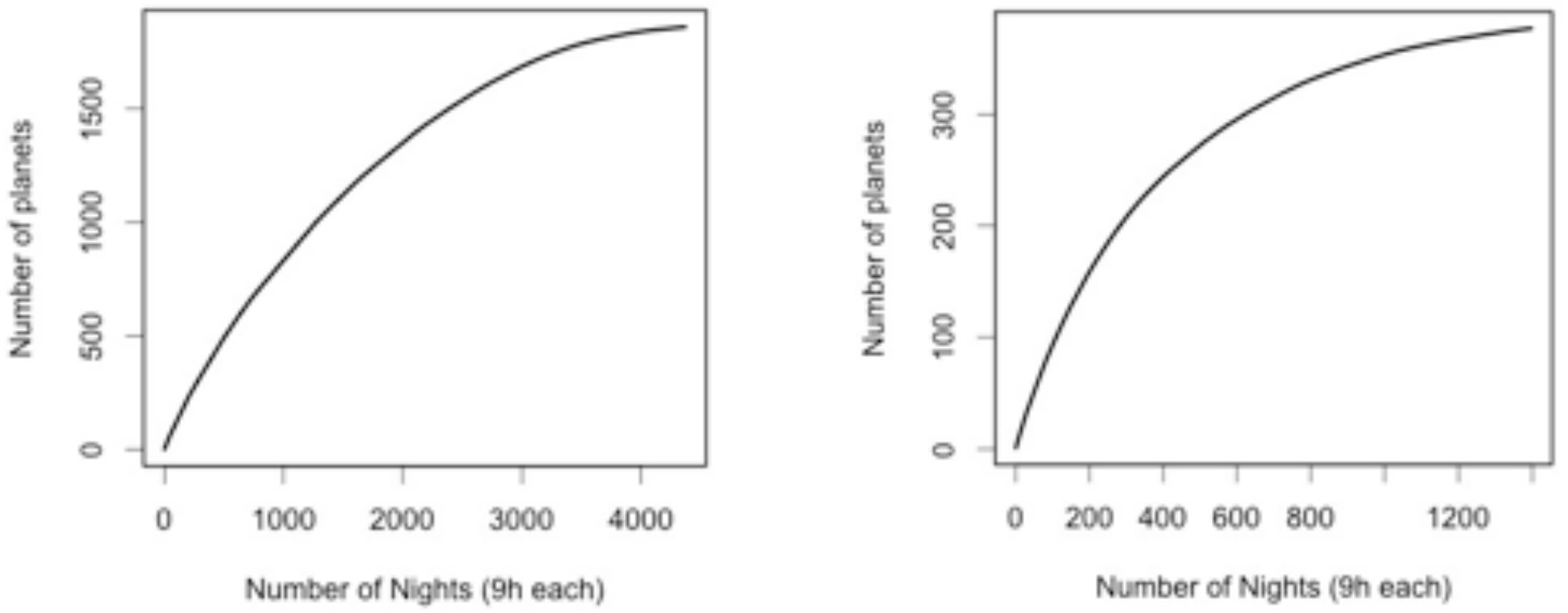

*Figure 7-9: Number of planets with measured mass as a function of the time (expressed in 9hr-nights) needed to measure the mass with a precision of 20% for the entire sample (left ) and for the planets with radius smaller than 4R⊕ (right ).*

The Mission Reference Sample (MRS) of the Ariel mission represents a possible example of the exoplanets that will be observed with Ariel to reach its objectives.

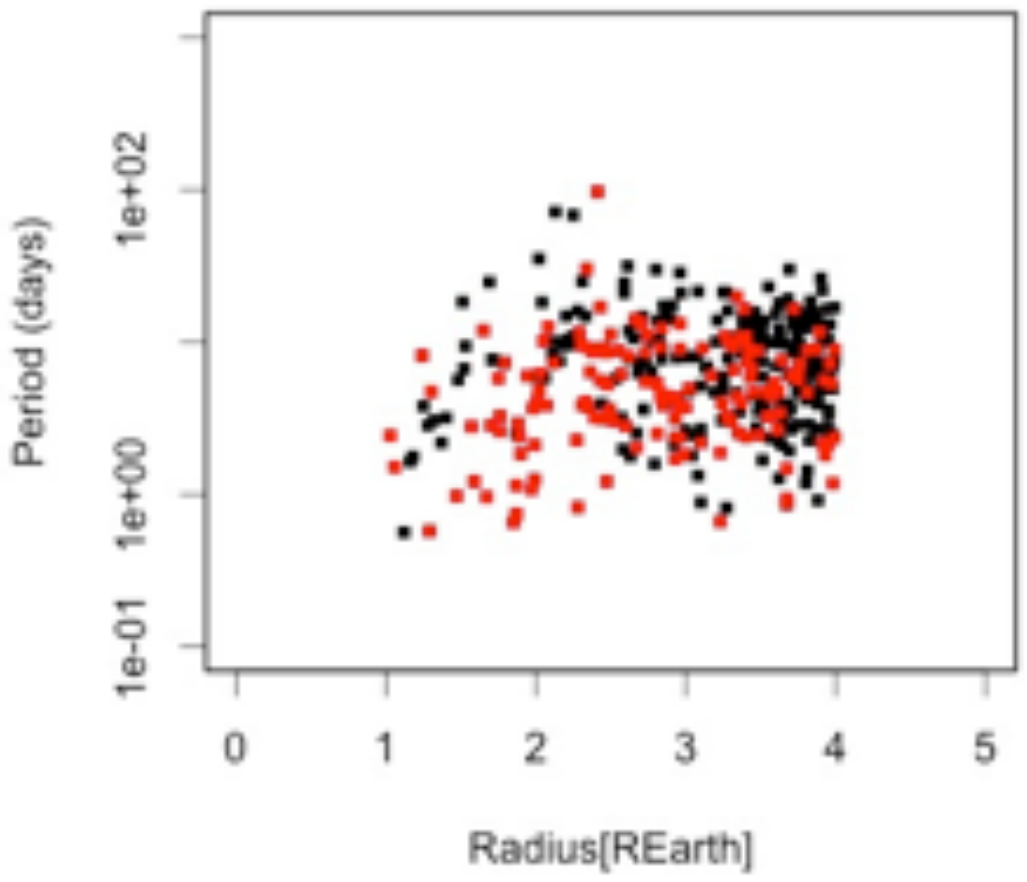

Let us focus on the smallest planets ($R_p$ < 4R⊕) for which the knowledge of the mass is an important requirement for a correct recovery of the atmospheric properties. As an example, we select a sample of six planets with radii between 1 and 1.5R⊕ , 20 planets with radii between 1.5 and 2R⊕ and 30 planets for each bin of 0.5R⊕ size in the range from 2 to 4R⊕, for a total of 146 planets. With this choice we need 303 observing nights and a uniform coverage of the period-radius diagram (Figure 7-10) if we require a precision of 50% the number of nights decreases to 283 nights.

*Figure 7-10: Left: Period-Radius distribution of the 166 planets selected as described in the text.*

To validate the reliability of the estimates provided above, we performed a comparison with three real cases: Kepler-1655 (Haywood et al. 2018), LTT 3870 (Cloutier et al. 2020) and TOI-132 (Diaz et al. 2020). We compared the results of our simulations with the actual number of RV data points used to measure the masses, as reported in the corresponding papers. These three works concern the RV follow-up of transiting exoplanets detected by Kepler and TESS by using the HARPS-N and HARPS spectrographs. We adapted our code to reproduce the same conditions as the ones presented in these three analyses, in particular:

- we considered the same temporal sampling
- we used the stellar and planetary parameters, usually well constrained in the papers, to generate the RV time series (e.g. the rotation period, the stellar age, mass and *v* sin*i*, the logR'$_{HK}$ index, the RV semi-amplitude)
- we used the average RV error and RV dispersion showed in the paper

This comparison shows that our code provides a realistic number of the radial velocities data points required to measure the planetary masses of transiting planets with a given significance.

Summary: as a final remark we note that with both approaches reported here the mass measurement of the small planets can be obtained before launch and/or during the Ariel operations. This implies that an affordable number of nights/year at 4-m class telescopes, from now to Ariel launch, are sufficient to achieve the mass measurements needed for Ariel targets. Furthermore, we have assumed here that we do not know the mass of any planets. Obviously this is not the case, as a large fraction (see previous section) of the known planets included in the CRS have known mass, and many of the candidates discovered by TESS are currently observed with RV instruments by different teams from different telescopes. It is difficult to estimate the number of nights today devoted to the mass determination of TESS (small) candidates, but likely is not too far from the needs for Ariel. Our result is therefore a very conservative estimate and can be considered an upper limit to the time actually necessary for the determination of the planetary masses for Ariel. Given the large number of facilities and the RV precision available (listed in Section 3.3.2), we are confident a mass precision of 50% or even better will be achievable for most of the targets in the Ariel target list.

## 7.6 Dealing with Systematic and Astrophysical Noise

The photometric stability of the Ariel payload is critical to achieve its science goals. The top level requirement is that all sources of uncertainties shall not add significantly to the photometric noise from the astrophysical scene (star, planet and zodiacal light), over the characteristic timescales of transit and eclipse events, typically from about 5 minutes to 10 hours.

*Table 7-4: Summary of noise sources and systematic errors.*

| Type of uncertainty | Source | Mitigation Strategy |
|---|---|---|
| **Detector noise** | Dark current noise | Choice of low-noise detectors |
| | Readout noise | |
| | Gain stability | Calibration, post- processing data analysis, choice of stable detectors. |
| | Persistence | Post-processing decorrelation. Continuously staring at a target for the whole duration of the observation. |
| **Thermal noise** | Emission from telescope, common optics and all optical elements | Negligible due to surface emissivity properties and in-flight temperatures of the payload. |
| | Temperature fluctuations in time | Negligible impact by design |
| **Astrophysical noise** | Photon noise arising from the target | Fundamental noise limit, choice of aperture size (M1 diameter). |
| | Photon noise arising from local zodiacal light | Negligible over ARIEL band |
| | Stellar variability with time | Multi-wavelength stellar monitoring, post-processing decorrelation |
| **Pointing jitter** | RPE and PDE effects on the position, Spectral Energy Distribution, and detector intra/inter pixel response | Small RPE and PDE, Nyquist sampling, post-processing decorrelation |
| | Slit losses | Spectrometer input slit sufficiently large |

We have designed the Ariel payload considering the lessons learned in measuring exoplanetary atmospheres using Spitzer, HST and ground-based instruments.

Ariel will obtain spectroscopic and photometric time series of transiting exoplanets with better than 20 to 100 ppm stability over a single transit observation, depending on the target brightness. Key aspects which allow Ariel to obtain its stable performance are:

1. Simultaneous observations of the same transit event by all photometric and spectroscopic channels.
2. Continuous observation of the transit event such that the measurement is conducted in a thermally and photometrically stable condition.
3. A payload design which makes Ariel resilient to major sources of systematics or makes it possible for their removal in post processing.

The most important disturbances of astrophysical and instrumental origin are listed in Table 7-4, along with the approach used to mitigate their impact on the detection, and on the photometric stability.

The overall noise budget is shown in Figure 7-11. It gives the best estimate of the complete instrument noise performance for three sizing targets: HD 219134, HD 209458 and GJ 1214. HD 219134 has spectral type K3V, magK = 3.25 and represents the brightest target observable by Ariel. HD 209458 is a G0V star with magK =

6.3; it is 10 times fainter than HD 219134, and represents a typical bright target Ariel will observe. GJ 1214 is a moderately faint target of spectral type M5V and magK = 8.8.

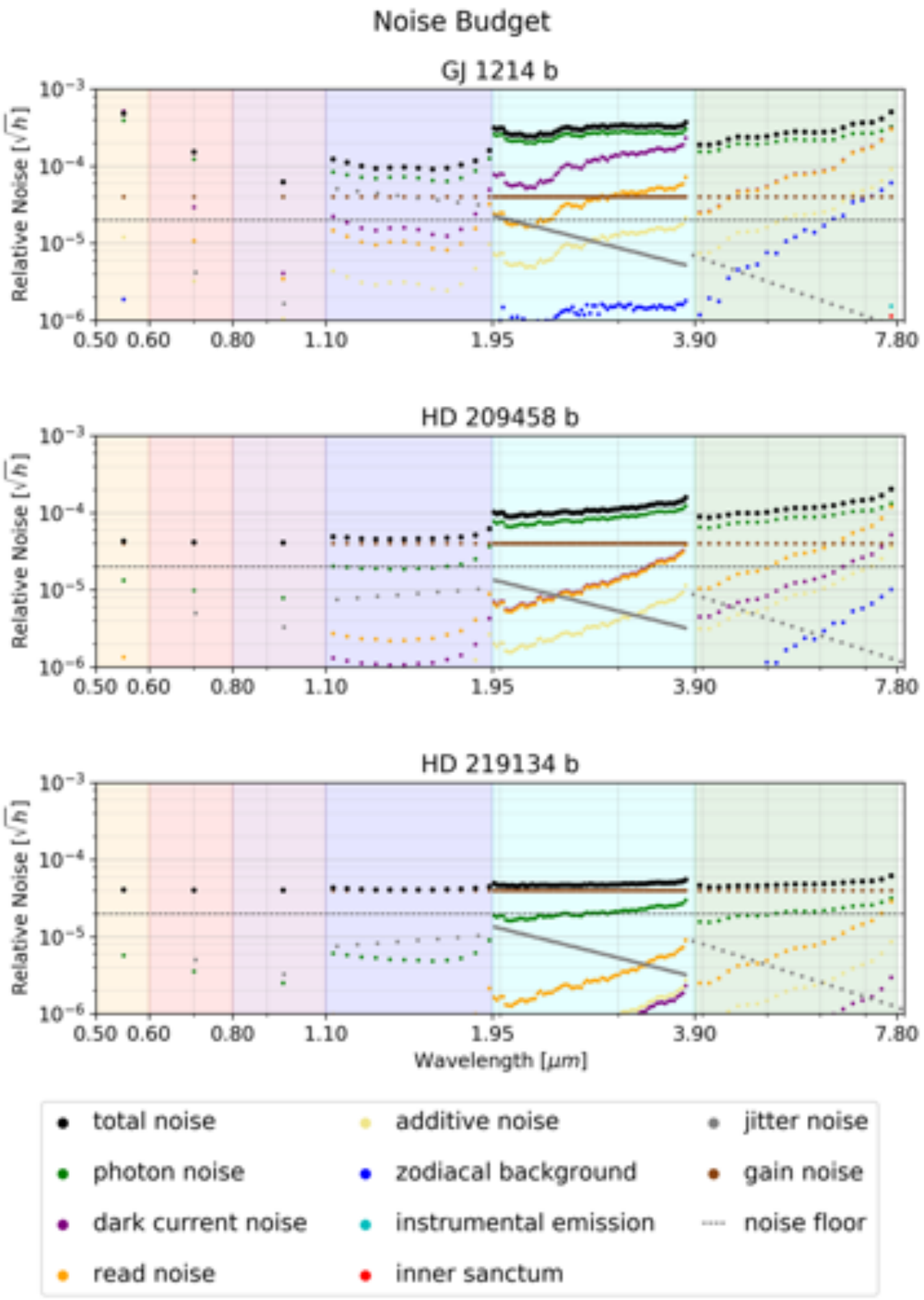

*Figure 7-11: Post-processing noise budgets. Panels from top to bottom show the budget when observing a faint target (GJ 1214) a typical bright target (HD 209458) and a very bright target (HD 219134), respectively.*

Figure 7-11 shows that, excluding gain noise effects, the dominant noise source is photon noise from the target star over most of the Ariel band. A gain noise of 40 ppm $\sqrt{hr}$, with a $1/f^2$ knee at 10hrs (equivalent to a noise floor of 20 ppm), has been assumed as a budget margin. A 20% photon noise margin is further assumed in the budget. Total margin carried is at least 30% across all MRS targets.

We used radiometric modelling (ArielRad, Mugnai et al. 2020) and dynamic simulations (ExoSim, Sarkar et al. 2020) to assess all aspects of the payload and satellite stability in order, among other things, to establish the optimum payload design, observing strategy and the calibration programme for the mission. Thermal and Zodiacal backgrounds are negligible in the Ariel bands. Detectors have measured stability upper limits in the range of 10 to 20 ppm.

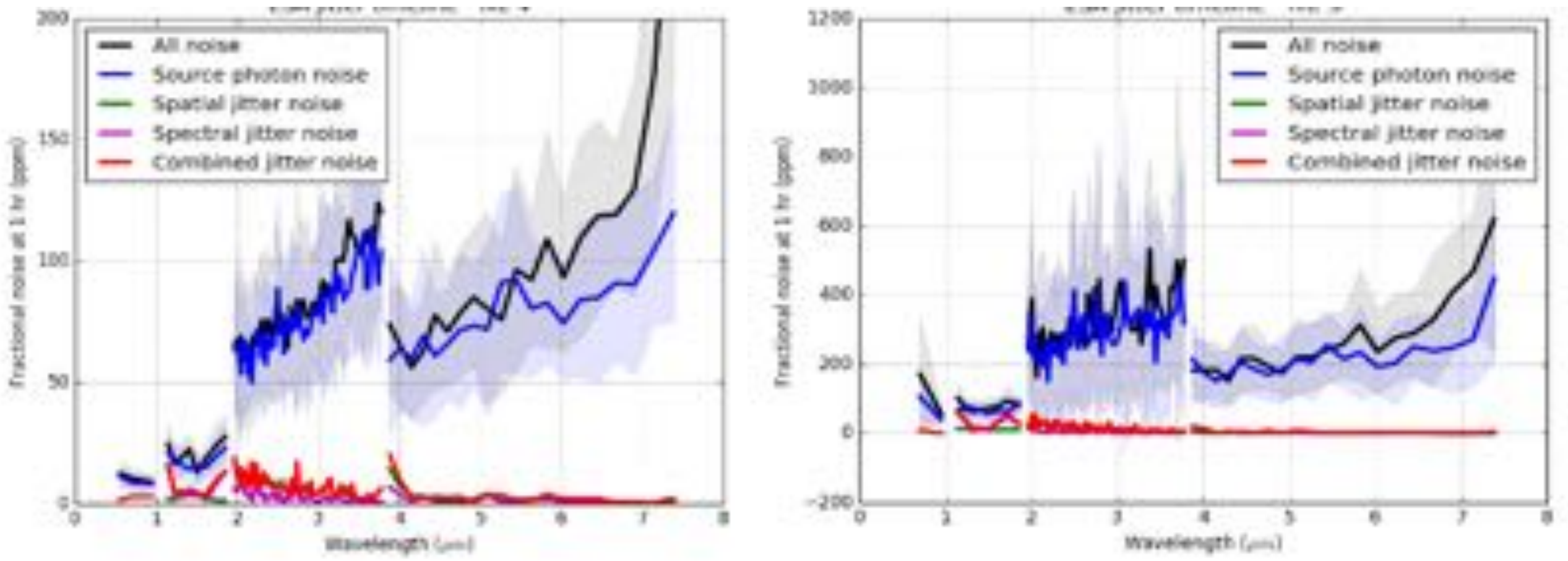

*Figure 7-12: Photometric uncertainties at one-hour integration evaluated for a typical bright target (HD 209458, left panel) and faint target (GJ 1214, right panel). Pointing jitter noise (red line) is the result of both spatial (green) and spectral (violet) jitter noise. Total noise (black) is dominated by the target photon noise across the whole band, with detector read noise (not shown in the figure) contributing the excess above target photon noise at the red end of AIRS CH1. Jitter noise is either negligible, or small in absolute terms when below a few 10s of ppm√hr, and, consequently, it is generally irrelevant.*

The observing strategy involving continuous staring at a target during a full observation makes effects such as detector persistence and thermoelastic induced instabilities negligible in the overall budget. While pointing jitter is a potential source of systematic, time-correlated noise, it is made negligible thanks to a careful instrument design where FoVs are un-vignetted, and optical signals are fully (Nyquist) sampled spatially across all focal planes. Figure 7-12 shows an example of dynamic simulations of Ariel observations that demonstrate the negligible impact of pointing jitter noise relative to the fundamental photon noise of the target source.

### 7.6.1 Activity Monitoring and Correction

The differential spectroscopy measurement strategy of Ariel (before/during/after the transit) may be affected by changes in the host star spectrum on the timescale of the transit. Changes in the host star spectrum are caused by magnetic activity (flares, co-rotating active regions and spots) and convective turbulence (granulation, pulsations).

The Ariel mission has been designed to be self-sufficient in its ability to correct for the effects of stellar activity. This is possible thanks to the instantaneous, broad-wavelength coverage and the strong chromatic dependence of light modulations caused by stellar variations. This is a unique capability of Ariel, unmatched by present and future space and ground observatories.

#### *7.6.1.1 ExoSim simulations of impact of stellar variability on Ariel observations*

We have simulated two sources of stellar variability:

*a)* convection-driven variability arising from *pulsations and granulations*

*b) star spot activity.*

These simulations have been coupled with the end-to-end simulator, ExoSim incorporating the Ariel instrument model, to obtain estimates of the impact of these sources of variability on Ariel observations.

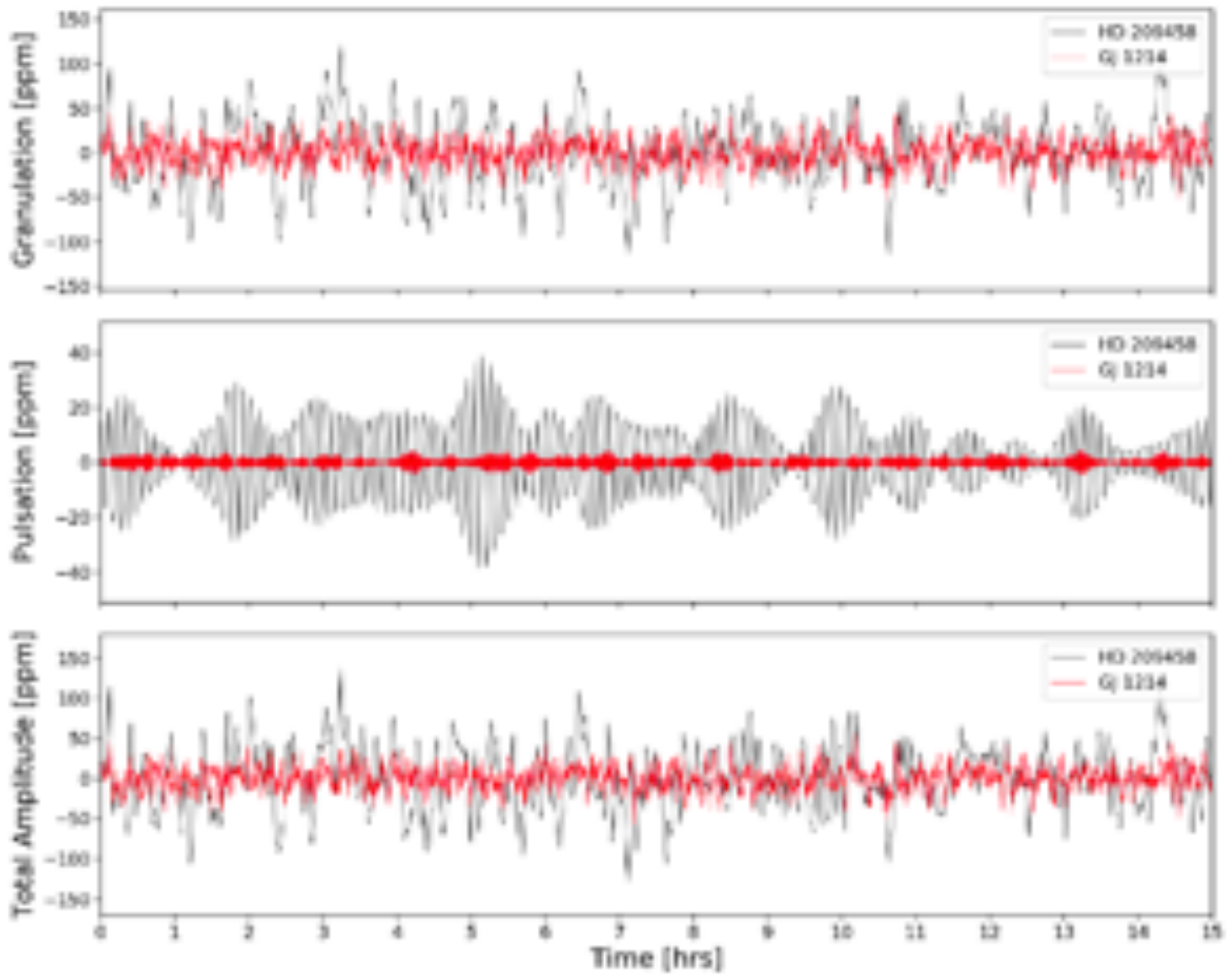

*Figure 7-13: Stellar variability model example time series for GJ1214 (M4.5V) and HD 209458 (G0V). One realization of 15 hours is shown for each star. Displayed are the variations to the bolometric luminosity with time due to granulation (top), pulsation (middle) and their combined effects (bottom). Adapted from Sarkar et al. (2018), their figure 1.*

a) <u>*Pulsations and granulations*</u> – Convection in the outer layer of stars induces stochastic pulsations and the emergence of granulation on the stellar surface. Both of these phenomena cause local temperature fluctuations and yield a stellar effective temperature that varies over time. Sarkar et al. (2018) presented simulations to assess the impact of pulsations and granulations on Ariel observations of GJ 1214b and HD 209458b (Figure 7-13). In the range 1.95-7.8 µm, stellar pulsation and granulation noise has insignificant impact compared to photon noise for both targets. In the visual range the contribution increases significantly but remains small in absolute terms and will have minimal impact on the transmission spectra of the targets studied. The impact of pulsation and granulation will be greatest for planets with low scale height atmospheres and long transit times around bright stars.

b) *Stellar spots* – A star spot model has been developed which is now fully integrated with ArielSim. The spot model can simulate different star classes, with spots and faculae of different sizes, temperatures and spatial distributions. The transit of a planet across the spotted star can be simulated at multiple wavelengths and the resulting light curves from the Ariel instrument model recovered, and used to obtain a reconstructed planet spectrum with the effects of spots in each channel. Specially important are the Ariel photometric bands in the visible, purposely chosen to correct for stellar activity. Spots may affect transit depths and the reconstructed spectrum in a number of ways:

- Unocculted spots may increase the transit depth, with the bias being wavelength dependent.
- Occulted spots may reduce the transit depth obtained through curve fitting, with the amplitude on the light curve being wavelength dependent.
- Spots with molecular absorption features can contaminate the planet spectrum.

The situation is simpler for eclipses, where the planetary emission follows directly from the depth measurement. In this case, only activity-induced variations on the timescale of the duration of the occultation need to be corrected to ensure that the proper stellar flux baseline is used.

Star spot simulations using the ArielSim was used to explore the effects of spots on transit spectra. We show the results of the spot simulator and we describe in the various decorrelation methodologies to mitigate the effects of spots.

#### *7.6.1.2 Methods to correct the effect of star spots on transit spectra*

We have developed two different methods based on stellar models to correct for effects occurring during transits.

***Method 1 –*** We have developed the StarSim code (Herrero et al. 2016; Rosich et al. 2020), which uses realistic stellar spectra to simulate light curves in various passbands and spectroscopic time series of rotating stars considering the effects of active regions. The simulator considers surface inhomogeneities in the form of (dark) starspots and (bright) faculae, implements limb darkening (or brightening in the case of faculae), and includes time-variable effects such as differential rotation and active region evolution. StarSim is based on surface integration techniques that take into account the particular properties of each element of a fine stellar surface grid such as the effective temperature, the limb darkening and convection effects. StarSim can estimate the surface structure of active stars through the inversion of photometric time-series, as well as representative activity-related parameters such as spot temperature contrast, relative coverage of facular regions and stellar rotation period. With the resulting model, one can investigate and correct out the influence of the chromatic effect produced by activity features for transmission spectroscopy.

As an illustrative example, we show here a test case based on BVRI-filter photometry of the young active planet-host star WASP-52 covering ~600 days of 2016 and 2017 (Rosich et al. 2020). The photometric data are inverted using a StarSim activity model yielding a heterogeneous stellar surface composed of dark spots with a temperature difference of 575±150 K with respect to the surrounding photosphere (see Figure 7-14). The fit also produces a probabilistic map of active regions, showing in this case two or three prominent spot complexes at stable stellar longitudes. To study the influence of the derived activity model on the observables of transmission spectroscopy, we produced simulated transits of WASP-52 b by avoiding spot-crossing events. The simulations reveal chromatic effects of the spot map resulting in relative uncertainties of ~10% in transit depth and of up to ~5% in the planetary radius, at visible wavelengths.

Table 7-5 provides the statistics of the solutions for two fiducial transits (TR-1 – low activity and TR-2 – high activity) at different wavelengths, encompassing the full wavelength range of Ariel. We computed the mean transit depth variation induced by spots and the corresponding planetary radius variation, along with the standard deviation induced by the uncertainty in the determination of the spot map. The statistics reveals that, even for the relatively low stellar activity level of WASP-52, with a spot filling factor of ~5%, the signature of stellar activity on the transit depth can be close to $10^{-3}$ of the flux in the visual bands and $\sim 2\times10^{-4}$ in the NIR bands. This translates into relative planetary radius changes from $\sim 2\times10^{-3}$ to $\sim 5\times10^{-4}$, depending on the wavelength considered, which are of the same order as the typical signature of exoplanet atmospheres on transmission spectra. Furthermore, at higher activity level such as that of TR-2, the effect of the starspots is about twice as large. After correcting for spot effect using the StarSim model, we are able to reduce residual

depth uncertainties down to ~$10^{-4}$ at 550 nm (VIS-Phot) and ~$3\times10^{-5}$ at 6 $\mu$m (AIRS-Ch1), regardless of the fraction of photosphere covered by spots (see error bars in the first two columns of Table 7-5). And we recall that we are analysing a star displaying a modulation of about 7% in flux in the visible band, and a spot filling factor of about 3-14%.

*Table 7-5: Mean values of transit depth variations and the corresponding relative radius variations due to spots on WASP-52 simulated transits. Values for the low-activity transit TR-1 and the high-activity transit TR-2 cases are provided. The error bars of the variations can be considered as the residual uncertainties of the corrections using the fitting procedure.*

| | Transit depth effect ($10^{-3}$) | | Planet radius effect ($10^{-3}$) | |
|---|---|---|---|---|
| **Instrument** | **Low activity (TR-1)** | **High activity (TR-2)** | **Low activity (TR-1)** | **High activity (TR-2)** |
| VIS-Phot | 0.73 ± 0.13 | 1.99 ± 0.18 | 2.30 ± 0.41 | 6.26 ± 0.57 |
| FGS-1 | 0.57 ± 0.10 | 1.51 ± 0.15 | 1.79 ± 0.31 | 4.75 ± 0.47 |
| FGS-2 | 0.43 ± 0.07 | 1.11 ± 0.11 | 1.35 ± 0.22 | 3.49 ± 0.35 |
| NIR-Spec | 0.22 ± 0.04 | 0.56 ± 0.06 | 0.69 ± 0.13 | 1.76 ± 0.19 |
| AIRS-Ch0 | 0.17 ± 0.03 | 0.43 ± 0.05 | 0.53 ± 0.10 | 1.35 ± 0.16 |
| AIRS-Ch1 | 0.15 ± 0.02 | 0.37 ± 0.04 | 0.47 ± 0.10 | 1.16 ± 0.13 |

The analysis of WASP-52 reveals that, without additional information allowing inference of the distribution and filling factor of spots during the transit, the planet-to-star radius cannot be determined with an accuracy better than a few percent. This is because in that case, we do not know at which rotation phase of the star the transit is observed, and therefore, we can only estimate a range of possible filling factors based on the amplitude of the photometric modulation. However, if we know the rotation phase at the time of transit, and we can estimate the spot properties and distribution from light curves, we can infer the correction that needs to be applied to the relative radius of the exoplanet (or transit depth).

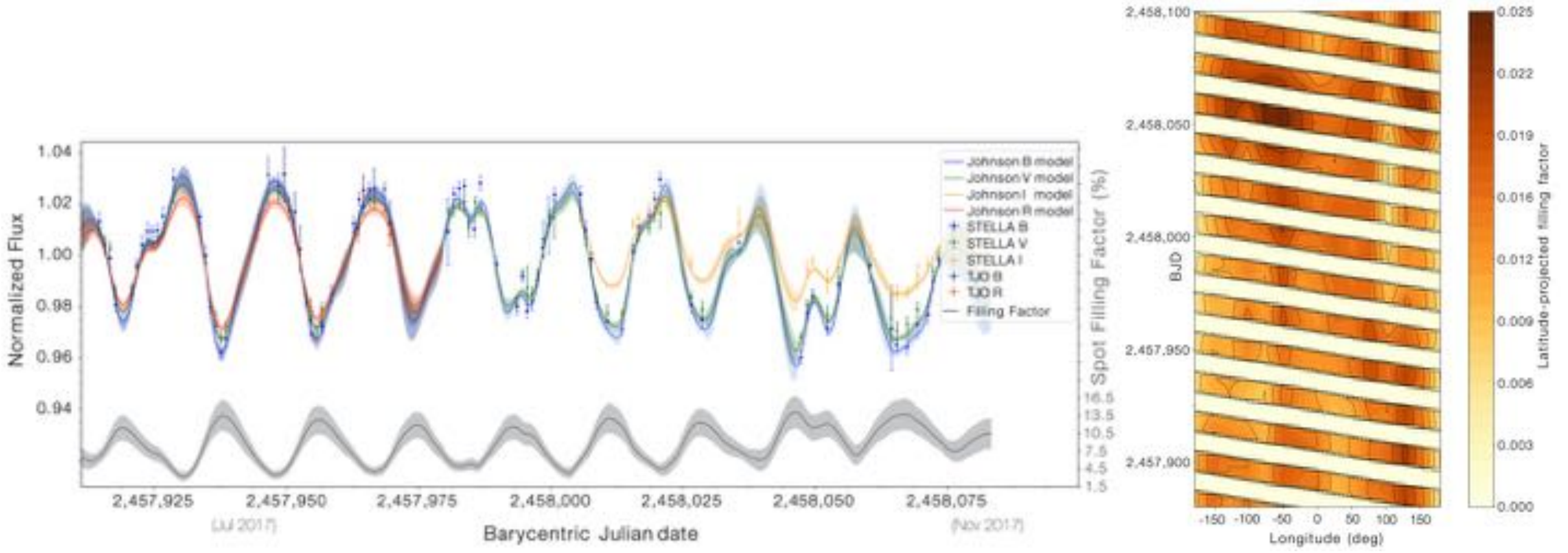

*Figure 7-14: Left: Light curve of WASP-52 in the BVRI bands together with the best-fitting StarSim photometric model. The shaded bands indicate the 1σ uncertainties and the grey line at the bottom is the projected spot filling factor of the maps. Right: Longitudinal spot filling factor projected on the stellar equator of WASP-52 as a function of time. Only stellar longitudes visible at each time are shown (hence the band structure). The colour scale indicates the fractional latitude-projected spot coverage for each 15-degree longitude bin. See Rosich et al. (2020).*

The main advantage of this approach to compute the chromatic effect on the depth of exoplanet transits is that it is based on an independent and consistent deterministic method allowing accurate determination of the stellar parameters, the filling factor and the distribution of spots. Thus, the effect of spots and their different positions on the stellar disk are also considered. We have only considered here the effects of unocculted spots, which produce a bias without any visible signature on the transit photometry. In the case of occulted starspots, the Ariel high-precision measurements could reveal their presence and allow a direct correction or the modelling of their properties. We show that contemporaneous ground-based photometric monitoring over a wide wavelength range can be used to model spot-driven stellar activity variations and correct out their chromatic effects on the transit depth measurements. We expect that stronger activity constraints could be achieved by optimizing the strategy of the photometric observations, i.e., making sure that multi-colour measurements are obtained before and after the transits to better define the model. Photometric observations from Ariel itself in the intervals before and after the transits will boost the accuracy in light curve modelling, and hence, the correction capabilities.

**Method 2 –** A complementary method (Cracchiolo et al. 2020) to extract the planetary spectrum in presence of stellar magnetic activity has been developed, based exclusively on Ariel spectra. Ariel-like spectra were simulated using the Ariel radiometric model (ArielRad; Mugnai et al. 2020). In the next paragraphs we show how we can extract the planetary signal from a transit observation in presence of stellar activity, instrumental and Poisson noise. We model the stellar activity assuming that the activity is triggered by the presence of a dominant spot colder than the surrounding photosphere. The approach can be applied in a more general case, including other photospheric inhomogeneities. The observed out-of-transit spectrum $F_{out}$ can be expressed as:

$$F_{out} = (1-ff) \cdot F_\lambda(T_*) + ff \cdot F_\lambda(T_s)$$

Where $F_\lambda(T_*)$ and $F_\lambda(T_s)$ are the fluxes radiated by the unspotted photosphere and the spot respectively, and $ff$ (filling factor) is the fraction of the stellar surface covered by the spot, $0 \le ff \le 1$.

During the transit, in the presence of a spot on the stellar surface, the planet can occult a fraction of the radiation coming from the spot and a fraction coming from the unperturbed photosphere. Since the planetary radius is wavelength dependent, the fraction $g$ of the planet that occults the spot at time $t$ is wavelength dependent too (with $0 \le g \le 1$). So, the measured flux of the system "spotted star+planet" $F_{in}$ can be expressed as

$$F_{in} = [1-ff-(1-g_\lambda)\cdot \varepsilon_\lambda] \cdot F_\lambda(T_*) + (ff - g_\lambda \cdot \varepsilon_\lambda) \cdot F_\lambda(T_s)$$

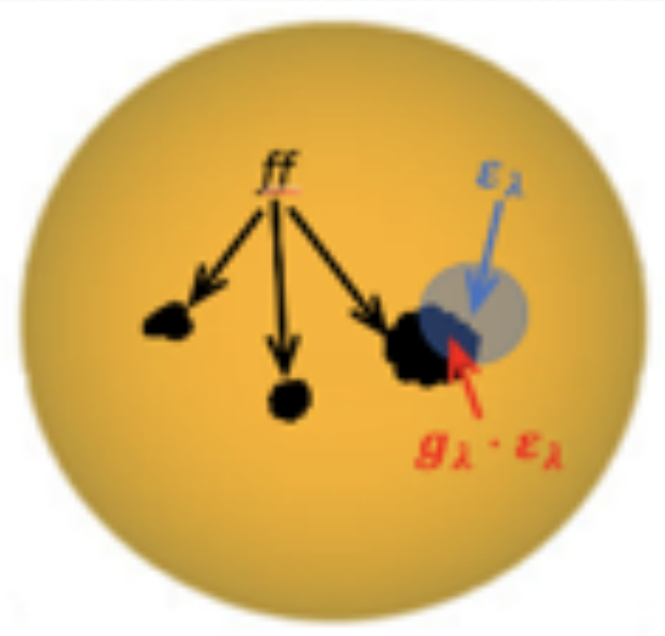

where $\varepsilon_\lambda$ is the transmission spectrum of the planetary atmosphere, see Figure 7-15 for a graphical representation.

*Figure 7-15: Scheme of the geometry of a transit on a spotted star. ff is the fraction of the stellar surface covered by the spot, $\varepsilon_\lambda$ is the transmission spectrum of the planetary atmosphere, and $g_\lambda$ is the fraction g of the planet that occults the spot at time t.*

Since $ff$ evolves on a timescale comparable with the rotation period, we expect that during a single transit the spot pattern does not change and the measured stellar flux variations can only be attributable only to the planet occultation.

$$F_{in} = F_{out} - \varepsilon_\lambda \cdot [ ( 1 - g_\lambda) \cdot F_\lambda(T_*) + g_\lambda \cdot F_\lambda(T_s)]$$

This equation can be used to derive the planetary spectrum $\varepsilon_\lambda$. If the planet transits out-of the spot, $g_\lambda = 0$ at every time $t$, and therefore

$$\varepsilon_\lambda = (F_{out} - F_{in})/F_{out} \cdot [ 1 + ff \cdot (F_\lambda(T_*)/ F_\lambda(T_s) - 1)]$$

A similar approach can be found in McCullough et al. (2014) and Zellem et al. (2017). The method has been tested on three potential Ariel targets taken from Edwards et al. (2019) catalogue. For each star, a grid of stellar spectra out-of-transit is generated where fluxes $F_\lambda(T)$ are generated with the Ariel Radiometric Model (ArielRad; Mugnai et al. 2020). For a given input temperature $T$(K) and surface gravity $g$ of a target star, ArielRad selects the input spectrum by using BT-Settl models (https://phoenix.ens-lyon.fr/Grids/ BT-Settl/, Baraffe et al. 2015). Spectra corresponding to temperatures not present in the Phoenix library are generated by interpolating spectra with the closest temperatures. Assuming a given magnitude of the star, we simulate the stellar flux $F_\lambda(T)$ (in *counts/sec*) in Ariel's photometric and spectral channels. Figure 7-16 shows the capability to recover the spot properties for three specific cases. In all cases the spot parameters are very well recovered, thanks to the Ariel bluest photometric bands. If we used only the Ariel spectroscopic channels at wavelengths longer than 1.1 μm, the uncertainty on the temperature of the spots becomes very large.

We simulate the observations of planetary transits in the presence of spots of three real transiting systems from Edwards et al. (2019) catalogue: HD 17156 b (a Jupiter), HAT-P-11 b (a Neptune) and K2-21 b (a super-Earth). We adopted the number of transits estimated by Edwards et al. (2019) for Tier 2 (n=2, 3, 184, respectively) for the three planets. For each planet, we generate the synthetic transmission spectrum by using TauRex code (Al-Refaie et al. 2020), by assuming that the three planets have atmospheres $H_2$-He dominated. We assume an isothermal temperature profile, that the planetary atmosphere absorbs the stellar radiation within five scale heights (Sing et al. 2018; Tinetti et al. 2013), and we set the pressure at the bottom of the atmosphere to 1 *bar*. In addition to the presence of $H_2$, He, we assume that each atmosphere contains traces of water (with mixing ratio $10^{-4}$). The results are illustrated in Figure 7-17, where we compare the extraction of the planetary signal without and with correction for two different activity levels. We find that in case of low stellar activity we do not need to correct for the presence of spots because their effect is smaller than the measurement errors, while in case of high activity we have to correct the spectrum and the proposed method works well in extracting the planetary signal.

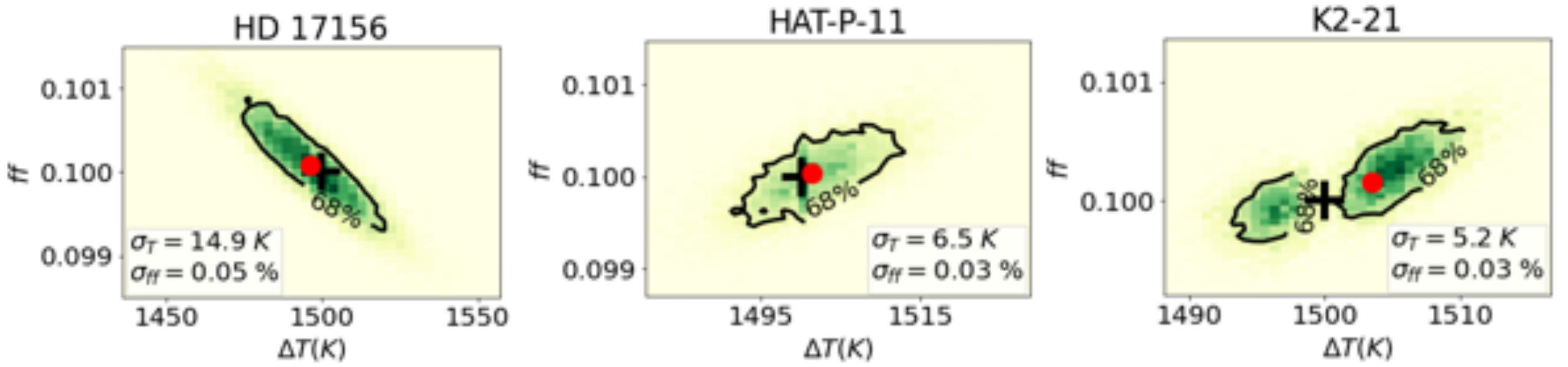

*Figure 7-16: 2D probability distribution functions of the derived parameters from 10 000 realizations of noisy spectra of V=9 stars with $T_{exp}$=100 s. We show the case with a spot temperature 1500.0 K cooler than the photosphere and with ff = 0.10, for a G0 star (left), a K4 star (centre) and a M0 star (right), respectively. The black cross represents the input parameters, while the red point marks the mode of the distribution. The black contour shows the confidence region around the mode containing the 68% of the sample.*

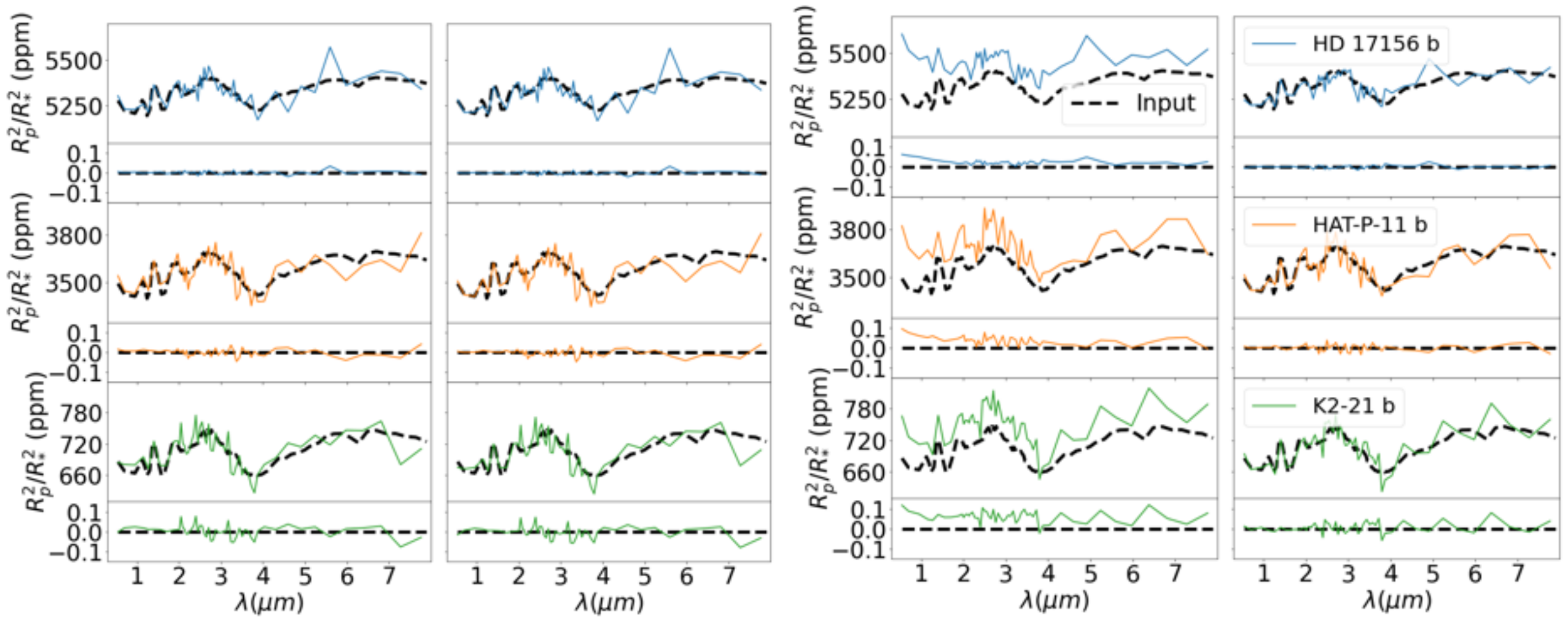

*Figure 7-17: Planetary spectrum extracted (with Tier 2 resolution) in presence of activity for three specific planets, as indicated in the panels. The dashed black line is the planetary signal $\varepsilon_\lambda$ in Ariel bins. Left: low stellar activity (ff = 0.03, ΔT = 300.0K). The coloured lines on the left panels show the planetary signals extracted without any correction, while the right panels show the corrected spectra. Right: the same in presence of high stellar activity (ff =0.10 and ΔT =1000.0K).*

### 7.6.1.3 Ariel Machine Learning Data Challenge to correct for stellar spots

In preparation for the Ariel Mission, a number of open problems related to analysing the data that will be collected from Ariel have been released to the public. Among these, the Ariel Mission's 1st Machine Learning Challenge was concerned with the task of correcting transiting exoplanet light curves for the presence of stellar spots. Both the astrophysics and machine learning communities were invited. To this end, the Challenge targeted both audiences by being officially organized in the context of the ECML-PKDD 2019 conference and also having a strong presence in the joint EPSC-DPS 20195 conference via a dedicated session. The Challenge ran from April to August 2019. More than 120 teams participated and it attracted the interest of researchers from both communities – as evidenced from the top-5 ranked teams co-authoring the Nikolaou et al. (2020)

article and the solutions they submitted. As such, we consider the secondary objective of the Ariel ML Challenge has been met successfully.

The main goal of the ML Challenge was to automate the extraction of useful parameters from transiting exoplanet light curves in the presence of stellar spots. For the purposes of the competition, we generated simulated Ariel-like light curves for 2097 candidate target planets of Ariel. For each planet, we included 10 different instances of stellar spot noise (details of the stellar spot model are provided below). This resulted in 20,970 light curves distorted by stellar spots. Finally, for each of these light curves 10 different instances of additive Gaussian photon noise were introduced (no additional instrument systematics were assumed) resulting in 209,700 total light curves distorted by both stellar spot and photon noise. The two sources of noise were treated as independent. These 209,700 light curves formed the dataset of the competition.

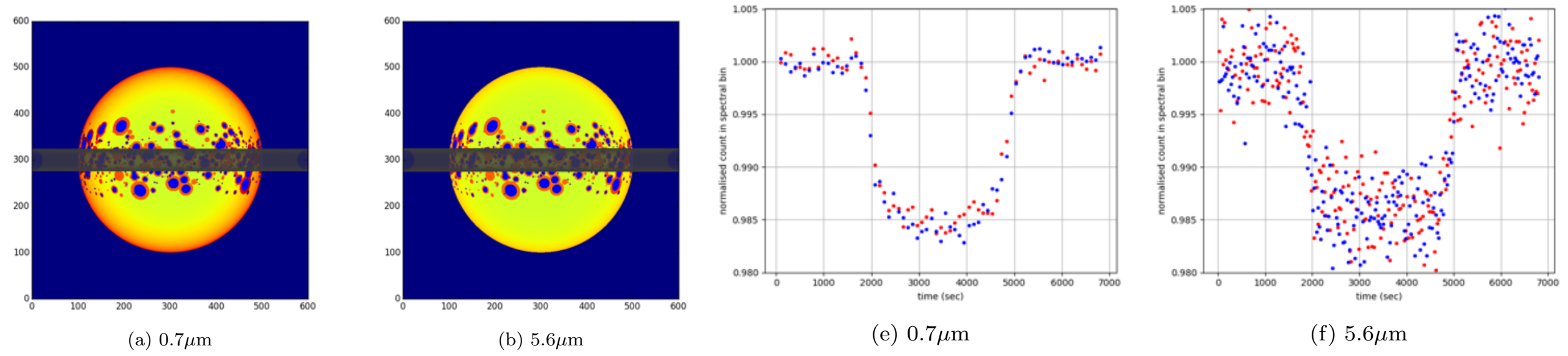

*Figure 7-18: Figures from Nikolaou et al. (2020). Examples of simulations for two of the 55 wavelength channels, 0.7µm and 5.6µm. (a) & (b), stellar surface simulations of a spotty star. (e) & (f) Normalised observed flux as the planet transits across the star, with stellar photon noise included. Blue shows the perfect transit across a spotless star; red shows the transit across a spotty star.*

A large number of solutions submitted achieved the desired precision of $10^{-5}$ for correctly predicting the relative transit depth per each wavelength from the noisy light curves. The solutions of the top-5 ranking teams that participated in the ML Challenge are presented in detail in Nikolaou et al. (2020). Most solutions amount to constructing highly non-linear (w.r.t. the light-curves) models with minimal pre-processing using deep neural networks and/or ensemble learning methods. However, there exist almost as good –in terms of the precision of the obtained predictions– approaches that involve obtaining meaningful (i.e. informed by physics) statistics from the light curves and then training models that are linear w.r.t. them.

#### *7.6.1.4 Flares*

Flares show up as stochastic stellar variations on short time scales on all solar type stars (A to BDs). They are frequent events, more prominent in the bluest bands than in the red and in infrared, since their emission is due to material heated by accelerated particles and conduction fronts after magnetic reconnection (leading to photospheric/chromospheric heating up to ten(s) thousand degrees). Analysis of the statistical properties of flares may be subject to several biases due to the quality of the data and the methods adopted to identify them, but all the analyses agree that that large (with regard to the non-flaring brightness) flares are rare events.

In Table 7-6 (adapted from Yang & Liu 2019) we compare results derived from different authors using Kepler data. Beside the detailed numbers, these numbers indicate that stars with large flares that can impact on the extraction of the planetary signal are rare, with a maximum of ~10% among the K-M stars. The average number of flares for each flaring star (~15) derived from Balona (2015) indicates that flaring/non-flaring stars belong to different classes, and that it is very unlikely that a star in the non-flaring class shows a flare. Flaring stars are more common among fast rotators and therefore among active young stars. Most of those will not be observed by Ariel.

The first step will be, therefore, the identification of  planets around flaring stars and either eliminate them from the sample, unless we want to study the effect of flares on atmospheres explicitly (Venot et al. 2016). In the eventuality that some flares occur during Ariel observations, it will be possible to recognize a typical flare from the shape of light curve with the photometric channels from the FGS. The typical sudden rise and slow decay signature of a flare over one/several hours allows it to be disentangled from a transit signature. The flare emission can also be discriminated by its colour temperature which will be measured via the FGS1, VGS2 and VisPhot optical channels. Data can be discarded if necessary. If some low intensity flares remain undetected in the optical channels, the contamination of the NIR-MIR spectra will be very limited. In fact, a recent detailed study of three M dwarf flare stars (Tofflemire et al. 2012) and M dwarf stars from SDSS and 2MASS surveys,

Davenport et al. (2012) did a multi-wavelength characterization of flares in low-mass stars. Using their results, it is possible to obtain the relation between the flare intensity in the u-band and in the IR. The flare intensity reduces towards the longer wavelengths nearly following a power-law, and it depends on the sub-spectral type. In Table 7-7 $\Delta m_u$ and $\Delta m_v$ are the expected magnitude changes in the U and V band, respectively, and produce effects in K band at levels as indicated (courtesy of J. Zoreck, IAP).

*Table 7-6: Fraction of flaring stars in SC Kepler data (adapted from Yang & Liu 2019).*

| Type | Teff | % (Yang & Liu 2019) | % Balona (2015) (Long Cadence data) | % Balona (2015) (Short Cadence data) | % Van Doorsselaere et al. 2017) |
|---|---|---|---|---|---|
| A | >7500 | 1.16 | 2.78 | 2.36 | 1.31 |
| F | 6000-7500 | 0.69 | 0.94 | 2.54 | 3.2 |
| G | 5000-6000 | 1.46 | 3.29 | 4.91 | 2.90 |
| K | 4000-5000 | 2.96 | 12.75 | 10.16 | 5.28 |
| M | < 4000 | 9.74 | | | |

To summarise, stars with intense flaring are uncommon and might not make it to the final Ariel sample. Isolated flares would be immediately identified with the FGS and the flare effect in the IR is very small.

*Table 7-7: expected variation in K band for a given variation in V-band. The reported case is relative to a very intense flare. A flare on a $M_0$ star with an amplitude in U of u=2.05, has an amplitude of V=0.21 in V and of $m_k$=0.01 in K band. Such a strong flare will be easily detected.*

| $\Delta m_k$ [mag] | M0 | | M3 | |
|---|---|---|---|---|
| | $\Delta m_u$ | $\Delta m_v$ | $\Delta m_u$ | $\Delta m_v$ |
| 0.01 | 2.05 | 0.21 | 3.20 | 0.53 |
| 0.001 | 0.48 | 0.02 | 1.15 | 0.07 |
| 0.0001 | 0.15 | 0.002 | 0.25 | 0.007 |

### *7.6.1.5 Conclusions*

The combination of a stable platform, operating in a stable thermal environment and with a highly integrated payload and systems design, will ensure the very high level of photometric stability required to record exoplanet atmospheric signals, i.e. 10-50 ppm relative to the star (post-processing). The broad, instantaneous wavelength range covered by Ariel will allow to detect many molecular species, probe the thermal structure, identify/characterize clouds and monitor/correct the stellar activity. Finally, requiring an agile, highly stable platform in orbit around L2, from which the complete sky is accessible within a year, will enable the observation of hundreds of planets during the mission lifetime.

# 8 Extended use of Ariel Observations

Thus far we have exclusively discussed the core scientific objectives of Ariel and how they will be achieved. In this short chapter, we now consider the scientific opportunities which Ariel will deliver within and in addition to the core science. Nothing in this chapter drives the design of the Ariel Mission, but nonetheless the science described herein will be significant. In Section 8.1 we describe some of the possible scientific advances which can be made using the Ariel Core Survey data for goals additional to the core science. This "Extended use" science is a bonus and requires no additional observations. Section 8.2 discusses the synergies between Ariel and other facilities. Section 8.3 discusses Complementary Science, i.e. that which can be achieved with non-time-constrained observations to be made in the gaps in the schedule after the Core Survey observations have been optimally scheduled. As the Core Survey is comprised entirely of time-constrained observations, it is estimated that there will be approximately 5-10% of the total time available for Complementary Science.

## 8.1 Extended use of Ariel Core Survey Observations

Ariel will produce exquisite optical and near-IR multi-band photometric data, and thermal IR spectrophotometry, all at 1 Hz sampling. For a generic target the total noise in each of VISPhot, FGS1 and FGS2 is less than 50 ppm √hr (Pascale 2020). With photometric data of this quality, Ariel is sensitive to the intrinsic variability of even very quiet stars. This section outlines a variety of science which can be executed with the exquisite, multi-band, high-speed sampling of Ariel's Core Survey data. Further details and more possibilities are given in Szabó et al. 2020; Borsato et al. 2020, and Haswell 2020.

### 8.1.1 Stellar Variability across the Herzsprung-Russell Diagram

At the precision of Ariel photometry, Sun-like stars are variable even at intervals with little magnetic activity. The Helioseismic and Magnetic Imager (HMI) instrument aboard the Solar Dynamics Observatory (SDO) produces full disc optical continuum intensity images and Stokes parameters for the Solar photosphere at 45 s time resolution (Schou 2012). These data were used to produce high precision disc-integrated light curves of the Solar optical continuum which show a ~100 ppm variability even for a day of low magnetic activity close to the Solar Minimum (Morris et al. 2020). This variability arises from *p*-mode oscillations, stochastic variability due to the basal level of magnetic activity which is present even in the quiet Sun, and temporal fluctuations in the convective granulation.

While Ariel's photometric precision will not approach that of the HMI solar continuum data, Ariel's out of transit data will produce multi-band photometric data capable of detecting the analogous variability for bright exoplanet host stars. Coadding VISphot, FGS1, and FGS2 data in 30 minute bins, Ariel can achieve optical intensity light curves with ~35 ppm standard deviation for typical bright targets (e.g. HD 209458 b). Numerical simulations suggest that the length scales of stellar photospheric granules scale inversely with stellar surface gravity. The granule size, and the granulation-induced stellar variability is thus expected to be largest for evolved stars (Trampedach et al. 2013). Ariel's diverse sample of planetary systems will allow this to be directly assessed for stars across the Hertzsprung-Russell (HR) diagram.

### 8.1.2 Stellar Activity and Stellar Flares

Stellar activity is a key issue which must be dealt with in the extraction of Ariel's spectra of planetary atmospheres. The strategies for this are discussed in Section 7.6, here we outline the benefits of Ariel observations to scientific issues connected with stellar activity. From this perspective, transit spectroscopy is a powerful new technique for probing stellar surface inhomogeneities, i.e. effectively spatially resolving other stars (Haswell 2010; Barnes et al. 2016; Dravins et al. 2018; Kowalski et al. 2019).

Currently the structural details of inhomogeneities on the surfaces of stars other than the Sun are challenging to constrain; three methods currently prevail: Doppler Imaging producing temperature maps and therefore providing spatially resolved maps of the large-scale brightness structure, showing the presence of dark, spotted areas or bright spots (e.g. Kuerster et al. 1994; Rice & Strassmeier 1998); Zeeman Doppler Imaging making use of polarization measurements and providing orientation and strength of magnetic fields in spatial resolution (see Section 7.3.2). And third, classical spot mapping using light modulation during stellar rotation has also often been used to map large spot structures (Notsu et al. 2013). Where direct measurements of stellar surface features have been made, for example for magnetically active M dwarfs, there are generally significant differences in comparison to the Sun (e.g. Reiners 2012; Barnes et al. 2015; Barnes et al. 2017a). Solar analogue stars at moderate activity levels may show spotted areas unlike any active region ever observed on the Sun (Notsu et al. 2013), or spot features close to the stellar poles also defy solar analogy (Hussain et al. 2002). Cegla et al. (2019) point out that photometry is a key diagnostic, so ground-based high-resolution spectroscopy simultaneous with Ariel photometry is likely to produce data able to challenge and drive significant advances in modelling stellar surface magnetoconvection. Transiting planets may occult some surface activity features such as active regions or large spots, thus not only producing bumps in the photometric transit light curve but also changing the part of the stellar spectrum that is modified by the planetary atmosphere; such observations may pinpoint location, size and type of magnetic features on the star and contribute to some partial mapping (Section 7.6.1).

Stellar observations in the optical and U-band range have uncovered classes of stars with conspicuous flare brightening for minutes to tens of minutes that are the analogues of solar white-light flares. They are impulsive events powered by magnetic reconnection. They are inherently energetic and rapid phenomena, and much of their optical and IR emission is generally likely to be due to neutral hydrogen line emission. Ariel IR data can reveal temporally varying Paschen, Brackett and Pfundt line and continuum emission. While some of Ariel's core survey data will be suffering from stellar-flare contamination, these same data will provide a rich source for the study of flares and super-flares. Although the detectability of such flares is highly biased toward small, late-type stars because of luminosity and temperature contrast issues, the most energetic super-flares are known from active solar analogues (e.g. Schaefer et al. 2002; Notsu et al. 2013), but the mentioned contrast issue makes flare statistics more challenging on this type of star; therefore, a survey of flares using the very stable photometric recordings on Ariel will be valuable for this purpose, building on previous studies (e.g. Kowalski et al. 2013).

Ariel's high precision optical photometry and IR spectrophotometry at 1s cadence for ~1000 Ariel core survey target stars opens up new parameter space for discovery. This rapid cadence and the precision of Ariel offers opportunities to simultaneously sample the short timescale behaviour across the optical, near-IR and thermal IR (c.f. Szabó et al. 2020). Of special interest will be statistics relating flare amplitudes to the overall stellar activity level (as determined by chromospheric, coronal, or spot indicators – active stars expected to produce the largest flares), to rotation rate (the prime determinant of magnetic activity), and spectral type (solar analogues expected to produce the highest-amplitude flares). The flare occurrence rate distribution in energy itself will be relevant to pinpoint the total energy released in flares. The high cadence will also allow us to record flares with secondary peaks, suggesting some complexity in flaring active regions, or look for optical precursor flares. Some relatively simple modelling will allow us to estimate the involved flare footpoint area impulsively heated to some 10000 K at photospheric/chromospheric levels (Hawley et al. 1995).

Of course, many of Ariel's core survey targets may also exhibit magnetic star-planet interactions (c.f. Section 2.2.2.5). These may be characterised at unprecedented precision through Ariel photometric data.

### 8.1.3 Transit Timing Data

The Ariel core survey will cover hundreds of transits, including multiple transits for the more challenging targets. While the core goals of Ariel centre on the spectroscopy of the exoplanet atmospheres, these data will also routinely furnish exquisite timing data (Figure 8-1). Timing data is invaluable for a range of scientific issues related to planetary system dynamics; a brief overview is given in Haswell 2010.

Simulations of transits of 55 Cnc e, a planet with radius 2 $R_\oplus$ orbiting a bright star (V=5.95) establish that transit timing precision of around 12 s can be achieved for bright targets with FGS1 and FGS2 data combined (Borsato et al. 2020.) For K2-24 b, a Neptune-sized planet transiting a fainter (V=11, K=9.18) star, Ariel photometry is photon-limited, and produces a transit timing error of 35 s. This improves on Kepler/K2 by a factor of two (Borsato et al. 2020; Petigura et al. 2018).

Transit timing data is valuable for constraining the dynamics of planetary systems, and can produce robust and precise values for a range of derived quantities. For multi-planet systems, the transit timing variations (TTVs) due to the mutual gravitational interactions between planets can be used to determine planet masses (Agol et al. 2005). Importantly, for targets where Ariel's survey will make multiple transit observations, Ariel can independently detect small TTV signals. With about 10 transit observations Ariel data alone will determine planet masses with a precision better than 20% in the Earth-Neptune regime. Analysis of the K2-24 system is an illustration of Ariel's capabilities in this area. With only two transit times from Ariel we get an improvement on the uncertainty of the mass of about 22% and 31% for planets b and c, respectively (Borsato et al. 2020). The K2-24 case study demonstrates that Ariel will inevitably extend TTV coverage, break degeneracies in orbital solutions, and improve the uncertainties on planet masses and orbital parameters. These fundamental measurements place strong constraints on formation, migration, and planetary evolution processes (see Section 2.2.4.1).

### 8.1.4 Transit Duration Variations

The transit duration can be measured by Ariel with similar precision to the transit mid-time. Transit duration variations (TDVs) arise from orbital precession changing the impact parameter of the transit. Orbital precession may arise either as a result of misaligned stellar spin and orbital angular momentum axes or due to mutual gravitational interactions between orbiting bodies. Boley et al. (2020) discuss the latter, finding orbital precession can cause ~10 minute changes in transit duration in compact multi-planet systems. TDV measurements can be coupled with TTVs to derive three-dimensional solutions for the planets' orbits. In some cases, these data might reveal the presence of additional, non-transiting planets.

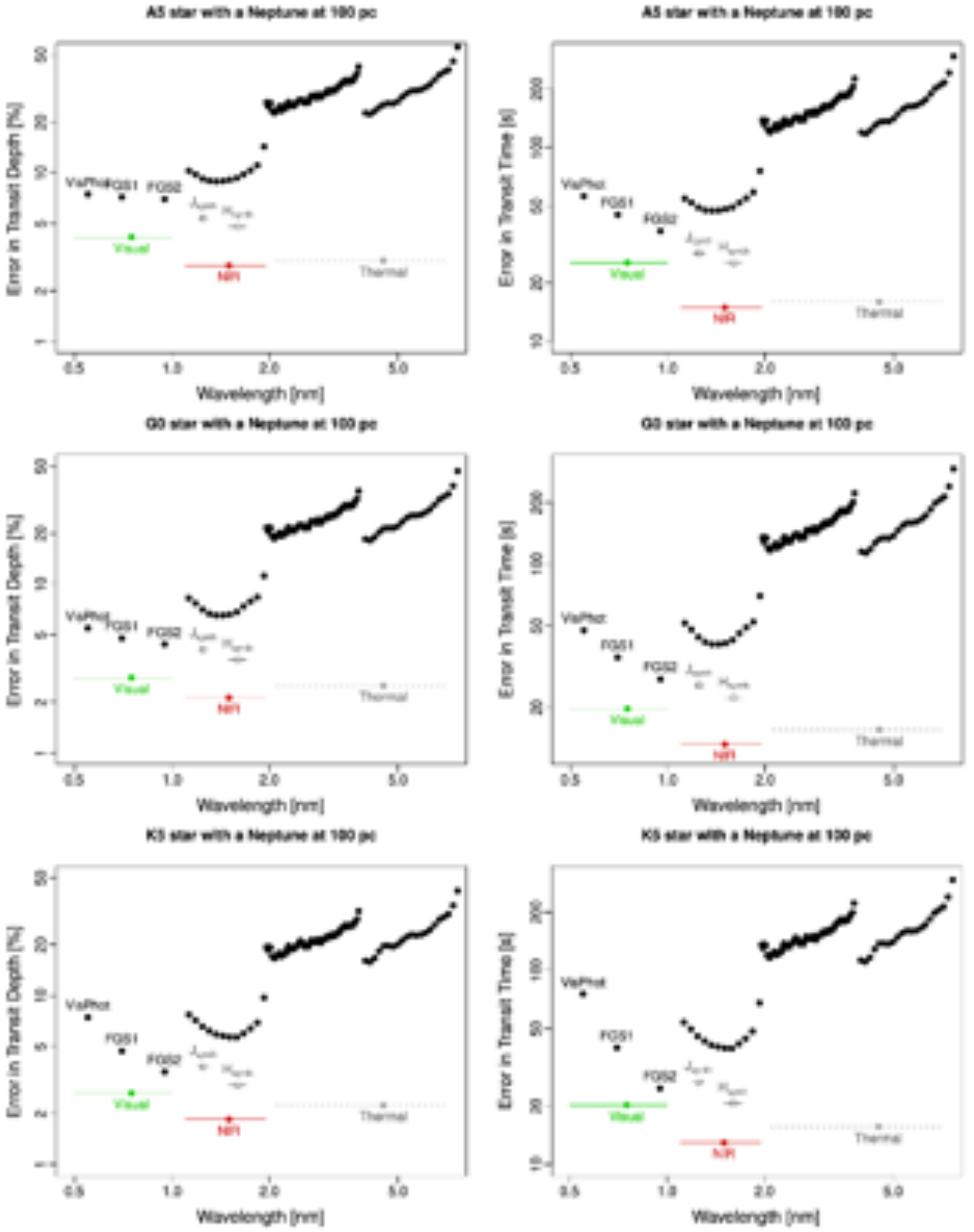

*Figure 8-1: Transit depth and timing precision attainable for Ariel core survey observations of three fiducial systems at a distance of 100 pc. By binning the data, timing precisions of 10-20 s are possible in the NIR, with independent measurements of similar precision possible in the visual and thermal IR. From Szabó et al. 2020.*

Observations of the Rossiter-McLaughlin effect in the RVs of close-orbiting giant planets have established that stellar spin-orbit misalignments are common (Fabrycky & Winn 2009; Muñoz & Perets 2018 and references therein). Asteroseismic inferences from TESS and PLATO will provide measurements of misalignments for many Ariel targets (Campante 2015; Campante et al. 2018). An oblate rotating star with a misaligned planetary orbit leads to transit duration variations. The second moment of the gravitational potential of the non-spherical star leads to precession of the ascending node of the orbit (Kaula 1966). This effect has been observed in Kepler-13 (Szabó et al. 2011; 2012) with recent TESS data confirming the predicted increase in transit duration and impact parameter. By the time Ariel observations are made, the planet will be transiting with $b \approx 0$; by the year 2100, Kepler-13Ab will no longer transit (Szabó et al. 2020).

### 8.1.5 Exomoons, exorings, exocomets and exotrojans

Exomoons can cause both TTVs and TDVs through their gravitational interactions with the transiting planet they orbit (Kipping 2009). Ariel's core survey will yield precise measurements of both. Exomoons can also be directly detected through their transits. The combination of these effects observed with Ariel's photometric and timing precision will produce data both sensitive to exomoons and powerfully able to determine the mass and size of any detected (Szabó et al. 2006; Simon et al. 2007). The best exomoon candidate to date, orbiting Kepler-1625 b, is still debated in the literature (Teachey & Kipping 2018; Heller et al. 2019; Kreidberg et al. 2019a; Teachey et al. 2020). As Figure 8-1 illustrates, Ariel' suite of instruments are all capable of producing sensitive, high cadence data. Systematic effects in their light curves should be largely uncorrelated. This will help enormously in promptly resolving debates like that on Kepler-1625 b's putative moon.

Heller (2018) gives a comprehensive review of the methods which can reveal both exomoons and exorings. In the Solar System, rings are ubiquitous around giant planets, and have even been detected around an asteroid (Braga-Ribas et al. 2014). Rings are unlikely to produce dynamical effects on transit timing, but can be directly detected through the flux dips due to their own transit across the host star (Tusnski & Valio 2011). The light curves of transiting ring systems might also exhibit forward-scattering (e.g. Barnes & Fortney 2004).

Trojan bodies, also known as co-orbital bodies, orbit at the L4 or L5 Lagrange points. L4 and L5 lie respectively at points 60° ahead of and behind a planet. None have yet been found in exoplanetary systems (Leleu et al. 2019), though they are present in the Solar System and predicted by theories of planetary system formation (Horner et al. 2020). Ford & Holman (2007) discussed identifying exotrojans through TTVs. Ariel's photometric precision also offers prospects for direct detection through transits of ~1000 km sized exotrojans, but within the core survey, the L4 and L5 points will only be covered for objects in Tier 4 (phase curves). The detection of Trojans accompanying a hot or warm giant planet would imply the planet had arrived in its present location via migration through a dissipative disc rather than via tidal circularisation of a highly eccentric orbit.

### 8.1.6 Characterisation of transiting dust clouds in CDEs

The catastrophic endpoint of hot rocky exoplanets' evolution is exemplified by Kepler-1520b (Rappaport et al. 2012), the prototype catastrophically disintegrating exoplanet (CDE). These objects are low mass bodies heated to ~2100 K at the sub-stellar point, where a thermal wind carries the vaporised rocky surface free of the planet's gravity. The Kepler/K2 mission identified a handful of CDEs through transits of dust clouds condensing out of the metal rich vapour ablated from the planet. The Kepler CDEs are all too faint to be targets in the Ariel core survey, but nearby analogues orbiting bright stars would be attractive targets. Jones et al. (2020) present a marginal detection in TESS data of a CDE orbiting a V=8 star. Croll et al. (2014) discuss the dust grain size, composition and wavelength-dependent behaviour of scattering in the context of CDEs. Bochinski et al. (2015) detected the colour dependence of the transits of Kepler-1520 b, concluding that the grain sizes are in the range 0.25-1μm. Ariel's multiband light curves of transits of any bright CDEs within the core survey will provide sensitive spatially-resolved dust grain size and composition measurements (Garai 2020).

## 8.2 Synergies with Other Facilities & the Uniqueness of Ariel

### 8.2.1 Overview

Ariel is the first space mission fully dedicated to the study of exoplanet atmospheres and has the unique ability to probe the atmosphere of warm and hot exoplanets with a statistical approach (~1000 exoplanet atmospheres observed during Tier 1). Other facilities in this decade will also contribute to the characterization of exoplanet atmospheres, Table 8-1 gives an overview of these facilities. The complementarities and synergies with Ariel are discussed in detail in Sections 8.2.1-8.2.6.

*Table 8-1: Various facilities with the capability of probing the atmospheres of exoplanets.*

| Facilities, telescope size | Wavelength coverage | Operation | Instruments | Main goal | Complementaries with Ariel |
|---|---|---|---|---|---|
| HST, 2.4 m | 0.1-1.7 µm | To be in operation for several more years (TBD) | WFC3, STIS | Multi-purpose observatory. | Unique facility to observe in the UV |
| JWST, 6.5 m | 0.6-28 µm | From 2022 up to at least 2027, potentially 2032 | 0.6-5 µm : NIRISS, NIRCAM, NIRSPEC, 5-28 µm: MIRI. Not observing the same field simultaneously | Multi-purpose observatory. | Large collecting area as required to study temperate rocky exoplanets. Limited number of exoplanets to be studied. |
| Roman Space Telescope (WFIRST), 2.4 m | 0.5-2 µm | Mid 2020s' for at least 5 years, potentially 10 years | Wide Field Camera, Coronagraph instrument | Dark Energy. Exoplanet detection by microlensing. Exoplanet study by direct imaging. | Potential to characterize exoplanet in transit |
| TESS, 4 cameras with an entrance pupil of 0.1 m | 0.6-1 µm | In operation since 2018 and up to at least 2021 | Broad-band photometer | Nearly all sky search for exoplanets orbiting bright stars. | Exoplanets Phase curves in the optical |
| CHEOPS, 0.32 m | 0.4-1.1 µm | In operation since 2020 and up to at least 2023 | Broad-band photometer | Precise radius determination of known exoplanets | Exoplanets Phase curves in the optical |
| PLATO, 24+2 cameras entrance pupil diameter of 0.12 m | 0.4-1.0 µm | From 2027 to at least 2030, potentially 2034 | Broad-band photometer | Exoplanet detection, with emphasis on temperate Earth mass planet around Sunlike stars. | Exoplanets Phase curves in the optical |
| Gaia, 2 telescopes of $1.46 \times 0.5$ m$^2$ | 0.3-1.0 µm | In operation since 2013 and up to at least 2022 | 3 instruments: Astrometric, Photometric, Radial velocity spectrometer | Complete census of large planets to 200–500 pc | |
| Athena | 0.3-10 keV | Launch scheduled early 2030s. | Wide Field Camera; X-ray Integral Field Unit | Multi-purpose observatory | X-ray irradiation from the hosts, flares. Star-Planet Interaction. |
| Current 3-10 m class ground-based telescopes | Visible – near IR | In operation | Various imagers and spectrometers | Multi-purpose observatories | Exoplanets detection and characterization by high contrast imaging and high spectral resolution observations. |
| ELTs (ELT, GMT, TMT), up to 39 m. | Visible - near and mid IR | To be in operation in mid-2020s. | Various imagers and spectrometers | Multi-purpose observatories | Exoplanets detection and characterization by high contrast imaging and high spectral resolution observations. |

### 8.2.2 Synergies between Ariel and JWST

Observations of exoplanet atmospheres with HST have been very successful. For example, UV observations have allowed to probe escape processes in the atmospheres of close-in exoplanets (e.g. Vidal Madjar et al. 2003); near-IR observations with the WFC3 camera have led to the detection of water in tens of atmospheres (e.g. Tsiaras et al. 2018). Once JWST is launched, HST will be obsolete for observations in the optical and infrared spectral range, but will remain a unique facility for UV observations.

#### *8.2.2.1 JWST and Ariel: complementary goals for exoplanet studies*

JWST and Ariel have been conceived with different science goals. JWST is an observatory which will serve many fields in astronomy; four themes have been highlighted: first light and reionisation, assembly of galaxies, birth of stars and planetary systems, planets and origin of life. The requirements for such science goals have led to the development of a large (6.5 m size) infrared telescope equipped with 4 instruments covering the 0.6-28 µm wavelength range. Based on JWST Guaranteed Time Observations (GTO) and Early Release Science (ERS) programs, we can expect that about 20% of the JWST observing time will be devoted to transiting exoplanet programs.

Ariel is a mission entirely devoted to the study of exoplanet atmospheres with a statistical approach; ~1000 transiting exoplanets will be first observed in a 'reconnaissance' survey (Tier 1); at least 500 of those will be characterized spectroscopically. The mission is optimized to observe transiting exoplanets: it features a 1 m class telescope equipped with two instruments covering *simultaneously* the 0.5-1.95 and 1.95-7.8 µm wavelength range.

JWST will observe in great detail a more limited number of exoplanets, down to temperate rocky planets. Ariel will enable the **statistical understanding** of warm and hot exoplanets. In Table 8-2, possible science portfolios for transiting exoplanet observations with the JWST are indicated, assuming a reasonable guess of 300 days of JWST observations dedicated to exoplanets (Cowan et al. 2015).

*Table 8-2: JWST Transiting Planet Observing Portfolios from Cowan et al. (2015).*

*JWST* TRANSITING PLANET OBSERVING PORTFOLIOS

| Portfolio name | Number of targets | Duration per target (days) | Total time (days) |
|---|---|---|---|
| Atmospheric structure | 150 | 2 | 300 |
| Atmospheric mapping | 25 | 12 | 300 |
| Temperate terrestrials | 3 | 100 | 300 |

NOTE.—Endmember portfolios for transiting planet science with *JWST*. Linear combinations of these are possible, and probably more scientifically productive: e.g., $70 \times 2 = 140$ days of structure, $5 \times 12 = 60$ days of mapping, and 1 terrestrial (100 days) also adds up to 300 days.

To understand the complementarity, it is first important to remember that Ariel can be more efficient than JWST to characterise exoplanets orbiting bright stars. Contrary to most fields in astronomy, the general statement: "the larger the telescope is, the smaller the time to observe an object is", does not hold for transiting exoplanets. To achieve the spectro-photometric precision needed to characterise exoplanet atmospheres (down to a few tens of ppm), the star-planet system has to be observed before, during and after the transit, adding up to an observation that should last at least twice the transit duration, independently of the telescope size. A larger telescope will bring a higher signal over noise ratio, which is not always needed to retrieve the information about exoplanet atmospheres (see Figure 8-2).

Given that JWST, contrary to Ariel, has not been optimized for transiting exoplanet observations, only a fraction of the Ariel wavelength coverage is obtained during the JWST observation of a single transit. Multiple transits observations (two to four, according to the star brightness) with different JWST instruments or instrumental configurations are needed to cover the full Ariel wavelength range. Edwards et al. (2019) have shown that ~800 exoplanets can be observed with a single transit during the Tier 1 reconnaissance survey (see Figure 3-3), making Ariel much more efficient than JWST for such a survey.

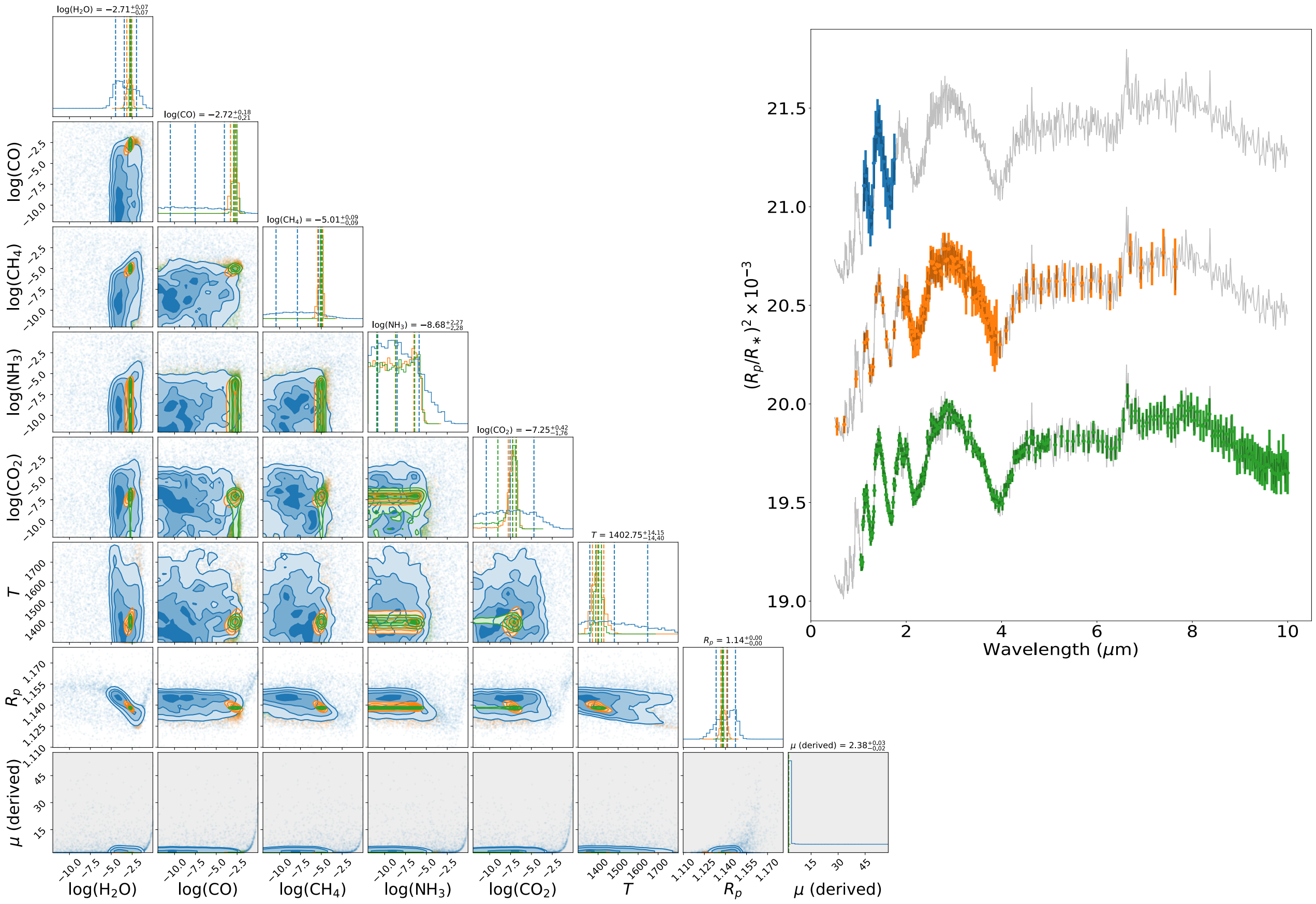

*Figure 8-2: posterior distributions of various atmospheric trace gases, temperature and top cloud pressure obtained with TauREx by retrieving the simulated spectra shown on the right, as observed by HST/WFC (blue), JWST (green) and Ariel (orange). Dashed lines in the histogram plots show the 1 sigma confidence intervals. The gases and atmospheric parameters retrieved include $H_2O$, $CO_2$, CO, $CH_4$ and $NH_3$, temperature, radius and derived mean molecular weight. There is still considerable degeneracy for retrievals from WFC3 spectra, but Ariel and JWST data are expected to be very constraining. The information content for JWST and Ariel in the case of bright sources is comparable, indicating that Ariel will be able to characterize atmospheres to a similar degree of accuracy.*

#### *8.2.2.2 Preparing Ariel with JWST*

a) Adjusting the Mission Reference Sample: the study of exoplanet atmospheres is still in its infancy and no doubt that the field will rapidly evolve with JWST observations. Such evolution will influence the Ariel MRS. The possibility to adjust the MRS, as mentioned in the Science Management Plan, is thus key to get out the best science from Ariel.

b) Providing calibration targets: some of the exoplanets observed by JWST will be used by Ariel for calibration purpose in the first months of observation and will give rapidly confidence in the Ariel observations.

c) Data reduction and retrieval techniques: an important effort has been made in Europe to develop sophisticated data reduction methods for HST data (Tsiaras et al. 2016a) or in preparation of JWST observations (Bouwman et al. 2020). The reduction of Ariel data will benefit from all the lessons learned during the reduction of JWST data.

A great effort has also been made in Europe, especially in the Ariel Mission Consortium, to develop retrieval codes (e.g. Ariel Spectral Retrieval WG report 2020). There is still room for developments and optimization of the retrieval methods and atmospheric models: JWST will provide a unique opportunity to do so before the launch of Ariel.

#### *8.2.2.3 Follow-up of Ariel observations with JWST observations.*

Given the expected start of JWST scientific observations (2022) and those of Ariel (2029) and given the expected duration of JWST operations (5+5 years), there is the possibility for JWST and Ariel to operate simultaneously for a few years. In this case, Ariel observations can be used in exploratory mode and particularly interesting targets identified can be re-observed with JWST, taking advantage of the wavelength coverage of the MIRI instrument (8-28 μm range containing dust features, $NH_3$ strong lines, Si-O feature, $CO_2$), or taking advantage of the higher spectral resolution (several thousands in resolution) provided by MIRI and NIRSPEC.

### 8.2.3 Synergies between Ariel and TESS, CHEOPS, PLATO

TESS is a NASA mission launched in April 2018 (Ricker et al. 2015): it has completed its two years near all sky survey looking for exoplanets around bright stars, with as primary goal to discover planets smaller than Neptune that transit stars bright enough to enable follow-up spectroscopic observations. CHEOPS, ESA's first S-class mission, launched in December 2019, is in its routine operations, measuring with high precision the size of planets orbiting bright stars (magV<12) already known to host planets (Benz et al. 2020). PLATO, ESA's M3 mission in the Cosmic Vision program, will survey the brightest stars over large (2250 deg²) fields for long continuous periods (years), detecting and measuring the size of thousands of exoplanets, also with long orbital periods (Rauer & Heras 2018).

The main exoplanet characterization potential of TESS, CHEOPS and PLATO is in the bulk properties of the planet. However, these missions also hold the potential to characterize exoplanet atmospheres via phase curve measurements. The photometric passbands of TESS, CHEOPS and PLATO will allow to make optical phase curves, complementary to the thermal phase curves Ariel will be able to observe. CHEOPS has reached a photometric precision of 10-17 ppm for a $mag_V$=6.6 star, using a one-hour binning (Lendl et al. 2020). TESS is currently reaching a precision of around ~85 ppm in 1 hour for a $mag_V$=9 star. PLATO will have a photometric precision of ~30 ppm in 1 hour for bright solar-like stars ($mag_V$<11). In addition, the PLATO fields will be observed for 1-3 years, cumulating a few hundreds of orbits for short period planets. Ariel will have a precision of around 50 ppm with VISPhot and FGS for stars brighter than $mag_V$ = 11. TESS, similarly to PLATO, stares at the same targets for at least 27 days and at best 1 year (without considering possible extensions of the mission) and can thus achieve exquisite phase-curves over time. CHEOPS, as a mono-target instrument like Ariel, will not be able to stare at a numerous targets for an extended period of time, but will still be able to make significant contributions for a handful of golden targets. PLATO, TESS and CHEOPS will thus provide strong constraints on the phase curve of giant planets prior to the launch of Ariel. Presently, more than 136 known giant planets would be suitable to measure Ariel phase curves. Prioritizing the phase curve targets to those with a good TESS, CHEOPS or PLATO optical phase curve would be a strong synergy, combining visible and infrared phase curves from these telescopes.

### 8.2.4 Potential synergies between Ariel and the Roman Telescope (WFIRST)

WFIRST, recently renamed as the Roman Space Telescope, is a 2.4 m telescope with two instruments: a Wide Field Camera (WFC) and a Coronagraph Instrument (CI) (https://www.nasa.gov/content/goddard/nancy-grace-roman-space-telescope). Both instruments have spectroscopic capabilities. The spectroscopic modes of the WFC cover the 1.0-1.9 μm wavelength range with a spectral resolving power of 450-850, and the 0.8-1.8 μm wavelength range with a spectral resolving power of 70-140. The Coronagraph Instrument has a spectroscopic imaging mode with a spectral resolving power R=50 in a 0.675-0.785 μm bandpass.

The main scientific goals of the observations with the Wide Field Camera are two folds: to study the dark universe (dark energy and dark matter) and to keep on the census of exoplanets by detecting exoplanets with masses down to Mars mass using the microlensing technique.

The Coronagraph Instrument is a technology demonstration instrument, it is expected to produce scientific results from high-contrast imaging and spectroscopic observations of nearby exoplanets. Thus Ariel and the Roman Telescope have distinct goals in exoplanetary science. The use of the spectroscopic mode of the Wide Field instrument to probe the atmosphere of transiting exoplanets could enable potential synergies.

### 8.2.5 Synergies between Ariel and Athena

Athena (Advanced Telescope for High Energy Astrophysics) is the second L-class mission in ESA's Cosmic Vision programme. It is devoted to the study of the hot and energetic universe and will bring significant advancements over current X-ray observatories. The mission has several scientific objectives including exoplanet science (Branduardi-Raymont et al. 2013), improving our knowledge of exoplanets and exploring how planetary atmospheres and magnetospheres interact with stellar winds and flares. Athena will be launched in 2031 with an expected lifetime of 4 years, with possible extensions. It will be equipped with a 40’×40’ camera with moderate spectral resolution (WFI) and with a high spectral resolution X-ray spectrograph (XIFU). According to current plans, it will be in orbit at the same time as Ariel for few years.

Beyond the common interests on exoplanets observed in X-rays, Athena will be of help for Ariel's observations because it will allow to obtain X-ray measurements reasonably close in time to Ariel's observations. This possibility will give us the opportunity to determine the activity of the host stars at the time of the planetary observations and in particular in which phase of the magnetic cycle they are. In fact, the stellar X-ray emission is variable even on long time scale with changes of coronal activity at least of one order of magnitude during cycles. Furthermore, in order to be able to point possible bright GRBs, one of Athena's requirements is to have a large fraction of the sky instantly accessible. This is a great opportunity, since it implies that it is likely that it will be possible to coordinate Athena and Ariel simultaneous observations of some planetary transits, giving us the unique possibility of observing, in the case of stellar flares, in real time the effects of high energy radiation on the chemistry of planetary atmospheres.

### 8.2.6 Synergies between Ariel and large ground-based Telescopes

Ground based observations are complementary to space observations by providing high contrast imaging and high resolution spectroscopy. The high contrast imaging allows to probe a class of exoplanets not accessible to Ariel: the exoplanet detected by direct imaging. The number of such exoplanets is, at the moment, limited to a few dozen and these exoplanets are mostly young giant exoplanets, but the situation will change with the advent of Extremely Large Telescopes. Ariel and the ELTs (ELT, GMT, TMT) are expected to be operational in the same time-frame (2025+), and their observations will have important synergies that, when used in combination, will allow the exploration of a wide parameter space of exoplanet atmospheres. Ariel has the fundamental advantage to cover simultaneously the whole spectral region from 0.5 to 7.9 μm. Thus, Ariel will be measuring the global planetary atmospheric parameters – temperature-pressure profile, cloud coverage and the bulk molecular abundances – while ground-based measurements will provide key information on elements to resolve the overall picture of a given planet.

Ground-based observations have many challenges and limitations: large parts of the electromagnetic spectrum are blocked due to absorption and scattering in the Earth’s atmosphere. A large fraction of the spectral range observed with Ariel is, in fact, inaccessible from the ground. In addition, the thermal background from the sky and telescope are strongly variable, making high-precision ground-based transit or eclipse spectroscopy practically impossible at wavelengths longer than 5 μm. However, the ELTs will be very valuable in specific ways. Indeed, all ELTs are expected to have both high-contrast imaging and high-dispersion spectroscopy instruments for exoplanet studies, sometimes even used in combination. High contrast imaging will probe a volume-limited population of young, non-transiting exoplanets. High-dispersion spectroscopy instruments, however, are designed to study the atmospheres of a sample of exoplanets that will overlap with that of Ariel (comprised of stars of magnitudes between 3-4 and 9-10 in the Ks band). At resolving power R>100,000, atomic and molecular bands in exoplanet spectra are resolved into hundreds to thousands of individual lines, whose signals can be combined to secure a detection. This technique has been successfully applied, for both exoplanet transmission (Snellen et al. 2010) and emission spectroscopy (Brogi et al. 2013), and will be more effective on the next-generation of extremely-large telescopes. These high-resolution measurements can also be used to measure accurate planet masses (Snellen et al. 2010), which can be combined with the bulk compositions, T-p profiles, and thermal inversions derived from Ariel data (Brogi et al. 2017). For the most favourable targets, the ELTs can even provide information on the rotation of the planet and high-altitude wind speeds using the absorption line profiles (Snellen et al. 2014). In particular, ELT exoplanet observations will provide the following complementary measurements to Ariel:

- Metallicity measurements via the detection of atomic and ionized species. Species such as Fe, Fe+, Ca, Ca+, etc, have already been detected in the atmospheres of ultra hot-Jupiter planets (Hoeijmakers et al. 2018; Casasayas-Barris et al. 2019; Yan et al. 2019).
- Detecting inhomogeneities and dynamics in the planetary atmosphere via time-resolved transmission measurements during the transit, which lead to probing the different planetary limbs subject to different wind/temperature regimes (Ehrenreich et al. 2020).

- Measuring atmospheric lines that might be crucial to understand evaporation and atmospheric evolution processes, and will help to put Ariel bulk composition measurements in context. Some of these lines are the Hα line and the Balmer series in the visible range (Yan & Henning 2018; Casasayas-Barris et al. 2019) and the He I triplet in the mid-infrared (Nortmann et al. 2018)
- Finally, although Ariel has been designed to be self-sufficient in its ability to mitigate the stellar activity induced noise in the retrieved transmission spectra, thanks to the instantaneous and broad wavelength coverage (Sarkar et al. 2018; Tinetti et al. 2018), having access to simultaneous high-resolution and high signal-to-noise spectra from ELTs can also provide a wealth of unique and detailed information about the properties of stellar active regions (Rackham et al. 2018; Boldt et al. 2020).

## 8.3 Complementary Science

The primary scientific objectives of Ariel will be addressed using the observations conducted in the Core Survey (Chapter 2). The primary scientific requirement of the mission is to perform this survey. All observations in the Core Survey require particular orbital phases of the target exoplanets. This means that each of these observations are time constrained. As explained in Section 7.2.1.3, building a schedule comprised entirely of time constrained observations inevitably creates intervals of unused time. These gaps will be used for calibration observations and other necessary activities, but unused time will still remain. It is anticipated to be between 5 and 10% of all the available science time, or hundreds of hours of observing time per year. This time will be used for conducting 'Complementary Science' observations. These observations cannot be time constrained; they will be inserted into the gaps in the schedule when the target is visible. Complementary Science observations must not drive any technical or operational requirements of the mission. The majority of the unscheduled time will be in relatively short time blocks, so that most Complementary Science observation windows are expected to have durations between 0.5 and 4 hours.

The selection of Ariel Complementary Science observations will be through an open ESA-run call. Below we list some examples of possible Complementary Science observations, but these are by no means exhaustive or prescriptive.

*Ultra-cool dwarfs* – Ultra-cool dwarf is a general term used to refer to low-mass stars, brown dwarfs and planetary-mass objects with temperatures below 2700 K or spectral types later than M6. They represent a rather abundant population in the solar neighbourhood. The warmest and more massive ultra-cool dwarfs can be brighter than magnitude 14.5 in the optical (depending on distance and age). According to theory, the thermodynamical condition of condensation is well met in their atmospheres (Tsuji et al. 1996), and some show photometric variability ascribed to the presence of patchy clouds of "dusty particles" (e.g. Crossfield et al. 2014). Their rotation periods and variability time scales are typically of hours (e.g. Zapatero Osorio et al. 2006). Ariel observations of these bright ultra-cool dwarfs can provide valuable information on the simultaneous optical and infrared variability with a clear insight on the onset of condensation in the upper atmospheres. These atmospheres may resemble those of the ultra-hot exoplanets orbiting stars, which Ariel will likely observe.

*Protoplanetary discs* – Exoplanets form from protoplanetary discs surrounding virtually every young star. Spectroscopy of these discs reveal the presence and size distribution of the dust component, which usually dominates the continuum opacity. In the Ariel spectral wavelength range there are several interesting and accessible dust components. Of particular interest is the carbonaceous component of the discs, thus far difficult to study. Nano-diamonds have features around 3.5 micron (Jones et al. 2004) and have been tentatively detected in one protoplanetary disc (Van Kerckhoven et al. 2002). Polycylic Aromatic Hydrocarbons (PAHs) also have features in the Ariel wavelength range and are broadly found in protoplanetary discs. The ratio of different PAH features reveals their ionization state and total size. In addition to these carbonaceous components, water ice is an important constituent of protoplanetary discs; it traces the temperature and thermal history of the disc (Min et al. 2016). The shape of the 3 μm water ice feature is a valuable tracer of the ice structure (crystalline or amorphous), revealing the thermal history of the disc material.

*Non-transiting Catastrophically Disintegrating Exoplanets* – Kepler-1520b is the prototype catastrophically disintegrating exoplanet (CDE), an object discovered through variable depth transits in the Kepler light curve. These transits are attributed to a dust cloud condensing from a metal-rich, thermal wind escaping from an ablating, approximately Mercury mass planet (Rappaport et al. 2012). CDEs are hot, rocky planets, with insolation ~1000 $S_\oplus$. This heats the sub-stellar point to ~2100 K, creating lava lakes, mineral atmospheres and winds. The Kepler CDEs are too faint for Ariel to observe, but we anticipate the discovery of nearby, bright analogues before Ariel launch (Haswell et al. 2019; Jones et al. 2020). These bright, nearby CDE systems will offer unique opportunities to probe the mineral content of rocky planets outside our Solar System. Croll et al. (2014) explored the effect of composition on the Mie scattering cross-section of the dust, and van Lieshout et al. (2014) used analytic modelling to show that the tails of Kepler-1520b and KOI-2700b are consistent with corundum ($Al_2O_3$) or iron-rich silicates such as fayalite ($Fe_2SiO_4$). Transiting CDEs may become part of the Ariel core survey, but non-transiting examples are statistically likely to be even closer and brighter (Garai 2020).

<u>Interacting binary stars</u> – The 1 Hz sampling of Ariel data allows a variety of investigations of accretion-powered binary star systems. The low space density of some classes of interacting binary stars means there may be few objects which are bright enough to ensure Ariel can acquire and guide on them. However, there are ~50 high mass X-ray binaries and ~110 cataclysmic variables brighter than magnitude 14 in the optical (Liu et al. 2006; Ritter & Kolb 2015). Ariel observations of these can probe their simultaneous optical and IR time variability, allowing e.g. eclipse mapping and echo-mapping of their accretion flows (e.g. Pratt et al. 1999a,b; Khangale et al. 2020; O'Brien et al. 2002).

<u>Stellar activity and flares</u> – See Section 8.1.2.

<u>Solar System Objects</u> – Although there is no requirement for Ariel to be able to track a moving target, it is possible that the telescope will be able to observe slowly moving objects. This opens the possibility of observing a variety of Solar System targets, particularly those in the outer Solar System.

Among the possible options, occultations by Solar System objects (in particular KB0, and possibly Uranus/Neptune) searching for atmospheres (e.g. Pluto's): with the Gaia catalogue, a prediction of occultation is now routinely achieved and a few targets could be accessible during Ariel lifetime ; the observing strategy would be identical to exoplanets, since pointing on the star and observing during occultation would be similar.

<u>Impact of YSO variability on the planetary cradle</u> – During their early, pre-main sequence phase young stellar objects (YSOs) are highly variable, although their variability becomes less and less violent with age. Optical variability has been a long-known defining characteristics of young stars, but the growing amount of multiepoch infrared data show that these systems exhibit also flux changes at infrared wavelengths. While the optical variability is caused either by hot or cold stellar spots, or by extinction changes along the line of sight, mid-infrared variability partly reflects changes in the thermal emission of the disc as a response to its varying irradiation by the central star. While photometric monitoring of young stars at optical and infrared wavelengths are available for relatively larger samples (e.g. Cody et al. 2014), multi-epoch infrared spectroscopic surveys are still very rare. A small but conspicious group of pre-main sequence stars is the class of FU Orionis- (FUor) or EX Lupi- (EXor) type young eruptive stars (Audard et al. 2014). Their luminosity bursts have a very strong impact on the circumstellar disc. During the large outburst of EX Lup, the prototypical EXor, our Spitzer mid-infrared spectra revealed observable changes in the mineralogy of solids and in the molecular abundances. Rab et al. (2017) and Molyarova et al. (2018) suggest FUor and EXor-type outbursts may play an important role in setting the chemical initial conditions for exoplanet formation.

<u>Extreme debris discs</u> – In the region around main-sequence stars not only planets, but also numerous smaller planetesimals, asteroids and comets can be present. The destruction of these minor bodies through mutual collisions produces fresh dust, that is removed from the system by stellar radiation pressure and/or stellar wind. While in most cases the dust forms a debris disc at several ten au from the star, similarly to our Kuiper-belt, in a small sample the dust is located within a few au, and emits at mid-infrared wavelengths, like the zodiacal dust in the solar system. Recently a few highly interesting objects, typically young (30-150 Myr), Sun-like stars were discovered, where the mid-infrared emission is stronger than average, and is highly variable (Meng et al. 2012, 2015). The current explanation for these extreme debris discs is that we witness the aftermath of a very recent collision between two larger planetesimals, that produced an immense amount of fresh dust in the inner part of the system system (Meng et al. 2015). Such dynamic events might have occurred in the early solar system (Wyatt & Jackson 2016) thus these extreme debris discs provide a glimpse into our past.

<u>Other Possibilities</u> – There are ~450 active galaxies with optical apparent magnitude brighter than 14 (Veron-Cetty & Veron 2010), all of which are possible targets. The rapid variability of a variety of classes of stars, including pulsating stars, T-Tauri stars and massive stars could be explored. Dust, ices, PAHs and other molecules might be studied in molecular clouds and star forming regions, though observations of such extended objects cannot impose any requirements on the mission.

# 9 Management

## 9.1 Overview of Mission Implementation

Following the selection of Ariel as the M4 mission by the SPC in 2018, ESA has conducted the Definition Phase (phase B1) study, encompassing all aspects of the mission:

- The spacecraft: two parallel industrial contracts were run in parallel, and were concluded by the Mission Adoption Review (MAR);
- The payload complement (PLM and warm payload units) and the IOSDC: under the responsibility of the Ariel Mission Consortium (AMC), this study was concluded by the payload and IOSDC System Requirements Review (pSRR);
- The technology readiness: all technologies that were identified to be below the TRL6 threshold are going through dedicated Technology Development Activities (TDA), with the objective to demonstrate sufficient maturity to proceed with the mission.
- The mission: all other aspects of the Ariel mission, including the consolidation of the launcher and ground segment aspects, were also the subject of dedicated studies.

Ariel, for both the space and ground segments, will be implemented and operated by ESA and the Ariel Mission Consortium (AMC) in collaboration. ESA will have overall responsibility for all aspects of the Ariel mission (see Chapter 5). While it is common that for ESA science directorate missions scientific instruments are provided by instrument consortia, it is noteworthy that for Ariel the AMC is responsible for the provision of the entire payload module (PLM) as well as payload related warm electronics (see Chapter 4) that will go into the service module (SVM). In addition, the AMC will also be part of the science ground segment (see Chapter 6) in the form of the Instrument Operations and Science Data Centre (IOSDC).

### 9.1.1 Project Management and Responsibilities

After mission adoption, the ESA Project Team led by the Project Manager (PM), will have overall responsibility to ESA for implementing the Ariel mission. The PM interfaces with the Project Scientist (PS) for all science-related matters. The ESA project team will build the data package for the Invitation To Tender (ITT) that will be issued just after the mission adoption for the selection of the industrial prime contractor.

Over the course of the implementation phase, the ESA project team will conduct reviews:

- for the space segment (Spacecraft and Payload): System Requirement Review (SRR), Preliminary Design Review (PDR), a Critical Design Review (CDR) and finally a Flight Acceptance Review (FAR);
- for the Science Ground Segment (SOC and IOSDC): SGS System Requirements Review (SGS-SRR) and the SGS Preliminary Design Review (SGS-PDR);
- for the overall ground segment (MOC, SOC and IOSDC): GS Design Review (GSDR), GS Implementation Review (GS-IR), GS Readiness Review (GSRR);
- for the Launcher: Preliminary Mission Analysis Review (PMAR), Final Mission Analysis Review (FMAR);
- for the overall Mission: Mission PDR, Mission CDR and Mission Commissioning Result Review (MCRR).

The responsibility for the Ariel mission will transfer from the PM to the Mission Manager, following the successful commissioning of the satellite and its scientific payload. The task of the PS will continue throughout the operations and post-operations phases.

### 9.1.2 Management of Operations

ESA will be responsible for the launch, checkout and operation of the Ariel spacecraft. ESA will establish a mission operations centre (MOC), and a science operations centre (SOC).

Definition of the MOC will commence at the beginning of the definition phase, under the responsibility of a Ground Segment Manager located at ESOC who will report to the Project Manager. The responsibility for the MOC will transfer from the Ground Segment Manager to the Ariel Spacecraft Operations Manager (SOM, located at ESOC), following the successful commissioning of the satellite and scientific payload.

Definition of the SOC will commence at the same point in time, and will be under the responsibility of a SOC Development Manager in the Operations Development Division at ESAC. The SOC Development Manager will work closely with the PS, but will formally report to the Project Manager.

Management of the Science Ground Segment will be transferred from the Operations Development Division to the Operations Division following successful commissioning of the satellite and scientific payload. As described in Chapter 6, the mission scientific operations will be under the overall control of the Ariel SOC in close collaboration with the IOSDC provided by the Ariel Mission Consortium.

## 9.2 Ariel Mission Consortium Management Plan Summary

The Ariel Mission Consortium (AMC) has distributed effort according to the skills and resources in the participating institutes and reflecting the areas of national interest in the science of Ariel. This also ensures that a broad spectrum of state-of-the-art knowledge and appropriate technical design expertise is brought to bear on the design, construction and test of the instrument. An overview of the planned division of work between the participating countries is shown in Figure 4-1, and the organisation of the AMC management bodies is shown in Figure 9-1. The detailed plans for the management of the large multi-national consortium are defined in the Ariel Mission Consortium Management Plan and its subservient reference documents (lower level plans).

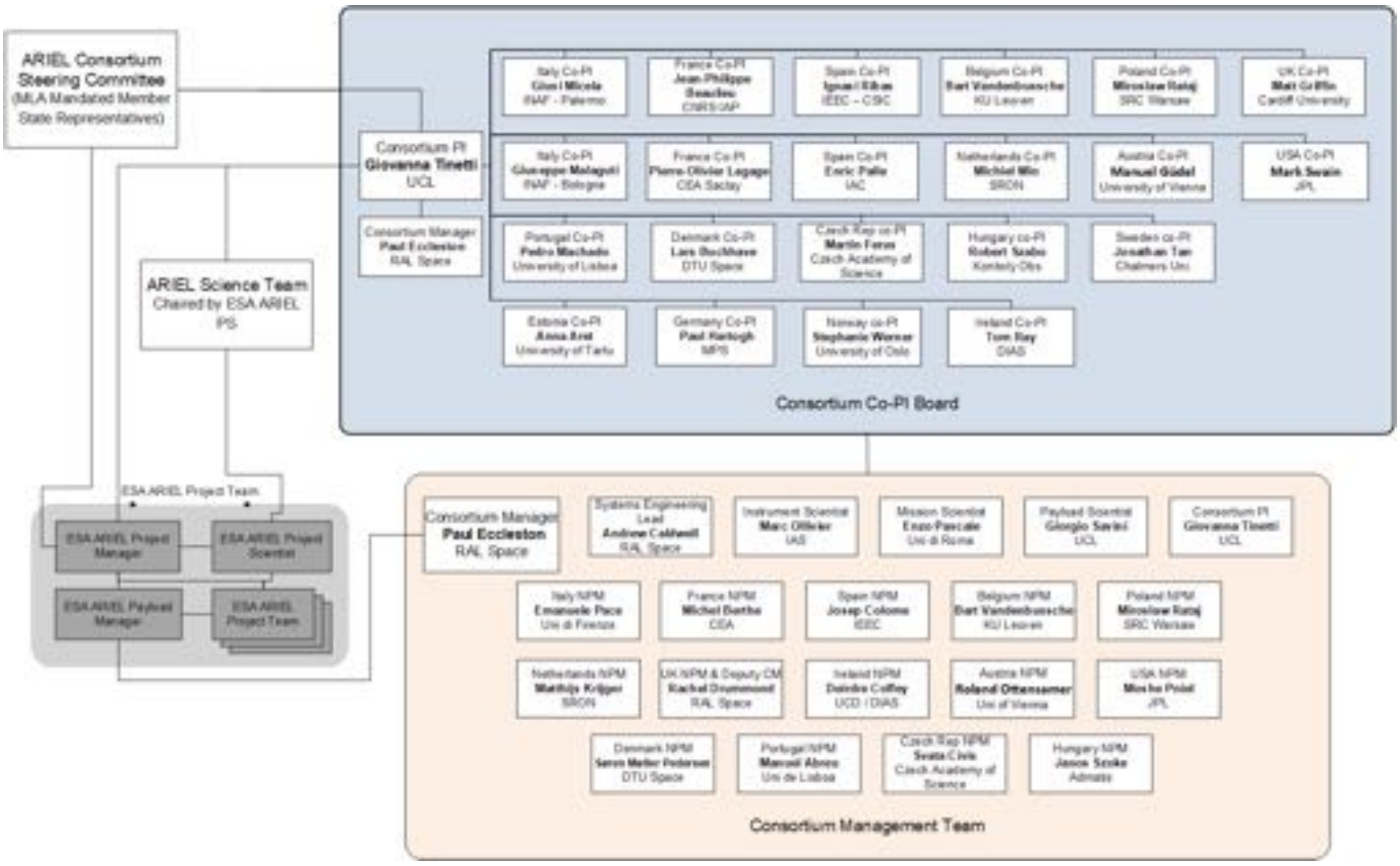

*Figure 9-1: Consortium Co-PIs Board and Consortium Management Team along with key ESA interfaces*

The responsibility for the work of the consortium rests with the Consortium PI and the Co-PIs, assisted by the Consortium Project Manager. In order to manage the work of the distributed consortium seven defined teams will assist the consortium management to coordinate the work. These are the Consortium Co-PIs Board, Consortium Management Team (CMT), the Consortium Science Coordination Group (CSCG) consisting of the leads of the Science Working Groups, the Consortium Engineering Leadership Team (CELT), the Consortium Instrument Scientist Team (CIST), the Instrument Operations and Science Data Centre Management Team and the PA Management Team. The key directional bodies (strategic from Co-PIs Board, functional from the Consortium Management Team) of the consortium and the foreseen interactions with ESA are illustrated in Figure 9-1.

## 9.3 Procurement philosophy

ESA will have overall responsibility for the following items:

- Overall design of the mission
- Provision of the spacecraft service module through an industrial contract, including integration of the SVM and PLM and environmental testing
- Procurement of the detector chains via a cooperation with NASA (CASE, see below) for the FGS instrument and via a direct procurement in the US for the AIRS instrument
- Procurement of the PLM V-grooves, in support of the AMC
- Provision of the launch (Arianespace)
- Mission and science operations (ESOC and ESAC)

The payload consortium team, funded by the national agencies and institutes, will have responsibility for developing and providing the following items:

- The cold payload module (PLM) including:
    - The Ariel telescope assembly including an optical bench, all mirrors, baffles and structures plus the M2M refocusing mechanism;
    - The AIRS and FGS instruments, common optics and the on-board calibration system;
    - The telescope assembly support structure (isolating bipods);
    - The thermal control system (instrument radiators, thermal straps, active cooler heat exchanger, and integration of the ESA provided V-grooves into the PLM);
- The warm payload units located in the SVM
    - The Instrument Control Unit (ICU) including the AIRS Warm Front End Electronics, Data Processing Unit and Power distribution;
    - The Active Cooling System (ACS) including the Ne JT cooler with associated Cooler Drive Electronics (including JT-pipes interconnecting the warm units and the PLM);
    - The Fine Guidance System Control Unit (FCU) electronics providing detector control, readout and centroiding to provide the input necessary for the AOCS system.
    - The Telescope Control Unit (TCU) for control of the telescope mechanism, calibration source and thermal control;
- The cryo-harnesses that connect between the warm SVM upper panel and the cold PLM units.
- Elements of science ground segment:
    - Provision of the Instrument Operations and Science Data Centre (IOSDC; note: instrument operations will be performed by MOC);
    - Payload-specific software, modules and processing blocks for processing up to Level 3 data;
    - Long-term mission planning tool;
    - Calibration and instrument-monitoring;
    - Support to the MOC and SOC (contingencies, expert advice, payload monitoring).

The following payload activities are planned to be managed through the ESA PRODEX program:

- Austrian contribution to FGS software
- Belgium contribution to the telescope assembly
- Czech contribution to the common optics
- Danish contribution to the PLM bipods
- Hungarian contribution to the PLM radiator and GSE
- Polish contribution of the FGS and FCU
- Portuguese contribution to the telescope assembly baffle, and OGSE

The baseline for the Ariel Mission Consortium is an European mission, with a few key technologies procured from US suppliers (detector chains for AIRS and FGS) where this is determined to be optimal for mission science return. The possibility of other external collaboration remains open, but only within the strict programmatic limits of the existing schedule.

The CASE (Contribution to Ariel Spectroscopy of Exoplanets) proposal, submitted under the 2016 Astrophysics call for Missions of Opportunity for US participation in Ariel, has been approved by NASA. This proposal is to supply the Sensor Chip Assemblies (SCAs) and Cold Front End Electronics (CFEE) for both of the detectors within the FGS. The US contribution to the AMC will be matter of an ESA / NASA agreement.

## 9.4 Baseline Schedule

The key dates for the Ariel project top level schedule are shown in Table below. The planned launch date for M4 is in 2029. With the kick-off of the Implementation Phase straight after adoption in Dec 2020 (initiation of payload B2 activities, and release of S/C ITT), this leaves ~8.5 years for the complete development, manufacturing, assembly, integration, testing and launch campaign.

The critical path lies in the payload activities, in particular all AIT activities leading to the payload CDR (including the PLM SM/EM test campaign and the S/C SM test campaign), and the subsequent release of the PL FM production and the S/C FM AIV activities (with some long lead items released ahead of the payload CDR, at the instruments / units CDRs). The payload FM is expected to be delivered to the spacecraft in Q4 2026.

*Table 9-1: Key dates (indicative) from Ariel top-level schedule.*

| Milestone | Schedule |
|---|---|
| Mission Adoption | Nov 2020 |
| Phase B2/C/D/E1 industrial KO | Sept 2021 |
| System SRR | Q1/2022 |
| System PDR | mid-2023 |
| System CDR | Q2/2025 |
| FAR | Q1/2029 |
| Launch (L) | 2029 |
| LEOP | L + 48 hrs |
| Start of Satellite and Payload commissioning phase | L + few days |
| Start of performance verification and science demonstration phase | L + < 3 months |
| Start of nominal in-orbit science operations phase | L + < 6 months |
| End of nominal in-orbit operations phase | L + 4 years |
| End of extended in-orbit operation phase (goal) | L + 6 years |
| S/c disposal (in parallel with post-operations phase) | L + 4/6 years + 3 months |
| End of post-operations phase | L + 6/8 years |

## 9.5 Science Management Overview

In this section we outline the principles currently foreseen for the science management responsibilities for the Ariel mission, as presented to the community in the ‘Ariel: Science, Mission & Community 2020’ conference as well as to the ESA Advisory Bodies (AWG and SSEWG) in January 2020. These are expected to form the basis of the ‘Science Management Plan’ (SMP) which is being written.

The SMP will be the top-level science management document for the mission, and will require approval by the Science Programme Committee (SPC) in the context of mission adoption. It is emphasized that this subsection is subject to change pending approval of the SMP.

### 9.5.1 Ariel Project Scientist and Science Team

After the mission adoption, ESA will appoint an Ariel Science Team (AST), chaired by an ESA Project Scientist (PS). The task of the AST will be detailed in the Science Management Plan. The AST will comprise the instrument consortium PI and scientists from both the AMC and the wider scientific community.

The AST will advise ESA on all aspects of the mission potentially affecting its scientific performance. It will assist the PS in maximising the overall scientific return of the mission within the established boundary conditions and will act as a focus for the interests of the scientific community in Ariel. The AST will monitor the implementation of the scientific requirements and priorities of the mission, and the preparation and execution of the scientific operations. The AST will be responsible to define the scientific priorities for the selection of the Ariel targets.

### 9.5.2 Target Lists

Ariel is a survey mission with the primary objective to observe a large diverse sample of known, transiting exoplanets as described in Chapters 2 and 3 and documented in the DRM. The targets to be observed in the Ariel core survey from the Mission Reference Sample (MRS – aka the 'core sample'). The choice of targets and associated observations shall meet the science requirements that govern the core sample (see Chapter 3), and will be made by the AST.

The MRS is derived from a list of all identified potential targets observable by Ariel, the Mission Candidate Sample (MCS), according to the adopted strategy. This is a two-stage process whereby the scientific priority of the targets in the MCS and the associated observations are defined by the AST in accordance with maximising the science return of the mission; followed by a scheduling exercise with an algorithm which accounts for this prioritisation. Inputs will be solicited from the wider community (e.g. through whitepapers, meetings, and other mechanisms) that will be kept informed about the status of the target lists, as will the ESA Advisory Bodies whose feedback will be solicited.

The MCS will be continuously updated, with the goal of having a consolidated version six months before launch. It is expected that over time many MRSs will be produced as the MCS and scientific priorities evolve. By the time of launch the MRS will have evolved into a description of the nominal pre-launch Ariel target list documented in the DRM. The initial MRS will have to be ready for the start of the nominal science operations. The MRS can – and should – be updated on a monthly basis throughout the in-flight mission given sound scientific reasons, such as e.g. including previously unknown particularly interesting targets and actual achieved in-flight mission performance.

### 9.5.3 Data Rights and Proprietary Periods

Ariel is an ESA mission with AMC lead that will characterise a large diverse sample of known transiting exoplanets. It is recognized that Ariel data and science will be of interest to a large general community of exoplanetary scientists. The intention is to provide high quality data products in a timely manner and to have a continuous dialogue with the wider community, thus ultimately maximising the science that can be achieved by the mission.

The last step (see Section 6.2) before commencing the 'routine science phase' (RSP) of the mission will be the 'science demonstration phase' (SDP). Data from the approximately 1-month duration of the SDP will be released soon after observing, and a public workshop will be organized in connection to the data release.

In the RSP the Ariel observing strategy (see Section 3.1) will be to observe the targets in the various 'Tiers' (see Section 3.1.2) in 6-month cycles. The data will be pipeline processed to different levels of data products (see Sections 6.3.3-6.3.4) labelled 'raw photometric or spectral science frames' (Level 1), 'calibrated photometric or spectral science exposures' (Level 1.5), 'photometric or spectral target (star + exoplanet) light-curves' (Level 2), which is the highest level output of the systematic pipeline processing performed at the SOC. Finally 'individual exoplanet planet spectra' (Level 3) will be produced by the AMC. All data products will be ingested into the Ariel Science Archive through which they will be made available to the scientific community, after certain proprietary period have elapsed as will be described in the SMP.

Preliminarily, data products up to and including Level 2 will be publicly released as follows:

- SDP (cycle 0): data public immediately after quality control has been completed
- Tier 1: data public immediately after quality control has been completed
- Tiers 2 & 3: data public 6 months after quality control has been completed
- Tier 4: data public 12 months after quality control has been completed

For the SDP (cycle 0) this includes all tiers. For subsequent cycles the above periods are applicable. The processing and quality control period is assumed to be 2 months for the SDP and RSP cycles 1 and 2. Early in the mission this delay may need to be longer, if the calibration needs and data processing/correction or systematics have not been fully mastered. The goal is to reduce this period to 1 month from cycle 3 onwards based on the assumption that as the mission progresses and a more complete understanding of the instrument characteristics, calibration needs and data processing/correction or systematics have been gained.

Datasets up to and including Level 2 products for Tier 2 and Tier 3 targets will be released after each semester where the required signal-to-noise (SNR) and spectral resolution for a particular target requiring multiple observations has been achieved.

Release of associated ancillary data, pipeline input files, and similar will be performed in connection with the above-mentioned releases.

In addition Level 3 products will be produced and publicly released. These are not pipeline products but involve manual processing, and require a good understanding of Ariel data, their instrumental signatures, and possible peculiarities that will only be obtained once in orbit. It is in the interest of the Ariel Mission Consortium to have Level 3 data processed, published and released soon and frequently. It is foreseen to provide Level 3 products at least on an annual basis, and sooner/more frequently when the knowledge to produce them is firmly in hand.

# 10 Communications and Outreach

A mission to characterize the atmospheres of diverse worlds beyond our Solar System provides an excellent opportunity to harness curiosity, imagination and philosophical reflection in people of all ages. The discovery of more than 4300 exoplanets in the last 25 years is possibly one of the most exciting developments of modern astronomy, awarded with the Nobel Prize in 2019. These discoveries resonate with the public, who have already shown very strong curiosity and interest in the exploration of the disparate worlds in our own Solar System, as well as the exotic worlds that are starting to be revealed around other stars. Closer to home, our own planet's atmosphere is so familiar that it is often taken for granted, but gives us constant reminders of the profound influence it has on our environment and very existence. The atmosphere provides the air we breathe; its presence is felt through the winds that drive it, and the colours of our sky – whether blue during the day or orange/red at sunset and sunrise – show the direct fingerprints of the Earth's atmosphere on the light arriving from the Sun.

Ariel's communication and outreach activities reach out to a wide audience that includes the public at large, as well as focused groups such as school students, amateur astronomers, politicians and artists. Details of the communications, outreach and education activity plan will evolve with the mission, with an outline for responsibilities of the different Ariel stakeholders detailed in the science management plan. In this chapter we describe some of the initiatives that are already being implemented or under consideration, with activities intended to be Member State-wide where possible.

ESA is responsible for planning and coordinating press release activities relating to the Ariel mission, with the support of the Ariel Mission Consortium and guidance from the Ariel Science Team. An open approach is adopted by the Ariel Mission Consortium teams – except where commercial confidentiality is at stake – they welcome media professionals into their institutions, laboratories and workshops during all phases of the mission. Bespoke programmes and documentaries that cover scientific and engineering aspects of Ariel have been and will be produced in the next phases of the mission (https://arielmission.space). These activities build on the strong record that many Ariel Ariel Mission Consortium scientists have with TV and radio interviews, including with leading European broadcasters such as the BBC and Euronews.

The Ariel Mission Consortium plans to execute an active programme to brief and inform policy makers at national and European levels on scientific and technological developments of Ariel. One-on-one meetings, seminars for politicians and stakeholders, exhibitions at venues such as the European Parliament, and public events that involve political figures as keynote speakers are being organized to keep policy makers abreast of developments which, although in the "blue-skies" field of space exploration, create indirect economic benefits to society.

Social media channels, including Twitter, Facebook and Instagram, are providing up-to-the minute engagement with all aspects of the mission (@ArielTelescope), with the opportunity for different instruments or exoplanet targets to have their own, individual feeds and "voices". This complements the dissemination of mission news and updates through existing ESA channels, allowing interested parties to stay informed about mission progress and performance, as well as to engage with the stories and experiences of the people behind the mission. Online outlets such as YouTube and blogs recorded in multiple languages are being used to post updates and interviews with Ariel scientists and engineers to illustrate the wide range of tasks that technical professionals engage in over the course of a space mission (https://arielmission.space).

The Ariel Mission Consortium plans to continue to work closely with space outreach and educational networks, including Europlanet, Space Awareness, Ecsite, European Schoolnet, Hands-on Universe, the Galileo Teacher Training Programme, ESA's own European Space Education Resource Offices, as well as national and more local networks, to achieve a wide engagement across Member States.

The excitement generated by the Ariel mission and its discoveries provides a topical platform around which to develop educational materials, helping to raise the profile of both the Ariel mission and ESA in general within schools. Many of the core concepts behind Ariel's science objectives and technological challenges are covered in school syllabi at different levels e.g. conditions for life, or power generation. Curriculum-linked resources are developed covering a broad range of scientific and technical topics, such as the study of exoplanets and their formation, exoplanet discovery techniques, spectroscopic signatures of atoms and molecules, and orbital mechanics. When Ariel data will become available, materials will be disseminated to school students Europe-wide through educational partner networks, and will be supported by Continued Professional Development courses to enable school teachers to use the discoveries of Ariel to enliven classroom lessons and activities.

Ariel offers many enquiry-based learning opportunities for school students. Involving students in state-of-the-art research from an early age eliminates the idea that science is 'only for scientists' and empowers young people to explore STEM subjects. In July 2018 ESA commissioned to PricewaterhouseCoopers France and its partners an impact assessment study to analyse the broader socio-economic impact of ESA's Scientific Programme. The study has been completed in May 2019. We show in Figure 10-1 the outcome of the study PwC performed by contacting students attending 100 universities in Europe. At the time of the study Ariel had just been selected, but was already known to 31% of STEM students and 21% of non-STEM students consulted.

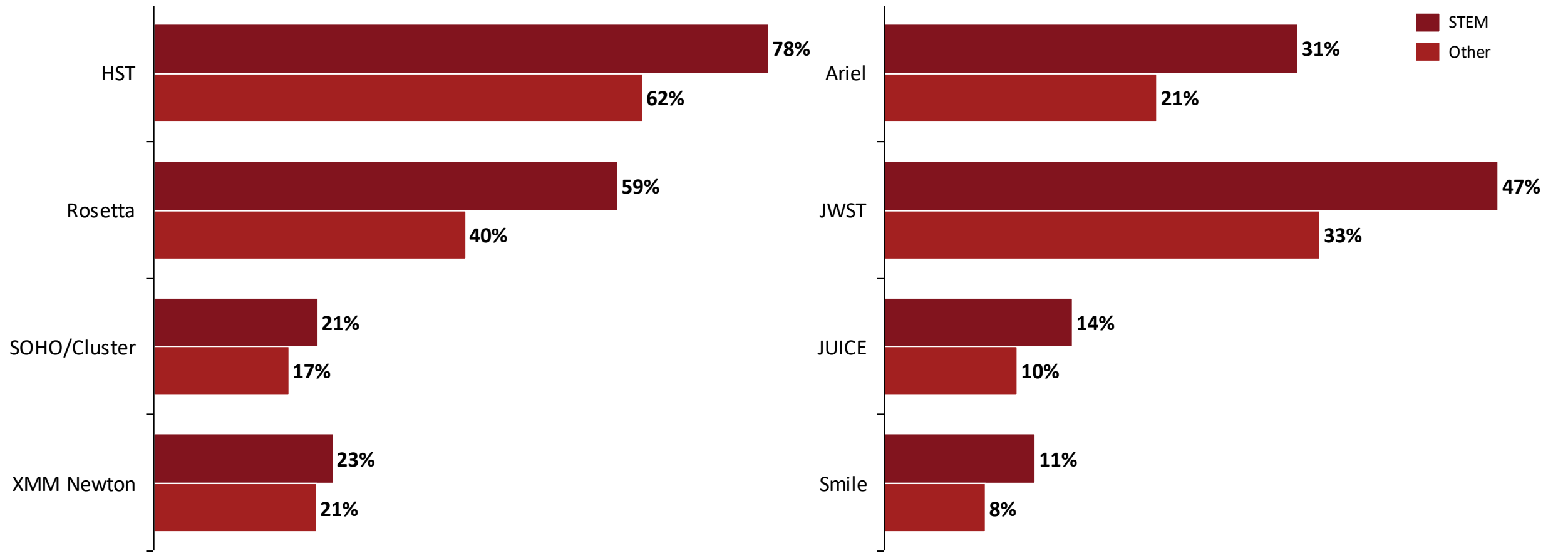

*Figure 10-1: Percentage of students who have heard about a certain space mission extracted from the PwC's report commissioned by ESA. PwC contacted 100 universities in Europe.*

Current PhD students and postdocs will be at the forefront of the scientific exploitation of Ariel after its launch. To prepare the future science community, Ariel summer schools will be organised every year including theory and hands-on exercises with experts of an Ariel-related theme for a group of about 20 PhD students, postdocs and young scientists. The first edition was organised in 2019 in Biarritz, ARES (Ariel Retrieval of Exoplanet School) and funded by CNES. The data for 9 planets observed with the Hubble-WFC3 camera have been analysed by the students and tutors and submitted to the Astronomical Journal. Each of the paper is ending by a paragraph showing Ariel and JWST simulated observations for the analysed planets. The lead authors are the students responsible for each analysis (e.g. Pluriel et al. 2020; Skaf et al. 2020; Edwards et al. 2020b).

Pilot programmes, already underway with secondary school groups in the UK (e.g. ORBYTS programme), have started to produce original, publishable scientific research associated with the characterisation of exoplanetary atmospheres, under the supervision of young PhD students and Post-Docs. Initial projects have focused on compiling data points for molecules of interest for modelling the atmospheres of cool stars and exoplanets (e.g. acetylene, titanium oxide, methane). The Ariel Mission Consortium supports the roll-out of these projects through ESA member states.

Schools will be actively engaged in the selection of the Ariel core sample. A competition will be run across ESA member states to choose a School's Target Exoplanet. Supporting material detailing potential Ariel candidate targets will be developed to enable students to make a scientifically-informed vote. Students will be able to follow observations of the chosen planet via a dedicated website, and participate in the analysis and interpretation of the data.

The public is also invited to participate in the science preparation and exploitation of the Ariel mission through access to Ariel data sets for analysis and interpretation. In 2019 Ariel, has launched a global competition series to find innovative solutions for the interpretation and analysis of exoplanet data. The first Ariel Data Challenge[3] invited professional and amateur data scientists around the world to use Machine Learning (ML) to remove noise from exoplanet observations caused by star-spots and by instrumentation. The Ariel ML contest has been selected as a Discovery Challenge by the European Conference on Machine Learning and

[3] https://arielmission.space/data-challenges/

Principles and Practice of Knowledge Discovery in Databases (ECMLPKDD). Over 100 international teams participated to the challenge. A second Ariel Data Challenge that focuses on the retrieval of spectra from simulations of cloudy and cloud-free super-Earth and hot-Jupiter data was also launched in April at the UK-Exom meeting in London. A further data analysis challenge to create pipelines for faster, more effective processing of the raw data gathered by the mission has been launched in June at the EWASS conference in Lyon.

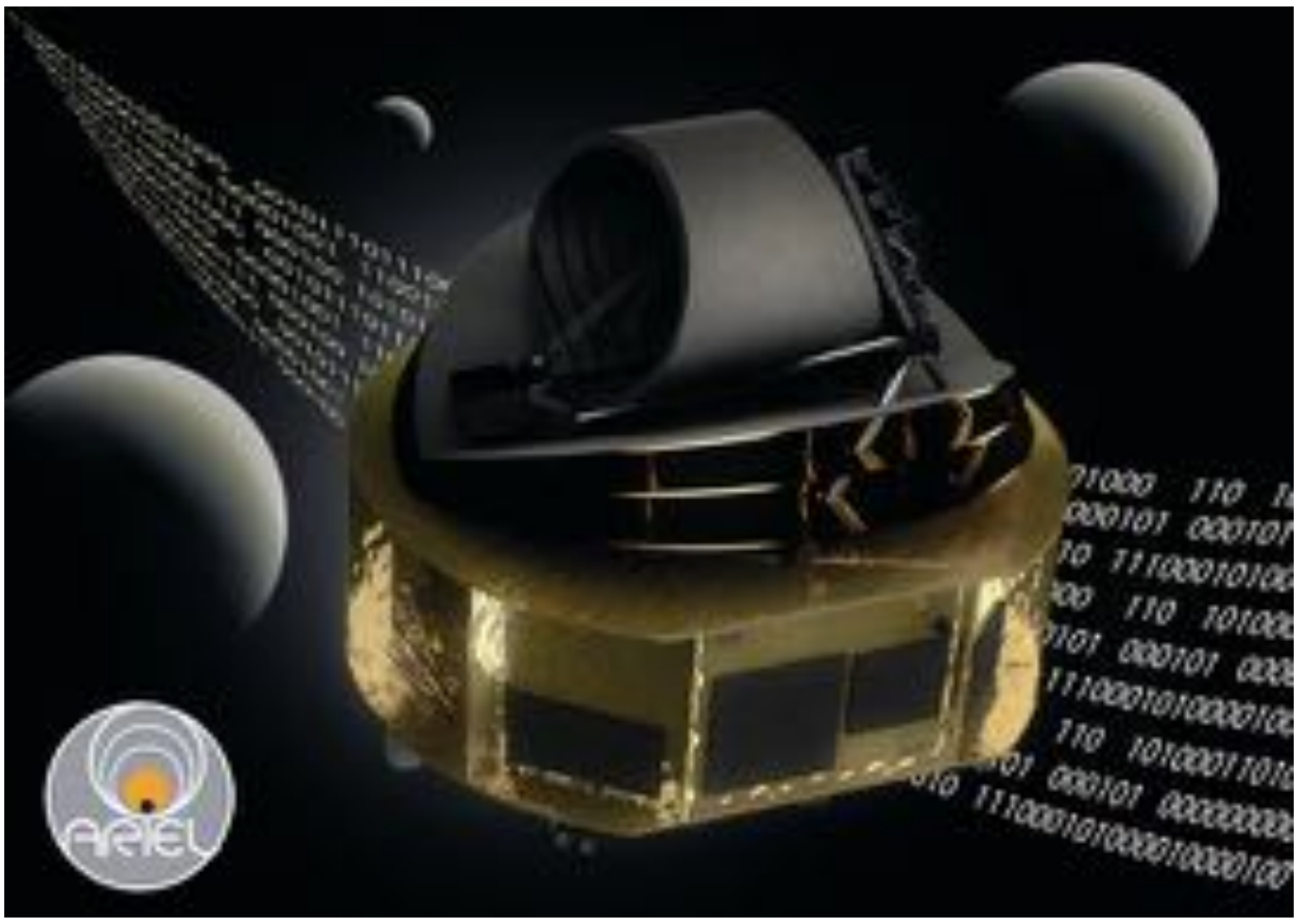

*Figure 10-2: The Ariel Data Challenges, https://arielmission.space/data-challenges/*

Outcomes from all three Ariel Data Challenges have been discussed at the ECMLPKDD in Würzburg 16-20 September 2019 and at the EPSC-DPS Joint Meeting 2019, which took place in Geneva during the same week. The winners were awarded at ECMLPKDD and Ariel Open Conference in ESTEC in January 2020.

Amateur astronomers play a crucial role, both in cascading the outreach efforts of professional scientists by providing a link with the broader general public, and by providing valuable scientific input. Some robotic telescope networks (LCOGT website; Telescope Live) are available to amateur astronomers and school groups. Since most of the targets of Ariel are relatively bright stars, target preparation and follow-up observations are feasible, profitable and exciting to both the highly experienced and relatively novice amateur astronomers and schools.

Ariel scientists have already started to work with the amateur astronomer community, citizen astronomers and schools encouraging them to undertake a programme of observations to support the Ariel ephemeris refinement. The ExoClock programme (https://www.exoclock.space, Kokori et al. 2020) aims to facilitate a coordinated programme of ground-based observations to maximise the efficiency of the Ariel mission. The programme also aims to stimulate engagement with citizen astronomers, allowing them to contribute to an upcoming ESA mission. The ExoClock initiative has the explicit rule that all those who upload data for a planetary system will be included on any subsequent publications. Edwards et al (2020a) published the refined ephemerides of eight exoplanets using a fully robotic telescope network, observations from citizen astronomers and data from TESS; a significant portion of the work has been completed by high school students via the ORBYTS programme.

The fascinating and exotic new worlds that will be revealed by Ariel will need visual support to capture the imagination of the public. Ariel scientists will work together with ESA to produce images, animations, and 3-D simulations suitable for a wide range of online and broadcast media formats. Artistic and musical collaborations will be fostered to spread the impact of the mission through artworks, compositions, writing and performance inspired by the Ariel mission and its findings. This continues and expands the tradition of the "Space Art" movement that was initiated in Europe a century ago (most notably by L. Rudaux, (IAAA website, 2013)), and the multiple artistic and musical tributes to Rosetta (Rosetta Art Tribute website, 2016) Cultural and visual arts programmes will be developed at school level also, facilitating cross-curriculum discussion and interpretation of the scientific, historical and philosophical contexts of Ariel.

# 11 References

All ESA documents and Consortium reports and Technical Notes that are referred to in the text are available on request. These include:

Cosmic Vision, ESA BR-247, 2005: http://www.esa.int/esapub/br/br247/br247.pdf
AIRS Instrument Detailed Design Description, ARIEL-CEA-INST-DD-003, Issue 4.0, Feb 2020
Ariel Calibration OGSE Design Definition, ARIEL-OXF-PL-DD-003, Issue 2.0, Feb 2020
Ariel Data Processing Pipeline Design Description Document, ARIEL-CRDF-GS-DD-001, Issue 4.0, Feb 2020
Ariel IOSDC Management Plan, ARIEL-INAF-GS-PL-002, Issue 3.8, April 2020
Ariel Mission Planning, Ariel-IEEC-GS-TN-001, 2020
Ariel Science Operations Concept Document, ESA-ARIEL-ESAC-SOC-OD-001, 2020
Ariel Payload Design Description, ARIEL-RAL-PL-DD-001, Issue 5.0, July 2020
Ariel Payload Budgets Report, ARIEL-RAL-PL-RP-003, Issue 2.0, Feb 2020
Sky Calibrators for ARIEL, ARIEL-INAF-SCI-TN-003, Issue 0.1, 2017
The parameter space of planetary systems explored by ARIEL, ARIEL-INAF-SCI-TN-002, Issue 0.1, 2017

Ariel Phase B1, Experimental Astronomy Special Issue manuscripts:

Barnes, J.R. et al. *Exoplanet mass estimation for a sample of targets for the Ariel mission;*

Barstow, J. et al. *A retrieval challenge exercise for the Ariel mission;*

Brucalassi, A. et al. *Determination of stellar parameters for Ariel targets: a comparison analysis between different spectroscopic methods;*

Caines, H. et al. *Simulation of Ephemeris Maintenance of Transiting Exoplanets'*

Changeat et al. *Disentangling Atmospheric Compositions of K2-18 b with Next Generation Facilities;*

Charnay, B. et al. *A survey of exoplanet phase curves with Ariel;*

Chioetto, P. et al. *Qualification of the thermal stabilization, polishing and coating procedures for the aluminum telescope mirrors of the Ariel mission;*

Danielski, C. et al. *The homogeneous characterisation of Ariel host stars;*

Demangeon, O. et al. *Need, Scale and Feasibility of an Ariel radial velocity campaign;*

Encrenaz, T. et al. *Observability of temperate exoplanets with Ariel;*

Focardi, M. et al. *The Ariel Instrument Control Unit its role within the Payload and B1 Phase design;*

Ferus M. et al. *Ariel – a window to the origin of life on early Earth?*

Garai, Z. et al. *Grazing, non-transiting disintegrating exoplanets observed with the planned Ariel space observatory A case study using Kepler-1520b;*

Garcia Perez, A. et al. *Thermoelastic evaluation of the Payload Module of the Ariel mission;*

Guilluy, G. et al. *On The Synergy Between Ariel and Ground-Based High-Resolution Spectroscopy;*

Haswell, C. A. *Extended Use of the Ariel Core Survey Data;*

Helled, R. et al. *Ariel Planetary Interiors White Paper;*

Ito, Y. et al. *Detectability of mineral atmospheres with Ariel;*

Kiss, C. et al. *Ancillary science with Ariel: Feasibility and scientific potential of young star observations;*

Kokori A. et al. *ExoClock Project: An open platform for monitoring the ephemerides of Ariel targets with contributions from the public;*

Morales, J.C. et al. *Ariel scheduling using Artificial Intelligence;*

Morello, G. et al. *The Ariel 0.6 - 7.8 μm stellar limb-darkening coefficients;*

Morgante, G. et al. *The thermal architecture of the ESA Ariel payload at the end of Phase B1;*

Moses, J.I. et al. *Chemical variation with altitude and longitude on exo-Neptunes: Predictions for Ariel phase-curve observations;*

Pearson C. et al. *The Ariel Ground Segment and Instrument Operations Science Data Centre;*

Seli, B. et al. *Stellar flares with Ariel;*

Szabó, G. et al. *High-precision photometry with Ariel;*

Turrini, D. et al. *Exploring the link between star and planetary formation with Ariel;*

Wolkenberg, P. et al. *Effect of clouds on emission spectra for Super Venus within Ariel;*

Most references are also available on ADS at:
https://ui.adsabs.harvard.edu/public-libraries/ZdDby7vCRW6XtW9dl1V6pg

Abel, M. et al. 2011, JPCA 115, 6805
Adibekyan, V. Z. et al. 2012, A&A 545, A32
Adibekyan, V. Z. et al. 2013, A&A 560, A51
Adibekyan, V. Z. et al. 2015, MNRAS 450, 1900
Agol, E. et al. 2005, MNRAS 359, 567
Agúndez, M. et al. 2014, A&A 564, A73
Airapetian, V. S. et al. 2016, ApJL 817, L24
Al-Refaie, A. F. et al. 2020, ApJ, arXiv:1912.07759
Allart, R. et al. 2018, Sci 362, 1384
Allen-Sutter, H. et al. 2020, PSJ 1, 39
ALMA Partnership, Vlahakis et al. 2015, ApJL 808, L4
Alonso, R. et al. 2004, ApJL 613, L153
Amiaux, J. et al. 2020, SPIE 2020, Paper 11443-63
Amyay, B. et al. 2018, JChPh 148, 169902
Andreasen, D. T. et al. 2017, A&A 600, A69
Andrews, S. M. et al. 2018, ApJL 869, L41
Angelo, I. et al. 2017, AJ 154, 232
Apai, D. et al. 2013, ApJ 768, 121
Arcangeli, J. et al. 2019, A&A 625, A136
Armstrong, D. J. et al. 2016, NatAs 1, 0004
Armstrong, D. J. et al. 2020, Natur 583, 39
Artigau, É. et al. 2009, ApJ 701, 1534
Audard, M. et al. 2014, prpl.conf 387
Auvergne, M. et al. 2009, A&A 506, 411
Azzam, A. A. A. et al. 2016, MNRAS 460, 4063
Babánková, D. et al. 2006, PQE 30, 75
Bagnasco, G. et al. 2007, SPIE 6692, 66920M
Bakos, G. et al. 2004, PASP 116, 266
Balona, L. A. 2015, MNRAS 447, 2714
Baraffe, I. et al. 2008, A&A 482, 315
Baraffe, I. et al. 2015, A&A 577, A42
Baranov, Y. I. et al. 2004, JMoSp 228, 432
Barber, R. J. et al. 2014, MNRAS 437, 1828
Barclay, T. et al. 2018, ApJS 239, 2
Barman, T. S. et al. 2015, ApJ 804, 61
Barnes, J. R. et al. 2015, ApJ 812, 42
Barnes, J. R. et al. 2017, MNRAS 471, 811
Barnes, J. R. et al. 2016, MNRAS 462, 1012
Barnes, J. W. et al. 2004, ApJ 616, 1193
Barstow, J. K. et al. 2015, ExA 40, 545
Barstow, J. K. et al. 2017, ApJ 834, 50
Barstow, J. K. et al. 2020, MNRAS 493, 4884
Batalha, N. M. 2014, PNAS 111, 12647
Batalha, N. E. et al. 2019, ApJL 885, L25
Batista, V. 2018, Handbook of Exoplanets, 120
Beatty, T. G. et al. 2020, arXiv:2006.10292
Benatti, S. et al. 2020, A&A 639, A50
Benneke, B. et al. 2019, ApJL 887, L14
Benz, W. et al. 2020, ExA, arXiv:2009.11633
Berdyugina, S. V. 2005, LRSP 2, 8
Bernath, P. F. 2020, JQSRT 240, 106687
Bianchi, E. et al. 2019, ECS 3, 2659
Biersteker, J. B. et al. 2019, MNRAS 485, 4454
Boccaletti, A. et al. 2020, arXiv:2003.05714
Bochinski, J. J. et al. 2015, ApJL 800, L21
Bogdan, T. J. et al. 1988, ApJ 327, 451
Bohn, A. J. et al. 2020, ApJL 898, L16
Boldt, S. et al. 2020, A&A 635, A123
Boley, A. C. et al. 2020, AJ 159, 207
Boro Saikia, S. et al. 2018, A&A 620, L11
Boro Saikia, S. et al. 2020, A&A 635, A178
Borucki, W. J. et al. 2009, Sci 325, 709
Borucki, W. J. 2016, RPPh 79, 036901
Borucki, W. J. et al. 2010, Sci 327, 977
Bosman, A. D. et al. 2019, A&A 632, L11
Bouchy, F. et al. 2001, A&A 374, 733
Bourgalais, J. et al. 2020, ApJ 895, 77
Bourrier, V. et al. 2013, A&A 557, A124
Bourrier, V. et al. 2020, MNRAS 493, 559
Braga-Ribas, F. et al. 2014, Natur 508, 72
Branduardi-Raymont, G. et al. 2013, arXiv:1306.2332
Brogi, M. et al. 2013, ApJ 767, 27
Brogi, M. et al. 2012, A&A 545, L5
Brogi, M. et al. 2017, ApJL 839, L2
Brogi, M. et al. 2012, Natur 486, 502
Brown, T. M. 2001, ApJ 553, 1006
Buchhave, L. A. et al. 2014, Natur 509, 593
Burrows, A. et al. 1999, ApJ 512, 843
Campante, T. L. et al. 2015, ESS 47, 503.03
Campante, T. L. 2015, IAUGA 29, 2250891
Campante, T. L. et al. 2018, ASSP 49,
Carmichael, T. W. et al. 2020, AAS 122.06
Carron, J. et al. 2017, SPIE 10378, 1037805
Casasayas-Barris, N. et al. 2019, A&A 628, A9
Casasayas-Barris, N. et al. 2019, yCat J/A+A/628/A9
Casewell, S. L. et al. 2018, MNRAS 481, 5216
Casewell, S. L. et al. 2020, MNRAS 497, 3571
Cassan, A. et al. 2012, Natur 481, 167
Cavalié, T. et al. 2017, Icar 291, 1
Cegla, H. M. et al. 2019, ApJ 879, 55
Changeat, Q., Al-Refaie, A. 2020, ApJ 898, 155
Changeat, Q. et al. 2019, ApJ 886, 39
Changeat, Q. et al. 2020a, AJ 160, 80
Changeat, Q. et al. 2020b, ApJ 896, 107
Changeat, Q. et al. 2020c, ApJ submitted.
Changeat, Q. et al. 2020d, ExA, arXiv:2003.01486
Changeat, Q. et al. 2020e, ApJ, arXiv:2010.01310
Charbonneau, D. et al. 2005, ApJ 626, 523

Charnay, B. et al. 2015, ApJL 813, L1
Charnay, B. et al. 2018, ApJ 854, 172
Chassefière, E. 1996, JGR 101, 26039
Chazelas, B. et al. 2012, SPIE 8444, 84440E
Chen, J. et al. 2017, ApJ 834, 17
Cho, J. Y.-K. et al. 2015, MNRAS 454, 3423
Chubb, K. L. et al. 2020a, MNRAS 493, 1531
Chubb, K. L. et al. 2020b, A&A, arXiv:2009.00687
Civiš, S. et al. 2004, CPL 386, 169
Civiš, S. et al. 2008, JPCA 112, 7162
Civiš, M. et al. 2016a, PCCP 18, 27317
Civiš, S. et al. 2016b, NatSR 6, 23199
Civiš, S. et al. 2017, NatAs 1, 721
Cloutier, R. et al. 2018, AJ 156, 82
Cloutier, R. & Menou K. 2020, AJ 159, 211
Cloutier, R. et al. 2020, AJ 160, 3
Cody, A. M. et al. 2014, AJ 147, 82
Cohen, O. et al. 2011a, ApJ 733, 67
Cohen, O. et al. 2011b, ApJ 738, 166
Cohen, O. et al. 2017, ApJ 834, 14
Coles, P. A. et al. 2019, MNRAS 490, 4638
Collier Cameron, A. 2001, LNP 573, 183
Cowan, N. B. et al. 2007, MNRAS 379, 641
Cowan, N. B. et al. 2009, ApJ 700, 915
Cowan, N. B. et al. 2013, MNRAS 434, 2465
Cowan, N. B. et al. 2015, PASP 127, 311
Cowan, N. B. et al. 2018, Handbook of Exoplanets, 147
Cranmer, S. R. et al. 2011, AAS 218, 205.03
Cridland, A. J. et al. 2019, A&A 632, A63
Croll, B. et al. 2014, ApJ 786, 100
Crossfield, I. J. M. et al. 2010, ApJ 723, 1436
Crossfield, I. et al. 2014, Natur 505, 654
Cubillos, P. et al. 2016, ascl.soft ascl:1608.004
Danielski, C. et al. 2014, ApJ 785, 35
Davenport, J. R. A. et al. 2012, ApJ 748, 58
Davis, T. A. et al. 2009, MNRAS 396, 1012
Dorn C. et al. 2018, MNRAS 484, 712
de Wit, J. et al. 2012, A&A 548, A128
de Wit, J. et al. 2016, Natur 537, 69
de Wit, J. et al. 2018, NatAs 2, 214
Debras, F. et al. 2019, ApJ 872, 100
Deming, D. et al. 2005, Natur 434, 740
Deming, D. et al. 2013, ApJ 774, 95
Demory, B.-O. et al. 2013, ApJL 776, L25
Demory, B.-O. et al. 2016, Natur 532, 207
Desch, S. J. et al. 2018, ApJS 238, 11
Donati, J.-F. et al. 2008, MNRAS 390, 545
Doyle, L. R. et al. 2011, Sci 333, 1602
Doyle, A. E. et al. 2019, Sci 366, 356
Dragomir, D. et al. 2019, ApJL 875, L7
Dransfield, G. et al. 2020, MNRAS 499, 505
Dravins, D. et al. 2018, A&A 616, A144
Drozdovskaya, M. N. et al. 2019, MNRAS 490, 50
Drummond, B. et al. 2020, A&A 636, A68
Ducrot, E. et al. 2018, AJ 156, 218
Dudás, E. et al. 2020, JChPh 152, 134201
Díaz, M. R. et al. 2020, MNRAS 493, 973
Edwards, B. et al. 2019, AJ 157, 242
Edwards, B. et al. 2020a, MNRAS, arXiv:2005.01684
Edwards, B. et al. 2020b, AJ 160, 8
Ehrenreich, D. et al. 2020, Natur 580, 597
Eistrup, C. et al. 2016, A&A 595, A83
Eistrup, C. et al. 2018, A&A 613, A14
Elkins-Tanton, L. T. et al. 2008, ApJ 688, 628
Elkins-Tanton, L. T. 2012, AREPS 40, 113
Encrenaz, T. et al. 2015, ExA 40, 523
Esteves, L. J. et al. 2015, ApJ 804, 150
Evans, T. M. et al. 2016, ApJL 822, L4
Fabrycky, D. C. et al. 2009, ApJ 696, 1230
Fares, R. et al. 2009, MNRAS 398, 1383
Fedele, D. et al. 2017, A&A 600, A72
Fedele, D. et al. 2018, A&A 610, A24
Ferus, M. et al. 2009, CPL 472, 14
Ferus, M. et al. 2011, JPCA 115, 12132
Ferus, M. et al. 2014, JPCA 118, 719
Ferus, M. et al. 2015a, PNAS 112, 657
Ferus, M. et al. 2015b, PNAS 112, 7109
Ferus, M. et al. 2017, NatSR 7, 6275
Ferus, M. et al. 2018a, A&A 616, A150
Ferus, M. et al. 2018b, A&A 610, A73
Ferus, M. et al. 2019, A&A 630, A127
Ferus, M. et al. 2020a, Icar 341, 113670
Ferus, M. et al. 2020b, Ex A, accepted
Fichtinger, B. et al. 2017, A&A 599, A127
Fisher, C. et al. 2018, MNRAS 481, 4698
Ford, E. B. et al. 2007, ApJL 664, L51
Fossati, L. et al. 2010, ApJL 714, L222
Fressin, F. et al. 2013, ApJ 766, 81
Fridlund, M. et al. 2006, ESASP 1306,
Fromang, S. et al. 2016, A&A 591, A144
Fulton, B. J. et al. 2017, AJ 154, 109
Fulton, B. J. et al. 2018, AJ 156, 264
Gaillard, F. et al. 2009, E&PSL 279, 34
Gallet, F. et al. 2013, A&A 556, A36
Gallet, F. et al. 2015, A&A 577, A98
Gandhi, S. et al. 2018, MNRAS 474, 271
Gandhi, S. et al. 2019, MNRAS 485, 5817
Gao, P. et al. 2017, AJ 153, 139

Gao, P. et al. 2018, ApJ 863, 165
Garhart, E. et al. 2020, AJ 159, 137
Georges, R. et al. 2019, RScI 90, 093103
Ghysels, M. et al. 2018, JQSRT 215, 59
Gillon, M. et al. 2016, Natur 533, 221
Gillon, M. et al. 2017, Natur 542, 456
Gillon, M. 2018, NatAs 2, 344
Ginzburg, S. et al. 2018, MNRAS 476, 759
Gordon, I. E. et al. 2017, JQSRT 203, 3
Goyal, J. M. et al. 2018, MNRAS 474, 5158
Gravity Collaboration, 2019, A&A 623, L11
Guillot, T. 2008, PhST 130, 014023
Gupta, A. et al. 2019, MNRAS 487, 24
Hakim, K. et al. 2018, A&A 618, L6
Hamano, K. et al. 2013, Natur 497, 607
Hammond, M. et al. 2017, ApJ 849, 152
Hargreaves, R. J. et al. 2020, ApJS 247, 55
Harrington, J. et al. 2006, Sci 314, 623
Haswell, C. A. 2010, Cambridge University Press
Haswell, C. A. et al. 2019, NatAs 4, 408
Hawley, S. L. et al. 1995, ApJ 453, 464
Haynes, K. et al. 2015, ApJ 806, 146
Haywood, R. D. et al. 2018, AJ 155, 203
Heller, R. 2018, Handbook of Exoplanets, 35
Heller, R. et al. 2019, A&A 624, A95
Herrero, E. et al. 2016, A&A 586, A131
Hoeijmakers, H. J. et al. 2018, Natur 560, 453
Horner, J. et al. 2020, PASP 132, 102001
Howard, A. W. et al. 2010, Sci 330, 653
Howard, A. W. et al. 2012, ApJS 201, 15
Howell, S. B. et al. 2014, PASP 126, 398
Huang, X. et al. 2013, JQSRT 130, 134
Huang, C. X. et al. 2018, ApJL 868, L39
Hussain, G. A. J. et al. 2002, ApJ 575, 1078
Hussain, G. A. J. 2012, AN 333, 4
Hörst, S. M. et al. 2018, NatAs 2, 303
Ikoma, M. et al. 2018, SSRv 214, 76
Ingersoll, A. P. 1969, JAtS 26, 1191
Irwin, P. G. J. et al. 2008, JQSRT 109, 1136
Isella, A. et al. 2016, PhRvL 117, 251101
Ito, Y. et al. 2015, ApJ 801, 144
Ito, Y. et al. 2020, Ex A, in press.
Iyer, A. R. et al. 2016, ApJ 823, 109
Jardine, M. et al. 2002, MNRAS 336, 1364
Jin, S. et al. 2018, ApJ 853, 163
Johnstone, C. P. et al. 2015a, A&A 577, A27
Johnstone, C. P. et al. 2015b, A&A 577, A28
Johnstone, C. P. et al. 2015c, ApJL 815, L12
Johnstone, C. P. et al. 2019, A&A 624, L10
Johnstone, C. P. et al. 2020, AJ 890, 79
Jones, A. P. et al. 2004, A&A 416, 235
Jones, M. H. et al. 2020, ApJL 895, L17
Kama, M. et al. 2016a, A&A 588, A108
Kama, M. et al. 2016b, A&A 592, A83
Kama, M. et al. 2019, ApJ 885, 114
Kataria, T. et al. 2014, ApJ 785, 92
Kawashima, Y. et al. 2019, ApJ 877, 109
Keating, D. et al. 2017, ApJL 849, L5
Khangale, Z. N. et al. 2020, MNRAS 495, 637
Kimura, T. et al. 2020, MNRAS 496, 3755
Kipping, D. M. 2009, MNRAS 392, 181
Kite, E. S. et al. 2016, ApJ 828, 80
Kite, E. S. et al. 2020, ApJ 891, 111
Knutson, H. A. et al. 2007, Natur 447, 183
Kochukhov, O. et al. 2002, A&A 388, 868
Komacek, T. D. et al. 2016, ApJ 821, 16
Komacek, T. D. et al. 2020, ApJ 888, 2
Koskinen, T. T. et al. 2013a, Icar 226, 1695
Koskinen, T. T. et al. 2013b, Icar 226, 1678
Kowalski, A. F. et al. 2013, ApJS 207, 15
Kowalski, A. F. et al. 2019, ApJ 878, 135
Kraft, R. P. 1967, ApJ 150, 551
Kreidberg, L. et al. 2014, Natur 505, 69
Kreidberg, L. et al. 2018, AJ 156, 17
Kreidberg, L. et al. 2019a, ApJL 877, L15
Kreidberg, L. et al. 2019b, Natur 573, 87
Krick, J. E. et al. 2016, ApJ 824, 27
Kuerster, M. et al. 1994, A&A 289, 899
Kurosaki, K. et al. 2014, A&A 562, A80
Lammer, H. et al. 2020, SSRv 216, 74
Lampón, M. et al. 2020, A&A 636, A13
Lanza, A. F. 2008, A&A 487, 1163
Lanza, A. F. 2013, A&A 557, A31
Laskar, J. et al. 2017, A&A 605, A72
Laughlin, G. et al. 2009, Natur 457, 562
Lavie, B. et al. 2017, AJ 154, 91
Lavvas, P. et al. 2017, ApJ 847, 32
Lavvas, P. et al. 2019, ApJ 878, 118
Leconte, J., Chabrier, G., A&A 540, A20
Leconte, J. et al. 2013a, A&A 554, A69
Leconte, J., Chabrier, G., 2013b, NatGe 6, 347
Leleu, A. et al. 2019, A&A 624, A46
Li, G. et al. 2015, ApJS 216, 15
Limbach, M. A. et al. 2015, PNAS 112, 20
Line, M. R. et al. 2013, ApJ 775, 137
Line, M. R. et al. 2016, AJ 152, 203
Linsky, J. L. et al. 2010, ApJ 717, 1291
Lissauer, J. J. et al. 2011, Natur 470, 53

Liu, Q. Z. et al. 2006, A&A 455, 1165
Lodders, K. 2010, ASSP 16, 379
Long, F. et al. 2018, ApJ 869, 17
Lopez, E. D. et al. 2013, ApJ 776, 2
Lopez, E. D. 2017, MNRAS 472, 245
Léger, A. et al. 2011, Icar 213, 1
Lüftinger, T. et al. 2010a, A&A 509, A43
Lüftinger, T. et al. 2010b, A&A 509, A71
Lüftinger, T. et al. 2020, IAUS 345, 181
MacDonald, R. J. et al. 2017a, ApJL 850, L15
MacDonald, R. J. et al. 2017b, MNRAS 469, 1979
Macintosh, B. et al. 2015, Sci 350, 64
Madhusudhan, N. et al. 2009, ApJ 707, 24
Madhusudhan, N. et al. 2012, ApJL 759, L40
Madhusudhan, N. et al. 2016, SSRv 205, 285
Madhusudhan, N. et al. 2020, ApJL 891, L7
Maggio, A. et al. 2015, ApJL 811, L2
Magrini, L. et al. 2013, A&A 558, A38
Majeau, C. et al. 2012, ApJL 747, L20
Maldonado, J. et al. 2020, A&A submitted.
Mamajek, E. E. et al. 2008, ApJ 687, 1264
Managadze, G. G. et al. 2003, GeoRL 30, 1247
Mant, B. P. et al. 2018, MNRAS 478, 3220
Marboeuf, U. et al. 2014a, A&A 570, A35
Marboeuf, U. et al. 2014b, A&A 570, A36
Marcq, E. et al. 2017, JGRP 122, 1539
Martins, J. H. C. et al. 2018, MNRAS 478, 5240
Matt, S. P. et al. 2015, ApJL 799, L23
Mayo, A. W. et al. 2018, AJ 155, 136
Mayor, M. et al. 1995, Natur 378, 355
Mazeh, T. et al. 2016, A&A 589, A75
McCullough, P. R. et al. 2005, PASP 117, 783
McCullough, P. R. et al. 2014, ApJ 791, 55
McIvor, T. et al. 2006, MNRAS 367, L1
McKay, C. P. et al. 1997, Sci 276, 390
McKemmish, L. K. et al. 2016, MNRAS 463, 771
McKemmish, L. K. et al. 2019, MNRAS 488, 2836
McMurty, C. et al. 2013, Optical Engineering 52, 9
Mendonça, J. M. et al. 2016, ApJ 829, 2
Mendonça, J. M. et al. 2018a, AJ 155, 150
Mendonça, J. M. et al. 2018b, ApJ 869, 107
Meng, H. Y. A. et al. 2012, ApJL 751, L17
Meng, H. Y. A. et al. 2015, ApJ 805, 77
Meunier, N. et al. 2010, A&A 512, A39
Miguel, Y. et al. 2011, ApJL 742, L19
Miguel, Y. 2019, MNRAS 482, 2893
Mikal-Evans, T. et al. 2020, MNRAS 496, 1638
Min, M. et al. 2016, A&A 593, A11
Min, M. et al. 2020, arXiv:2006.12821
Miozzi, F. et al. 2018, JGRE 123, 2295
Miyagoshi, T. et al. 2018, EP&S 70, 200
Mizus, I. I. et al. 2017, MNRAS 468, 1717
Modirrousta-Galian, D. et al. 2020, ApJ 891, 158
Mollière, P. et al. 2015, ApJ 813, 47
Mollière, P. et al. 2017, A&A 605, C3
Molyarova, T. et al. 2018, ApJ 866, 46
Morbidelli, A. et al. 2016, Icar 267, 368
Mordasini, C. et al. 2016, ApJ 832, 41
Morley, C. V. et al. 2012, ApJ 756, 172
Morris, B. M. et al. 2020, MNRAS 493, 5489
Mugnai, L. V. et al. 2020, Ex A, 50, 303
Mugnai, L. V. et al. 2021, ApJ submitted
Muñoz, D. J. et al. 2018, AJ 156, 253
Navarro-González, R. et al. 2019, JGRE 124, 94
Nielsen, E. L. et al. 2019, AJ 158, 13
Nikolaou, A. et al. 2019, ApJ 875, 11
Nikolaou, N. et al. 2021, AJ, arXiv:2010.15996
Nortmann, L. et al. 2018, Sci 362, 1388
Notsu, Y. et al. 2013, ApJ 771, 127
Nutzman, P. et al. 2008, PASP 120, 317
Öberg, K. I. et al. 2019, AJ 158, 194
O'Brien, K. et al. 2002, MNRAS 334, 426
Owen, J. E. et al. 2013, ApJ 775, 105
Owen, J. E. et al. 2017, ApJ 847, 29
Palme, H. et al. 2014, Treatise on Geochemistry 2, 15
Parmentier, V. et al. 2013, A&A 558, A91
Parmentier, V. et al. 2016, ApJ 828, 22
Parmentier, V., Crossfield, I., 2018, Handbook of Exoplanets 116
Parmentier, V. et al. 2018, A&A 617, A110
Pascale, E. et al. 2015, ExA 40, 601
Pepe, F. et al. 2018, Handbook of Exoplanets, 190
Perryman, M. et al. 2014, ApJ 797, 14
Petigura, E. A. et al. 2018, AJ 156, 89
Petralia, A. et al. 2020, ExA 49, 97
Pinhas, A. et al. 2019, MNRAS 482, 1485
Pirani, S. et al. 2019, A&A 623, A169
Pirri, A. N. 1977, PhFl 20, 221
Piskunov, N. et al. 2002, A&A 381, 736
Plavchan, P. et al. 2020, Natur 582, 497
Pluriel, W. et al. 2019, Icarus 317, 583
Pluriel, W. et al. 2020, AJ 160, 112
Pollacco, D. et al. 2006, Ap&SS 304, 253
Pollack, J. B. et al. 1994, ApJ 421, 615
Polyansky, O. L. et al. 2018, MNRAS 480, 2597
Pont, F. et al. 2008, MNRAS 385, 109
Powell, D. et al. 2018, ApJ 860, 18
Powell, K. G. et al. 1999, JCoPh 154, 284

Pratt, G. W. et al. 1999a, MNRAS 309, 847
Pratt, G. W. et al. 1999b, MNRAS 307, 413
Quintana, E. V. et al. 2014, ApJ 786, 33
Rab, C. et al. 2017, A&A 604, A15
Rackham, B. V. et al. 2018, ApJ 853, 122
Rappaport, S. et al. 2012, ApJ 752, 1
Rauer, H., Heras A, 2018, Handbook of Exoplanets 86
Rauscher, E. et al. 2007, ApJ 664, 1199
Raymond, S. N. et al. 2017, Icar 297, 134
Redfield, S. et al. 2008, ApJL 673, L87
Reiners, A. et al. 2012, AJ 143, 93
Reiners, A. 2012, LRSP 9, 1
Rey, M. et al. 2014a, ApJ 789, 2
Rey, M. et al. 2014b, JChPh 141, 044316
Rey, M. et al. 2016a, A&A 594, A47
Rey, M. et al. 2016b, JMoSp 327, 138
Rey, M. et al. 2017, ApJ 847, 105
Rey, M. et al. 2018a, Icar 303, 114
Rey, M. et al. 2018b, PCCP 20, 21008
Rice, J. B. et al. 1998, A&A 336, 972
Ricker, G. R. et al. 2014, SPIE 9143, 914320
Ricker, G. R. et al. 2015, JATIS 1, 014003
Ricker, G. R. et al. 2016, SPIE 9904, 99042B
Ritter, H. et al. 2015, AcPPP 2, 21
Robertson, P. et al. 2015, ApJL 805, L22
Roble, R. G. et al. 1987, JGR 92, 8745
Rogers, L. A. 2015, ApJ 801, 41
Rogers, T. M. 2017, NatAs 1, 0131
Rosich, A. et al. 2020, A&A 641, A82
Réville, V. et al. 2015, ApJ 798, 116
Sainsbury-Martinez, F. et al. 2019, A&A 632, A114
Santos, N. C. et al. 2000, A&A 361, 265
Santos, N. C. et al. 2004, A&A 415, 1153
Santos, N. C. et al. 2013, A&A 556, A150
Sarkar, S. et al. 2016, SPIE 9904, 99043R
Sarkar, S. et al. 2018, MNRAS 481, 2871
Sarkar, S. et al. 2020, ExA, arXiv:2002.03739
Saumon, D. et al. 2008, ApJ 689, 1327
Scattergood, T. W. et al. 1989, Icar 81, 413
Schaefer, B. E. 2012, Natur 485, 456
Schaefer, J. J. et al. 2002, AAS 201, 119.01
Schaefer, L. et al. 2009, ApJL 703, L113
Schaefer, L. et al. 2012, ApJ 755, 41
Schlaufman, K. C. 2015, ApJL 799, L26
Schou, J. et al. 2012, SoPh 275, 327
Seager, S. et al. 2000, ApJ 537, 916
Shallue, C. J. et al. 2018, AJ 155, 94
Shibata, S. et al. 2020, A&A 633, A33
Showman, A. P. et al. 2002, A&A 385, 166
Showman, A. P. et al. 2009, ApJ 699, 564
Showman, A. P. 2016, Natur 533, 330
Shulyak, D. et al. 2020, A&A 639, A48
Simon, A. et al. 2007, A&A 470, 727
Sing, D. K. et al. 2016, Natur 529, 59
Sing, D. K. 2018, arXiv:1804.07357
Skaf, N. et al. 2020, AJ 160, 109
Snellen, I. A. G. et al. 2010, Natur 465, 1049
Snellen, I. A. G. et al. 2014, Natur 509, 63
Solanki, S. K. 1999, ASPC 158, 109
Sousa-Silva, C. et al. 2015, MNRAS 446, 2337
Sousa, S. G. et al. 2008, A&A 487, 373
Sousa, S. G. et al. 2018, A&A 620, A58
Sozzetti, A. et al. 2013, EPJWC 47, 03006
Spake, J. J. et al. 2018, Natur 557, 68
Šponer, J. E. et al. 2016, PCCP 18, 20047
Stevenson, K. B. et al. 2014, Sci 346, 838
Swain, M. R. et al. 2009, ApJL 690, L114
Szabó, G. M. et al. 2006, A&A 450, 395
Szabó, G. M., Kiss, L. L., 2011, ApJL 727, L44
Szabó, G. M. et al. 2011, ApJL 736, L4
Szabó, G. M. et al. 2012, MNRAS 421, L122
Szabó, G. M. et al. 2020, MNRAS 492, L17
Tackley, P. J. et al. 2013, Icar 225, 50
Teachey, A. et al. 2020, AJ 159, 142
Teachey, A. et al. 2018, SciA 4, eaav1784
Terrile, R. J. et al. 2006, AGUSM.A21 2007, A21A-07
Tessenyi, M. et al. 2013, Icar 226, 1654
Thiabaud, A. et al. 2014, A&A 562, A27
Thorngren, D. P. et al. 2016, ApJ 831, 64
Thorngren, D. et al. 2019, ApJL 874, L3
Tian, F. et al. 2015, NatGe 8, 177
Tinetti, G. et al. 2007a, ApJL 654, L99
Tinetti, G. et al. 2007b, Natur 448, 169
Tinetti, G. et al. 2013, A&ARv 21, 63
Tinetti, G. et al. 2018, ExA 46, 135
Trampedach, R. et al. 2013, ApJ 769, 18
Triaud, A. H. M. J. et al. 2014, MNRAS 444, 711
Tsantaki, M. et al. 2018, MNRAS 473, 5066
Tsiaras, A. et al. 2016a, ApJ 832, 202
Tsiaras, A. et al. 2016b, ApJ 820, 99
Tsiaras, A. et al. 2018, AJ 155, 156
Tsiaras, A. et al. 2019, NatAs 3, 1086
Tsuji, T. et al. 1996, A&A 308, L29
Turbet, M. et al. 2020, A&A 638, A41
Turrini, D. et al. 2014, Life 4, 4
Turrini, D. et al. 2018, ExA 46, 45
Turrini, D. et al. 2020a, ExA submitted
Turrini, D. et al. 2020b, A&A 636, A53

Turrini, D. et al. 2020c, ApJ submitted
Tusnski, L. R. M. et al. 2011, ApJ 743, 97
Underwood, D. S. et al. 2016, MNRAS 459, 3890
Valencia, D. et al. 2013, ApJ 775, 10
van der Holst, B. et al. 2014, ApJ 782, 81
Van Kerckhoven, C. et al. 2002, A&A 384, 568
Van Lieshout, R. et al. 2014, A&A 572, A76
Vanderspek, R. et al. 2019, ApJL 871, L24
Venot, O. et al. 2012, A&A 546, A43
Venot, O. et al. 2013, A&A 551, A131
Venot, O. et al. 2015, A&A 577, A33
Venot, O. et al. 2016, ApJ 830, 77
Venot, O. et al. 2018, A&A 609, A34
Venot, O. et al. 2019, A&A 624, A58
Venot, O. et al. 2020a, A&A 634, A78
Venot, O. et al. 2020b, ApJ 890, 176
Vidal-Madjar, A. et al. 2003, Natur 422, 143
Vidotto, A. A. et al. 2015, MNRAS 449, 4117
von Essen, C. et al. 2019, A&A 622, A71
Voronin, B. A. et al. 2010, MNRAS 402, 492
Véron-Cetty, M.-P. et al. 2010, A&A 518, A10
Wakeford, H. R. et al. 2017, Sci 356, 628
Wakelam, V. et al. 2015, ApJS 217, 20
Waldmann, I. P. et al. 2015a, ApJ 802, 107
Waldmann, I. P. et al. 2015b, ApJ 813, 13
Wang, D. et al. 2016, Icar 276, 21
Way, M. J. et al. 2020a, JGRE 125, e06276
Welsh, W. F. et al. 2012, Natur 481, 475
Werner, M. W. et al. 2004, ApJS 154, 1
Williams, D. M. et al. 2002, IJAsB 1, 61
Wilson, R. F. et al. 2018, AJ 155, 68
Wolszczan, A. et al. 1992, Natur 355, 145
Wong, A. et al. 2019, ApJS 240, 4
Wong, I. et al. 2014, ApJ 794, 134
Wood, B. E. 2004, LRSP 1, 2
Wright, G. et al. 2015, PASP 127, 595
Wyatt, M. C. et al. 2016, SSRv 205, 231
Yan, F. et al. 2018, NatAs 2, 714
Yan, F. et al. 2019, A&A 632, A69
Yang, H. et al. 2019, ApJS 241, 29
Yip, H. K. et al. 2020a, ApJ, arXiv:1811.04686
Yip, H. K. et al. 2020b, ApJ, arXiv:2009.10438
Yurchenko, S. N. et al. 2014, MNRAS 440, 1649
Yurchenko, S. N. et al. 2020, MNRAS 496, 5282
Zhang, Ke et al. 2020, ApJ 891, L16
Zapatero Osorio, M. R. et al. 2006, ApJ 647, 1405
Zechmeister, M. et al. 2009, A&A 496, 577
Zellem, R. T. et al. 2017, ApJ 844, 27
Zellem, R. T. et al. 2019, PASP 131, 094401
Zellem, R. T. et al. 2020, PASP 132, 054401
Zeng, L. et al. 2019, PNAS 116, 9723
Zingales T. et al. 2018, ExA 46, 67
Zinzi, A. et al. 2017, A&A 605, L4

# 12 List of Acronyms

| | |
|---|---|
| ACS | Active Cooler System |
| ADaRP | Ariel Data Reduction Pipeline |
| AIRS | Ariel IR Spectrometer |
| ALMA | Atacama Large Millimetre Array |
| AMC | Ariel Mission Consortium |
| AOCS | Attitude and Orbit Control System |
| APE | Absolute Pointing Error |
| ArielRad | Ariel Radiometric model |
| AST | Ariel Science Team |
| ATHENA | Advanced Telescope for High Energy Astrophysics |
| AVM | Avionics Model |
| CASE | Contribution to Ariel Spectroscopy of Exoplanet |
| CBE | Current Best Estimate |
| CDE | Catastrophically Disintegrating Exoplanet |
| CDE | Cooler Drive Electronics |
| CDF | Concurrent Design Facility |
| CDPU | Command and Data Processing Unit |
| CDR | Critical Design Review |
| CEA | Centre d'Energie Atomique |
| CEE | Centroiding Estimation Error |
| CELT | Consortium Engineering Leadership Team |
| CFC | Cryogenic Flex Cable |
| CFEE | Cold Front End Electronics |
| CFRP | Carbon Fibre Reinforced Plastic |
| CHEOPS | CHaracterising ExOPlanet Satellite |
| CILAS | Compagnie Industrielle des Lasers |
| CIST | Consortium Instrument Scientist Team |
| CMT | Consortium Management Team |
| CoP | Commissioning Phase |
| CSCG | Consortium Science Coordination Group |
| CURE | CUbesat for Refining Ephemerides |
| DCU | Detector Control Unit |
| DLS | Dual Launch Structure |
| DMS | Data Management System |
| DPU | Data Processing Unit |
| DRM | Design Reference Mission |
| ECMLPKDD | European Conference on Machine Learning and Principles and Practice of Knowledge Discovery in Databases |
| ECSS | European Cooperation for Space Standardisation |
| EGSE | Electrical Ground Support Equipment |
| EM | Engineering Model |
| EMC | Electro-Magnetic Compatibility |
| ERS | Early Release Science |
| ESA | European Space Agency |
| ESAC | European Space Astronomy Centre |
| ESOC | European Space Operations Centre |
| ESTEC | European Space Research and Technology Centre |
| FAR | Flight Acceptance Review |
| FCU | Fine Guidance System Control Unit |
| FDIR | Failure Detection Isolation and Recovery |
| FGS | Fine Guidance System |
| FMAR | Final Mission Analysis Review |
| FPA | Focal Plane Assemblies |
| FPGA | Field Programmable Gate Array |
| FSM | Finite State Machine |
| G/S | Ground Stations |
| GFRP | Glass Fibre Reinforced Plastic |
| GS | Ariel Ground Segment |
| GSIR | Ground Segment Implementation Review |
| GSRR | Ground Segment Readiness Review |
| GTO | Guaranteed Time Observations |
| HK | Housekeeping |
| HMI | Helioseismic and Magnetic Imager |
| HMS | Health Monitoring and trend analysis System |
| HST | Hubble Space Telescope |
| IC | Integrated Circuit |
| ICU | Instrument Control Unit |
| IML | Infrared Multilayer Laboratory (at University of Oxford) |
| IOCR | In-Orbit Commissioning Review |
| IOSDC | Instrument Operations and Science Data Centre |
| IRAC | InfraRed Array Camera (on Spitzer) |
| ITT | Invitation To Tender |
| IVOA | International Virtual Observatory Alliance |
| IWS | Instrument Work Station |
| JPL | Jet Propulsion Laboratory |
| JT | Joule-Thomson |
| JUICE (ESA) | JUpiter ICy moons Explorer |
| JWST-NIRSpec | James Webb Space Telescope – Near-InfraRed Spectrometer |
| LED | Light Emitting Diodes |
| LEOP | Launch and Early-orbit OPerations |
| LoS | Line of Sight |
| LVDS | Low Voltage Differential Signal |
| M2M | M2 Mirror |
| MAR | Mission Adoption Review |
| MCRR | Mission Commissioning Result Review |
| MCS | Mission Candidate Sample |
| MCS | Mission Candidates Sample |
| MHD | Magneto-HydroDynamics |
| MIR | Mid-InfraRed |
| MIRI | Mid-InfraRed Instrument (on JWST) |
| MOC | Mission Operations Centre |
| MRS | Mission Reference Sample |
| NASA | National Aeronautics and Space Administration |
| NDR | Non-Desructive Read |
| NEOCam | Near Earth Object Camera (proposed NASA mission) |
| NGTS | Next Generation Transit Survey |
| NIR | Near infrared |
| NIRSpec | Near-IR Spectrometer |
| OBCU | On-Board Calibration Unit |
| OBSW | On-Board Software |
| OGS | Operational Ground Segment |
| PCDU | Power Conditioning and Distribution Unit |
| PDE | Pointing Drift Error |
| PDR | Preliminary Design Review |
| pFM | proto Flight Model (also on P96 referred to as PF) |
| PIP | Payload Interface Panel |
| PLATO | PLAnetary Transits and Oscillations |
| PLM | PayLoad Module |
| PMAR | Preliminary Mission Analysis Review |
| PS | Project Scientist |
| PSU | Power Supply Unit |
| PTM | Pathfinder Telescopic Model |
| PVP | Performance Verification Phase |
| QLA | Quick Look Analysis |
| RAL | Rutherford Appleton Laboratory |
| ROIC | Read Out Integrated Circuit |
| RSP | Routine Science Phase |
| RV | Radial Velocity |
| SAA | Solar Aspect Angle |
| SCA | Sensor Chip Assembly |
| SDO | Solar Dynamics Observatory |
| SDP | Science Demonstration Phase |
| SF | Science Frame |
| SGS | Science Ground Segment |
| SIDECAR ASIC | System for Image Digitization, Enhancement, Control And Retrieval-Application Specific Integrated Circuit |
| SM | Structural Model |
| SNR | Signal to Noise Ratio |
| SOC | Science Operations Centre |
| SPC | Science Programme Committee |
| SPI | Serial Peripheral Interface |
| SRR | System Requirement Review |
| SSCE | Sun-S/C-Earth |
| SVM | Spacecraft Service Module |
| SW | Software |
| TAS | Thales Alenia Space |
| TC | Telecommand |
| TCM | Transfer Correction Manoeuvres |
| TCU | Telescopic Control Unit |
| TDA | Technology Development Activities |
| TDVs | Transit Duration Variations |
| TESS | Transiting Exoplanet Survey Satellite |
| TIS | Teledyne Imaging Sensors |
| TLM | Telemetry |
| TOI | TESS Objects of Interest |
| TTVs | Transit Timing Variations |
| TVAC | Thermal Vacuum Chamber |
| VGs | V-Grooves |
| VISPhot | Visible Photometer |
| VM | Virtual Machine |
| VUV | Vacuum Ultraviolet |
| WASP | Wide Angle Search for Planets |
| WFE | Wave Front Error |
| WG | Working Group |
| XUV | eXtreme Ultraviolet Radiation |
| YSO | Young Stellar Objects |
| ZDI | Zeeman Doppler Imaging |